\documentclass[11pt]{article}
\usepackage{pifont}

\usepackage{dsfont}
\usepackage{tikz}

\usepackage{amsmath,amssymb,graphicx}
\usepackage{hyperref}
\usepackage{stackengine}

\usepackage[nosort]{cite}
\usepackage[Symbol]{upgreek}%for \upmu
\usepackage[vcentermath]{youngtab}%for Young tableaux

\DeclareMathAlphabet{\mathpzc}{OT1}{pzc}{m}{rm}

\usepackage{multicol,xcolor,longtable}

\usepackage[T1]{fontenc}
\usepackage{hyphenat}
\usepackage{amsfonts}
\usepackage{mathrsfs}
\usepackage{mathdots}
\definecolor{darkred}{rgb}{0.65,0.15,0}
\definecolor{newgreen}{rgb}{0.2,0.62,0.14}
\hypersetup{pdfborder={0 0 0},colorlinks=true,urlcolor=blue,citecolor=blue,linkcolor=darkred,linktocpage=true}

\usepackage{array}
\newcolumntype{P}[1]{>{\centering\arraybackslash}p{#1}}
\usepackage{lscape}
\usepackage{multirow}

\usepackage{empheq}

\newcommand{\be}{\begin{equation}}
\newcommand{\ee}{\end{equation}}
\newcommand{\bea}{\setlength\arraycolsep{2pt} \begin{eqnarray}}
\newcommand{\eea}{\end{eqnarray}}
\def\cR{\mathcal{R}}

\def\<{\langle}
\def\>{\rangle}
\newcommand{\colvec}[2][.8]{\scalebox{#1}{\renewcommand{\arraystretch}{.8}$\begin{matrix}#2\end{matrix}$}}

\newcommand{\grad}[3]{{\scriptscriptstyle (#1 , #2, #3 )}}
\newcommand{\gra}[2]{{\scriptscriptstyle (#1 , #2 )}}
\newcommand{\ord}[1]{{\scriptscriptstyle (#1)}}

\newcommand{\nn}{\nonumber}

\newcommand{\cV}{\mathcal{V}}
\newcommand{\cE}{\mathcal{E}}

\newcommand{\cF}{\mathcal{F}}
\newcommand{\cL}{\mathcal{L}}

\newcommand{\mf}[1]{{\mathfrak{#1}}}

\newcommand{\eprint}[1]{{\href{http://arxiv.org/abs/#1}{[\texttt{#1}]}}}
\newcommand{\eprintN}[1]{{\href{http://arxiv.org/abs/#1}{[\texttt{#1 [hep-th]}]}}}
\newcommand{\eprintNT}[1]{{\href{http://arxiv.org/abs/#1}{[\texttt{#1 [math.NT]}]}}}

\newcommand{\doi}[2]{{\hypersetup{urlcolor=darkred}\href{http://dx.doi.org/#2}{#1}\hypersetup{urlcolor=blue}}}

\newcommand{\CR}{\nonumber \\*}

\newcommand{\cN}{\mathcal{N}}

\newcommand{\cD}{\mathcal{D}}

\def\a{\alpha}

\newcommand{\Scal}[1]{\Bigl ({#1} \Bigr )}

\makeatletter

\@addtoreset{equation}{section}
\makeatother

\def\a{\alpha}		\def\b{\beta}				
			\def\h{\eta}			
\def\i{\iota}						
						\def\p{\pi}				
		\def\t{\tau}

						\def\C{\Chi}

\newcommand{\calE}{{\mathcal E}}
\newcommand{\calF}{{\mathcal F}}
\newcommand{\calG}{{\mathcal G}}

\newcommand{\calI}{{\mathcal I}}

\newcommand{\calN}{{\mathcal N}}

\newcommand{\calY}{{\mathcal Y}}

\def\gfour{g_{\scalebox{0.5}{4}}}
\def\gFour{g_{\scalebox{0.6}{4}}}
\def\gfive{g_{\scalebox{0.5}{5}}}
\def\gFive{g_{\scalebox{0.6}{5}}}
\def\Rs{R_{\scalebox{0.5}{s}}}

\def\RS{R_{\scalebox{0.6}{s}}}

\newcommand{\Cong}[2]{\Gamma^{\scalebox{0.6}{#1}_{\scalebox{0.5}{#2}}}}

\def\cL{{\cal L}}
\def\cN{\mathcal{N}}

\newcommand{\sLambda}{I \hspace{-1mm}I}

\def\adt{\dot{\alpha}}
\def\tr{{\rm tr}}

\def\cL{{\cal L}}
\def\cN{\mathcal{N}}

  \def\xmin{1}
 \def\ymin{0}

\begin{document}

\begin{flushright} CPHT-RR019.032022 \end{flushright} 
 \vspace{8mm}

\begin{center}

{\LARGE \bf \sc Perturbative higher-derivative terms \\
in ${\cal N}=6$ asymmetric orbifolds}

\vspace{6mm}
\normalsize
{\large  Massimo Bianchi${}^{1}$, Guillaume Bossard${}^{2}$, Dario Consoli${}^1$}

\vspace{10mm}

${}^1${\it Dipartimento di Fisica, Universit\`a di Roma “Tor Vergata” \& Sezione INFN Roma2,\\
Via della ricerca scientifica 1, 00133, Roma, Italy}\\
${}^2${\it Centre de Physique Th\'eorique, CNRS,  Institut Polytechnique de Paris\\
91128 Palaiseau cedex, France}
\vskip 1 em
\vspace{20mm}

\hrule

\vspace{5mm}

 \begin{tabular}{p{14cm}}

We analyse the perturbative four-point amplitudes in the simplest string theory examples of  T-fold backgrounds, which enjoy ${\cal N}=6$ supersymmetries in four dimensions. There are two theories defined as asymmetric orbifolds of order $2$ and $3$, respectively. The perturbative spectrum and the one-loop four-point amplitudes are shown to be invariant under an arithmetic symplectic group defined over the Hurwitz (respectively Eisenstein) quaternions. The supersymmetry constraints on the low energy effective couplings are determined and we conjecture the U-duality group of the theory and the non-perturbative completion of the six-derivative coupling as a rank one theta series. We discuss the non-perturbative spectrum of BPS states in the light of our analysis.  

\end{tabular}

\vspace{6mm}
\hrule
\end{center}

\vspace{5mm} 
\begin{flushright} {\it We wish to dedicate this paper to the memory of Costas Kounnas} \end{flushright} 

\thispagestyle{empty}

\newpage

\setcounter{page}{1}

\setcounter{tocdepth}{2}
\tableofcontents

%%%%%%%%%%%%%%%%%%%%%%%%
%%%%%%%%%%%%%%%%%%%%%%%%
%%%%%%%%%%%%%%%%%%%%%%%% %%%%%%%%%%%%%%%%%%%%%%%%
\section{Introduction}
\label{intro}
Supergravity theories with $\mathcal{N}$ extended supersymmetry admit different ultra-violet completions in string theory. For $\mathcal{N}\ge 5$, all the massless fields are in the graviton super-multiplet and there is no lower-spin `matter' super-multiplet.  The two-derivative action is then uniquely determined by supersymmetry and the scalar fields are moduli that parametrise a symmetric space $G/K$ \cite{Cremmer:1978ds}. There is a unique string theory ultra-violet completion of (ungauged) maximal $\mathcal{N}=8$ supergravity in four dimensions.\footnote{The issue of its non-perturbative completion was addressed in \cite{Bianchi:2009wj, Bianchi:2009mj}.} This theory has been very useful in understanding dualities and non-perturbative effects in string theory \cite{Hull:1994ys,Witten:1995ex}, including D-brane and NS-five-brane instantons. Supersymmetry and U-duality give very stringent constraints on the low-energy effective action \cite{Green:1981yb,Green:1997tv,Green:1997di,Berkovits:1997pj,Pioline:1998mn,Green:1998by,Obers:1999um,Green:1999pv,Obers:2001sw,Kazhdan:2001nx,Basu:2008cf,Green:2005ba,Pioline:2010kb,Green:2011vz,Bossard:2014lra,Bossard:2014aea,Gustafsson:2014iva,Wang:2015jna,Bossard:2015uga,Gourevitch:2019knu}, whose exact couplings are known up to fourteen-derivative terms \cite{Green:1997as,Kiritsis:1997em,Pioline:1997pu,Green:1999pu,Basu:2007ru,Green:2008uj,Green:2010wi,Green:2010sp,Green:2010kv,Green:2014yxa,Bossard:2020xod}. Such results offer an invaluable window into the non-perturbative regime of string theory, like supersymmetric D-branes and their bound states. Similar results have been obtained for theories with sixteen supercharges, i.e. $\mathcal{N}=4$ in four dimensions \cite{Harvey:1996ir,Bachas:1997mc,Gregori:1997hi,Kiritsis:1997hf, Bianchi:1998vq,Bossard:2018rlt}. Another interesting example is $\mathcal{N}=6$ supersymmetry in four dimensions. Such theories provide the simplest examples of compactifications on T-folds \cite{Hull:2004in}, and can therefore be used to study D-branes and NS-five-branes on non-geometric backgrounds. One may wonder: how many independent string theories give (ungauged)  $\mathcal{N}=6$ supergravity at low energy? What is their group of U-duality symmetries as an arithmetic subgroup of the non-compact global $SO^*(12)$ symmetry group? Can one determine supersymmetry protected couplings exactly using constraints from supersymmetry and U-duality? What is the spectrum of supersymmetric D-branes in these theories? 

The first construction of a string theory with $\mathcal{N}=6$ supersymmetry was proposed by Ferrara and Kounnas in \cite{Ferrara:1989nm} as a model of free fermions. This model can also be formulated as an asymmetric orbifold of $(T^2\times T^{4})/ \mathds{Z}_2$ in which $\mathds{Z}_2$ acts on the left moving sector of $T^4$ as for the K3 orbifold conformal field theory and as a shift on one additional circle in $T^2$. The D-branes and the four-graviton amplitude have been analysed in \cite{Bianchi:2008cj, Bianchi:2010aw}. In this paper we refine this analysis, with particular emphasis on the dependence on the NS-NS moduli (internal metric and B-field)\footnote{The NS-NS moduli are the internal metric and $B$ fields preserved by the orbifold that parametrise the Grassmannian $SO(2,6)/ (SO(2)\times SO(6))$ and the axio-dilaton  that parametrise $SL(2)/SO(2)$. We will sometimes call the Grassmannian coordinates `Narain moduli' \cite{Narain:1986am}. The perturbative amplitudes only depend on the dilaton and the Narain moduli, while the NS-NS axion and the R-R moduli do not contribute due to shift symmetries resulting from their `axionic' nature.}  of the couplings and their automorphic symmetry under the T-duality group. We also demonstrate that there is a similar string theory defined as an asymmetric orbifold of $(T^2\times T^{4})/ \mathds{Z}_3$.  The obvious generalisation to $\mathds{Z}_4$ and $\mathds{Z}_6$ orbifold theories does not work, contrary to naive expectations. Another string theory with $\mathcal{N}=6$ supersymmetry was identified in \cite{Dabholkar:1998kv} as an asymmetric orbifold in which $\mathds{Z}_3$ acts on the left moving sector of $T^6$ as for the corresponding Calabi--Yau orbifold conformal field theory. We find that this theory has exactly the same (perturbative) spectrum as  the $(T^2\times T^{4})/ \mathds{Z}_3$ and therefore conclude that they are dual to each other. It seems therefore that there are only two inequivalent string theories with $\mathcal{N}=6$ supersymmetry.

The bosonic sector of $\mathcal{N}=6$ supergravity is identical to the bosonic sector of the $\mathcal{N}=2$ supergravity theory associated to the Jordan algebra of Hermitian three by three matrices over the quaternions \cite{Gunaydin:1983bi}.\footnote{The string version of the `magic' $\mathcal{N}=2$ supergravity associated to the octonionic algebra with $E_{7(-25)}$ global symmetry was constructed in \cite{Bianchi:2007va}.} In particular the scalar fields parametrise the special K\"{a}hler symmetric space $SO^*(12)/U(6)$, whose coordinates can be defined as  complexified ${3\times 3}$ Hermitian matrices $T$ over the quaternions. The duality symmetry acts on this space by linear fractional transformations and the theory of modular forms on $SO^*(12)/U(6)$ involves the choice of a discrete ring over the quaternions admitting prime factorisation \cite{Krieg,Vigneras}. We find that the relevant ring for the $\mathds{Z}_2$ orbifold theory is the ring $\mathds{H}(2)$ of Hurwitz quaternions isomorphic to the $D_4$ root lattice, while the relevant ring for the $\mathds{Z}_3$ orbifold theory is the ring $\mathds{H}(3)$ of Eisenstein quaternions (obtained through the Cayley--Dickson construction from a pair of Eisenstein integers) isomorphic to the $A_2\oplus A_2$ root lattice. 

Following the discussion of \cite{Green:2010wi} for the four-graviton amplitude in the maximally supersymmetric theory, the type II superstring four-graviton amplitude for the $\mathds{Z}_K$ asymmetric orbifolds with $K=2,3$ can be decomposed at low energy as 
\be \mathcal{M}_{\scalebox{0.6}{type II}}(T,s,t,u) =\mathcal{M}^{\scalebox{0.6}{sugra}}_{\scalebox{0.6}{type II}}(T,s,t,u,\mu) +   \mathcal{M}^{\scalebox{0.6}{Wilson}}_{\scalebox{0.6}{type II}}(T,s,t,u,\mu) \label{Split} \ee
where $\mu$ is a regulating mass scale. As usual we factor out  $(2\pi)^4 \delta( \sum_a k_a)$ of the amplitude. The analytic term $\mathcal{M}^{\scalebox{0.6}{Wilson}}_{\scalebox{0.6}{type II}}$ is interpreted as defining the Wilsonian effective action obtained by integrating out all the massive fields in string theory, whereas the first component $\mathcal{M}^{\scalebox{0.6}{sugra}}_{\scalebox{0.6}{type II}}$ is not analytic in the Mandelstam variables and corresponds to the supergravity amplitude computed with the Wilsonian effective action. The former expands in the Mandelstam variables as
\be  i \mathcal{M}^{\scalebox{0.6}{Wilson}}_{\scalebox{0.6}{type II}} =  \frac{\pi}{2^{10}}  \gFour^8 \alpha^{\prime 4} \hspace{-3.8mm} \sum_{m,n,p\in \mathds{N}} \hspace{-3mm}\mathcal{E}_\grad{m}{n}{p}(T) \hspace{-0.1mm} \bigl(\scalebox{1.0}{$ \frac{ \gfour^2 \alpha^\prime}{4}$}\bigr)^{\! \scalebox{0.6}{$m{+}2n{+}3p$}} \hspace{-0.6mm} (s^2 {+} t^2 {+} u^2)^n (s^3 {+} t^3 {+} u^3 )^p ( s^m t^s_8  + t^m t_8^t + u^m t_8^u ) t_8 R^4 \ee
where we included the Weyl rescaling to Einstein frame that involves the four-dimensional effective string coupling constant 
\be \gFour^2  = \frac{ K (2\pi)^6 \alpha^{\prime 3}}{V(T^6)} g_{\rm s}^{\; 2} \ee
with $g_{\rm s}$ the string coupling constant in ten dimensions and $V(T^6)$ the volume of the six-torus, so much so that  the volume of the orbifold is $V(T^6/\mathds{Z}_K) = V(T^6)/K$. Denoting by $\RS$ and $\tilde{\RS}$ the dimensionless radii of $T^2$ one finds that 
\be \gFour  = \frac{ 2}{\sqrt{\RS \tilde{\RS} }} g_{\rm s} \; . \ee
The gravitational coupling $\kappa$ is given in these conventions by 
 \be \kappa^2=\pi  \alpha^{\prime } \gFour^{2}   \; .\label{kappaalpha} \ee
The effective couplings $ \mathcal{E}_\grad{m}{n}{p}(T)  $ are U-duality invariant functions of the moduli. For maximal supersymmetry, only the terms with $m=0$ appear in the low energy expansion and 
\be t_8 = t^s_8  +  t_8^t +  t_8^u  \ee
is the Lorentz invariant tensor, ubiquitous in 4-point amplitudes \cite{GreenSchwWitt}, while $t_8^{r}$ for $r = s$, $t$ and $u$ are the projections to the respective channels defined in equation \eqref{t8s}. The non-analytic component expands similarly as
\be \mathcal{M}^{\scalebox{0.6}{sugra}}_{\scalebox{0.6}{type II}}(T,s,t,u,\mu)  =  -\frac{i}{2^{4}} \kappa^2 t_8 t_8 R^4 \frac{1}{s t u} +  \kappa^4  \mathcal{M}^{\scalebox{0.6}{1-loop}}_{\scalebox{0.6}{sugra}}(s,t,u, \mu) + \kappa^6  \mathcal{M}^{\scalebox{0.6}{2-loop}}_{\scalebox{0.6}{sugra}}(s,t,u, \mu)  + \dots \ee
The onset of moduli dependence appears at order $\alpha^{\prime 5}$ with the supergravity form-factor of the $\mathcal{E}_\grad{0}{0}{0}(T)  R^4$ supersymmetric completion. The fiducial scale $\mu$ is introduced to keep track of the ambiguity in the split \eqref{Split} associated to the presence of logarithmic divergences in the supergravity amplitudes and form factors. Its counterpart in the Wilsonian action gives an ambiguity in the definition of some couplings $ \mathcal{E}_\grad{m}{n}{p}(T) $ in terms of lower order ones, in which case we write $ \mathcal{E}_\grad{m}{n}{p}(T,\mu)$. There is no ultra-violet divergence in $\mathcal{N}=6$ supergravity up to four-loop included \cite{Bossard:2010bd}, and comparison with $\mathcal{N}=4$ and $5$ suggests that the five-loop amplitude should be finite as well \cite{Bern:2013uka,Bern:2014sna,Bossard:2011tq}. We will find however that the $\mathcal{E}_\grad{0}{0}{0}(T)  R^4$ form factor diverges at one-loop.

In $\mathcal{N}=6$ supergravity in four dimensions, there are two super-multiplets of 4-point amplitudes, the $U(1)$ preserving super-multiplet of the four-graviton amplitude and the $U(1)$ violating super-multiplet of the  two-graviphoton (of negative helicities) two-graviton (of positive helicities) amplitude. There is a similar analytic component of this two-graviphoton two-graviton amplitude and factorising out the helicity structure for simplicity one has
\be i \mathcal{A}^{\scalebox{0.6}{Wilson}}_{\scalebox{0.6}{type II}} =  \pi  \gFour^6 \alpha^{\prime 3} \sum_{m,n\in \mathds{N}} \hspace{-1mm}\mathcal{F}_\gra{m}{n}(T) \hspace{-0.1mm} \bigl(\scalebox{1.1}{$ \frac{ \gfour^2 \alpha^\prime}{4}$}\bigr)^{\! \scalebox{0.7}{$2m{+}3n$}} (s^2 {+} t^2 {+} u^2)^m (s^3 {+} t^3 {+} u^3 )^n \; , \ee
where the permutation symmetry in the Mandelstam variables follows by supersymmetry. The effective couplings $ \mathcal{F}_\gra{m}{n}(T)  $ are $U(1)$ weight  $2$ modular forms of the U-duality group.

The perturbative expansions of $ \mathcal{M}^{\scalebox{0.6}{Wilson}}_{\scalebox{0.6}{type II}} $ and $\mathcal{A}^{\scalebox{0.6}{Wilson}}_{\scalebox{0.6}{type II}} $ are obtained from the expansion of the string amplitude in $\alpha^\prime$ before integration over the genus $h$ moduli space of Riemann surfaces with four punctures. Although the string theory amplitude is finite, these couplings must then be regularised in order to tame the supergravity divergences due to the massless fields.

In the maximally supersymmetric theory, $\mathcal{E}_\grad{0}{0}{0}(\phi)$, $\mathcal{E}_\grad{0}{1}{0}(\phi)$, $\mathcal{E}_\grad{0}{0}{1}(\phi,\mu)$ are protected by supersymmetry and only receive perturbative corrections in string theory up to one, two and three-loop respectively. They have been the subject of extensive studies and have been determined as automorphic functions of $E_7(\mathds{Z})$ in \cite{Obers:1999um,Pioline:2010kb,Green:2011vz,Green:2010kv,Bossard:2020xod}.

Similarly we demonstrate that $\cF_\gra{0}{0}(T)$,  $\mathcal{E}_\grad{0}{0}{0}(T)$,  $\cF_\gra{1}{0}(T)$ and $\mathcal{E}_\grad{1}{0}{0}(T,\mu)$ are protected by supersymmetry in $\mathcal{N}=6$ supergravity. The first two only receive perturbative corrections in string theory up to one loop, and the last two up to two loops. Moreover $\cF_\gra{1}{0}(T)  = \det \! \cD \, \mathcal{E}_\grad{1}{0}{0}(T,\mu)$ by supersymmetry, where $\det \cD$ is an $SU(6)$ invariant third order differential operator defined in \eqref{detD}.\footnote{In many respects  $\cF_\gra{0}{0}(T)$,  $\mathcal{E}_\grad{0}{0}{0}(T)$,  and $\mathcal{E}_\grad{1}{0}{0}(T,\mu)$ are respectively analogues of $\mathcal{E}_\grad{0}{0}{0}(\phi)$, $\mathcal{E}_\grad{0}{1}{0}(\phi)$ and $\mathcal{E}_\grad{0}{0}{1}(\phi,\mu)$  in the maximally supersymmetric theory.}

The main purpose of this paper is to compute the one-loop contribution to these four BPS protected couplings $\cF_\gra{0}{0}(T)$,  $\mathcal{E}_\grad{0}{0}{0}(T)$,  $\cF_\gra{1}{0}(T)$ and $\mathcal{E}_\grad{1}{0}{0}(T)$ in the $\mathds{Z}_K$ orbifold theory for $K=2$ and $3$. Using `chiral' factorisation of the one-loop intergrands and relying on the results of \cite{Bianchi:2015vsa} for 4-point open-string amplitudes of gauge bosons in supersymmetric D-brane worlds, we derive rather compact expressions for the 4-graviton and 2-graviton 2-graviphoton amplitudes. We then find that the one-loop four-point amplitudes can be simplified by  partly  unfolding the genus one modular domain to the fundamental domain of the congruent subgroup $\Gamma_0(K)\subset SL(2,\mathds{Z})$. In this form they can be  written as integrals of $\Gamma_0(K)$ modular forms times the partition function for the Narain lattice $\mathds{F}_2(K)$ defined as
\be \mathds{F}_2(2)  =\sLambda_{1,1}\oplus \sLambda_{1,1}[ 2]\oplus D_4\; , \qquad  \mathds{F}_2(3)= \sLambda_{1,1}\oplus \sLambda_{1,1}[ 3]\oplus A_2\oplus A_2 \; , \label{PertF2} \ee
where $\sLambda_{1,1}[ K]$ is the Lorentzian circle Narain lattice with a momentum $m\in K \mathds{Z}$. Therefore the symmetries of the four-point one-loop amplitude are the automorphisms ${\rm Aut}(\mathds{F}_2(K))$ of the lattice $\mathds{F}_2(K)$. We also find that the character valued partition function that determines the perturbative spectrum in string theory\footnote{Similar techniques apply to theories with higher spin fields \cite{Bianchi:2010mg}.} \cite{Kiritsis:1997hj} can be written in terms of the Narain partition function of $\mathds{F}_2(K)$ and its dual lattice $\mathds{F}_2(K)^*$, such that the group ${\rm Aut}(\mathds{F}_2(K))$ is a symmetry of the full perturbative spectrum of the theory. In particular we find that for each charge of vanishing norm in $\mathds{F}_2(K)$ there is a single massive spin two 1/2 BPS super-multiplets and for each charge of vanishing norm in $\mathds{F}_2(K)^*\smallsetminus \mathds{F}_2(K) $ there are respectively two and one  massive spin three-half 1/2 BPS super-multiplets for $K=2$ and $3$. The true perturbative symmetry of these string theories may only be a proper subgroup of the group of automorphisms ${\rm Aut}(\mathds{F}_2(K))$, but for brevity we will nevertheless refer to ${\rm Aut}(\mathds{F}_2(K))$ as the T-duality group. 

The perturbative T-duality group ${\rm Aut}(\mathds{F}_2(K))$ will be identified as a group of ${4\times 4}$ symplectic matrices over the quaternions $\mathds{H}(K)$ that we call $\Cong{D}{4}_{0*}(\alpha)$, where $\alpha$ is a prime in $\mathds{H}(K)$ of norm square $K$. In this notation $D_4$ refers to $SO(2,6)$ and not to the lattice $\mathds{H}(2)$. For brevity we write  $\Cong{D}{4}_{0*}(\alpha)$ for both $K=2$ and $3$ such that $\Cong{D}{4}_{0}(\alpha) \subset \mathds{H}(K)$, is the subgroup with a lower left two by two matrix divisible by $\alpha$ in $\mathds{H}(K)$. We will moreover conjecture that the U-duality group is defined similarly as $\Cong{D}{6}_{0*}(\alpha) = {\rm Aut}(\mathds{F}_3(K))$ for the lattice of non-perturbative states 
\be \mathds{F}_3(K) = K  \mathds{F}_2(K)^* \oplus [ \mathds{H}(K) \oplus\mathds{H}(K) \oplus  \alpha \mathds{H}(K) \oplus  \alpha \mathds{H}(K) ]\oplus  \mathds{F}_2(K) \ee
where the middle component $ \mathds{H}(K)^2 \oplus ( \alpha  \mathds{H}(K))^2$ is the lattice of D-brane R-R charges and $K  \mathds{F}_2(K)^*$ the lattice of NS5-branes and KK(6,1)-branes. Here KK(6,1) refers to the Kaluza--Klein monopole brane wrapping $T^6$ with a Taub-NUT fibration along one circle in $T^6$ \cite{Bergshoeff:1997gy}. We will identify these lattices as the set of brane charges invariant under the $\mathds{Z}_K$ orbifold action, while non-perturbative states will generally be in 
\be \mathds{F}^*_3(K) = \mathds{F}_2(K) \oplus [ \tfrac{1}{\alpha} \mathds{H}(K) \oplus \tfrac{1}{\alpha}  \mathds{H}(K) \oplus   \mathds{H}(K) \oplus   \mathds{H}(K) ]\oplus  \mathds{F}^*_2(K) \; . \ee
Because of the asymmetric action of $\mathds{Z}_K$ on $T^4$, the D-branes are necessarily bound states of branes wrapping a cycle and its Poincar\'e dual in $T^4$ \cite{Bianchi:2008cj}. The D-brane states of the orbifold theory therefore preserve at most one third of the supersymmetries, i.e. eight supercharges.

We shall argue in particular that  NS5-branes instantons have integer charges as a winding number over $ T^6 / \mathds{Z}_K$. Assuming T-duality and some minimal assumptions about the non-perturbative symmetry we shall identify the non-perturbative coupling $ \cF_\gra{0}{0} $ as a the theta series 
\be \cF_\gra{0}{0} = - \frac{\pi}{3} \det T_2  \sum_{{\bf q} \in  \mathds{H}(K)^2 \oplus \frac{1}{\alpha}\mathds{H}(K) }  \hspace{-2mm} e^{2\pi i    {\bf q}^\dagger T {\bf q}}\; . \label{NonPerturbativeF00} \ee
This is the unique rank one theta series invariant under the conjectured U-duality group  $\Cong{D}{6}_{0*}(\alpha)$. This 1/2 BPS protected coupling does not receive corrections from  D-instantons, consistently with the property that there is no 1/2 BPS (Euclidean) D-brane.

The organisation of the paper is as follows. In Section \ref{StringOrbifolds} we review asymmetric orbifolds that allow for string embeddings of ${\cal N}=6$ supergravity. We prove that only $K=2,3$ give rise to perturbatively consistent models in $D=4,5$ while $K=4,6$ don't, contrary to what one would have naively expected.  We study the perturbative BPS states and emphasise the role of Hurwitz quarternions in the definition of the BPS charges and the T-duality group. We then turn our attention on  supersymmetry constraints on threshold corrections to the low-energy effective action in Section \ref{SusyLEEA} and prove non-renormalisation theorems in string theory for the threshold corrections to the $\mathcal{R}^4, D^2\mathcal{R}^4$ and $\bar{\mathcal{F}}^2\mathcal{R}^2$ couplings. In Section \ref{String1-loop}, we compute the one-loop threshold corrections to these couplings. In order to achieve our goals, we exploit chiral factorisation of the integrands and rely on the results of \cite{Bianchi:2015vsa} for 4-point open-string amplitudes of vector bosons at one-loop in vacuum configurations with (unoriented) intersecting and magnetised D-branes preserving ${\cal N}\le 4$ supersymmetry in $D=4$. In Section \ref{MinTheta} we rewrite the perturbative $\bar{\mathcal{F}}^2\mathcal{R}^2$ coupling as a rank one theta series and conjecture its nonpertubative completion \eqref{NonPerturbativeF00}. We discuss in more detail the D-brane and NS5-brane  instantons in Section \ref{NonPerturbativeBranes}. We conclude with a discussion of our results and an anticipation of forthcoming papers on the non-perturbative $\mathcal{R}^4, D^2\mathcal{R}^4$ couplings. 

Several appendices contains technical details on Hurwitz quaternions, helicity super-traces, elliptic functions, Eisenstein series, one-loop computations in string theory, matching of normalisations with supergravity and equivalence with another orbifold theory.

%%%%%%%%%%%%%%%%%%%%%%%% %%%%%%%%%%%%%%%%%%%%%%%%

\section{${\cal N}=6$ supergravity in diverse dimensions}

\label{StringOrbifolds}

Maximal supergravity with 32 supercharges (i.e. ${\cal N}=8$ in $D=4$) can be defined in any dimension $D\le 11$. It is the low energy effective theory of either M-theory on $\mathds{R}^{1,D-1}\times T^{11-D}$ or type IIA/B string theory on $\mathds{R}^{1,D-1}\times T^{10-D}$. In $D=10$ type IIA and IIB are respectively the non-chiral $({\cal N}_{ L}, {\cal N}_{ R})=(1,1)$  and chiral $({\cal N}_{ L}, {\cal N}_{ R})=(2,0)$ supergravity theory. The consistency of $({\cal N}_L, {\cal N}_R)=(2,0)$ supergravity inherited  from type IIB string theory relies on gravitational anomaly cancellation \cite{Alvarez-Gaume:1983ihn}. 

The maximal number of supercharges below 32 is 24 (i.e. ${\cal N}=6$ in $D=4$). The classical theory can be defined in $D=6$ dimensions  with  $({\cal N}_L, {\cal N}_R)=(1,2)$ as the consistent truncation of maximal supergravity corresponding to the subgroup $SU(2)_{ L} \times_{\mathds{Z}_2} SU(2)_{ R} \times SO(1,5)\subset SO(5,5)$, where one sets to zero all the fields with a non-trivial $SU(2)_{L}$ weight. The $SU(2)_{L}$ singlet sector can be understood as the untwisted sector of the orbifold by a discrete subgroup $\mathds{Z}_K \subset SU(2)_{L}$ that can be realised as an asymmetric $\mathds{Z}_K$ orbifold of $T^4$ in type II string theory. The $({\cal N}_L, {\cal N}_R)=(1,2)$ supergravity theory  is plagued by gravitational anomalies and is inconsistent at the quantum level in six dimensions. String theory `cures' the problem  thanks to the presence of a gravitino multiplet in the twisted sector of the asymmetric $\mathds{Z}_K$ orbifold leading to supersymmetry enhancement and to maximal non-chiral $({\cal N}_L, {\cal N}_R)=(2,2)$ supergravity. Only in dimension $D\le 5$, one can define a freely acting orbifold with the common asymmetric action of $\mathds{Z}_K$ on $T^4$ and a shift on an additional circle $S^1$.

The theory in five dimensions is ${\cal N}=3$ supergravity\footnote{Here as above, in $D=6$ and $D=10$, we count the number of super-symmetries in terms of the relevant spinorial representation in $D=5$ with 8 real (4 complex) components.} and includes: one graviton, 6 gravitini, 15 vectors (after dualizing all massless 2-forms),  
10 dilatini and 14 scalars. The latter parameterize the non-compact symmetric space ${\cal M}^{D=5}_{\mathcal{N}=6} = SU^*(6)/Sp(3)$.\footnote{We write $Sp(3) = U\hspace{-0.4mm}Sp(6)$ the subgroup of $SU(6)$ preserving a symplectic form.}

The $\mathcal{N}=6$ theory in four dimensions  can be obtained by dimensional reduction from $D=5$, the massless content is: one graviton, 6 gravitini, 16 vectors,  
26 dilatini and 30 scalars. The latter parameterize the non-compact symmetric space ${\cal M}^{D=4}_{\mathcal{N}=6} = SO^*(12)/U(6)$ \cite{Andrianopoli:2001zh}. Quite remarkably ${\cal M}^{D=4}_{\mathcal{N}=6}$ is a special K\"ahler manifold, which allows $\mathcal{N}=6$ supergravity to have the same bosonic sector as a peculiar $\mathcal{N}=2$ supergravity coupled to 15 vector multiplets \cite{Gunaydin:1983bi}. 

Further reduction to $D=3$ produces a (non-propagating) graviton, 6 (non-propagating) gravitini, 64 dilatini and 64 scalars (after dualizing all 17 vectors). The latter parameterize the non-compact quaternionic-K\"ahler symmetric space ${\cal M}^{D=3}_{\mathcal{N}=6} = E_{7(-5)} / ( SU(2) \times_{\mathds{Z}_2} Spin(12))$.

%%%%%%%%%%%%%%%%%%%%%%%%
\subsection{String Theory on (freely acting) asymmetric orbifolds}

Due to the large amount of supersymmetry, only Type II strings admit perturbative vacuum configurations with $\mathcal{N}=6$ supersymmetry \cite{Ferrara:1989nm}. The easiest way to achieve the breaking of $\mathcal{N}=8$ to $\mathcal{N}=6$ is to rely on asymmetric orbifolds \cite{Narain:1986qm,Narain:1990mw} that preserve all  supercharges in one sector (say R-moving) and break half of them in the other sector (say L-moving). One can accomplish the task with a combination of `asymmetric' twists $\tau_{L,R}$ and shifts $\sigma_{L,R}$ \cite{Anastasopoulos:2009kj, Bianchi:2012xz}. In the simplest ${\mathds{ Z}}_2$ case, $\tau_L$ reflects 4 internal Left-moving coordinates $\tau_L X_L^i = - X^i_L$ as well as their (world-sheet) fermionic  partners $\tau_L \Psi_L^i = - \Psi^i_L$, while $\sigma$ acts `geometrically' by an order 2 shift on another (compact circle) direction: $\sigma X = X + \pi \sqrt{\alpha^\prime} \tilde{\RS}$ where $2\pi \sqrt{\alpha^\prime} \tilde{\RS}$ is the periodicity.\footnote{In general $\sigma$ can act in a non  L-R symmetric fashion, compatibly with modular invariance. All these cases are however related by the T-duality of the original theory and can be obtained by field redefinitions. It will be convenient to consider a geometric shift for which the Narain moduli are all in the untwisted sector.} In this way the twisted sector does not contribute new massless states and supersymmetry is broken from $\mathcal{N}=4=2_L+2_R$ (maximal) to $\mathcal{N}=3=1_L+2_R$ in $D=5$. The resulting construction is the simplest possible (supersymmetric) T-fold \cite{Hull:2004in}. This construction straightforwardly generalises to $D=4$ or $D=3$ by compactification on additional circles.

For later use, let us describe the construction in more details, relying on the characters of $SO(2n)$ current algebra at level one 
\be 
O_{2n} = {\vartheta_3^n + \vartheta_4^n \over 2 \eta^n} \, , \quad 
V_{2n} = {\vartheta_3^n - \vartheta_4^n \over 2 \eta^n} \, ,\quad
S_{2n} = {\vartheta_2^n + i^n \vartheta_1^n \over 2 \eta^n} \, ,\quad
C_{2n} = {\vartheta_2^n - i^n\vartheta_1^n \over 2 \eta^n} \; , 
\ee
where $\vartheta_\alpha$ are Jacobi (elliptic) theta functions and $\eta$ is the 
Dedekind function.

One defines the weight $(d/2,d/2)$ modular form $\mathcal{Z}$ associated to a partition function $Z$ via an integral 
\be Z = {\rm Tr} [(-1)^F ] =   \int_{-\frac12}^{\frac12} \hspace{-1.8mm}d\tau_1 \; \mathcal{Z} \; , \ee
that implements the level-matching condition.\footnote{The vacuum energy would be 
$V_{{\rm eff}} \propto \int_{\cF} {d^2 \tau} \tau_2^{\frac{d}{2}-2} \mathcal{Z} $.}

The Type II one-loop partition functions on $T^d$ then compactly read \footnote{Here we do not include the divergent $D$-dimensional zero mode contribution.}
\be
\mathcal{Z}_{{\rm IIA}} = \frac{\overline{V}_8-\overline{C}_8}{\overline{\eta}^8} \frac{V_8-S_8}{\eta^8} \Lambda_{\sLambda_{d,d}}  \quad , \quad \mathcal{Z}_{{\rm IIB}} =\frac{\overline{V}_8-\overline{S}_8}{\overline{\eta}^8} \frac{V_8-S_8}{\eta^8}\Lambda_{\sLambda_{d,d}}  \; , 
\ee
where one identifies the contribution from the fermions ${\cal Q} = V_8-S_8$ and ${\cal Q}'=V_8-C_8$ as the supersymmetric characters in $D=10$, whereby $SO(8)$ acts as the little group. Here $ \Lambda_{\sLambda_{d,d}} $ is the Narain partition function for the Lorentzian lattice $\sLambda_{d,d}$.

The $\mathds{Z}_2$ breaking of supersymmetry in $D=6$ is achieved on the fermionic component via \cite{Bianchi:1990yu, Bianchi:1990tb}  
\be 
{\cal Q}=V_8-S_8 = V_4O_4 + O_4V_4 - S_4S_4 - C_4C_4 \rightarrow 
V_4O_4 - O_4V_4 - S_4S_4 + C_4C_4 = {\cal Q}_o - {\cal Q}_v\; , 
\ee 
with the `untwisted' supersymmetric characters given by 
\be 
{\cal Q}_o = V_4O_4 - S_4S_4 \quad , \quad {\cal Q}_v = O_4V_4 - C_4C_4\; . 
\ee
Performing an $S$-modular transformation ($\tau\rightarrow-1/\tau$) on ${\cal Q}_o$ produces two `twisted' supersymmetric characters
\be 
{\cal Q}_s = O_4S_4 - C_4O_4 \quad , \quad {\cal Q}_c = V_4C_4 - S_4V_4\; , 
\ee
while ${\cal Q}_s$ accommodates massless states, ${\cal Q}_c$ doesn't.

As already mentioned, in $D=6$ the twisted sector produces two gravitini multiplets that enhance supersymmetry back to 32 supercharges, {\it i.e.} $(\mathcal{N}_L,\mathcal{N}_R)=(2,2)$. In $D\le 5$ one can prevent the appearance of twisted  massless states by introducing the order 2 `shift' $\sigma$ defined above. In particular starting with $T^{4+d}= T^4\times T^d$ with $d=6-D$ and $T^4$ the maximal torus of $SO(8)=D_4$, one can immediately determine the invariant lattice and the perturbative charge lattice of the model.

Other possibilities may be envisaged to break $\mathcal{N}=8$ to $\mathcal{N}=6$ in $D<6$ that rely on other asymmetric $\mathds{Z}_K$ orbifolds with $K=3,4,6$, that act on the supersymmetric character ${\cal Q}$ according to 
\be 
{\cal Q} \rightarrow \sum_{k=0}^{K-1} \omega^k {\cal Q}_k \quad {\rm with} \quad \omega=e^{2\pi i\over K} \; . 
\ee
For this purpose one further decomposes the second $SO(4)$ current algebra character into $SU(2)\times SU(2)^\prime$ characters as
\be \mathcal{Q} = V_4 \chi_0 \chi_0^\prime  + O_4 \chi_\frac12 \chi_\frac12^\prime - S_4 \chi_\frac12 \chi_0^\prime - C_4 \chi_0 \chi_\frac12^\prime = \mathcal{V} \chi_0^\prime  + {\cal H}  \chi_\frac12^\prime\; ,  \ee
where one identifies the vector multiplet super-character ${\cal V}= V_4\chi_0 - S_4\chi_{1/2}$, the (half) hypermultiplet  super-character ${\cal H} = O_4\chi_{1/2} - C_4\chi_{0}$, and $SU(2)$ as the R-symmetry group. The orbifold acts on the $SU(2)^\prime$ characters $\chi_s^\prime$ only. For instance for $K=3$ the $\mathds{Z}_3$ orbifold acts on the $SU(2)'$ characters as
\be 
\chi_0' \rightarrow \xi_0 +\bar\omega \xi_{+4} +\omega \xi_{-4}, \quad \chi_{1/2}' \rightarrow \omega \xi_{+2} +\bar\omega \xi_{-2} + \xi_{6} \quad ,
\ee
where $\xi_{p}$ are the characters of a free boson with radius $R=\sqrt{6\alpha'}$ and ${\cal Q}_0 = {\cal V} \xi_0 + {\cal H} \xi_6$, ${\cal Q}_1 = {\cal H} \xi_2 + {\cal V} \xi_{-4}$ and ${\cal Q}_2 = {\cal H} \xi_{-2} + {\cal V} \xi_{+4}$. Starting from the boson partition function on $T^4$ at the  $SU(3)\times SU(3)$ symmetric point 
\be \frac{\Lambda_{\sLambda_{4,4}}}{\eta^2 \bar \eta^2} =\bigl(|\chi_{[00]}|^2 +  |\chi_{[10]}|^2 +  |\chi_{[01]}|^2\bigr)^2\; , 
\ee
 and acting with $\mathds{Z}_3$ only on the Left-movers one finds 
\be
 \frac{\Lambda_{\sLambda_{4,4}}}{\eta^2 \bar \eta^2}  \rightarrow  \bigl(|\chi_{[00]}|^2 + \omega |\chi_{[10]}|^2 + \bar\omega |\chi_{[01]}|^2\bigr)^2\; . 
\ee

In general one may consider a subgroup $\mathds{Z}_K\subset SU(2)$ acting on $T^4$. Consistency of the asymmetric orbifold requires that the number of `chiral' fixed points $N_{\rm f} = 4 \sin(\pi/K)^2$ be an integer, which is the case only for $K=2,3,4,6$. For $K=4$ one starts with the maximal torus of $SU(2)^4$. For $K=6$ from $SU(3)_1\times SU(3)_2$, with the second $SU(3)$ current algebra at level $2$. The lattices for $K=2,3,4,6$ are therefore $D_4$, $A_2\oplus A_2$, $A_1\oplus A_1 \oplus A_1 \oplus A_1$ and $A_2\oplus A_2[2]$, respectively. Quite remarkably, these lattices can be realised as integral quaternions and admit in this way a non-abelian ring structure. We will therefore denote these lattices as $\mathds{H}(K)$, following the definition \eqref{HKquaternions}. The ring $\mathds{H}(2)$, that we will often abreviate as $\mathds{H}$, is the ring of Hurwitz quaternions, $\mathds{H}(3)$ is a generalisation of the Eisenstein integers that we call the Eisenstein quaternions, while $\mathds{H}(4)$ and $\mathds{H}(6)$ are subrings of the Hurwitz quaternions. This is consistent with the property that the duality group in $\mathcal{N}=6$ is naturally defined over the quaternions, and the T-duality group is a theta congruent subgroup of $SL(2,\mathds{H}(K))$ and $Sp(4,\mathds{H}(K))$  in five and four dimensions dimensions. 

Yet the condition that $N_{\rm f}$ be an integer is not enough for the asymmetric orbifold to be consistent, and one needs to check that the number of bosonic ground states in the twisted sectors be an integer too \cite{Narain:1986qm,Narain:1990mw}. Indeed, the twisted sectors of the asymmetric orbifold are only defined abstractly by modular invariance, and one must verify that this construction can be realised on a Hilbert space. The number of bosonic ground states is determined by   $N_{\rm f}=\sqrt{\det(1-g)}$ (where $g$ is the L-R symmetric extension of the twist $\tau_L$) divided by the square root of the index $| \Lambda^* / \Lambda |$ 
 of the $\tau_L$ invariant lattice $\Lambda \subset \sLambda_{4,4}$ \cite{Narain:1986qm,Narain:1990mw}. For the $\mathds{Z}_K$ asymmetric orbifold discussed above, $\Lambda=\mathds{H}(K) $ and $|\mathds{H}(K) ^* / \mathds{H}(K)  | = K^2 $. One obtains  \be \sqrt{\frac{\det(1-g)}{|\mathds{H}(K) ^* / \mathds{H}(K)  |}} = \frac{4 \sin(\pi/K)^2}{K} \underset{K=2,3,4,6}{=} \{ 2, 1 , \tfrac12 , \tfrac16 \} \label{NumberGravitini}  \ee
 and the asymmetric orbifold is only consistent for $K=2$ and $3$. We will find more explicitly below that $ 4 \sin^2(\pi/K) /K $ is the number of massive 1/2 BPS gravitini multiplets for a given charge of vanishing norm in the Narain lattice in the first twisted sector. To this end we will now turn our attention onto the identification and counting of (perturbative) BPS states via character-valued partition functions and helicity super-traces \cite{Kiritsis:1997em, Pioline:1997pu, Kiritsis:1997hj}.
  
\subsection{Character-valued partition functions and helicity super-traces}
Thanks to Jacobi {\it aequatio satis abstrusa}, the 10-D super-characters ${\cal Q}$ and ${\cal Q}'$ vanish, which implies equal number of bosons and fermions at each mass level and ensures the  absence of tachyons \cite{GreenSchwWitt}. As a consequence, the one-loop contribution to the vacuum energy is zero, which guarantees quantum stability of flat ten-dimensional space-time and of toroidal compactifications thereof. Actually supersymmetry is more powerful in that also 1-, 2- and 3-point amplitudes are zero for massless external states both in Type IIA and B as well as in Type I and Heterotic strings. The first non-trivial amplitude with only massless external states at one-loop is a 4-point amplitude.  In type II, the prototypical case is the 4-graviton amplitude.
After partial supersymmetry breaking, lower-point amplitudes with (non-conserved) complex momenta are non-zero but some are still protected. The easiest way to identify the special amplitudes and derive the associated (on-shell) effective action is to introduce the concept of `character valued partition functions' and helicity super-traces \cite{Kiritsis:1997em, Pioline:1997pu, Kiritsis:1997hj}.

Recalling that the (Left-moving) Lorentz generators are given by 
\be 
J^{MN}= \Psi^M\Psi^N + X^M\partial X^N -   
X^N\partial X^M \ee
 one can consider including a coupling to the zero-modes of $J^{MN}$ in the Cartan subalgebra, {\it i.e.} replace $Z={\rm Tr} [ (-)^F]$ with $Z={\rm Tr} [(-)^F\prod_a e^{2\pi  i v_aJ_a}]$. For the theory in four dimensions, we only need to include the zero mode $J_3 = J^{12}$ in the four-dimensional little group and that does not act on the compact bosons, so we consider $Z={\rm Tr}[(-)^F e^{2\pi  i v J_3}]$.  In practice, this amounts to including a fugacity in the partition function. Writing the maximally supersymmetric Left-mover partition function as a sum over spin structures  
\be \frac{V_8-S_8}{\eta^8} = \frac{1}{2} \Bigl[ \frac{\vartheta_3(0)^4}{\eta^{12}} - \frac{\vartheta_4(0)^4}{\eta^{12}}- \frac{\vartheta_2(0)^4}{\eta^{12}}- \frac{\vartheta_1(0)^4}{\eta^{12}} \Bigr]\; .  \ee
and including the fugacity produces for each spin structure $\alpha = 1,2,3,4$ \footnote{In general one would have $$
{\vartheta_\alpha(0)^4 \over \eta^{12}} \rightarrow \prod_a 2\sin\pi v_a {\vartheta_\alpha(v_a)\over \vartheta_1(v_a)} = \prod_a {\vartheta_\alpha(v_a)\over \eta^3}{\sin\pi v_a \vartheta_1'(0) \over \pi \vartheta_1(v_a)} = \prod_a {\xi(v_a)  \vartheta_\alpha(v_a)\over \eta^3}
$$
where 
$
\xi(v_a) = {\sin\pi v_a \vartheta_1'(0) /\pi \vartheta_1(v_a)} \; . 
$}
\be
{\vartheta_\alpha(0)^4 \over \eta^{12}} \rightarrow  2\sin\pi v {\vartheta_\alpha(v)\over \vartheta_1(v)} {\vartheta_\alpha(0)^3 \over \eta^{9}} = {\sin\pi v \vartheta_1'(0) \over \pi \vartheta_1(v)}  {\vartheta_\alpha(v)\over \eta^3}{\vartheta_\alpha(0)^3 \over \eta^{9}}  =\xi(v)  {\vartheta_\alpha(v)\vartheta_\alpha(0)^3  \over \eta^{12}} \label{spinChZ} \ee
where 
\be
\xi(v) = {\sin\pi v \, \vartheta_1'(0) \over \pi \vartheta_1(v)} \; .
\ee
Summing over the spin structures with the help of Riemann identities yields
\bea  \frac{V_8-S_8}{\eta^8} &\rightarrow& \frac{\xi(v)}{2} \Bigl( \frac{\vartheta_3(v)\vartheta_3(0)^3 }{\eta^{12}} - \frac{\vartheta_4(v)\vartheta_4(0)^3 }{\eta^{12}}- \frac{\vartheta_2(v)\vartheta_2(0)^3 }{\eta^{12}}- \frac{\vartheta_1(v)\vartheta_1(0)^3 }{\eta^{12}} \Bigr) \CR
&=& \frac{\xi(v) \vartheta_1(\frac{v}{2})^4}{\eta^{12}}\; , \label{TypeIIChZ} \eea
that correctly reproduces the (vanishing) partition function in the limit $v\rightarrow 0$, using $\vartheta_1'(0) = 2\pi \eta^3$. 

Taking derivatives with respect to $v$ and then setting it to zero one brings down powers of $J_3$ and obtains what are known as helicity super-traces, denoted by $B_{n}$.   Indeed recall that the four Cartan generators of $SO(8)$ represent the `helicities' in the four planes transverse to the light-cone in ten dimensions, and $J_3$ determines the four-dimensional helicity. For massive states in four dimensions $J_3$ is the Cartan generator of the $SU(2)$ little group. In five dimensions one must distinguish the Left- and Right-mover $J_{3 L,R}$ that define the Cartan generators of the $SU(2)_L\times SU(2)_R$ little group of massive states. One defines accordingly  the `character-valued' partition function $\mathcal{Z}(v,\bar v)$ and introduces the differential operators 
\be 
{{\cal H}} = {1\over 2\pi i} {\partial \over \partial v} \;  , \quad \bar{{{\cal H}}} = - {1\over 2\pi i} {\partial \over \partial \bar{v}} \; , 
\ee
such that 
\be
{\cal B}_{2n} =  ({{\cal H}}+\bar{{{\cal H}}})^{2n}{\cal Z}(v,\bar{v})\vert_{v=\bar{v}=0}  \; . 
\ee
The helicity supertrace is then defined by implementing the level matching condition as
\be B_{2n} = {\rm Tr} [ (-1)^{2J_3} J_3^{\, 2n}] = \int_{-\frac12}^{\frac{1}{2}} \hspace{-1.8mm} d\tau_1 \; {\cal B}_{2n}\; . \ee
$B_{2n+1}=0$ by CPT symmetry.  By construction $B_0=0$ in all supersymmetric theories with a supersymmetric ground state, since BPS multiplets have the same number of bosonic and fermionic degrees of freedom. For $\mathcal{N}$ extended supersymmetry with $\mathcal{N}$ even in four dimensions, one finds that $B_{2n}=0$ for $2n< \mathcal{N}$, and $B_{\mathcal{N}}$ only receives contributions from 1/2 BPS states in the theory. The non-BPS long supermultiplets only contribute to $B_{2n}$ for $n \ge \mathcal{N}$, and the helicity supertraces for $\mathcal{N}/2\le n<\mathcal{N}$ are BPS protected observables of the theory. The corresponding `modular forms' $\mathcal{B}_{2n}$ are associated to BPS protected threshold corrections. 

In particular the leading one-loop $\mathcal{R}^4$ type supersymmetry invariant in the low-energy effective action of Type II strings in  $D=10$ and in toroidal compactifications is associated to the helicity super-trace ${\cal B}_{8}$, the ${D}^4 \mathcal{R}^4$ to ${\cal B}_{12}$ and the ${D}^6 \mathcal{R}^4$ to ${\cal B}_{14}$.\footnote{Similarly the one-loop $F^4$ supersymmetry invariant in Type I and Heterotic strings is related to ${\cal B}_{4}$  and ${D}^2 F^4$ to ${\cal B}_{6}$ \cite{Bachas:1997mc, Kiritsis:1997hf, Bianchi:1998vq,Bossard:2018rlt}.}

\subsection{The asymmetric orbifold helicity supertraces}
\label{CharZK} 
The `character-valued' partition function can be obtained similarly for the Type II asymmetric $\mathds{Z}_K$ orbifolds. We will concentrate on the relevant theories with $K=2,3$, although the first part of this section applies to the `inconsistent' cases of $K=4,6$, too. 

The contribution from the bosonic zero modes in the untwisted sector with the insertion of $(\tau_L,\sigma)^r \in \mathds{Z}_K$ for $r=1$ to $K-1$ gives the Narain lattice partition function  
\be   \Lambda_{\sLambda_{2,2}\oplus \mathds{H}(K)}[^0_r] = \Lambda_{\sLambda_{2,2}\oplus \mathds{H}(K)}[e^{\frac{2\pi i r \tilde{m}}{K}}] = \sum_{m,n,\tilde{m},\tilde{n}\in \mathds{Z}} \sum_{q\in  \mathds{H}(K)} e^{\frac{2\pi i r \tilde{m}}{K}} e^{\pi i \tau p_L(m,n,\tilde{m},\tilde{n},q)^2 - \pi i \bar \tau p_R(m,n,\tilde{m},\tilde{n},q)^2 } \label{UntwistNarain} \ee
where $\tilde{m}$ is the momentum along the circle on which $\sigma$ acts as a geometric shift $\sigma X= X + \frac{2\pi}{K}\sqrt{\alpha^\prime} \tilde{\RS}$. There is no loss of generality in choosing the geometric shift along a particular circle in $T^2$ since the other cases could be obtained by T-duality. 

Since the current $J_3$ does not act on the fields affected by the orbifold, the contribution from the Left-mover oscillators in each spin structure $\alpha$ is modified as in \eqref{spinChZ} as
\be  4\sin^2(\tfrac {\pi r} K)  \frac{\vartheta_\alpha(0)^2 \vartheta_\alpha(\frac{r}{K})^2}{\vartheta_1(\frac{r}{K})^2 \eta^6} \rightarrow 4\sin^2(\tfrac {\pi r} K)  \frac{ \xi(v) \vartheta_\alpha(\frac v 2) \vartheta_\alpha(0) \vartheta_\alpha(\frac{r}{K})^2}{\vartheta_1(\frac{r}{K})^2 \eta^6} \ee
and they recombine accordingly  as in \eqref{TypeIIChZ}
\bea  &&    \frac{\xi(v)}{2} \Bigl( \frac{ \vartheta_3(\frac v 2) \vartheta_3(0) \vartheta_3(\frac{r}{K})^2}{\vartheta_1(\frac{r}{K})^2 \eta^6} -  \frac{ \vartheta_4(\frac v 2) \vartheta_4(0) \vartheta_4(\frac{r}{K})^2}{\vartheta_1(\frac{r}{K})^2 \eta^6} - \frac{ \vartheta_2(\frac v 2) \vartheta_2(0) \vartheta_2(\frac{r}{K})^2}{\vartheta_1(\frac{r}{K})^2 \eta^6} - \frac{ \vartheta_1(\frac v 2) \vartheta_1(0) \vartheta_1(\frac{r}{K})^2}{\vartheta_1(\frac{r}{K})^2 \eta^6} \Bigr) \CR
&=& \frac{\xi(v) \vartheta_1(\frac{v}{2})^2\vartheta_1(\frac{v}{2} + \frac{r}{K})\vartheta_1(\frac{v}{2}-\frac{r}{K})}{\vartheta_1(\frac{r}{K})^2 \eta^6}\; , \label{LChZ} \eea
in such a way that the total contribution from the untwisted sector with  insertion of $(\tau_L,\sigma)^r \in \mathds{Z}_K$ is
\be 4\sin^2(\tfrac {\pi r} K)  \frac{\xi(v) \vartheta_1(\frac{v}{2})^2\vartheta_1(\frac{v}{2} + \frac{r}{K})\vartheta_1(\frac{v}{2}-\frac{r}{K})}{\vartheta_1(\frac{r}{K})^2 \eta^6}   \overline{\frac{\xi(v) \vartheta_1(\frac{v}{2})^4}{\eta^{12}}} \Lambda_{\sLambda_{2,2}\oplus \mathds{H}(K)}[^0_r]\; . \ee
As usual the twisted sector is determined by modular invariance. We define the contribution from the bosonic zero modes for the twisted sector $s\ne 0$ mod $K$ dividing $r$ as \footnote{For $K=2,3$ this includes all $s\ne 0$ mod $K$, while this excludes $(r,s)= (1,2)$ and $(3,2)$ for $K=4$ and  $(r,s)= (1,2), \, (3,2),\, (5,2),\, (1,3),\, (2,3),\, (4,3),\, (5,3),\, (1,4), \, (3,4),\, (5,4)$ for $K=6$.} 
\be \Lambda_{\sLambda_{2,2}\oplus \mathds{H}(K)}[^s_r]  =- \frac{1}{K} \sum_{\substack{m,n,\tilde{m}\in \mathds{Z}\\ \tilde{n} \in \mathds{Z}+ \frac{s}{K}}} \sum_{q\in  \mathds{H}(K)^*} e^{\frac{2\pi i r \tilde{m}}{K} - 2\pi i r |q|^2} e^{\pi i \tau p_L(m,n,\tilde{m},\tilde{n},q)^2 - \pi i \bar \tau p_R(m,n,\tilde{m},\tilde{n},q)^2 }  \label{TwistNarain}\ee
where we included the factor of $-\frac{1}{K}$ appearing in the Poisson summation over $\mathds{H}(K)$ in the definition in such a way that 
\be    \Lambda_{\sLambda_{2,2}\oplus \mathds{H}(K)}[^0_s](- \tfrac{1}{\tau}, - \tfrac{1}{\bar \tau}) =  \tau \bar \tau^3  \Lambda_{\sLambda_{2,2}\oplus \mathds{H}(K)}[^{\scalebox{0.6}{-}s}_0](\tau,\bar\tau)\; .\ee
The complete `character-valued' partition function can finally be written with these definitions as
 \begin{multline}
{\cal Z}^{\mathds{Z}_K}_{D=4}(v,\bar{v})=\frac{\xi(v)}{K} \overline{ \frac{\xi({v})\vartheta_1({v}/2)^4}{\eta(\tau)^{12}} }\Biggl[  \frac{{\vartheta_1({v\over 2})^4}}{   \eta(\tau)^{12} } \Lambda_{\sLambda_{6,6}}\\ +  \sum_{\substack{r , s\, {\rm mod} K\\ (r,s)\ne (0,0)} } 4 \sin^2\Bigl( \frac {\pi {\rm gcd}(r,s)} K\Bigr)    \frac{  {\vartheta_1({v\over 2})^2 \vartheta_1({v\over 2}+\frac{r + s \tau}{K})\vartheta_1({v\over 2}-\frac{r + s \tau}{K})
}}{  \vartheta_1(\frac{r + s \tau}{K})^2 \eta(\tau)^6}  \Lambda_{\sLambda_{2,2}\oplus \mathds{H}(K)}[^s_r] \Biggr] \label{ChaPart} \end{multline}
with the sum over $r$ and $s$ from $0$ to $K-1$ excluding $r=s=0$. Recall that we have fixed the external four-dimensional momenta to zero.

Setting $i\pi v= i \pi\bar{v}=\log(y)$ with $y$ the fugacity for the angular momentum $J_3$, one can define the partition function as a trace over the Hilbert space of perturbative string states  as 
\be  
Z^{\mathds{Z}_K}_{D=4}(y) = {\rm Tr}[e^{2\pi i \tau L_0- 2\pi i \bar \tau \bar L_0} (-y)^{2J_3} ]  = \int_{-\frac12}^{\frac12} \hspace{-1.8mm} d\tau_1 {\cal Z}^{\mathds{Z}_K}_{D=4}(v,\bar{v}) \; ,
\ee
where the integral over $\tau_1$ implements level matching, as above. The partition function $Z^{\mathds{Z}_K}_{D=4}(y) $ expands as a sum over $\frac{k}{6}$-BPS supermultiplets $\psi \in \mathcal{E}_j^{k/6}$ of  highest spin $j+3 -\frac k 2$  with $k=1,2,3$ and $k=0$ fo non-BPS states as 
\bea Z^{\mathds{Z}_K}_{D=4}(y) &=&  \sum_{j\in \mathds{N}/2} (-1)^{2j}   \frac{y^{2j+1}-y^{-2j-1}}{y-y^{-1}}\Biggl[- (y^{\frac12}- y^{-\frac12})^{6} \sum_{\psi \in \mathcal{E}_j^{\scalebox{0.5}{$1/2$}} } e^{- \pi \tau_2\, \alpha^\prime \! M_\psi^2} \label{PartSMult} \\
&& \hspace{-18mm}  + (y^{\frac12}- y^{-\frac12})^{8} \hspace{-1mm} \sum_{\psi \in \mathcal{E}_j^{\scalebox{0.5}{$1/3$}} } e^{- \pi \tau_2\, \alpha^\prime \! M_\psi^2}  -(y^{\frac12}- y^{-\frac12})^{10} \hspace{-1mm} \sum_{\psi \in \mathcal{E}_j^{\scalebox{0.5}{$1/6$}} } e^{- \pi \tau_2\, \alpha^\prime \! M_\psi^2} + (y^{\frac12}- y^{-\frac12})^{12} \hspace{-1mm}\sum_{\psi \in \mathcal{E}_j^{\scalebox{0.5}{$0$}} } e^{- \pi \tau_2\, \alpha^\prime \! M_\psi^2} \Biggr] \; , \nonumber\eea
where $M_\psi$ is the mass of the supermultiplet $\psi$. The large $\tau_2$ expansion of \eqref{ChaPart} (not expanding the lattice partition functions) is determined by 
\bea  &&  \overline{ \frac{\xi({v})\vartheta_1(\frac{v}{2})^4}{\eta(\tau)^{12}} } \frac{ \xi(v) {\vartheta_1(\frac{v}{2})^2 \vartheta_1(\frac{v}{2}+u)\vartheta_1(\frac{v}{2} -u)
}}{  \vartheta_1(u)^2 \eta(\tau)^6}  \CR
&=& \frac{(1-\zeta y)(1-\zeta y^{-1})}{(1-\zeta)^2}  (y^{\frac12}- y^{-\frac12})^{6}  \Bigl( 1 +  (y^{\frac12}- y^{-\frac12})^{4} e^{-2\pi i \bar \tau} + \dots \Bigr) \CR
&& \hspace{40mm} \times \Bigl( 1-\frac{(1-\zeta y)(1-\zeta y^{-1})}{\zeta } (y^{\frac12}- y^{-\frac12})^{2} e^{2\pi i  \tau}  + \dots \Bigr) \label{ZThetaExpand} \eea
where $\zeta = e^{2\pi i u}$ for $u = \frac{r+s \tau}{K}$. This exhibits that the corresponding constant term determines the spectrum of 1/2 BPS supermultiplets, the holomorphic part the spectrum of  1/3 BPS supermultiplets, the anti-holomorphic part the spectrum of  1/6 BPS supermultiplets while the generic terms determine the spectrum of  non-BPS supermultiplets. This is consistent with the  BPS black hole solitons in supergravity \cite{Andrianopoli:2006ub}. The 1/2 BPS black holes have a rank one electromagnetic charge that is a vector ${Q}$ with ${Q}^2=0$ in the perturbative Narain lattice. The 1/3 BPS black holes have a rank two electromagnetic charge that is a vector ${Q}$ with ${Q}^2<0$ in the perturbative Narain lattice, consistently with the level matching condition in the Left-moving sector. The 1/6 BPS black holes with an  electromagnetic charge in the perturbative Narain lattice can only be rank two,  with a  vector ${Q}$ with ${Q}^2>0$, consistently with the level matching condition in the Right-moving sector.

Using \eqref{ChaPart} and  \eqref{ZThetaExpand} one obtains the partition function for 1/2 BPS states 
\begin{multline}  Z^{\scalebox{0.6}{$\frac12$-BPS}}_{D=4}(y)  = (y^{\frac12}- y^{-\frac12})^{6} \int_{-\frac12}^{\frac12} \hspace{-2mm} d\tau_1 \Biggl[( y + y^{-1} ) \sum_{m,n, \tilde{m},\tilde{n}\in \mathds{Z}} \sum_{q\in  \mathds{H}(K)}  e^{\pi i \tau p_L({Q})^2 - \pi i \bar \tau p_R({Q})^2 }\\
-\Biggl( \sum_{\substack{m,n,\tilde{n}\in \mathds{Z}\\\tilde{m} \in K\mathds{Z}+1 }} +\sum_{\substack{m,n,\tilde{n}\in \mathds{Z}\\\tilde{m} \in K\mathds{Z}-1 }} \Biggr) \sum_{q\in  \mathds{H}(K)}  e^{\pi i \tau p_L({Q})^2 - \pi i \bar \tau p_R({Q})^2 } \\ 
-  \frac{4 \sin^2( \frac\pi K)  }{K} \sum_{\substack{m,n,\tilde{m}\in \mathds{Z}\\\tilde{n} \in \mathds{Z}+\frac{1}{K} }}  \sum_{q\in  \mathds{H}(K)}  e^{\pi i \tau p_L({Q})^2 - \pi i \bar \tau p_R({Q})^2 }
\\+ \sum_{s=2}^{K-1} \sum_{r=0}^K  \frac{4 \sin^2\bigl( \frac {\pi {\rm gcd}(r,s)} K\bigr)  }{K}    \Lambda_{\sLambda_{2,2}\oplus \mathds{H}(K)}[^s_r]\Biggr] \; . \end{multline}
The first two lines correspond to the untwisted sector, with one spin two 1/2 BPS supermultiplet for each null charge ${Q} \in \sLambda_{2,2}\oplus \mathds{H}(K)$ with $\tilde{m}=0$ mod $K$, and one spin three half 1/2 BPS supermultiplet for each null charge ${Q} \in \sLambda_{2,2}\oplus \mathds{H}(K)$ with $\tilde{m} = \pm 1 $ mod $K$ (two supermultiplets for $K=2$). For the first twisted sector, one gets for each charge of vanishing norm with $\tilde{n} \in \mathds{Z} + \frac{1}{K}$ and $q\in \mathds{H}(K)^*$, ${ 4 \sin(\pi/K)^2}/{K}$ massive spin three half 1/2 BPS supermultiplets.  This confirms that \eqref{NumberGravitini} must be an integer and therefore that the asymmetric orbifold theory is only consistent for $K=2,3$. 

From now on we will therefore consider $K=2,3$ only. In these cases \eqref{ChaPart} simplifies because $ \sin^2(\pi {\rm gcd}(r,s)/ K) =\sin^2(\pi/ K) $ for all non-zero $r$ and $s$. In order to make the physical lattice of charges manifest, it is convenient to write the Lorentzian lattice partition function as a sum over orbits
\be   \Lambda_{\sLambda_{6,6}} =  4 \sin^2(\tfrac \pi K)  \Biggl(  \Lambda_{\sLambda_{2,2} \oplus  \mathds{H}(K)} \frac{\wp(\frac{1}{K})}{4\pi^2}- \frac{1}{K} \sum_{r=0}^{K-1} \Lambda_{\sLambda_{2,2} \oplus \mathds{H}(K)^*}[e^{2\pi i r|q|^2}]  \frac{\wp(\frac{r-\tau}{K})}{4\pi^2} \Biggr)   \; .  \ee
Using identity \eqref{WeierDiffTheta} one gets 
\be  \frac{ \xi(v) {\vartheta_1({v\over 2})^2 \vartheta_1({v\over 2}+\frac{r + s \tau}{K})\vartheta_1({v\over 2}-\frac{r + s \tau}{K})
}}{  \vartheta_1(\frac{r + s \tau}{K})^2 \eta(\tau)^6}  =   \frac{ \xi(v) \vartheta_1({v\over 2})^4
}{\eta(\tau)^{12}} \frac{\wp(\frac{r + s \tau}{K})-\wp(\frac{v}{2})}{4\pi^2}\; . \label{WeierTheta}\ee
The symmetries of the Weierstrass function imply that the distinct values of $\wp(\frac{r + s \tau}{K})$ reduce to $\wp(\frac{1}{K})$ and $\wp(\frac{\tau+r}{K})$ for $r=0$ to $K-1$. Moreover  using 
\be   \Lambda_{\sLambda_{2,2} \oplus  \mathds{H}(K)} - \frac{1}{K} \sum_{r=0}^{K-1} \Lambda_{\sLambda_{2,2} \oplus \mathds{H}(K)^*}[e^{2\pi i r |q|^2}]  = 0 \; , \ee
the `character-valued'  partition function \eqref{ChaPart} can be rewritten as
\bea {\cal Z}^{\mathds{Z}_K}_{D=4}(v,\bar{v}) &=&4 \sin^2(\tfrac \pi K)  \Biggl| \frac{ \xi(v) \vartheta_1({v}/2)^4}{ \eta( \tau)^{12}} \Biggr|^2 \Biggl[  \frac{\wp(\frac{1}{K})-\wp(\frac{v}{2})}{4\pi^2} \frac{1}{K} \sum_{r=0}^{K-1}  \Lambda_{\sLambda_{2,2} \oplus  \mathds{H}(K)}[e^{\frac{2\pi i r \tilde{m}}{K}}]   \CR
&& \hspace{15mm} - \frac{1}{K^2} \sum_{r=0}^{K-1} \frac{\wp(\frac{\tau+r}{K})-\wp(\frac{v}{2})}{4\pi^2} \Lambda_{\sLambda_{1,1}\oplus \sLambda_{1,1}[\frac 1 K] \oplus  \mathds{H}(K)^*}[e^{\pi i r  {Q}^2}]  \Biggr] \; ,\; \label{calZ23} \eea
where 
\be \Lambda_{\sLambda_{1,1}\oplus \sLambda_{1,1}[\frac 1 K] \oplus  \mathds{H}(K)^*}[e^{\pi i r  {Q}^2} ]   =  \sum_{s=0}^{K-1} \sum_{\substack{m,n,\tilde{m}\in \mathds{Z}\\ \tilde{n} \in \mathds{Z}+ \frac{s}{K}}} \sum_{q\in  \mathds{H}(K)^*} \hspace{-1.5mm} e^{\pi i (\tau +r)p_L(m,n,\tilde{m},\tilde{n},q)^2 - \pi i (\bar \tau+r) p_R(m,n,\tilde{m},\tilde{n},q)^2 } \; .  \ee
One finally arrives to the very compact expression for the character-valued partition function 
\bea \label{PartitionHelicity}
Z^{\mathds{Z}_K}_{D=4}(y) 
&=& 4 \sin^2(\tfrac \pi K)  \int_{-\frac12}^{\frac12}\hspace{-2.5mm}d\tau_1   \overline{ \frac{\xi({v})\vartheta_1(\frac{v}{2})^4}{\eta(\tau)^{12}} }  \frac{  \xi(v){\vartheta_1(\frac{v}{2})^2 \vartheta_1(\frac{v}{2}{+}\frac{1}{K})\vartheta_1(\frac{v}{2}{-}\frac{1}{K})
}}{  \vartheta_1(\frac{1}{K})^2 \eta(\tau)^6}  \Lambda_{\sLambda_{1,1}\oplus \sLambda_{1,1}[K]\oplus \mathds{H}(K)} \\
&& - \frac{4 \sin^2(\tfrac \pi K)  }{K^2}  \int_{-\frac K2}^{\frac K2}\hspace{-2mm}d\tau_1    \overline{ \frac{\xi({v})\vartheta_1(\frac{v}{2})^4}{\eta(\tau)^{12}} }  \frac{  \xi(v){\vartheta_1(\frac{v}{2})^2 \vartheta_1(\frac{v}{2}{+}\frac{\tau}{K})\vartheta_1(\frac{v}{2}{-}\frac{\tau}{K})
}}{  \vartheta_1(\frac{\tau}{K})^2 \eta(\tau)^6} \Lambda_{\sLambda_{1,1}\oplus \sLambda_{1,1}[\frac 1 K]\oplus \mathds{H}(K)^*}   \; , \nonumber \eea
where we have traded the sum over $r$ from $0$ to $K-1$ with the extension of the integration domain of $\tau_1$ and used \eqref{WeierTheta} reversely. The main advantage of this formula is that it makes manifest that the physical lattice of bosonic zero modes is the lattice $\mathds{F}_2(K)^* $ dual to 
\be \mathds{F}_2(K) = \sLambda_{1,1}\oplus \sLambda_{1,1}[ K]\oplus \mathds{H}(K)\; . \label{LatticeUntwisted} \ee
Therefore one finds that the spectrum of supermultiplets with a given charge ${Q} \in \mathds{F}_2(K)^* $ depends only of its norm square ${Q}^2$ and the property that it lies or not in the sublattice $\mathds{F}_2(K)$. All the charges in $\mathds{F}_2(K)$ correspond to states in the untwisted sector, but charges ${Q} \in  \mathds{F}_2(K)^* \smallsetminus \mathds{F}_2(K) $ are in the twisted sector if and only if $\tilde{n}$ is not an integer. However, \eqref{PartitionHelicity} exhibits that the physical distinction between states relies on the condition of being or not in $\mathds{F}_2(K)$ and we will therefore sometimes refer by abuse of language to all the states with charge ${Q} \in  \mathds{F}_2(K)^* \smallsetminus \mathds{F}_2(K) $ as twisted states, including the ones with $\tilde{n}  $ an integer. 

In particular the constant term in the large $\tau_2$ expansion (not expanding the lattice partition function) gives the partition function of 1/2 BPS supermultiplets 
\begin{multline} Z^{\scalebox{0.6}{$\frac12$-BPS}}_{D=4}(y)  =\bigl( y^{\frac12}-y^{-\frac12}\bigr)^6 \bigl(y+y^{-1}\bigr) \hspace{-2mm} \sum_{\substack{{Q}\in \mathds{F}_2(K) \\{Q}^2=0 }}\hspace{-2mm}e^{-2\pi \tau_2  p_R({Q})^2} \\
-\frac{4\sin^2(\frac{\pi}{K})}{K} \bigl( y^{\frac12}-y^{-\frac12}\bigr)^6 \hspace{-4mm} \sum_{\substack{{Q}\in  \mathds{F}_2(K)^* \smallsetminus   \mathds{F}_2(K) \\ {Q}^2=0 }}\hspace{-4mm}e^{-2\pi \tau_2  p_R({Q})^2} \; . \end{multline}
One has therefore a single spin two 1/2 BPS supermultiplet for each untwisted charge  ${Q}\in \mathds{F}_2(K) $ with ${Q}^2=0$, including the massless gravity multiplet for ${Q}=0$, and respectively two (for $\mathds{Z}_2$) and one (for $\mathds{Z}_3$)  spin three half 1/2 BPS supermultiplets for each `twisted' charge $ {Q}\in  \mathds{F}_2(K)^* \smallsetminus   \mathds{F}_2(K)$ with ${Q}^2=0$. Note that an $\mathcal{N}=8$ spin two 1/2 BPS supermultiplet decomposes into one $\mathcal{N}=6$ spin two 1/2 BPS multiplet and two spin three half 1/2 BPS multiplets, so we understand that the two spin three half 1/2 BPS supermultiplets are in the single `twisted sector' for $K=2$, and each spin three half 1/2 BPS multiplet is in one of the two (conjugate) `twisted sectors' for $K=3$.

The character valued partition function \eqref{PartitionHelicity} is not protected in that it receives contribution from all kinds of supermultiplets. Yet the helicity supertraces
\be
B^{\mathds{Z}_K}_{2n} = \int_{-\frac12}^{\frac12} \hspace{-2mm} d\tau_1 \; {\cal B}^{\mathds{Z}_K}_{2n} =  \int_{-\frac12}^{\frac12} \hspace{-2mm} d\tau_1 \; ({{\cal H}}+\bar{{{\cal H}}})^{2n}{\cal Z}^{\mathds{Z}_K}_{D=4}(v,\bar{v})\vert_{v=\bar{v}=0}  \; , 
\ee
with $n\le 5$ are protected in that they only receive contribution from short BPS multiplets.  

So far as the perturbative threshold contributions are concerned, due to the splitting between Left- and Right-movers on the word-sheet, we are interested in the decomposition of BPS representations of (space-time) ${\cal N}=6$ supersymmetry into $({\cal N}_L,\mathcal{N}_R)=(2,4)$. To this end we introduce the notations 
\bea \langle ({{\cal H}}+\bar{{{\cal H}}})^{2n} \rangle_{u} &=&   ({{\cal H}}+\bar{{{\cal H}}})^{2n} \Biggl(4 \sin^2(\tfrac{\pi}{K})  \frac{  \xi(v){\vartheta_1(\frac{v}{2})^2 \vartheta_1(\frac{v}{2}{+}u)\vartheta_1(\frac{v}{2}{-}u)
}}{  \vartheta_1(u)^2 \eta(\tau)^6} \overline{ \frac{\xi({v})\vartheta_1(\frac{v}{2})^4}{\eta(\tau)^{12}} } \Biggr) \Bigg|_{v=\bar{v}=0} \; , \CR
\langle {{\cal H}}^{2n} \rangle^{u}_L &=&   {{\cal H}}^{2n}  \Biggl(4 \sin^2(\tfrac{\pi}{K})  \frac{  \xi(v){\vartheta_1(\frac{v}{2})^2 \vartheta_1(\frac{v}{2}{+}u)\vartheta_1(\frac{v}{2}{-}u)
}}{  \vartheta_1(u)^2 \eta(\tau)^6}  \Biggr) \Bigg|_{v=0} \; , \CR
 \langle \bar{{{\cal H}}}^{2n} \rangle^{u}_R &=&   \bar{{{\cal H}}}^{2n} \Biggl( \overline{ \frac{\xi({v})\vartheta_1(\frac{v}{2})^4}{\eta(\tau)^{12}} } \Biggr) \Bigg|_{\bar{v}=0}\; ,\eea 
where $u=(r-s\tau)/K$ accounts for the sector and  
 \be {\cal B}^{\mathds{Z}_K}_{2n} = \langle ({{\cal H}}+\bar{{{\cal H}}})^{2n} \rangle_{\! \frac{1}{K}} \Lambda_{\mathds{F}_2(K)}- \frac{1}{K^2} \sum_{r=0}^{K-1}  \langle ({{\cal H}}+\bar{{{\cal H}}})^{2n} \rangle_{\! \frac{\tau + r }{K}} \Lambda_{\mathds{F}_2(K)^*}[e^{\pi i r  {Q}^2}] \; .  \label{BfromH}  \ee
In particular, $\langle {{\cal H}}^{0} \rangle^{u}_L=0$  and $ \langle \bar{{{\cal H}}}^{0} \rangle^{u}_R=\langle \bar{{{\cal H}}}^{2} \rangle^{u}_R=0$ and one obtains that 
\be
{\cal B}^{\mathds{Z}_K}_{0}  ={\cal B}^{\mathds{Z}_K}_{2}  ={\cal B}^{\mathds{Z}_K}_{4} = 0\; . 
\ee
The first non-vanishing contribution is 1/2 BPS protected with 
\be
 \langle ({{\cal H}}+\bar{{{\cal H}}})^{6}\rangle_{u}  =15  \langle  {{\cal H}}^{2}\rangle^{u}_L \langle \bar{{{\cal H}}}^4 \rangle^{u}_R = {45\over 4}\times 4\sin^2(\tfrac{\pi}{K}) \; , 
\ee
and preserves $({\cal N}_L,\mathcal{N}_R)=(1,2)$ supersymmetry. The 1/3 BPS protected contribution 
\bea
 \langle ({{\cal H}}+\bar{{{\cal H}}})^{8}\rangle_{u}  &=& 28 \langle  {{\cal H}}^{2} \rangle_L^{u} \langle \bar{{{\cal H}}}^6 \rangle_R^{u} +  70 \langle  {{\cal H}}^{4} \rangle_L^{u} \langle \bar{{{\cal H}}}^4\rangle^{u}_R \CR 
 &=& 28 \times 4 \sin^2(\tfrac{\pi}{K}) \frac{1}{2}\times  \frac{15}{8} + 70 \times 4 \sin^2(\tfrac{\pi}{K}) \Bigl( \frac{3 \wp(u)}{8 \pi^2} + \frac14\Bigr) \times \frac{3}{2} \CR
& =& {315 \over 2} \times 4 \sin^2(\tfrac{\pi}{K})  \Bigl( \frac{ \wp(u)}{4\pi^2}  + \frac{1}{3} \Bigr) 
\eea
preserves $({\cal N}_L,\mathcal{N}_R)=(0,2)$ supersymmetry. The insertion of four Left-moving currents and four Right-moving currents gives  the 1/3 BPS ${\cal R}^4$ coupling, and indeed we will find that the holomorphic modular form $\wp(\frac{1}{K}) $ determines this coupling in the low-energy effective action. 

The 1/6 BPS protected contribution is
\bea \langle ({{\cal H}}+\bar{{{\cal H}}})^{10}\rangle_{u}  &=&45  \langle  {{\cal H}}^{2}\rangle^{u}_L \langle \bar{{{\cal H}}}^8 \rangle^{u}_R + 210 \langle {{\cal H}}^{4}\rangle_L^{u} \langle \bar{{{\cal H}}}^6 \rangle^{u}_R  + 210\langle  {{\cal H}}^{6}\rangle_L^{u} \langle \bar{{{\cal H}}}^4\rangle^{u}_R  \CR
&=& 45\times  4 \sin^2(\tfrac{\pi}{K}) {1\over 2} \times \frac{21}{32} ( \overline{E_4} + 2) +210\times 4 \sin^2(\tfrac{\pi}{K}) \Bigl( \frac{3 \wp(u)}{8 \pi^2} + \frac14\Bigr) \times  \frac{15}{8} \CR
&& \hspace{10mm} + 210 \times  4 \sin^2(\tfrac{\pi}{K})\frac{3}{32} \Bigl( E_4 +\frac{5 \wp(u)}{\pi^2} + 1\Bigr)   \times \frac{3}{2}  \CR
&=& \frac{945}{64} \times 4 \sin^2(\tfrac{\pi}{K})\Bigl( \overline{E_4}+ 2E_4 + \frac{20\wp(u)}{\pi^2} + \frac{32}{3} \Bigr) \eea
where the first (anti-holomorphic) term preserves $({\cal N}_L,\mathcal{N}_R)=(1,0)$ supersymmetry, while the two other (holomorphic) terms preserve $({\cal N}_L,\mathcal{N}_R)=(0,2)$ supersymmetry. 

Substituting these expressions in \eqref{BfromH} one obtains the helicity supertraces 
\bea  \label{HelicityBn}  B^{\mathds{Z}_K}_6 &=& \frac{45}{4} \times 4 \sin^2(\tfrac \pi K) \Biggl( \int_{-\frac12}^{\frac12}\hspace{-2mm}d\tau_1\, \Lambda_{\mathds{F}_2(K)} -\frac{1}{K^2}  \int_{-\frac K2}^{\frac K2}\hspace{-2mm}d\tau_1  \, \Lambda_{\mathds{F}_2(K)^*}\Biggr)\\
B^{\mathds{Z}_K}_8 &=& \frac{315}{8} \times 4 \sin^2(\tfrac \pi K)\Biggl( \int_{-\frac12}^{\frac12}\hspace{-2mm}d\tau_1\, \Lambda_{\mathds{F}_2(K)}\Bigl( \frac{4}{3}+ \frac{\wp(\frac{1}{K})}{\pi^2} \Bigr)\CR
&& \hspace{40mm} -\frac{1}{K^2}  \int_{-\frac K2}^{\frac K2}\hspace{-2mm}d\tau_1  \, \Lambda_{\mathds{F}_2(K)^*}\Bigl( \frac{4}{3}+ \frac{\wp(\frac{ \tau}{K})}{\pi^2} \Bigr)\Biggr) \CR
B^{\mathds{Z}_K}_{10} &=& \frac{945}{64}\times 4 \sin^2(\tfrac \pi K)\Biggl[ \int_{-\frac12}^{\frac12}\hspace{-2mm}d\tau_1   \, \Lambda_{\mathds{F}_2(K)} \Biggl(\frac{32}{3}\! +\! E_4(-\bar \tau) \!+\! 2  E_4(\tau) \! +\!  \frac{20\wp(\frac{1}{K})}{\pi^2} \Biggr) \CR
&& \hspace{20mm} -\frac{1}{K^2}  \int_{-\frac K2}^{\frac K2}\hspace{-2mm}d\tau_1  \, \Lambda_{\mathds{F}_2(K)^*} \Biggl(\frac{32}{3}\! +\! E_4(-\bar \tau) \!+\! 2  E_4(\tau) \! +\!  \frac{20\wp(\frac{\tau}{K})}{\pi^2} \Biggr)  \Biggr] 
\; . \nn \eea
To interpret these helicity supertraces as counting BPS supermultiplets up to sign, it is useful to decompose them as a sum over supermultiplets using \eqref{PartSMult}
\bea B^{\mathds{Z}_K}_6 &=&-\frac{45}{4} \sum_{j\in \mathds{N}/2} (-1)^{2j} (2j+1) \sum_{\psi \in \mathcal{E}_j^{\scalebox{0.5}{$1/2$}} } e^{- \pi \tau_2\, \alpha^\prime \! M_\psi^2}  \; , \\
B^{\mathds{Z}_K}_8 &=&\frac{315}{8} \sum_{j\in \mathds{N}/2} (-1)^{2j} (2j+1) \Biggl( - \frac{2(2j+1)^2+1}{3}\sum_{\psi \in \mathcal{E}_j^{\scalebox{0.5}{$1/2$}} } e^{- \pi \tau_2\, \alpha^\prime \! M_\psi^2}  +4\sum_{\psi \in \mathcal{E}_j^{\scalebox{0.5}{$1/3$}} } e^{- \pi \tau_2\, \alpha^\prime \! M_\psi^2}    \Biggr)  \; , \CR
B^{\mathds{Z}_K}_{10} &=&\frac{945}{64} \sum_{j\in \mathds{N}/2} (-1)^{2j} (2j+1)  \Biggl( - \frac{6(2j+1)^4+10(2j+1)^2+5}{3} \sum_{\psi \in \mathcal{E}_j^{\scalebox{0.5}{$1/2$}} } e^{- \pi \tau_2\, \alpha^\prime \! M_\psi^2}   \CR
&& \hspace{45mm} +40 \bigl((2j+1)^2+1\bigr)\sum_{\psi \in \mathcal{E}_j^{\scalebox{0.5}{$1/3$}} } e^{- \pi \tau_2\, \alpha^\prime \! M_\psi^2}   -240 \sum_{\psi \in \mathcal{E}_j^{\scalebox{0.5}{$1/6$}} } e^{- \pi \tau_2\, \alpha^\prime \! M_\psi^2}  \Biggr)  \;  . \nonumber \label{BSMult} 
\eea
The explicit formulae one obtains from these equations are displayed in Appendix \ref{HelicitySupertraces23}. 
We will find that $\mathcal{B}^{\mathds{Z}_K}_6$ is associated to the 1/2 BPS protected $\mathcal{F}^2 \mathcal{R}^2$ coupling, $\mathcal{B}^{\mathds{Z}_K}_8$  to the 1/3 BPS protected $\mathcal{R}^4$ coupling, and $\mathcal{B}^{\mathds{Z}_K}_{10}$  to the 1/6 BPS protected ${D}^2 \mathcal{R}^4$ coupling.

\subsection{Preliminaries on U-duality}

As suggested by the coset structure $G/K$ of the scalar manifolds, ${\cal N}=6$ supergravity admits global non-compact symmetries $G$ acting linearly on the vector (field-strengths) and non-linearly on the scalars. As usual the fermions transform under R-symmetry (isotropy group $K$) that acts by local transformations on the coset representative $\mathcal{V} \in G$. More specifically for $D=4$, one has the coset representative $( \cV_A, \cV_{ij A} , \bar{\cV}^{kl}{}_A, \bar \cV_A)$  in the Majorana--Weyl representation of $Spin^*(12)$ that transforms as
\be \cV_A(x) \rightarrow \det \! k(x)\, \cV_B(x) g^B{}_A \; , \qquad \cV_{ij A}(x) \rightarrow k_i{}^k(x) k_j{}^l(x) \cV_{kl B}(x) g^B{}_A \; , \ee
with $g^B{}_A \in Spin^*(12)$ a constant $32\times 32$ real matrix and $k_i{}^j(x) \in U(6)$ a $6\times 6$ complex matrix. The sixteen vector fields field strengths  and their dual define thirty-two field strengths transforming as
\be
F_{\mu\nu}^A(x)\rightarrow g^{-1 A}{}_B F_{\mu\nu}^B(x) \; , 
\ee
while the Weyl fermions transform under $U(6)$ as
\bea \psi_{\mu \alpha i}(x) &\rightarrow& k_i{}^j(x) \psi_{\mu \alpha j}(x)\; , \CR
 \chi_{\alpha ijk}(x) &\rightarrow& k_i{}^l(x) k_j{}^p(x) k_k{}^q(x) \chi_{\alpha lpq}(x)\; , \quad \lambda_{\alpha i}(x) \rightarrow \det \! k(x)^{-1} k_i{}^j(x) \lambda_{\alpha j}(x) \; . \eea
The $Spin^*(12)$ invariant symplectic form $\omega_{AB}$ is normalised such that 
\bea \omega_{AB} &=& i \bar{\mathcal{V}}_A \mathcal{V}_B   + \frac{i}{2} \cV_{ij A} \bar{\cV}^{ij}{}_B - \frac{i}{2} \bar{\cV}^{ij}{}_A \cV_{ij B} - i \cV_A \bar{\cV}_B \; , \\
\omega^{AB} \cV_A \bar \cV_B &=& i \; ,  \quad \omega^{AB} \cV_A \bar{\cV}^{ij}{}_B = 0 \; , \quad  \omega^{AB} \cV_{ij A} \bar{\cV}^{kl}{}_B = - 2i \delta_{ij}^{kl} \; , \quad \omega^{AB} \cV_{ij A} {\cV}_{kl B} = 0 \; . \nonumber\eea
One decomposes as usual the Maurer-Cartan form into $Q_H\in \mathfrak{u}(6)$ and $P$ in the coset component such that 
\bea d \cV_A &=& (Q_H)^k{}_k \cV + \frac12 P^{ij} \cV_{ij A} \CR
d \cV_{ij A} &=& - 2 (Q_H)^k{}_{[i} \cV_{j]k A} + \frac14 \varepsilon_{ijklpq} P^{kl} \bar{\cV}^{pq}{}_A + \bar{P}_{ij} \bar{\cV}_A \label{QP}\eea
and $P^{ij}$ is holomorphic in the fifteen complex scalar fields $T^M$, i.e. 
\be P^{ij} = P_M{}^{ij} dT^M \; , \qquad \bar{P}_{ij} = P_{\bar{M} ij} d \bar T^{\bar M}\; . \ee

Due to charge quantisation and to the absence of continuous symmetries in (super)gravity, only a discrete U-duality subgroup of $SO^*(12)$ may survive as an exact symmetry of the quantum theory, such as string theory. The U-duality group is defined as the non-perturbative duality symmetry of string theory and the T-duality group as the perturbative symmetry. In this paper we determine the T-duality group as the symmetry of the four-point one-loop amplitudes and of the character valued partition function that counts states for a given charge in the Narain lattice $\mathds{F}_2(K)^* $.  The T-duality group defined in this way does not preserve the untwisted sector, and mixes twisted states with untwisted states with charge $\mathcal{Q}\in \mathds{F}_2(K)^* \smallsetminus \mathds{F}_2(K)$. It is possible that the true T-duality group is restricted to the automorphisms of the perturbative lattice preserving the untwisted sector, if for example the scattering of massive twisted states differs from the scattering of unwtisted states with a charge in $\mathds{F}_2(K)^* \smallsetminus \mathds{F}_2(K)$. We may nevertheless conjecture that this is indeed the T-duality group, i.e.  the  symmetry of the complete perturbative theory, and for simplicity we shall refer to the group of automorphisms of $\mathds{F}_2(K)$ as the T-duality group in the following.

We shall start with the five-dimensional theory, with global non-compact symmetry $SU^*(6)$.  The computation of \eqref{calZ23} applies directly to five dimensions with $\mathds{F}_2(K)$ replaced by 
\be \mathds{M}_2(K) = \sLambda_{1,1}[ K]\oplus \mathds{H}(K) \; . \ee
The perturbative states with bosonic zero modes $ Q\in \mathds{M}_2(K) $ are untwisted and the other states with $Q\in  \mathds{M}_2(K)^* \smallsetminus  \mathds{M}_2(K) $ contribute in the same way to the  the four-point one-loop amplitudes and of the character valued partition function. The T-duality group in five dimensions as defined above is therefore the group of automorphisms of the lattice $\mathds{M}_2(K) $. To determine this group of automorphisms it is convenient to define $\mathds{M}_2(K) $ as the set of Hermitian two by two matrices over the integral quaternions $\mathds{H}(K)$  with a top left component $\tilde{m}=0$ mod $K$, i.e.
\be Q = \left( \begin{array}{cc} \tilde{m}\,  &\,  q\\ q^* & \tilde{n} \end{array}\right) \in \mathds{M}_2(K) \qquad \mbox{with}\;  \tilde{n}\in \mathds{Z}\, ,\;  \tilde{m}\in K\mathds{Z}\, , \;  q\in \mathds{H}(K)\; .  \ee
Then $Q^2 = 2 \det Q$ and the action of $SO(1,5)$ on this charge is defined as 
\be Q\rightarrow \gamma^\dagger Q \gamma \; , \ee
where $\gamma = (^a_c{}^b_d)$ is a two by two matrix over the quaternions  of unit modulus determinant  $\bigl| a  d -  a c  a^{-1} b  \bigr|=1$. The condition that the lattice $\mathds{M}_2(K) $ is preserved implies that $a,d \in \mathds{H}(K)$, $b \in \frac{1}{\alpha} \mathds{H}(K)$ and $c \in \alpha \mathds{H}(K)$ where $\alpha$ is a quaternion of norm square $|\alpha|^2 = K$ in $\mathds{H}(K)$. For both $K=2,3$, $\alpha$ is defined uniquely up to left multiplication by a unit $u\in \mathds{H}(K)$ (with $|u|^2 =1$). $\alpha \mathds{H}(K)$ is therefore a two-sided ideal in $\mathds{H}(K)$ and $\mathds{H}(K)^* = \frac{1}{\alpha} \mathds{H}(K)$. We denote this arithmetic group $\Cong{A}{3}_{0*}(\alpha)$ and will refer to it as the  T-duality group of the $K=2$, $3$ asymmetric orbifold theories in five dimensions.\footnote{It is related to $SL(2,\mathds{H}(K))$ by a similarity transformation, but the notation indicates that it is generated from $\Cong{A}{3}_{0}(\alpha)\subset SL(2,\mathds{H}(K))$ with $c=0$ mod $\alpha$ and $\gamma_{\rm F} = (^0_\alpha{}^{1\!/\!\alpha}_{\; 0})$.}

In four dimensions the group of automorphisms of $\mathds{F}_2(K)$ is defined similarly (up to a similarity transformation) as the subgroup $\Cong{D}{4}_{0*}(\alpha)$ of symplectic matrices $(^A_C{}^B_D)$ with $A,B,C,D$ two by two rational matrices over $\mathds{H}(K)$ such that 
\be D^\dagger A - B^\dagger C = \mathds{1}\; , \qquad D^\dagger B = B^\dagger D\; , \qquad A^\dagger C = C^\dagger A \; , \ee
that is generated by the group $\Cong{D}{4}_{0}(\alpha) \subset Sp(4,\mathds{H}(K))$ of integral matrices $(^A_C{}^B_D)$ with $C = 0 $ mod $\alpha$ and the symplectic group element 
  \be \gamma_{\rm F} =  \left(\begin{array}{cccc} 0\, &\, 0\,  & - \frac{1}{\alpha^*}  \, & \, 0 \\ 0\, &\, 0 \, &\, 0 \,  &- \frac{1}{\alpha^*}   \\\;  \alpha  \, & \, 0\,  &\,  0\, &\, 0 \\ 0 \, &\, \alpha  \, & \, 0\,  & \, 0 \, \end{array}\right) \label{FrickeA} \; . \ee 
  This additional generator $\gamma_{\rm F}$ is not an element of the original T-duality group $O(6,6,\mathds{Z}) \supset Sp(4,\mathds{H}(K))$ prior to the orbifold and is similar to the Fricke duality introduced in \cite{Persson:2015jka}. 
We shall refer accordingly to $\Cong{D}{4}_{0*}(\alpha)$ as the T-duality group of the $K=2$, $3$ asymmetric orbifold theories in four dimensions.

We will find strong evidence in \cite{GBinprep} that the U-duality group in five dimensions is defined similarly as $\Cong{A}{5}_{0*}(\alpha)$: the group preserving $\alpha \mathds{H}(K)\oplus \alpha \mathds{H}(K) \oplus \mathds{H}(K)$ by left action, and which is related to $SL(3,\mathds{H}(K))$ by similarity transformation. The U-duality group in four dimensions will be conjectured to be $\Cong{D}{6}_{0*}(\alpha)$, the group generated by $\Cong{D}{6}_{0}(\alpha)\subset Sp(6,\mathds{H}(K))$ and $\gamma_{\rm F} \in \Cong{D}{4}_{0*}(\alpha)\subset \Cong{D}{6}_{0*}(\alpha)$, where $(^A_C{}^B_D) \in \Cong{D}{6}_{0}(\alpha)$ is such that $C = 0 $ mod $\alpha$ as a three by three matrix. This is the group of automorphisms of the 32-dimensional lattice 
\be \mathds{F}_3(K) = K  \mathds{F}_2(K)^* \oplus [ \mathds{H}(K) \oplus\mathds{H}(K) \oplus  \alpha \mathds{H}(K) \oplus  \alpha \mathds{H}(K) ]\oplus  \mathds{F}_2(K) \ee
in the Majorana--Weyl spinor representation of $Spin^*(12)$ of `untwisted states' in the sense explained below \eqref{LatticeUntwisted}. We will find strong evidence that one-half BPS states are associated to rank one charges $\Gamma$ in 
\be \mathds{F}^*_3(K) =   \mathds{F}_2(K) \oplus [\tfrac{1}{\alpha}  \mathds{H}(K) \oplus \tfrac{1}{\alpha}  \mathds{H}(K) \oplus   \mathds{H}(K)   \oplus  \mathds{H}(K) ]\oplus  \mathds{F}^*_2(K) \ee
for which the projection $\Gamma \times \Gamma$ to the adjoint representation of $\mathfrak{so}^*(12)$ vanishes. For each $\Gamma \in  \mathds{F}_3(K)$ there is one spin two supermultiplet and for each $\Gamma \in \mathds{F}^*_3(K)\smallsetminus  \mathds{F}_3(K) $ there are $\frac{2}{K-1}$ spin three half supermultiplets.

One may then consider the decomposition of the spectrum under T-duality and U-duality orbits. Perturbative one-half BPS states have a charge $\mathcal{Q}\in \mathds{F}_2^*$ of vanishing norm. We will show in Appendix \ref{OrbitsSp4}  that such charges are either in the $\Cong{D}{4}_{0*}(\alpha)$ orbit of a charge with only non-zero component the momentum $\tilde{m} = {\rm gcd}_{\mathds{F}^*_2}(\mathcal{Q})$ along the twisted circle  or with only non-zero component the momentum $m= {\rm gcd}_{\mathds{F}_2}(\mathcal{Q})$ along the additional circle. A charge associated to a spin 3/2 supermultiplet in $\mathds{F}_2^*\smallsetminus \mathds{F}_2$ is therefore in the $\Cong{D}{4}_{0*}(\alpha)$ orbit of a Kaluza--Klein mumentum charge along the twisted circle, with mass 
\be \sqrt{\alpha^\prime} M(\mathcal{Q}) = \frac{|\tilde{m}|}{\tilde{\RS}}\; .  \ee
For $\tilde{\RS}>\hspace{-1.8mm}>1$ there are $\frac{4}{K-1}$ gravitino supermultiplets with the minimal mass ${1}/{\tilde{\RS}}$ for $\tilde{m}=\pm1$. The two derivative low energy effective theory truncated to the massless graviton supermultiplet and two such gravitino supermultiplets  should be the corresponding Cremmer--Scherk--Schwarz 
 gauging of $\mathcal{N}=8$ supergravity \cite{Cremmer:1979uq}. This  gauged supergravity theory may be a consistent truncation, but it is not a good effective theory since there are also massive spin two supermuliplets for each Kaluza--Klein momentum of mass $\frac{|\tilde{m}|}{\tilde{\RS}}$ with $\tilde{m}=0$ mod $K$ and massive spin three-half supermuliplets for each Kaluza--Klein momentum of mass $\frac{|\tilde{m}|}{\tilde{\RS}}$ with $\tilde{m}\ne 0$ mod $K$ that are of the same order. For $K=2$ there are moreover four gravitino supermultiplets with the same minimal mass. The analysis for the non-perturbative one-half BPS states is very similar, and all the rank one charges $\Gamma \in \mathds{F}^*_3(K)\smallsetminus  \mathds{F}_3(K) $ are in the   $\Cong{D}{6}_{0*}(\alpha)$ orbit of a perturbative Kaluza--Klein charge along the twisted circle.

%%%%%%%%%%%%%%%%%%%%%%%%
%%%%%%%%%%%%%%%%%%%%%%%%
\section{Low-energy effective action and threshold corrections}
\label{SusyLEEA}

Different $\cN=6$ realisations such as the $K=2$ and $K=3$ asymmetric orbifold theories considered in this paper give rise to different low-energy effective actions, notwithstanding the uniqueness of the 2-derivative action. Comparison is easier for special couplings that are BPS-saturated  and whose dependence on the scalar fields is tightly constrained by differential equations and U-duality. Once again the prototypical example is the $\mathcal{R}^4$ term in Type IIB, whose threshold function is the real analytic Eisenstein series  $E_{\frac32,0}(S)$ of the $SL(2,\mathds{Z})$ S-duality group \cite{Green:1997tv}. Supersymmetry constrains this threshold function to be an eigen-function of the Laplace operator on the upper-complex half plane with eigenvalue $\frac{3}{4}$ \cite{Green:1998by}, which together with $SL(2,\mathds{Z})$ invariance determines the function uniquely up to normalisation \cite{Pioline:1998mn}. The maximal supergravity $\mathcal{R}^4$  threshold function is more generally understood in dimension $D\ge 3$ to be associated to the minimal automorphic representation of the (split real form) exceptional group $E_{11-D}$ \cite{Kazhdan:2001nx,Green:2010kv}, which is unique for $D\le 6$. This analysis extends to higher derivative couplings and the ${D}^4 \mathcal{R}^4$ threshold function is associated to the next to minimal automorphic representation \cite{Green:2011vz}. The fact that these threshold functions are attached to small automorphic representations follows from differential equations implied by supersymmetry  \cite{Bossard:2014lra,Bossard:2014aea,Bossard:2015uga}, which can be derived in four dimensions by classifying all the linearised supersymmetry invariants using the techniques developed in  \cite{Drummond:2003ex,Drummond:2010fp}.  In this section we extend this analysis to the BPS protected couplings of $\cN=6$ supergravity.

%%%%%%%%%%%%%%%%%%%%%%%%
\subsection{Supersymmetry constraints}
The field content of $D=4$ $\cN=6$ supergravity includes fifteen complex scalar fields $T^M$ parametrising the special K\"{a}hler symmetric space $SO^*(12)/U(6)$, where $U(6)$ is the R-symmetry group, the metric field $g_{\mu\nu}$ of Weyl tensor $C_{\alpha\beta\gamma\delta}$,  six Rarita--Schwinger fields $\psi_{\mu \alpha i }$ of field strength $\rho_{\alpha\beta\gamma i}$ in the fundamental of $U(6)$, sixteen vector fields and their Gaillard--Zumino duals transforming together as a 32-dimensional Majorana--Weyl spinor $A_\mu^A$ of $Spin^*(12)$, and twenty plus six Weyl spinors $\chi_{\alpha ijk}$ and $\lambda_\alpha^i$. One defines as usual the scalar dressed Gaillard--Zumino field strengths 
\be F_{\alpha\beta ij} =\frac12  \sigma^{ab}{}_{\alpha\beta} e_a{}^\mu e_b{}^\nu \cV_{ij A} F_{\mu\nu}^A \; , \qquad F_{\alpha\beta} = \frac12  \sigma^{ab}{}_{\alpha\beta} e_a{}^\mu e_b{}^\nu \bar{\cV}_{ A} F_{\mu\nu}^A\; , \ee
where $\cV(T,\bar T) = (  \bar{\cV}_{ A} ,  \bar{\cV}^{ij}{}_{ A},  \cV_{ij A} , \cV_A)$ define the coset representative in the $Spin^*(12)$ Majorana-Weyl spinor representation. The $\cN=6$ supermultiplet decomposes according to the increasing $U(1)$ weight \footnote{Here we call `weight' the charge with respect to $U(1)\subset U(6)$ R-symmetry / isotropy group.} from $w=0,\frac16,\frac{1}{3},\frac{1}{2},\frac{2}{3},\frac{5}{6}, 1$ as
\be \label{LinearisedFields} C_{\alpha\beta\gamma\delta}\; , \quad \rho_{\alpha\beta\gamma i}\; , \quad F_{\alpha\beta ij} \; , \quad \chi_{\alpha ijk} \; , \quad P_{M}{}^{ij} dT^M \; , \quad \bar{ \lambda}_{\dot{\alpha}}^i \; , \quad \bar F_{\dot{\alpha}\dot{\beta}} \; , \ee
and its CPT conjugate. In the linearised approximation, one can identify the linearised complex superfield $W^{ij}$
\be P_{M}{}^{ij} dT^M \approx d W^{ij} \; , \ee
which defines the ultra-short supersymmetry representation determined by the action of the superspace derivative as \cite{Howe:1981gz}
\be \begin{split}  D_\a^i  W^{jk} &=  \frac{1}{2} \varepsilon^{ijkpqr} \chi_{\a pqr}  \; , \\
D_\a^i  \chi_{\beta jkl} &= 3 \delta^i_{[j} F_{\alpha\beta kl]}  \; , \\
D_\a^i F_{\beta\gamma jk} &= 2\delta^i_{[j} \rho_{\alpha\beta\gamma k]} \; , \\
D_\a^i \rho_{\beta\gamma\delta j} &= \delta^i_{j} C_{\alpha\beta\gamma\delta}  \; .
\end{split}
\hspace{10mm}
\begin{split}
 \bar{D}_{\dot{\a} i} W^{jk} &= \delta_i^{[j} \bar \lambda_{\dot{\a}}^{k]}\; ,\\
  \bar{D}_{\dot{\a} i} \bar \lambda_{\dot{\beta}}^{j}  &= \delta_i^{j} \bar F_{\dot{\alpha}\dot{\beta}}\; , \\
  &\\
  &
 \end{split}\ee
We will need to define differential operators on the symmetric space $SO^*(12)/U(6)$. The moduli space metric is by construction
\be
2 G_{M \bar N} dT^M d\bar{T}^{\bar N} = \tr_{{\scriptscriptstyle \bf 12}} P \bar{P}  =\frac14   \tr_{{\scriptscriptstyle \bf 32}} P \bar{P}\; ,
 \ee
with components
\be G_{M \bar N}  =P_M{}^{ij}   \bar P_{\bar N ij}  \ee
and we define the covariant derivative
\be \cD_{ij} = \bar P_{\bar N ij} G^{M \bar N}  ( \partial_M  + Q_M) \ ,  \qquad \bar \cD^{ij} =  P_{M}{}^{ij}  G^{M\bar N} ( \bar{\partial}_{\bar M}  + Q_{\bar M}) \ , \label{Dso12} \ee
where $Q_{M}{}^i{}_j$ defined in \eqref{QP} acts on $U(6)$ tensors in the corresponding representation. These covariant derivatives acting on a $U(6)$ scalar function $f(T,\bar T)$ satisfy the commutation relation 
\be [ \cD_{ij} , \bar \cD^{kl} ] \cD_{pq} \, f(T,\bar T) = 2 \delta_{j][p}^{kl} \cD_{q][i}\,  f(T,\bar T) \ .  \label{ComSO12} \ee
The superspace derivative of a function of the scalars gives 
\be D_\a^i \cE(T,\bar T) = \Scal{   \frac{1}{2} \varepsilon^{ijkpqr} \chi_{\a pqr} \cD_{jk} + \lambda_{\a \, j} \bar \cD^{ij} }\cE(T,\bar T) \; .  \ee

One  classifies the BPS protected couplings using superconformal representations of $SU(2,2|6)$ according to \cite{Drummond:2003ex}. The superfield $W^{ij}$ defines an ultra-short representation of the superconformal algebra  $\mathfrak{su}(2,2|6)$ with the dilatation weight $L[W^{ij}] = 1$ and the R-symmetry weight $R[W^{ij}]=- \frac13$, with the convention that $R[D_\alpha^i]=- \frac16$ and $L[D_\alpha^i]=\frac12$ \cite{Dobrev:1985qv}.\footnote{The $SU(2,2|6)$ R-charge is not the $U(1)$ R-charge of the super-Poincar\'e algebra, but we have instead the $U(1)$ weight $w= -2 R-h$ where $h$ is the helicity of the field.} The BPS protected couplings are the top components of short multiplets, which must be Lorentz invariant but not necessarily $U(6)$ invariants. The polynomials in $W^{ij}$ defining short multiplets with a Lorentz invariant top component are represented in table \ref{Primaries}.
\begin{table}[h!]
\centering
\begin{tabular}{c|c|c|c|c}
BPS degree  & Polynomial & $SU(6)$ & $L$ & $R$  \\[2mm]
\hline\hline && && 
\\[-3mm]
$(2,4)$ &  $W^n$, $n\ge 4$ &$ [0,n,0,0,0] $& $n$ & $- \frac{n}{3}$ \\[2mm]
\hline&& && 
\\[-3mm]
$(2,2) $ & $W^{n} \bar W^{m}$, $n\ge 2$, $m\ge 2$ & $[0,n,0,m,0] $& $n+m$ & $ \frac{m-n}{3} $ \\[2mm]
\hline&& && 
\\[-3mm]
$(1,1) $ & $W^{n+k} \bar W^{m+k} $, $k\ge 2$, & $[k,n,0,m,k] $& $n+m+2k$ & $ \frac{m-n}{3} $\\[2mm]
\hline&& && 
\\[-3mm]
$(2,0)$ & $W^{n} \bar W^{m+2k} $, $n \ge 2$, $k\ge 2$ & $[0,n+k,0,m,0] $& $n+m+2k$ & $ \frac{m+2k-n}{3} $
\end{tabular}
\caption{\label{Primaries}{\small Short multiplets from polynomials in the superfield $W$.}}
\end{table}
A  BPS multiplet has $(p,q)$ degree if the chiral primary is annihilated by $D_\alpha^i$ for $i=1$ to $p$ and $\bar D_{\alpha j}$ for $j=7-q$ to $6$, preserving therefore a fraction $\frac{p+q}{12}$ of the supersymmetries. The $(p,q)$ BPS short multiplets are conveniently represented in harmonic superspace using the harmonic variables in $U(6) / ( U(p) \times U(6-p-q) \times U(q))$ \cite{Ferrara:1999zg}. The $(2,2)$ BPS primary with $n=m=2$ defines the $\mathcal{R}^4$ linearised invariant while the $(1,1)$ primary with $n=m=0$ and $k=2$ defines the ${D}^2 \mathcal{R}^4$ linearised invariant  \cite{Bossard:2010bd}. The representations with $m$, $n$, $k$ generic allow to determine the differential equations satisfied by the threshold function as in \cite{Bossard:2015uga}.

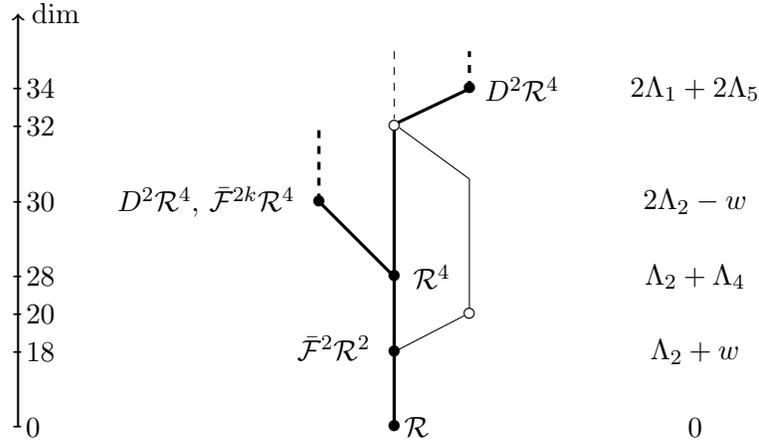
\begin{figure}[htbp]
\begin{center}
 \begin{tikzpicture}
 \draw (\xmin,\ymin) node{\textbullet};
 \draw (\xmin,\ymin - 1) node{\textbullet};
  \draw (\xmin,\ymin - 2)  node{\textbullet};
  \draw (\xmin,\ymin + 2)  node{$\circ$};
  \draw (\xmin + 1,\ymin + 2.5)  node{\textbullet};
    \draw (\xmin - 1 ,\ymin + 1) node{\textbullet};
     \draw (\xmin+1,\ymin -1/2)  node{$\circ$};
  \draw (\xmin + 0.3,\ymin - 2) node{$ \mathcal{R}$};
  \draw(\xmin + 4,\ymin - 2) node{$ 0 $};
  \draw (\xmin - 0.8,\ymin - 0.95) node{$\bar {\mathcal{F}}^2 \mathcal{R}^2 $};
  \draw (\xmin +4 ,\ymin - 1) node{$\Lambda_2 + w$};
  \draw (\xmin + 0.5,\ymin) node{$ \mathcal{R}^4$};
    \draw (\xmin + 4,\ymin) node{$ \Lambda_2 + \Lambda_4 $};
  \draw (\xmin - 2.5,\ymin + 1) node{$ {D}^2 \mathcal{R}^4,\, \bar{\mathcal{F}}^{2k} \mathcal{R}^4$};
   \draw (\xmin +4 ,\ymin +1) node{$2\Lambda_2-w$};
  \draw (\xmin + 0.7 + 1,\ymin + 2.5) node{$ {D}^2 \mathcal{R}^4$};
    \draw (\xmin + 4,\ymin + 2.5) node{$2 \Lambda_1 +2 \Lambda_5 $};
  \draw[-,draw=black,very thick](\xmin,\ymin) -- (\xmin,\ymin + 1.95 );
   \draw[dashed,draw=black,very thick](\xmin - 1,\ymin + 1) -- (\xmin - 1,\ymin + 2);
    \draw[dashed,draw=black, ](\xmin ,\ymin + 3) -- (\xmin,\ymin + 2.1);
    \draw[,draw=black,very thick](\xmin+0.05,\ymin + 2.05) -- (\xmin + 1,\ymin + 2.5);
    \draw[dashed,draw=black,very thick](\xmin + 1,\ymin + 2.5) -- (\xmin+1,\ymin + 3);
  \draw[-,draw=black,very thick](\xmin,\ymin) -- (\xmin - 1,\ymin + 1);
\draw[-,draw=black,very thick] (\xmin,\ymin - 1) -- (\xmin,\ymin);
\draw[-,draw=black,very thick] (\xmin,\ymin - 2) -- (\xmin,\ymin - 1);
    \draw[-,draw=black,](\xmin + 1-0.07,\ymin -1/2-0.02) -- (\xmin,\ymin -1);
        \draw[-,draw=black,](\xmin +0.07,\ymin +2-0.02) -- (\xmin+1,\ymin +1.3);
        \draw[-,draw=black,](\xmin + 1,\ymin -1/2+0.08) -- (\xmin+1,\ymin+1.3);
\draw[<-,draw=black,thick] (\xmin - 3-2,\ymin + 3.5) -- (\xmin - 3 - 2,\ymin - 2);
\draw (\xmin - 3 + 0.2 - 2,\ymin - 2) node{$0$};
\draw (\xmin - 3- 2,\ymin - 2) node{-};
\draw (\xmin - 3- 2,\ymin - 1) node{-};
\draw (\xmin - 3- 2,\ymin - 1/2) node{-};
\draw (\xmin - 3- 2,\ymin) node{-};
\draw (\xmin - 3- 2,\ymin + 2) node{-};
\draw (\xmin - 3- 2,\ymin + 1) node{-};
\draw (\xmin - 3- 2,\ymin + 2.5) node{-};
\draw (\xmin - 3 + 0.3 - 2,\ymin - 1) node{$18$};
\draw (\xmin - 3 + 0.3 - 2,\ymin-1/2) node{$20$};
\draw (\xmin - 3 + 0.3 - 2,\ymin) node{$28$};
\draw (\xmin - 3 + 0.3 - 2,\ymin + 1) node{$30$};
\draw (\xmin - 3 + 0.3 - 2,\ymin + 2) node{$32$};
\draw (\xmin - 3 + 0.3 - 2,\ymin + 2.5) node{$34$};
\draw (\xmin - 3 + 0.5 - 2,\ymin + 3.5) node{dim};
\end{tikzpicture}
\end{center}
\caption{\small Closure Hasse diagram of the small nilpotent orbits  of $SO(12,\mathds{C})$. The normal triple Dynkin label is given in Bourbaki convention for the fundamental weights $\Lambda_i= A^{-1 ij} \alpha_j$ of $SU(6)$ and $w$ the $U(1)$ weight.}
\label{ClosureDiag}
\end{figure}

We will see that each of these $(p,q)$ BPS degrees corresponds to an automorphic representation of $Spin^*(12)$. Functions and tensors on the symmetric space that are eigen functions of all Casimir operators, {\it i.e.} the three independent invariant operators 
 \bea \Delta &=& 2 \bar \cD^{ij} \cD_{ij}\; , \quad \Delta^{\!\times} = \frac{1}{32} \varepsilon^{ijklpq}\varepsilon_{rstupq} \bar \cD^{rs} \bar \cD^{tu} \cD_{ij} \cD_{kl}\; , \CR
  \Delta^{\! \rm det} &=& \frac{1}{48^2} \varepsilon^{ijklpq}\varepsilon_{rstuvw} \bar \cD^{rs} \bar \cD^{tu} \bar \cD^{vw}  \cD_{ij} \cD_{kl} \cD_{pq} \; ,  \label{Casimirs} \eea 
 for $Spin^*(12)$, are associated to nilpotent orbits of $Spin^*(12)$. To be more precise, the  variety associated to their annihilator in the universal enveloping algebra (realised as differential operators) is the closure of a union of nilpotent orbits \cite{CollingwoodMcGovern}. In particular the differential equations implied by supersymmetry correspond to small nilpotent orbits. In general, a nilpotent orbit of the complex simple Lie algebra $\mf{g}$ is determined by its normal $\mf{sl}_2$ triple such than the semi-simple element $h\in \mf{sl}_2$ is an integral weight that defines a $\mathds{Z}$ grading of the Lie algebra: $\mf{g}=\bigoplus_{n=-n_{\scalebox{0.5}{max}}}^{n_{\scalebox{0.5}{max}}} \mf{g}^\ord{n}$, and the nilpotent element $e\in \mf{sl}_2$ is generic in $\mf{g}^\ord{2}$, {\it i.e.} its $G^\ord{0}$ orbit is open in  $\mf{g}^\ord{2}$ for the Zariski topology.\footnote{Zariski topology is defined in such a way that the closed sets are solutions to algebraic equations. The closure of a nilpotent orbit is the set of Lie algebra elements that satisfy the same algebraic equations, {\it e.g.} they have the same nilpotency degree in all irreducible representation.} The dimension of the nilpotent orbit $G \cdot e$ is ${\rm dim}(\mf{g}^\ord{1})+2\sum_{n=2}^{n_{\scalebox{0.5}{max}}}{\rm dim}( \mf{g}^\ord{n})$ and is always even, because $G \cdot e$ is a symplectic variety. The space of functions / tensors  associated to the nilpotent orbit corresponds to a quantification of the Zariski closure $\overline{G \cdot e}$ of $G \cdot e$, and can be represented in principle by functions supported on the Lagrangian subspace of dimension $d_{\rm GK}=\frac12 {\rm dim}(\mf{g}^\ord{1})+\sum_{n=2}^{n_{\scalebox{0.5}{max}}}{\rm dim}( \mf{g}^\ord{n})$. The tensors  we are interested in for the string theory low-energy effective action are automorphic forms invariant under the U-duality group, and the associated representations are called automorphic representations. One defines the  Gelfand--Kirillov dimension $d_{\rm GK}$ of the associated automorphic representation, which defines the dimension of its Fourier support.  The relation between BPS couplings and nilpotent orbits of $SO^*(12)$ is summarised in Figure \ref{ClosureDiag}, where the white dotes are nilpotent orbits of $SO(12,\mathds{C})$ that do not intersect with $\mf{so}^*(12)$ and are therefore irrelevant. The lines between two nilpotent orbits mean that the smaller orbit is in the Zariski closure of the bigger one \cite{CollingwoodMcGovern}.  The Dynkin label is related to the BPS degree so that $U(p) \times U(6-p-q) \times U(q)$ is the stabiliser of the Cartan representative of the normal triple in $U(6)$ and the $U(1)$ weight coefficient is such that the G-analytic superfields are of weight $2$. 

\subsection{(2,4) BPS $\bar{\mathcal{F}}^2 \mathcal{R}^2$ type invariant}
\label{secF2R2}
To describe the class of supersymmetry invariants obtained from the $(2,4)$ chiral primaries in Table \ref{Primaries}, it is convenient to introduce the associated harmonic superspace, with harmonic variables $u^{\hat{r}}{}_i$ for $\hat{r}=1,2$ and $u^{r}{}_i$ for $r=3,4,5,6$ parametrising  $U(6) / ( U(2) \times U(4))$. The details of this construction can be found in  \cite{Ferrara:1999zg}. In practice the  harmonic variables simply allow to  restrict the set of indices to define the corresponding short multiplets. The G-analytic superfield $W^{12} = u^1{}_i u^2{}_j W^{ij}$ satisfies 
\be D_\a^{\hat{r}} W^{12} = 0 \ , \qquad \bar D_{\adt r} W^{12} = 0 \ . \ee
The chiral primaries are therefore all the harmonics of $(W^{12})^n$, that combine into the irreducible representation $[0,n,0,0,0]$ of $SU(6)$. The top-component can be shown to include the graviphoton--graviton coupling 
\be [D_\alpha^r]^8 [\bar D_{\dot{\alpha} \hat{r}}]^4 (W^{12})^{4+n} \sim (W^{12})^{n} \bar F_{\dot{\alpha}\dot{\beta}} \bar F^{\dot{\alpha}\dot{\beta}} C_{\alpha\beta\gamma\delta} C^{\alpha\beta\gamma\delta} + \dots  \ee
which is the only term with two Weyl tensors. This coupling is therefore determined by the amplitude with two gravitons of polarisation $+2$ and two  graviphotons of polarisation $-1$ (in the convention in which all particles are incoming). Following the same analysis as in \cite{Bossard:2014lra,Bossard:2014aea,Bossard:2015uga}, one deduces that the non-linear supersymmetric density $\cL_\gra{2}{4}[\Theta] $ must be determined by a form $\Theta(T,\bar T)$ of $U(1)$ weight $2$ such that 
\be \cL_\gra{2}{4}[\Theta]  = \cL_\gra{2}{4}^\ord{0} +  \cD_{ij}  \Theta(T) \cL_\gra{2}{4}^{\ord{1} ij} +  \cD_{ij} \cD_{kl}  \Theta(T)  \cL_\gra{2}{4}^{\ord{2} ij,kl} + \dots + \cD_{ij} \cD_{kl} \cdots  \cD_{pq} \Theta(T)  \cL_\gra{2}{4}^{\ord{8} ij,kl,\dots,pq} \; , \label{halfBPSExpand}\ee
 where $\cL_\gra{2}{4}^{\ord{n}}$ are $Spin^*(12)$ invariant densities that do not depend explicitly on the scalar fields $T^M$ and transform in the $[0,n,0,0,0]$ irreducible representation of $SU(6)$. More precisely, $\cL_\gra{2}{4}^{\ord{n}}$ are functions of $\hat{C},\hat{\rho},\hat{F},\hat{P},\chi,\lambda$ and their super-covariant derivatives, as well as the naked gravitino fields that only appear in wedge products. They are determined in the linearised approximation 
by the linearised invariant  as
 \be  [D_\alpha^r]^8 [\bar D_{\dot{\alpha} \hat{r}}]^4 (W^{12})^{4+n} =  (W^{12})^n \cL_{\gra{2}{4} \scalebox{0.6}{lin}}^\ord{0} + n  (W^{12})^{n-1}   \cL_{\gra{2}{4} \scalebox{0.6}{lin}}^{\ord{1} 12} + n(n-1)  (W^{12})^{n-2}   \cL_{\gra{2}{4} \scalebox{0.6}{lin}}^{\ord{2} 12,12} + \dots \ee
 For example 
 \be  \cL_\gra{2}{4}^\ord{0} \sim \hat{\bar F}^2 \hat{C}^2 + \dots \; , \qquad \cL_\gra{2}{4}^{\ord{8}[0,8,0,0,0] } \sim \bar \lambda^4_{[0,2,0,0,0]} \chi^8_{[0,6,0,0,0]} \; . \ee
 More schematically, we shall write \eqref{halfBPSExpand} as
 \be \cL_\gra{2}{4}[\Theta] = \sum_{n=0}^8 \bigl( \cD^n \Theta(T)\bigr)|_{[0,0,0,n,0]} \, \cL_\gra{2}{4}^{[0,n,0,0,0]} \ . \label{halfBPSExpandSchem} \ee
The fact that the supersymmetry invariant expands in this way implies that the derivative expansion of the form $\Theta(T)$ must only involve these specific irreducible representations, {\it i.e.} 
\be \bar \cD^{ij}  \Theta(T) = 0 \ , \qquad \varepsilon^{ijklpq} \cD_{kl} \cD_{pq} \Theta(T) = 0  \ .\label{F2R2}  \ee
Note that $\Theta(T)$ is a form of $U(1)$ weight $2$, so its covariant derivative involves the K\"{a}hler connexion, and $\frac{\Theta(T)}{\det (T-\bar T)}$ is a holomorphic function of $T$. For a form $\Theta_w$ of generic weight  $w$ one has
\be [ \cD_{ij} , \bar \cD^{kl} ] \Theta_w = \frac{w}{2} \delta_{ij}^{kl} \Theta_w\ , \qquad     [ \cD_{ij} , \bar \cD^{kl} ] \cD_{pq} \Theta_w = 2 \delta_{j][p}^{kl} \cD_{q][i}  \Theta_w +\frac{w}{2} \delta_{ij}^{kl} \cD_{pq}\Theta_w  \ , \ee
therefore 
\be \bar \cD^{kl} \cD_{[ij} \cD_{pq]}  \Theta_w - \cD_{[ij} \cD_{pq]}  \bar \cD^{kl}\Theta_w = ( 2-w) \delta^{kl}_{[ij} \cD_{pq]} \Theta_w\ , \ee
and equation \eqref{F2R2} is only consistent for $w=2$. This is indeed the $U(1)$ weight of $ \Theta(T)$, since the field strength $\bar F$ has $U(1)$ weight $w=-1$.

The $Spin^*(12)$ representation associated to the differential equation \eqref{F2R2} is the unique representation attached to the minimal nilpotent orbit of $\mf{so}^*(12)$. The minimal nilpotent orbit is represented by an element of degree $2$ in the Heisenberg parabolic decomposition 
\be \mf{so}^*(12) = {\bf 1}^\ord{-2} \oplus ({\bf 2}, {\bf 8})^\ord{-1} \oplus \bigl( \mf{gl}_1 \oplus \mf{su}(2)\oplus \mf{so}^*(8)\bigr)^\ord{0} \oplus  ({\bf 2}, {\bf 8})^\ord{1} \oplus {\bf 1}^\ord{2}\; , \ee
and is of dimension $18$. One can understand this from the property that a Lie algebra element in the coset component $\mf{so}(12,\mathds{C})\ominus \mf{gl}(6,\mathds{C})$ parametrised by a complex element $(\bar W_{ij},W^{ij})$ with only $W^{12}\ne 0$ is in the complex minimal nilpotent orbit of $\mf{so}(12,\mathds{C})$. The Cartan element of $\mf{u}(6,\mathds{C})$ defining the normal triple parametrising this orbit, {\it i.e.} $\Lambda_2+w$ in Bourbaki convention,\footnote{We define the fundamental weights of $SU(6)$ such that $\sum_in^i \Lambda_i$ is the highest weight for the representation of Dynkin label $[n^1,n^2,n^3,n^4,n^5]$, with $[1,0,0,0,0]={\bf 6}$, $[0,1,0,0,0]={\bf 15}$, $[0,0,1,0,0]={\bf 20}$, etc. We define the fundamental weights of $SO^*(12)$ such that $ \Lambda_1$ is the highest weight for the vector representation, $ \Lambda_2$ the adjoint, $ \Lambda_5$ the pseudo-real Weyl spinor and $ \Lambda_6$ the real Weyl spinor.} defines the harmonic coset space as its semi-simple $U(6)$ orbit.

The Gelfand--Kirillov dimension of the minimal automorphic representation is $9=18/2$, and corresponds to holomorphic forms of $T$ with Fourier expansion 
\be \Theta_2 =- \tfrac{i}{8}\det(T-\bar T)  \sum_{\substack{Q \in L \subset \mathds{R}^{15} \\ Q \times Q=0}} \tilde{c}(Q) e^{2\pi i Q_M T^M} \ee 
where $Q$ is in the minimal $SU^*(6)$ obit inside the real antisymmetric rank two tensor representation of dimension $9$, and  $\tilde{c}(Q)$ are constants. Supersymmetry allows for the sum to be replaced by an integral, but the axion shift symmetry of string theory implies that $Q$ only takes value in a lattice $L$  inside $\mathds{R}^{15}$.

\subsection{(2,2) BPS $\mathcal{R}^4$ type invariant}
\label{22Harmonic}
The $\mathcal{R}^4$ type invariant can be defined in the linearised approximation from the (2,2) chiral primaries in Table \ref{Primaries}. It is convenient to introduce the associated harmonic superspace, with harmonic variables $u^{\hat{r}}{}_i$ for $\hat{r}=1,2$, $u^{\check{r}}{}_i$ for $\check{r}=3,4$  and $u^{r}{}_i$ for $r=5,6$ parametrising  $U(6) / ( U(2) \times U(2)\times U(2))$. The associated G-analytic superfields $W^{12} = u^1{}_i u^2{}_j W^{ij}$ and $\bar W_{56} = u^i{}_5 u^j{}_6 \bar W_{ij}$ satisfy
\be D_\a^{\hat{r}} W^{12} = 0 \ , \qquad \bar D_{\adt r} W^{12} = 0 \;, \quad D_\a^{\hat{r}} \bar W_{56} = 0 \ , \qquad \bar D_{\adt r} \bar W_{56} = 0 \ . \ee
The chiral primaries are therefore all the harmonics of $(W^{12})^n(\bar W_{56})^m$, that combine into the irreducible representation $[0,n,0,m,0]$ of $SU(6)$. The top-component  can be shown to include the $\mathcal{R}^4$ Bell--Robinson square term
\be  [ D_\a^{\check{r}}]^4  [ D_\a^{{r}}]^4 [ \bar D_{\dot{\alpha} \hat{r}} ]^4[ \bar D_{\dot{\alpha}\check{r}}]^4  (W^{12})^{2+n}(\bar W_{56})^{2+m} \sim  (W^{12})^{n}(\bar W_{56})^{m} C^2 \bar C^2 + \dots   \ee
As in the previous section, one therefore concludes that the non-linear invariant is determined by a function $\cE(T,\bar T)$, and expands schematically as
\be \cL_\gra{2}{2}[\cE] = \sum_{\substack{n,m=0\\ 0\le n+m\le 12}}^{10} \bigl(\cD^n \bar \cD^{m} \cE(T,\bar T) \bigr)|_{[0,m,0,n,0]}    \cL_\gra{2}{2}^{[0,n,0,m,0]} \ee
where for example 
\bea    \cL_\gra{2}{2}^{[0,0,0,0,0]} &\sim&  C^2 \bar C^2 + \dots  \; , \\
\quad \cL_\gra{2}{2}^{[0,6,0,6,0]} &\sim&  \chi^{8\,  [0,0,0,6,0]} \bar \chi^{8\, [0,6,0,0,0]} \; , \qquad \cL_\gra{2}{2}^{[0,2,0,10,0]} \sim  \bar \lambda^{4\, [0,0,0,2,0]} \chi^{8\,  [0,0,0,6,0]} \bar \chi^{4\, [0,2,0,2,0]} \; . \nn \eea
The restriction to the irreducible  $U(6)$ representations appearing in the supersymmetry invariant implies the differential equations 
\be \varepsilon^{ijklpq} \cD_{kl} \cD_{pq} \cE = 0 \ , \qquad \cD_{ik} \bar \cD^{jk} \cE = \frac{1}{6} \delta_i^j \cD_{kl} \bar \cD^{kl} \cE  \ , \qquad   \varepsilon_{ijklpq} \bar \cD^{kl} \bar \cD^{pq} \cE = 0 \ .   \label{R4Eq0} \ee 
Using the commutation relations \eqref{ComSO12} one finds that 
\be \cD_{[ij} \bar \cD^{lp} \cD_{k]p} \cE = \bar \cD^{lp} \cD_{[ij} \cD_{kp]} \cE - \frac{3}{2} \delta^l_{[k} \cD_{ij]} \cE \ . \ee
Using then \eqref{R4Eq0} one gets 
\be \Delta  \cE = 2 \cD_{ij} \bar \cD^{ij} \cE =- 18 \cE \ . \ee
The function $\cE_\gra{2}{2}$ defining the $\mathcal{R}^4$ type invariant therefore satisfies the equations 
\be \varepsilon^{ijklpq} \cD_{kl} \cD_{pq}\cE_\gra{2}{2}= 0 \ , \qquad \cD_{ik} \bar \cD^{jk}\cE_\gra{2}{2} = - \frac{3}{2}  \delta_i^j  \cE_\gra{2}{2}  \ , \qquad   \varepsilon_{ijklpq} \bar \cD^{kl} \bar \cD^{pq} \cE_\gra{2}{2} = 0 \ .  \label{R4Eq} \ee 
This implies in particular 
\be \Delta^{\! \times } \cE_\gra{2}{2}= 0\; , \qquad \Delta^{\! \rm det } \cE_\gra{2}{2}= 0\; . \ee 
where $ \Delta^{\! \times }$ and $\Delta^{\! \rm det }$ were defined in \eqref{Casimirs}. 
The associated representation is attached to the next to minimal nilpotent orbit of dimension 28 that is represented by an element of degree $2$ in the parabolic decomposition 
\be \mf{so}^*(12) = ({\bf 6},{\bf 1})^\ord{-2} \oplus (\overline{\bf 4}, {\bf 4})^\ord{-1} \oplus \bigl( \mf{gl}_1 \oplus \mf{su}^*(4)\oplus \mf{so}^*(4)\bigr)^\ord{0} \oplus  ({\bf 4}, {\bf 4})^\ord{1} \oplus ({\bf 6},{\bf 1})^\ord{2}\; . \ee
The Gelfand--Kirillov dimension of the next to minimal automorphic representation is therefore $14$.  The complex element $(\bar W_{ij},W^{ij})$ with only $\bar W_{56}$ and $W^{12}$ non-zero parametrises a generic complex vector  in the  $({\bf 6},{\bf 1})^\ord{2}$.

\subsection{(1,1) BPS ${D}^2 \mathcal{R}^4$ type invariants}
\label{11Harmonics} 
A ${D}^2 \mathcal{R}^4$ type invariant can be defined in the linearised approximation from the (1,1) chiral primaries in Table \ref{Primaries}. It is convenient to introduce the associated harmonic superspace, with harmonic variables $u^{1}{}_i$, $u^{r}{}_i$ for ${r}=2,3,4,5$  and $u^{6}{}_i$ parametrising  $U(6) / ( U(1) \times U(4)\times U(1))$. The associated G-analytic superfields $W^{1r} = u^1{}_i u^r{}_j W^{ij}$ and $\bar W_{r6} = u^i{}_r u^j{}_6 \bar W_{ij}$ satisfy
\be D_\a^{1} W^{1r} = 0 \ , \qquad \bar D_{\adt 6} W^{1r} = 0 \;, \quad D_\a^{1} \bar W_{r6} = 0 \ , \qquad \bar D_{\adt 6} \bar W_{r6} = 0 \ . \ee
In this case the measure can be defined in the non-linear theory \cite{Bossard:2011tq}, and one finds that the class of supersymmetry invariants defined in this section can be identify with the set of non-linear G-analytic integrands. It will nonetheless be more convenient to use the linearised analysis. The chiral primaries are all the harmonics of $(W^{1r})^n(\bar W_{6s})^m (W^{1t} \bar W_{6t})^k$ with $k\ge 2$, that combine into the irreducible representation $[k,n,0,m,k]$ of $SU(6)$. The top-component can be shown to include the ${D}^2\mathcal{R}^4$  term \footnote{the only terms that contribute to the four graviton in the linearised approximation are where $[ D_\a^{r}]^8  [ D_\a^{6}]^2 [ \bar D_{\dot{\alpha} 1}]^2[ \bar D_{\dot{\alpha} r }]^8   (W^{1r } \bar W_{6r})^{2} \sim [ D_\a^{r_1}]^4   D_\a^{6}  \bar D_{\dot{\alpha} s_1 }  W^{1r} [ D_\a^{r_2}]^4   D_\a^{6}  \bar D_{\dot{\alpha} s_2 }  W^{1s}  [ \bar D_{\dot{\alpha} 1}] [ \bar D_{\dot{\alpha} s_1 }]^3 \bar W_{r6} [ \bar D_{\dot{\alpha} 1}] [ \bar D_{\dot{\alpha} s_2 }]^3 \bar W_{s6}  \sim \partial C \partial C \bar C \bar C $.}
\be  [ D_\a^{r}]^8  [ D_\a^{6}]^2 [ \bar D_{\dot{\alpha} 1}]^2[ \bar D_{\dot{\alpha} r }]^8  (W^{1r})^{n}(\bar W_{s6})^{m}  (W^{1t} \bar W_{6t})^{2+k} \sim (W^{1r})^{n}(\bar W_{s6})^{m}  (W^{1t} \bar W_{6t})^{k}  \nabla C^2  \nabla \bar C^2 + \dots   \label{GradientExpansion11}\ee
One therefore concludes as in the previous section that the non-linear invariant is determined by a function $\cE(T,\bar T)$, and expands schematically as
\be \cL_\gra{1}{1}[\cE] = \sum_{\substack{n,m,k\\ 0\le n+m+2k\le 16}} \bigl(\cD^{n+k} \bar \cD^{m+k} \cE(T,\bar T) \bigr)|_{[k,m,0,n,k]}    \cL_\gra{1}{1}^{[k,n,0,m,k]} \ee
where for example 
\be    \cL_\gra{1}{1}^{[0,0,0,0,0]} \sim \nabla  C^2\nabla  \bar C^2 + \dots  \; ,
\quad \cL_\gra{1}{1}^{[2,4,0,8,2]} \sim   \lambda^2_{[0,0,0,0,2]} \chi^8_{[0,0,0,6,0]} \bar \chi^{10}_{[2,4,0,2,0]}  \; . \ee
The restriction of the $U(6)$ irreducible representations appearing in the supersymmetry invariant implies the differential equations 
\be \varepsilon^{ijklpq} \cD_{kl} \cD_{pq} \cE = 0  \ , \qquad   \varepsilon_{ijklpq} \bar \cD^{kl} \bar \cD^{pq} \cE = 0 \ .   \label{DiffCons11}  \ee 
These constraints fix the eigenvalues of $\Delta^{\! \times}$ and $\Delta^{\! \rm det}$ to be zero. The eigenvalue of the Laplace operator is determined by supersymmetry, but does not follow directly from representation theory via \eqref{DiffCons11}. We shall find indirectly that $\cE$ satisfies the equation 
\be\Delta \cE= 2 \cD_{ij} \bar \cD^{ij} \cE = - 18 \cE \ . \ee
In principle this condition could be worked out explicitly using the same method as in \cite{Green:1998by,Bossard:2014lra}. 

The associated representation is attached to the nilpotent orbit of dimension 34  and is represented by a generic element of degree $2$ in the Heisenberg parabolic decomposition 
\be \mf{so}^*(12) = {\bf 1}^\ord{-4} \oplus ({\bf 2}, {\bf 8})^\ord{-2} \oplus \bigl( \mf{gl}_1 \oplus \mf{su}(2)\oplus \mf{so}^*(8)\bigr)^\ord{0} \oplus  ({\bf 2}, {\bf 8})^\ord{2} \oplus {\bf 1}^\ord{4}\; . \ee
The Gelfand--Kirillov dimension of the associated automorphic representations is therefore $17$. There is a one-parameter family of such representations determined by the eigen-value of the $SO^*(12)/U(6)$ Laplacian that corresponds to the maximal parabolic character of the $GL(1) $ factor, i.e. $\gFour^{-2s}$ with $\gFour$ the effective string coupling in our parametrisation.

As for the ${D}^6 \mathcal{R}^4$ correction in type IIB theory \cite{Green:2005ba}, one expects that  ${D}^2 \mathcal{R}^4$  will get a contribution depending quadratically in the $\bar F^2 C^2$ threshold function $\Theta$ in the $\cN=6$ supergravity effective action. This is consistent with power counting, as $\bar F^2 C^2$ has dimension $2+4$ and ${D}^2 \mathcal{R}^4$  dimension $2+8$.  Moreover one can check that this is consistent with the automorphic representation of Gelfand--Kirillov dimension $17$ as \eqref{F2R2} implies 
\be \varepsilon^{ijklpq} \cD_{kl} \cD_{pq}\bigl(  \Theta(T) \bar \Theta(\bar T) \bigr)  = 0  \ , \qquad   \varepsilon_{ijklpq} \bar \cD^{kl} \bar \cD^{pq} \bigl(  \Theta(T) \bar \Theta(\bar T)  \bigr) = 0\; , \ee
consistently with \eqref{DiffCons11}. 
We conjecture therefore that the Wilsonian effective action at this order includes a term in $\cE_\gra{1}{1}(T,\bar T) {D}^2 \mathcal{R}^4$ with the function $\cE_\gra{1}{1}(T,\bar T)$ satisfying to
\be \varepsilon^{ijklpq} \cD_{kl} \cD_{pq} \cE_\gra{1}{1} = 0  \ , \qquad   \varepsilon_{ijklpq} \bar \cD^{kl} \bar \cD^{pq} \cE_\gra{1}{1} = 0 \ , \quad 2 \cD_{ij} \bar \cD^{ij} \cE_\gra{1}{1} = - 18 \cE_\gra{1}{1} - \Theta(T) \bar \Theta(\bar T) \;.  \label{D2R4Source} \ee 
We shall further argue in \cite{GBinprep} that such a source term is indeed present in perturbative string theory. It originates from the singularity of the two-loop amplitude integrand at the  separating degeneration locus, as for the ${D}^6 \mathcal{R}^4$ threshold function in $\mathcal{N}=8$ \cite{DHoker:2014oxd} and ${D}^2 \mathcal{F}^4$ in $\mathcal{N}=4$ \cite{Bossard:2018rlt}.

\subsection{(2,0) BPS $ \mathcal{F}^{2k} \mathcal{R}^4  $ and ${D}^2 \mathcal{R}^4$ type invariants}
\label{20Harmonic}
The  class of supersymmetry invariants one can construct from the $(2,0)$ chiral primaries in Table~\ref{Primaries} is conveniently described using the associated harmonic superspace, with harmonic variables $u^{\hat{r}}{}_i$ for $\hat{r}=1,2$ and $u^{r}{}_i$ for $r=3,4,5,6$ parametrising  $U(6) / ( U(2) \times U(4))$.  The G-analytic superfields $W^{12} = u^1{}_i u^2{}_j W^{ij}$ and $\bar W_{rs} = u^i{}_r u^j{}_s \bar W_{ij}$ satisfy
\be
D_\a^{\hat{r}} W^{12} = 0 \ , \qquad  D_\a^{\hat{r}}  \bar W_{rs} = 0 \ .
\ee
The chiral primaries are therefore all the harmonics of $(W^{12})^n (\tfrac12 \varepsilon^{rstu}\bar W_{rs} \bar W_{tu})^k (\bar W_{rs})^m $ with $n\ge 2$ and $k\ge 2$,  that combine into the irreducible representation $[0,n+k,0,m,0]$ of $SU(6)$. The top-component includes the graviphoton--graviton coupling 
\bea &&  [D_\alpha^r]^8 [\bar D_{\dot{\alpha} i}]^{12} (W^{12})^{2+n}  (\tfrac12 \varepsilon^{rstu}\bar W_{rs} \bar W_{tu})^{k+2} (\bar W_{rs})^{m} \CR
 &\sim& (W^{12})^{n} (\tfrac12 \varepsilon^{rstu}\bar W_{rs} \bar W_{tu})^{k} (\bar W_{rs})^{m}  ( \bar F^{12} )^2 C^2 \bar C^2 + \dots  \eea
where the indices of  $(\bar W_{rs})^{m}$ are all meant to be uncontracted and projected to the traceless component. The linearised analysis therefore suggests that the non-linear supersymmetry invariant takes the form
\be \cL_\gra{2}{0}[\cE_{ij,kl}] = \sum_{n,m,k} \bigl(\cD^{n} \bar \cD^{m+2k} \cE_{ij,kl}(T,\bar T) \bigr)|_{[0,m,0,2+n+2k,0]}    \cL_\gra{2}{0}^{[0,2+n+2k,0,m,0]} \ee
where $\cE_{ij,kl}(T,\bar T)$ is a form of $U(1)$ weight $\frac{2}{3}$ in the $[0,0,0,2,0]$ of $SU(6)$ satisfying to 
\be \varepsilon^{ijrstu} \cD_{rs} \cD_{tu} \cE_{kl,pq} = 0  \ ,  \qquad \cD_{ir} \bar \cD^{jr} \cE_{kl,pq} = \frac{1}{6} \delta_i^j \cD_{rs} \bar \cD^{rs} \cE_{ij,kl} \ .\label{Cons20}  \ee
However, this linearised invariant only contributes to six and higher point amplitudes whereas we expect the non-linear supersymmetry invariant to contribute to four-point amplitude. Although it is always possible to linearised a non-linear supersymmetry invariant, one cannot neglect the possibility that a given non-linear invariant produce different types of linearised invariant in the linear approximation.  This is precisely the situation encountered  in $\mathcal{N}=8$ supergravity for the $(2,0)$-BPS  ${D}^6 \mathcal{R}^4$ type invariant \cite{Bossard:2015uga}.

The weight of $\cE_{ij,kl}(T,\bar T)$ is compatible with the requirement that 
\be \cE_{ij,kl}(T,\bar T) = \bigl(  \varepsilon_{ijpqrs}\cD_{kl}+ \varepsilon_{klpqrs}\cD_{ij} - 2\varepsilon_{pqrs[ij} \cD_{kl]}\bigr) \bar \cD^{pq} \bar \cD^{rs}    \cE(T,\bar T) \; , \label{Pot20}  \ee
and we cannot neglect the possibility that there be other linearised invariants in the expansion of the non-linear invariant in derivatives of $\cE(T,\bar T)$. To construct these linearised invariants one can use the $(2,2)$ harmonic superspace introduced in Section \ref{22Harmonic}, which is by construction consistent with the $(2,0)$-BPS condition. The first natural candidate is
\be   [ D_\a^{\check{r}}]^4  [ D_\a^{{r}}]^4 [ \bar D_{\dot{\alpha} \hat{r}} ]^4[ \bar D_{\dot{\alpha}\check{r}}]^4  (W^{12})^{2+n}(\bar W_{56})^{m} (\partial \bar W_{56} \cdot \partial \bar W_{56})  \sim  (W^{12})^{n}(\bar W_{56})^{m} C^2 (\nabla \bar C)^2 + \dots   \ee
which is compatible with the gradient expansion 
\be  \sum_{n,m} \bigl(\cD^{n} \bar \cD^{m} \cE(T,\bar T) \bigr)|_{[0,m,0,n,0]}    \cL_\gra{1}{1}^{[0,n,0,m,0]}\ee
such that $\cL_\gra{1}{1}^{[0,0,0,0,0]} \sim C^2 (\nabla \bar C)^2 $ plus supersymmetric completion. Up to integration by part, these densities are the same as in \ref{GradientExpansion11} for $k=0$. Another natural candidate is
\be   [ D_\a^{\check{r}}]^4  [ D_\a^{{r}}]^4 [ \bar D_{\dot{\alpha} \hat{r}} ]^4[ \bar D_{\dot{\alpha}\check{r}}]^4  (W^{12})^{2+n}(\bar W_{56})^{m} (F_{56} F_{56})  \sim  (W^{12})^{n}(\bar W_{56})^{m}  C^2 (\nabla \bar F)^2 + \dots  \label{C2D2F2}  \ee
which is of $U(1)$ weight $2$. It is therefore compatible with the gradient expansion 
\be
\sum_{n,m} \bigl(\cD^{n} \bar \cD^{m} \det\! \cD\, \cE(T,\bar T) \bigr)|_{[0,m,0,n,0]}    \cL_{\gra{1}{1}+2}^{ [0,n,0,m,0]}
\ee
where 
\be
\det\! \cD\,  \cE(T,\bar T)    = \frac{1}{48} \varepsilon^{ijklpq} \cD_{ij} \cD_{kl} \cD_{pq} \,  \cE(T,\bar T) \label{detD}
\ee
and $  \cL_{\gra{1}{1}+2}^{ [0,0,0,0,0]}\sim  C^2 ({\nabla} \bar {\mathcal{F}})^2 $ plus supersymmetric completion. There is no other available candidate $(2,2)$ harmonic superspace integral. We expect therefore the complete invariant to expand in all these structures with 
\bea \cL_{\gra{2}{0}+\gra{0}{2}}[\cE] &=& \sum_{n,m,k} \bigl(\cD^{n+1} \bar \cD^{m+2k+2} \cE(T,\bar T) \bigr)|_{[0,m,0,2+n+2k+2,0]}    \cL_\gra{2}{0}^{[0,2+n+2k,0,m,0]}  \CR
&& + \sum_{n,m} \bigl(\cD^{n} \bar \cD^{m} \cE(T,\bar T) \bigr)|_{[0,m,0,n,0]}    \cL_\gra{1}{1}^{[0,n,0,m,0]} \CR
&&  +  \sum_{n,m} \bigl(\cD^{n} \bar \cD^{m} \det\! \cD\, \cE(T,\bar T) \bigr)|_{[0,m,0,n,0]}    \cL_{\gra{1}{1}+2}^{ [0,n,0,m,0]} \CR
&& +  \sum_{n,m} \bigl(\cD^{n} \bar \cD^{m} \det\! \bar \cD\, \cE(T,\bar T) \bigr)|_{[0,m,0,n,0]}    \cL_{\gra{1}{1}-2}^{ [0,n,0,m,0]} \CR
&& + \sum_{n,m,k} \bigl(\bar \cD^{n+1}  \cD^{m+2k+2} \cE(T,\bar T) \bigr)|_{[0,2+n+2k+2,0,m,0]}    \cL_\gra{0}{2}^{[0,m,0,2+n+2k,0]}\; ,    \eea
where $ \cL_\gra{0}{2}^{[0,m,0,2+n+2k,0]}$ is the complex conjugate of $ \cL_\gra{2}{0}^{[0,2+n+2k,0,m,0]} $ and  $\cL_{\gra{1}{1}-2}^{ [0,n,0,m,0]} $ the complex conjugate of $ \cL_{\gra{1}{1}+2}^{ [0,m,0,n,0]}$. 

Using the constraints \eqref{Cons20} on \eqref{Pot20} one concludes that the real function $\cE(T,\bar T)$ must satisfy the constraint 
\be\varepsilon^{ijrstu} \varepsilon_{klpqvw} \cD_{rs} \cD_{tu} \bar \cD^{pq} \bar \cD^{vw} \cE(T,\bar T) = 0   \, , \qquad \cD_{ik} \bar \cD^{jk}  \cE(T,\bar T) =- \frac{3}{2} \delta_i^j  \cE(T,\bar T) \ , \label{Diff20} \ee
and in particular the three Casimir operators give \eqref{Casimirs} 
\be \Delta  \cE(T,\bar T) = -18 \cE(T,\bar T) \; , \qquad  \Delta^{\! \times}  \cE(T,\bar T) =0  \; , \qquad  \Delta^{\! \rm det}  \cE(T,\bar T) =0 \; . \ee
Note that the structure in \eqref{C2D2F2} must necessarily be present, because one does not find any solution $\cE(T,\bar T)$ to these equations such that $\det \!\cD\, \cE(T,\bar T)=0$ without satisfying moreover \eqref{R4Eq0} as for the $(2,2)$-BPS invariant. 

The existence of this invariant will be confirmed by the one-loop amplitude computation in string theory. 
The associated representation is attached to the nilpotent orbit of dimension 30 that is realised by an element of degree $2$ in the abelian parabolic decomposition 
\be \mf{so}^*(12) = \overline{\bf 15}^\ord{-2} \oplus \bigl( \mf{gl}_1 \oplus \mf{su}^*(6)\bigr)^\ord{0} \oplus  {\bf 15}^\ord{2} \; . \ee
The Gelfand--Kirillov dimension of the associated automorphic representations is therefore $15$. 

There is a one-parameter family of such representations fixed by the eigenvalue of the $SO^*(12)/U(6)$ Laplace operator that is represented by the maximal parabolic character associated to the $GL(1)$ factor, i.e. $R^{3s}$ in our conventions for $R$ the circle radius. 

According to this structure we find that $\cE(T,\bar T)$  appears in the four-graviton amplitude $\cE_\grad{1}{0}{0}$ at order $s$, while $\det \!\cD\,  \cE(T,\bar T)$ appears in the two-graviphoton two-graviton amplitude at order $s^2$. 

\medskip

 This class of supersymmetry invariants generalises to higher derivative couplings by including the additional G-analytic superfields $\bar F_{\dot{\alpha}\dot{\beta}}^{12} = u^1{}_i u^2{}_j \bar F_{\dot{\alpha}\dot{\beta}}^{ij}$ satisfying to 
\be D_\a^{\hat{r}} \bar F_{\dot{\alpha}\dot{\beta}}^{12} = 0 \ . \ee
The chiral primaries are in one-to-one correspondance with the harmonics of \footnote{The conformal primaries can be obtained as $ [S_{\dot{\alpha}}^r]^8  (W^{12})^n (\tfrac12 \varepsilon^{rstu}\bar W_{rs} \bar W_{tu})^{k-2} (\bar W_{rs})^m (\bar F^{12})^{2\ell+4}$ with the special supersymmetry $S_{\dot{\alpha}}^r = u^r{}_i S_{\dot{\alpha}}^i$. They only differ from \eqref{CPF2R4}, by a total superspace derivative and therefore define the same supersymmetry invariants.}
\be (W^{12})^n (\tfrac12 \varepsilon^{rstu}\bar W_{rs} \bar W_{tu})^k (\bar W_{rs})^m (\bar F^{12})^{2\ell} \; , \label{CPF2R4} \ee
that combine into the irreducible representation $[0,n+k+2\ell,0,m,0]$ of $SU(6)$. The top-component  includes the graviphoton--graviton coupling 
\bea &&  [D_\alpha^r]^8 [\bar D_{\dot{\alpha} i}]^{12} (W^{12})^{2+n}  (\tfrac12 \varepsilon^{rstu}\bar W_{rs} \bar W_{tu})^{k+2} (\bar W_{rs})^{m}  (\bar F^{12})^{2\ell-2}  \CR
 &\sim& (W^{12})^{n} (\tfrac12 \varepsilon^{rstu}\bar W_{rs} \bar W_{tu})^{k} (\bar W_{rs})^{m}  ( \bar F^{12} )^{2\ell} C^2 \bar C^2 + \dots  \eea
The linearised analysis therefore suggests that the non-linear supersymmetry invariant takes the form
\be \cL_\gra{2}{0}^\ord{\ell}[\cE_{12^{2\ell}}] = \sum_{n,m,k} \bigl(\cD^{n} \bar \cD^{m+2k} \cE_{12^{2\ell}}(T,\bar T) \bigr)|_{[0,m,0,2\ell +n+2k,0]}    \cL_\gra{2}{0}^{[0,2\ell+n+2k,0,m,0]} \ee
where $\cE_{12^{2\ell}}(T,\bar T)$ is a form of $U(1)$ weight $w=2\ell/3$ in the $[0,0,0,2\ell,0]$ of $SU(6)$ satisfying to 
\be \varepsilon^{ijrstu} \cD_{rs} \cD_{tu} \cE_{12^{2\ell}} = 0  \ ,  \qquad \cD_{ir} \bar \cD^{jr} \cE_{12^{2\ell}}  = \frac{1}{6} \delta_i^j \cD_{rs} \bar \cD^{rs} \cE_{12^{2\ell}}  \ . \ee
It is associated to a discrete series representation attached to the same nilpotent orbit corresponding the the maximal parabolic character $R^{6+3\ell}$.

\subsection{BPS invariants in five dimensions}
In five dimensions the linearised superfields do not define superconformal multiplets and linearised supersymmetry invariants cannot all be written as superspace integrals of appropriate G-analytic integrands. To classify five-dimensional supersymmetry invariants one must instead rely on the four-dimensional ones, and check if they consistently lift to five dimensions. In particular one finds that there is no 1/2 BPS $\mathcal{F}^2 \mathcal{R}^2$ invariant in five dimensions, whereas the $\mathcal{R}^4$ and ${D}^2 \mathcal{R}^4$ type invariants consistently lift. In this subsection we introduce the linearised superfield in five dimensions to determine the structure of the four-graviton amplitude for generic polarisations. 

The fourteen real scalar fields $\phi^M$ parametrise the symmetric space $SU^*(6) / Sp(3)$ and one defines the coset Maurer--Cartan form 
\be P_{M}{}^{ij} d\phi^M\, ,  \ee
in the real antisymmetric symplectic traceless representation of $Sp(3)$, with $i=1$ to $6$ and for the symplectic form $\Omega_{ij}$ satisfying $\Omega^{ik} \Omega_{jk} = \delta^i_j$ 
\be P^*_{M ij}  = \Omega_{ik} \Omega_{jl} P_M{}^{kl} \; , \qquad \Omega_{ij} P_M{}^{ij} = 0 \; \; . \ee
The $\cN=3$ supermultiplet ($\mathcal{N}=6$ in $D=4$) involves the supercovariant fields
\be P_{M}{}^{ij} d\phi^M \; , \quad \chi_{\alpha}^{ijk} \; , \;  \lambda_\alpha^i\; , \quad F_{\alpha\beta}^{ij} \; , \;  F_{\alpha\beta} \; ,  \quad \rho_{\alpha\beta\gamma}^i \; , \quad C_{\alpha\beta\gamma\delta}\;  , \ee
where $\alpha=1$ to $4$ are $Sp(1,1) = Spin(1,4)$ indices that are symmetrised, while all the $Sp(3)$ indices are antisymmetrised and symplectic-traceless. One defines the fifteen field strengths and the Weyl tensor using $Spin(1,4)$ gamma matrices as
\bea  F_{\alpha\beta}^{ij}  &=&  \frac12 \gamma^{ab}{}_{\alpha\beta} e_a{}^\mu e_b{}^\nu  \cV^{ij}{}_{IJ} F_{\mu\nu}^{IJ} \; , \qquad  F_{\alpha\beta} = \frac12 \gamma^{ab}{}_{\alpha\beta} e_a{}^\mu e_b{}^\nu  \cV_{IJ} F_{\mu\nu}^{IJ} \; , \CR
 C_{\alpha\beta\gamma\delta}  &=&   \frac14 \gamma^{ab}{}_{(\alpha\beta}  \gamma^{cd}{}_{\gamma\delta)}e_a{}^\mu e_b{}^\nu  e_c{}^\sigma e_c{}^\rho R_{\mu\nu\sigma\rho} \; . \eea
In the linearised approximation, one can identify the  real and symplectic-traceless linearised superfield $W^{ij}$
\be P_{M}{}^{ij} d\phi^M \approx d W^{ij} \; , \ee
which defines the ultra-short supersymmetry representation that is acted on by the superspace derivative according to
\bea  D_\a^i  W^{jk} &=&  \chi_{\a}^{ijk} + 2 \Omega^{i[j} \lambda_\alpha^{k]} + \frac13 \Omega^{jk} \lambda_\alpha^i   \; . \CR
D_\a^i \chi_\b^{jkl}  &=& 3 \Omega^{i[j} ( F_{\alpha\beta}^{kl]} - i \partial_{\alpha\beta} W^{kl]})- \frac32\Omega^{[kl} ( F_{\alpha\beta}^{j]i} - i \partial_{\alpha\beta} W^{j]i})  \; , \CR
 D_\a^i \lambda_\b^j &=& \Omega^{ij} F_{\alpha\beta} + \frac12 F^{ij}_{\alpha\beta} +  \frac{3i}{2} \partial_{\alpha\beta} W^{ij} \; ,  \eea
where we define $\partial_{\alpha\beta} = \frac12 \gamma^{a}{}_{\alpha\beta} \partial_a$. 
To describe the linearised invariants relevant to the four-graviton amplitude it is convenient to introduce the 1/3 BPS harmonic variables $(u^r{}_i , u^{\hat{r}}{}_i , u_{r i} )\in Sp(3)/ ( U(2) \times Sp(1))$ with $r=1, 2$ and $\hat{r}=3,4$ that satisfy 
\be \Omega^{ij} u^r{}_i u^s{}_j = 0 \; , \quad \Omega^{ij} u^r{}_i u_{ sj} = 2 \delta^r_s\; , \quad \Omega^{ij} u^r{}_i u^{\hat{s}}{}_j =0 \; , \quad \Omega^{ij} u^{\hat{r}}{}_i u^{\hat{s}}{}_j =\varepsilon^{\hat{r}\hat{s}}\; . \ee
One then has the two G-analytic superfields $W^{12} = u^1{}_i u^2{}_j W^{ij}$ and $F_{\alpha\beta}^{12} = u^1{}_i u^2{}_j F_{\alpha\beta}^{ij}$, satisfying  $D_\alpha^r W^{12} = D_\alpha F_{\beta\gamma}^{12}=0$ and one can define the measure 
\be [D^{16}] = [ D^{\hat{r}}_\alpha]^8  [ D_{{r\alpha}}]^8 \; .  \ee
From the four-dimensional analysis we know that the four-point $\mathcal{R}^4$ invariant is unique, and it can be realised in five dimensions as
\be [ D^{16}] (W^{12} )^4 \sim \prod_{a=1}^4 \varepsilon^{\alpha_a\beta_a\gamma_a\delta_a} C_{\alpha_1\alpha_2\alpha_3\alpha_4} C_{\beta_1\beta_2\beta_3\beta_4} C_{\gamma_1\gamma_2\gamma_3\gamma_4} C_{\delta_1\delta_2\delta_3\delta_4} + \dots = \frac{1}{256} t_8 t_8 \mathcal{R}^4 + \dots \ee
where we introduce the standard $t_8$ tensor 
\be t_8 F^4 = 24 \tr_{\bf 5} F^4 - 6 (\tr_{\bf 5} F^2)^2 =-96 \tr_{\bf 4} F^4 +48 (\tr_{\bf 4} F^2)^2 \; , \ee
with the notation \footnote{One raises $Sp(1,1)$ indices with the symplectic form as $F_{\alpha}{}^\beta = F_{\alpha\gamma} \Omega^{\gamma\beta}$ and $\varepsilon^{\alpha\beta\gamma\delta} = 3 \Omega^{[\alpha\beta} \Omega^{\gamma\delta]}$.}
\be \tr_{\bf 5} F^{2} = F_{ab} F^{ba} \; , \quad \tr_{\bf 5} F^{4} = F_{ab} F^{bc} F_{cd} F^{da} \;, \quad \tr_{\bf 4} F^{2} = F_{\alpha}{}^\beta F_\beta{}^\alpha \;  , \quad \tr_{\bf 4} F^{4} = F_{\alpha\beta} F^{\beta\gamma} F_{\gamma\delta} F^{\delta\alpha} \; . \ee
We also know from the four-dimensional analysis that there is a unique linearised ${D}^2 \mathcal{R}^4$ invariant, that can be written in five dimensions as
\bea [ D^{16}] (W^{12} )^2  F_{\alpha\beta}^{12} F^{\alpha\beta 12} &\sim& \prod_{a=1}^4 \varepsilon^{\alpha_a\beta_a\gamma_a\delta_a} \partial_{\epsilon_1 ( \alpha_1} C_{\alpha_2\alpha_3\alpha_4\epsilon_2} \partial^{(\epsilon_1}{}_{(\beta_1} C_{\beta_2\beta_3\beta_4)}{}^{\epsilon_2)}  C_{\gamma_1\gamma_2\gamma_3\gamma_4} C_{\delta_1\delta_2\delta_3\delta_4} + \dots \CR
&=& \frac{1}{3072} t_8^{(s)} t_8 \partial_a R \partial^a R R^2 + \dots \eea
where one uses integration by part in the last step, and the $t_8^{(s)}$ tensor is defined as \footnote{$(t_8^{(s)}){}^{a_1a_2b_1b_2c_1c_2d_1d_2}$ is antisymmetric under the permutations of $a_1a_2$, $b_1b_2$, $c_1c_2$ and $d_1d_2$ and symmetric under the permutations of the pairs $(a_1a_2)$ and $(b_1b_2)$ and the pairs $(c_1c_2)$ and $(d_1d_2)$. Therefore it is determined by its contraction with two independent rank two tensors $F_{1ab}$ and $F_{2 ab}$.} 
\be t_8^{(s)} F_1^{\, 2} F_2^{\; 2} =8\tr_{\bf 5} F_1 F_2 F_1 F_2 -4(\tr_{\bf 5} F_1 F_2)^2 + 2  \tr_{\bf 5} F_1^{\; 2}  \tr_{\bf 5} F_2^{\; 2} = - 32 \tr_{\bf 4} F_1^{\; 2} F_2^{\; 2} + 16  \tr_{\bf 4} F_1^{\; 2}  \tr_{\bf 4} F_2^{\; 2} \, . \ee
For the amplitude it is convenient to introduce the three permutations of the tensor $t_8^{(s)}$
\bea&&  t_8^{(s)} (F_1 F_2 F_3 F_4) =t_8^{(t)}(F_3 F_1 F_2 F_4) = t_8^{(u)} (F_2 F_3 F_1 F_4 ) \label{t8s} \\
&=& 4\tr_{\bf 5} F_1 F_3 F_2 F_4 + 4\tr_{\bf 5} F_2 F_3 F_1 F_4- 2 \tr_{\bf 5} F_1 F_3 \tr_{\bf 5} F_2 F_4 - 2 \tr_{\bf 5} F_2 F_3 \tr_{\bf 5} F_1 F_4 + 2 \tr_{\bf 5} F_1 F_2 \tr_{\bf 5} F_3 F_4 \nonumber\eea
that satisfy $t_8 = t_8^{(s)}+t_8^{(t)}+t_8^{(u)}$. The notation is justified because for four-dimensional polarisations, $t_8^{(s)} F_1 F_2 F_3 F_4$ vanishes if $F_1$ and $F_2$ do not have the same helicity. This structure of the supersymmetry invariant agrees with the six-dimensional four-graviton amplitude \cite{Berg:2016wux} reduced to five dimensions.

\subsection{Consequences for string perturbation theory}
\label{StringNonRenormalisation}
The solutions to the differential equations implied by supersymmetry derived in this section severely constrain possible perturbative corrections. We shall find in particular that the BPS saturated couplings can only receive corrections at specific loop orders in string perturbation theory.
 In this subsection we analyse these equations for the perturbative threshold functions that only depend on the 13 NS-NS moduli, excluding the axion dual to the external Kalb-Ramond two-form. These include the dilaton defining the effective string coupling $\gFour$, and the Narain lattice moduli $t^m,\bar t^{\bar m}$ that parametrise $SO(2,6)/ (SO(2)\times SO(6))$, corresponding to the internal components of the metric and B-field that survive the orbifold projection. We define the tangent frame differential operator on the Narain lattice moduli space
\be \cD_{ab} =\bar P_{\bar n \, ab} G^{m \bar n}  ( \partial_m  + Q_m) ,  \qquad \bar \cD^{ab} =  P_{m}{}^{ab}  G^{m\bar n} ( \bar{\partial}_{\bar n}  + Q_{\bar n}) \ , \ee  
as for $SO^*(12)/U(6)$ in \eqref{Dso12}, where $a=1,2,3,4$ of $SU(4)$. The Laplace operator decomposes as
\be 2 \cD_{ij} \bar \cD^{ij} \cE^{\scalebox{0.6}{per}}(\gFour,t,\bar t) = \biggl( \frac14 \gFour \frac{\partial\; }{\partial \gfour} \Bigl(\gFour \frac{\partial\; }{\partial \gfour} +18 \Bigr) +2 \cD_{ab} \bar \cD^{ab} \biggr) \cE^{\scalebox{0.6}{per}}(\gFour,t,\bar t) \; . \ee
By gauge invariance, the non-perturbative couplings only depend on the R-R moduli $C$ and the axio-dilaton axion $b$ through periodic functions. Expanding in Fourier series, the Laplace equation on a function of type $\cE_Q(\gFour,t,\bar t) e^{2\pi i Q C}$ is then exponentially suppressed in $\frac{|V(Q)|}{\gfour}$ at weak coupling, and equivalently $\cE_k(\gFour,t,\bar t,C) e^{2\pi i kb}$ is exponentially suppressed in $\frac{|k|}{\gfour^2}$. One interprets physically that the former correspond to Euclidean Dp-brane instantons \cite{Green:1997tv,Harvey:1996ir,Bianchi:2012ud,Bianchi:2012kt} and the latter to Euclidean NS5 brane instantons. It follows in particular that the perturbative couplings do not depend on the R-R and the external Kalb-Ramond two-form axions.

The relevant components of \eqref{DiffCons11} are 
\be \varepsilon^{abcd} \cD_{ab} \cD_{cd} \cE^{\scalebox{0.6}{per}}(\gFour,t,\bar t) = 0 \; , \qquad  \cD_{ab}  \Bigl(\gFour \frac{\partial\; }{\partial \gfour} +4 \Bigr)  \cE^{\scalebox{0.6}{per}}(\gFour,t,\bar t) = 0\; , \label{perBPS11} \ee
whereas the relevant components of \eqref{Diff20} are 
\be\varepsilon_{ef\hspace{-0.2mm}gh} \varepsilon^{abcd} \bar \cD^{ef} \bar \cD^{gh}  \cD_{ab} \cD_{cd} \cE^{\scalebox{0.6}{per}}(\gFour,t,\bar t) = 0 \; , \quad \Bigl(   \cD_{ac} \bar{\cD}^{bc} + \tfrac{1}{4} \delta_a^b \,   \gFour \frac{\partial\; }{\partial \gfour} + \tfrac32 \delta_a^b \Bigr) \cE^{\scalebox{0.6}{per}}(\gFour,t,\bar t) = 0\; . \label{perBPS20} \ee

\medskip

One obtains directly that the 1/2 BPS threshold function $\Theta(T)$ can only receive corrections at one-loop, and that the one-loop contribution is in the minimal automorphic representation of $SO(2,6)$ of Gelfand--Kirillov dimension $5$, i.e. must be a (linear combination of) rank one theta series $\theta(t)$ of the form
 \be \theta(t) =- \tfrac{1}{4}\det(t-\bar t)  \sum_{\substack{Q \in L_{1,5}^* \\ (Q , Q)=0}} \tilde{c}(Q) e^{2\pi i Q_n t^n} \; . \ee 
 Here we have introduced the notation $L_{p,q} $ for a lattice of signature $(p,q)$ and $L^*_{p,q} $ for its dual. For supersymmetry $L_{p,q}$ could be any discrete set of vectors, but it will turn out to be the lattice of perturbative untwisted states in string theory. To make this relation more explicit we anticipate that the Narain lattice partition function 
 \be \Gamma_{\Lambda_{2,6}} = \tau_2^{\; 3} \sum_{{Q}\in L_{2,6}} e^{ i \pi \tau p_L({Q})^2 - i \pi \bar \tau p_R({Q})^2} \ee
 is a modular form of weight $-2$ for a congruent subgroup $\Gamma \in SL(2,\mathds{Z})$.

\medskip

The (2,2) BPS equation solved by the $\mathcal{R}^4$ threshold function satisfies both \eqref{perBPS11} and \eqref{perBPS20}. The second equation in \eqref{perBPS11} gives  
\be \gFour^4 \cE^{\scalebox{0.6}{per}}_\gra{2}{2} = f_0(\gFour) + f_{1,2}(t,\bar t)  \ee
while the second in  \eqref{perBPS20} then implies that $f_0(\gFour) = c_0 / \gfour^2$ for a constant $c_0$. The function $f_{1,2}(t,\bar t)$ must moreover satisfy 
\be  \varepsilon^{abcd} \cD_{ab} \cD_{cd} f_{1,2}(t,\bar t)=0\; , \qquad   \qquad \bigl(   \cD_{ac} \bar{\cD}^{bc} + \tfrac12  \delta_a^b \bigr) f_{1,2}(t,\bar t)= 0 \; . \label{WardoneloopR4} \ee
The second implies that $f_{1,2}(t,\bar t)$ is a genus one theta lift consistently with string perturbation theory, while the first requires that the seed $\varphi_{0,2}(\tau)$ be a holomorphic function of the torus modulus $\tau$, so that 
\be \gFour^4 \cE^{\scalebox{0.6}{per}}_\gra{2}{2} = \frac{c_{0}}{\gFour^2} + \int_{\mathcal{H}^+\!/ \Gamma} \frac{d^2\tau}{\tau_2^{\, 2}} \varphi_{0,2}(\tau) \Gamma_{L_{2,6}} \ .  \label{R4twotwo}  \ee
As a consequence the $\mathcal{R}^4$ correction is therefore one-loop exact in string theory as it only receives contributions at tree-level and one-loop.

For the ${D}^2 \mathcal{R}^4$ threshold function we must distinguish the two supersymmetry invariants that correspond to distinct automorphic representations. The (2,0)-BPS invariant must be a solution of \eqref{perBPS20}. The second equation implies that all perturbative  terms can be written as genus one theta lifts, and the first 
\be \varepsilon_{efgh} \varepsilon^{abcd} \bar \cD^{ef} \bar \cD^{gh}  \cD_{ab} \cD_{cd} \int_{\mathcal{H}^+\!/ \Gamma} \frac{d^2\tau}{\tau_2^{\, 2}} \varphi_{i,2}(\tau,\bar \tau) \Gamma_{L_{2,6}} = 4 \int_{\mathcal{H}^+\!/ \Gamma} \frac{d^2\tau}{\tau_2^{\, 2}} \Delta( \Delta-2) \varphi_{i,2}(\tau,\bar \tau) \Gamma_{L_{2,6}} =0\ee
that the seeds $\varphi_{i,2}$ must satisfy 
\be \Delta( \Delta-2) \varphi_{i,2}(\tau,\bar \tau) = 0\; . \ee
Using moreover 
\be 2 \cD_{ab} \bar \cD^{ab} \int_{\mathcal{H}^+\!/ \Gamma} \frac{d^2\tau}{\tau_2^{\, 2}} \varphi_{i,2}(\tau,\bar \tau) \Gamma_{L_{2,6}} = 2 \int_{\mathcal{H}^+\!/ \Gamma} \frac{d^2\tau}{\tau_2^{\, 2}} ( \Delta -2)\varphi_{i,2}(\tau,\bar \tau) \Gamma_{L_{2,6}}  \; , \ee
one fixes the Laplace eigenvalue of the Narain lattice moduli space function and therefore the possible powers of the string coupling using again  \eqref{perBPS20}.  This gives 
\be \gFour^6 \cE^{\scalebox{0.6}{per}}_\gra{2}{0} =  \int_{\mathcal{H}^+\!/ \Gamma} \frac{d^2\tau}{\tau_2^{\, 2}} \varphi_{1,2}(\tau,\bar \tau) \Gamma_{L_{2,6}}  + \gFour^2   \int_{\mathcal{H}^+\!/ \Gamma} \frac{d^2\tau}{\tau_2^{\, 2}} \tilde{\varphi}_{0,2}(\tau) \Gamma_{L_{2,6}}   \label{D2R4twozero}  \ee
with 
\be  \Delta {\varphi}_{1,2}(\tau,\bar \tau) =2  {\varphi}_{1,2}(\tau,\bar \tau)\; , \label{E12LaplaceEquation} \ee
and $\tilde{\varphi}_{0,2}(\tau) $ harmonic. However, the derivation of these differential equations is based on the local Lagrangian while the effective action includes logarithmic divergences in the split of the amplitude into the supergravity amplitude and its analytic component in the Mandelstam variable. We will find in \cite{GBinprep} that the differential equation is modified and reads 
\be   \Delta \tilde{\varphi}_{0,2}(\tau)  =  -\frac14 \varphi_{0,2}(\tau)\; , \ee
where $\varphi_{0,2}(\tau)$ is the seed for the one-loop threshold function \eqref{R4twotwo}.  The first term in \eqref{D2R4twozero} is indeed a genus one correction while the second will be part of the regularised genus two Narain lattice theta lift.  

The (1,1)-BPS invariant solves instead the differential equations \eqref{perBPS11}. Using the second one, one obtains 
\be \gFour^6 \cE^{\scalebox{0.6}{per}}_\gra{1}{1} = f_2(\gFour)+ \gFour^2  f_{2,2}(t,\bar t) \ee
while the Laplace eigenvalue equation \eqref{D2R4Source} gives 
\be f_2(\gFour)=  \frac{c_{-2}}{\gFour^6} + c_1\; , \ee
and 
\be \bigl( 2 \cD_{ab} \bar \cD^{ab}  + 4\bigr) f_{2,2}(t,\bar t)   = -  \theta(t) \overline{\theta(t)}\; . \label{PoissonTwoLoop}  \ee
The coefficient $c_{-2}$ must vanish for consistency with string perturbation theory, $c_1$ is a constant and $f_{2,2}$ can be realised as a sum of a genus one and a genus two Narain lattice theta lift that we write as
\be f_{2,2} = \int_{\mathcal{H}_2^+\!/ \Gamma} \frac{d^6 \Omega}{\det \Omega_2^{\; 3}} {\varphi}_2(\Omega) \Gamma_{L_{2,6}}^{\scalebox{0.6}{2-loop}}- \int_{\mathcal{H}^+\!/ \Gamma} \frac{d^2\tau}{\tau_2^{\, 2}} \tilde{\varphi}_{0,2}(\tau) \Gamma_{L_{2,6}}    \; , \ee
with $\Omega = \Omega_1 + i \Omega_2$ the period matrix of the genus two surface and ${\varphi}_2(\Omega)$ a weight two modular form. Here we anticipate that the total two-loop contribution is the Narain lattice theta lift with seed function ${\varphi}_2(\Omega)$. This modular form satisfies differential equations that are discussed in detail in \cite{GBinprep}. We conclude therefore that 
\be \gFour^6 \cE^{\scalebox{0.6}{per}}_\gra{1}{1} = c_1 + \gFour^2 \Bigl(  \int_{\mathcal{H}_2^+\!/ \Gamma} \frac{d^6 \Omega}{\det \Omega_2^{\; 3}} {\varphi}_2(\Omega) \Gamma_{L_{2,6}}^{\scalebox{0.6}{2-loop}} - \int_{\mathcal{H}^+\!/ \Gamma} \frac{d^2\tau}{\tau_2^{\, 2}} \tilde{\varphi}_{0,2}(\tau) \Gamma_{L_{2,6}}   \Bigr)  \label{D2R4oneone} \; . \ee
The quadratic source to the Poisson equation \eqref{D2R4Source} corrects the Laplace equation for $ f_{2,2}$, consistently with the property that the genus 2 integrand ${\varphi}_2(\Omega) $ should be singular at the separating degeneration locus.\footnote{The fact that $c_{-2}=0$ in string theory implies that $ \cE_\gra{1}{1}$ satisfy an  inhomogeneous differential equation like \eqref{D2R4Source}, because the automorphic homogeneous solution would necessarily have  $c_{-2}\ne 0$. This situation is identical to the one for the  ${D}^6 \mathcal{R}^4$  threshold function in type IIB, for which the homogeneous solution $E_{4,0}(S)$ is compatible with supersymmetry but not with perturbation theory \cite{Green:2005ba}.}  The two-loop correction is discussed in more detail in  \cite{GBinprep}. The  ${D}^2 \mathcal{R}^4$ correction is therefore two-loop exact in string theory.

The complete one-loop ${D}^2 \mathcal{R}^4$ threshold function is the sum of \eqref{D2R4twozero} and \eqref{D2R4oneone}. As we shall see one must take into account that the ${D}^2 \mathcal{R}^4$ threshold function diverges as a worldsheet moduli space integral and must therefore be regularised. This is due to the logarithmic divergence of the supergravity  $\int  \mathcal{R}^4$ form factor at one-loop that affects $\int {D}^2 \mathcal{R}^4$ form factor. This is consistent with supersymmetry because $\cE_\gra{2}{2} $ is a solution to the differential equations satisfied by $\cE_\gra{2}{0} $ and $\cE_\gra{1}{1} $. Defining a cutoff $\tau_2\le L$ in the fundamental domain $\mathcal{F} = {\mathcal{H}^+\!/ SL(2,\mathds{Z})}$ one obtains the regulated integral
\be  \int_{\mathcal{H}^+\!/ \Gamma,L} \frac{d^2\tau}{\tau_2^{\, 2}} \varphi_{1,2}(\tau,\bar \tau) \Gamma_{L_{2,6}}  = \frac{c_0}{2\pi }  \log \frac{L}{\mu} +  \int_{\mathcal{H}^+\!/ \Gamma,\mu} \frac{d^2\tau}{\tau_2^{\, 2}} \varphi_{1,2}(\tau,\bar \tau) \Gamma_{L_{2,6}} \ee
where $c_0$ is the coefficient of the $\mathcal{R}^4$ threshold function \eqref{R4twotwo}. We find therefore that the constant $c_1$ in \eqref{D2R4oneone} can be reabsorbed in a redefinition of the renormalisation scale $\mu$. By consistency, one must find the same behaviour at two-loop. In this case this is the non-separating degeneration limit of the genus two surface that diverges and the appropriate cut-off integral defined in \cite{GBinprep} diverges in
\be   \int_{\mathcal{H}_2^+\!/ \Gamma,L} \frac{d^6 \Omega}{\det \Omega_2^{\; 3}} {\varphi}_2(\Omega) \Gamma_{L_{2,6}}^{\scalebox{0.6}{2-loop}} = \frac{1}{2\pi }  \log \frac{L}{\mu} \int_{\mathcal{H}^+\!/ \Gamma} \frac{d^2\tau}{\tau_2^{\, 2}} \varphi_{0,2}(\tau) \Gamma_{L_{2,6}} + \int_{\mathcal{H}_2^+\!/ \Gamma,\mu} \frac{d^6 \Omega}{\det \Omega_2^{\; 3}} {\varphi}_2(\Omega) \Gamma_{L_{2,6}}^{\scalebox{0.6}{2-loop}}\; .  \label{LogDivD2R4}  \ee

To conclude we briefly describe the higher derivative BPS protected threshold functions. The  $\bar {\mathcal{F}}^{2\ell} \mathcal{R}^4$ couplings realised as  (2,0) BPS invariants give rise to  solutions of the form
\be \gFour^{4+2\ell}  \cE^{\ord{\ell}\scalebox{0.6}{per} }_\gra{2}{0} \sim f_{1,2+\ell}(t,\bar t) + \gFour^{4\ell-2}  f_{1,3-\ell}(t,\bar t)  \ee
where $ f_{1,2+\ell}$ can be realised as a genus 1 Narain lattice theta lift and  $f_{1, 3-\ell}$ should correspond to a specific `genus $2\ell$ Narain lattice theta lifts', consistently with string perturbation theory.

\medskip

%%%%%%%%%%%%%%%%%%%%%%%%
\section{Perturbative results in String Theory: tree level and one-loop}
\label{String1-loop}

As we have just seen, ${\cal N}=6$ supersymmetry constrains the perturbative contributions to the higher derivative terms ${\cal R}^4$ and ${\cal F}^2{\cal R}^2$ to be one-loop exact and $D^2{\cal R}^4$ to be two-loop exact in string theory. Aim of the present section is to extract the perturbative threshold functions, that  depend only on the NS-NS moduli (the dilaton, internal metric and $B$-field excluding the axion dual to the external $B$-field) from one-loop string amplitudes and show that they perfectly match the expectations based on ${\cal N}=6$ supersymmetry and the ensuing T-duality. Non-perturbative corrections, that involve D-brane and NS5 instantons and thus introduce a dependence on the R-R moduli and the external $B$-field axion, are tightly constrained by the full U-duality group.  

After writing down the relevant vertex operators for the graviton and (one of) the gravi-photons, we will exploit `chiral' factorisation to reduce the computation to the computation in each of the two sectors (L- and R-movers) that in turn are the same as those for open strings \cite{Bianchi:2015vsa}. The correct normalisation of the vertex operators can be derived using tree-level amplitudes, factorisation and the field-theory limit as well as the embedding of the `trivial' contribution ($r=s=0$) in ${\cal N}=8$ supersymmetric compactifications on tori, rather than on asymmetric orbifolds. 

We will keep the discussion as general as possible so that it can be easily adapted to the two relevant asymmetric orbifold cases: $\mathds{Z}_2$ and $\mathds{Z}_3$. In fact the (open string) formulae are valid much more generally \cite{Bianchi:2015vsa,Bianchi:2006nf}

%%%%%%%%%%%%%%%%%%%%%%%%

\subsection{Vertex operators}

The vertex operators (VO's) for the graviton (NS-NS sector) in the canonical $(-1,-1)$ super-ghost picture reads
\begin{equation}
W^{(-1,-1)}= e^{-\varphi-\tilde{\varphi}} \, \psi^\mu h_{\mu \nu} \tilde{\psi}^\nu e^{i \frac{k}{2} (X+\tilde{X})}
\end{equation}
with $k_L^\mu=k_R^\mu = k^\mu/2$, while in the non-canonical (0,0) super-ghost picture it reads
\begin{equation}
W^{(0,0)}= h_{\mu \nu}(\partial X^\mu +i \tfrac{k_\rho}{2} \psi^\rho \psi^\mu)
(\bar{\partial} \tilde{X}^\nu +i \tfrac{k_\sigma}{2} \tilde{\psi}^\sigma \tilde{\psi}^\nu) e^{i \frac{k}{2} (X+\tilde{X})}
\end{equation}

The normalizations of the vertex operators on the sphere as well as on the torus  can be fixed by studying the field-theory limit using 3- and 4-graviton amplitudes from supergravity, that in 4-D read
\begin{align}
\mathcal{M}^{\scalebox{0.6}{tree}}_{\scalebox{0.6}{sugra}}(1^-,2^-,3^+) &= 
\kappa \frac{\langle12\rangle^6}{\langle23\rangle^2 \langle31\rangle^2}\\
\mathcal{M}^{\scalebox{0.6}{tree}}_{\scalebox{0.6}{sugra}}(1^-,2^-,3^+,4^+) &=
-i\, \kappa^2  \frac{\langle 12\rangle^4 [34]^4}{s  t u}
\end{align}
where $\kappa= \sqrt{8\pi G}$, with $G$  the Newton constant and the 3-pt function is defined with complex momenta.

Moreover it is very convenient to exploit chiral factorisation and relate closed-string graviton or gravi-photon amplitudes to open-string gauge boson amplitudes. At tree-level Kawai-Lewellen-Tye (KLT) have gone so far as writing the former as squares of the latter \cite{Kawai:1985xq}, pioneering the so-called double-copy construction. At one-loop and beyond, so far nobody has been able to generalise KLT relations for the integrals but at least the `integrands' can be related .

To this end one writes the (physical) polarization
tensors for gravitons in terms of photon polarization vectors 
\be h^{(2\sigma)}_{\mu\nu} = a_\mu^L
a_\nu^R \quad \rightarrow \quad r_{\mu\nu\rho\sigma} = f^L_{\mu\nu} f^R_{\rho\sigma} \ee where $r_{\mu\nu\rho\sigma}$ denotes the linearised Riemann tensor that on-shell coincides with the Weyl tensor. This will prove useful in order to process and re-cycle open string amplitudes at one-loop with (partially) broken supersymmetry \cite{Bianchi:2015vsa} in the present context. A similar story applies to the graviphotons, as we will see later on.

\subsection{`Chiral' factorisation: closed-strings from open-strings at one-loop}

Exploiting `factorization' of world-sheet correlation functions in closed-string amplitudes one has
\be
{\cal M}_{4} = {1\over K} \sum_{r,s}^{0,K-1}
\int {d^2\tau\over \tau_2^2} \prod_i {d^2z_i\over \tau_2} \tilde\Gamma[^r_s](\tau, \bar\tau)  \Pi_{\rm KN} (s,t)
\mathbb{G}_{4}^L[^r_s](\tau, z_i) \mathbb{G}_{4}^R[^r_s](\bar\tau, \bar{z}_i)
\ee
where \be
\Pi_{\rm KN} (s,t)= \exp[-{\alpha^\prime\over 2}  \sum_{a>b} k_a {\cdot} k_b\, \mathcal{G}(z_{ab}) ]
\ee
 is the one-loop Koba--Nielsen factor and $\tilde\Gamma[^r_s](\tau, \bar\tau)$ is a (modified) lattice sum (with two momentum insertions when gravi-photons are inserted). 

According to \cite{Bianchi:2006nf} the (un-)normalised one-loop amplitude for binary scattering of vector  bosons with helicity\footnote{As usual we consider all momenta as in-coming.} $[--++]$ in (unoriented) D-brane worlds preserving $\calN$ supersymmetry in $D=4$ reads
\begin{multline}\label{OpenAmpNsusy}
\mathcal{A}_{\calN}^{[--++]} = C_{A} \langle12\rangle^2 [34]^2  \int_0^\infty \frac{dT}{T} \int_0^{i T/2} dz_i \Pi_A(s,t) \times\\
\Bigl( 4\calF_\calN +\calE_\calN
\left[\calY(z_{12}){+}\calY(z_{34}){-}\calY(z_{13}){-}\calY(z_{24}){-}\calY(z_{14}){-}\calY(z_{23}) \right]\Bigr)
\end{multline}
where $\tau_\mathcal{A}= i T/2$, $\Pi_A(s,t)= \exp[-{2\alpha^\prime}  \sum_{a>b} k_a \cdot  k_b \mathcal{G}_A(z_{ab}) ]$ denotes the 1-loop (open-string) Koba--Nielsen factor and $\mathcal{G}_A$ the free-boson propagator on the annulus \eqref{annulus_prop}, while
\begin{equation}
\calF_{\mathcal{N}}= {1\over 4} \sum_\alpha c_\alpha {e}_{\alpha-1}^2 \mathcal{Z}_{\alpha}
\quad, \quad
\calE_{\mathcal{N}}= -{1\over 2} \sum_\alpha c_\alpha {e}_{\alpha-1} \mathcal{Z}_{\alpha}
\end{equation}
where $\mathcal{Z}_\alpha$ is the (unintegrated) partition function in the spin structure $\alpha=1,2,3,4$, the signs $c_\alpha$   
\begin{equation}
c_3= 1\; , \qquad c_2=c_4=c_1= -1\; , 
\end{equation}
are dictated by the GSO projection \cite{GreenSchwWitt} and $-e_{\alpha-1}$ are the  values of Weierstrass function at the non-zero half periods \eqref{eiJacobi}.\footnote{Note that in \cite{Bianchi:2015vsa} the $e_\alpha$ where defined as one-half the standard ones that we use here.} 

Last but not least
\begin{equation}
\calY(z) = 2 S(z)^2 - 2 \wp(z)
\end{equation}
is an ubiquitous combination that has no singularity. Henceforth we will often write for short  
$z_{ij}$, $\calY_{ij}$, $S_{ij}$  and $\wp_{ij}$ instead of $z_i{-}z_j$, 
$\calY(z_i{-}z_j)$, $S(z_i{-}z_j)$  and $\wp(z_i{-}z_j)$, respectively.

The maximal helicity violating (MHV) kinematic factor is given by
\bea
t_8 f_1 f_2 f_3 f_4 &=& 4(f_1 f_2 f_3 f_4)+4(f_1 f_2 f_4 f_3)+4(f_1 f_3 f_2 f_4)\CR
&& \quad +4(f_1 f_3 f_4 f_2)+4(f_1 f_4 f_1 f_2) + 4 (f_1 f_4 f_2 f_1) \nn \\ &&\quad - 2(f_1 f_2)( f_3 f_4) - 2(f_1 f_3)
(f_4 f_2) - 2(f_1 f_4)( f_2 f_3) 
\eea
where $f_{\mu\nu} = k_\mu a_\nu - k_\nu a_\mu$ and $(...) = {\rm Tr}(...)$ denotes the trace over 4-vector indices. Choosing for definiteness $[--++]$ helicities one has
\bea
&&t_8 f^-_1 f^-_2 f^+_3 f^+_4  = 4 (f^-_1 f^-_2)( f^+_3 f^+_4) =4  \langle 12 \rangle^2 [34]^2
\eea
since $(f^-_{1,2}f^+_{3,4}) = 0$, while $(f^-_1 f^-_2)=-\langle 12\rangle^2$ and $(f^+_3 f^+_4)=-[34]^2$.

Let us first focus on the R-mover part. Since the orbifold projection only acts by a shift on the lattice of the R-movers, {\it i.e.} preserves all four space-time supersymmetries, only terms with 4 fermion bilinears contribute in the terminology of \cite{Bianchi:2006nf}. After summing over spin structures one simply gets \begin{equation}
\calE_4 =0 
\quad, \quad
\calF_4 = C_A \frac{(2 \pi)^4 \Lambda^{(6)}}{4 K} 
\end{equation}
therefore, dropping the lattice and Koba--Nielsen factor, one simply has  
\begin{equation}
\mathbb{G}_{\rm R}^{[--++]}[^r_s] = \langle 12 \rangle^2 [34]^2
\end{equation}

Similarly for the L-movers we have
\begin{equation}
\mathbb{G}_{\rm L}^{[--++]}[^0_0] = \langle 12 \rangle^2 [34]^2
\end{equation}
in the trivial `untwisted' sector.

In the non-trivial $[^0_1]$ L-mover sector and its `modular orbit', the orbifold projection breaks 1/2 of the original four space-time supersymmetries, thus preserving ${\mathcal{N}}=2$ supersymmetry. The relevant (open-string) functions then are  
\begin{equation}
\calE_2 = C_A \frac{(2 \pi)^2 \Lambda^{(2)} I^{(4)}}{4 K}  
\quad, \quad
\calF_2 = - {1\over 2} \calE_2\, \wp(u) 
\end{equation}
with $u=u_{r,s} = (r\,\tau-s)/K$, depending on the orbifold group ($K=2,3$) and the `sector' $r,s=0,...K-1$, while $\Lambda^{(d)}$ and $I^{(6-d)}$ denote the relevant lattice sums and `intersections', that eventually get replaced by $\Gamma[^0_1]$ and $N^L_{\rm fp}$ the number of `chiral' fixed points.

For completeness let us briefly describe how to derive \eqref{OpenAmpNsusy} in the ${\mathcal{N}}=2$ case of interest here. More details can be found in \cite{Bianchi:2015vsa} or in Appendix \ref{open_details}. Given the form of the vector boson vertex operators, in principle there are five kinds of terms. Those of the form $\langle \partial X^3 \psi\psi \rangle$ or $\langle \partial{X}^4\rangle$, that involve contractions of only one world-sheet fermion bilinear or none, vanish thanks to `normal ordering' of the fermionic current or after summing over the spin structures, because of (residual) space-time supersymmetry. In 'our' case, with the given choice of helicities $[--++]$,  the terms with 3 fermion bilinears $\langle \partial X (\psi\psi)^3 \rangle$ vanishes too, since $(f_1^-f_2^-f_{3,4}^+)=0$ and $(f_{1,2}^-f_3^+f_4^+)=0$. 

Only three contributions then survive: two from connected and disconnected contractions of four fermion bilinears $\langle (\psi\psi)^4\rangle_{\rm conn}$ and  $\langle (\psi\psi)^4\rangle_{\rm disc} = \langle (\psi\psi)^2\rangle\langle (\psi\psi)^2\rangle$
and one from the contraction of 2 fermion bilinears $\langle (\psi\psi)^2\partial {X}^2\rangle$. Retracing the steps in \cite{Bianchi:2006nf}, in light of the simplifications found in \cite{Bianchi:2015vsa}, one finds
\begin{align}
\mathbb{G}^L_\textup{4-bil,conn}[^r_s]&={-}
\sum_\textup{conn} (f_1 f_2 f_3 f_4) 
\Bigl( {\mathcal{E}[^r_s]}(\wp_{13}+\wp_{24}+\omega_{123} \omega_{341}+\omega_{234} \omega_{412})+ 4 \mathcal{F}[^r_s]\Bigr)
\label{G4conn}
\\
\mathbb{G}^L_\textup{4-bil,disc}[^r_s]&=
\sum_\textup{disc} (f_1 f_2) (f_3 f_4)
\Bigl( -\frac{1}{2}\mathcal{E}[^r_s](\wp_{12}+\wp_{34})+\mathcal{F}[^r_s]\Bigr) 
\label{G4disc}
\\
\mathbb{G}^L_\textup{2-bil}[^r_s]&= 
\sum_\textup{pairs} (f_3 f_4) 
{\mathcal{E}[^r_s]} 
a_1 {\cdot} {\cal P}_1\, a_2 {\cdot} {\cal P}_2
\label{G2}
\end{align}
where\footnote{Having stripped off the lattice sum and the Koba--Nielsen factor, $\mathcal{E}[^r_s]=1$ and $\mathcal{F}[^r_s] $ are only kept for easier comparison with the open-string formulae in \cite{Bianchi:2015vsa}.} $\mathcal{E}[^r_s]=1$ and $\mathcal{F}[^r_s] = -\wp(u_{r,s}) /2$ and 
\be
{\cal P}_i = \sum_{j\neq i} k_j S_{ij} \quad , \quad S_{ij} = - \partial_i G_{ij} \quad , \quad 
\omega_{ijk} = S_{ij} + S_{jk} + S_{ki}\; . 
\ee
Note that the terms with $a^-_1a^-_2$ and $a^+_3a^+_4$ have been set to zero, thanks to the freedom of choosing the `reference' spinors to be the same for polarizations with the same helicity. 

In Appendix \ref{open_details} we manipulate the above contributions from the correlators and obtain 
\begin{equation}
\mathbb{G}^{(4)}_L = 
\langle 12 \rangle^2 [34]^2 \,
\mathcal{E}[^r_s] \left[\wp(u_{r,s}) + \frac{1}{8} (\mathcal{Y}_{12}+\mathcal{Y}_{34}-\mathcal{Y}_{13}-\mathcal{Y}_{14}-\mathcal{Y}_{23}-\mathcal{Y}_{24})\right]\; . \label{G4L} 
\end{equation}

Let us notice that this expression first obtained in \cite{Bianchi:2015vsa} matches precisely the one obtained in \cite{Berg:2016wux}, notwithstanding the different approach used there (relaxing momentum conservation at 4-points).\footnote{This is easily checked using $\frac{X_{23,4}}{s_{12} s_{23} } +  \frac{X_{24,3}}{ s_{12} s_{13} } = - f_{34}^{(2)} -  f_{24}^{(2)}- f_{23}^{(2)}$, see (4.40) in \cite{Berg:2016wux}.}

Combining L- and R-movers one finds \footnote{This amplitude was first analysed for the $\mathds{Z}_2$ orbifold in \cite{Bianchi:2010aw}.}
\begin{multline}  \label{M4grav1-loop} \mathcal{M}^{\scalebox{0.6}{1-loop}}_{\scalebox{0.6}{type II}}(1^-,2^-,3^+,4^+) =  i \frac{\alpha^{\prime\, 4}g_{\scalebox{0.5}{4}}^{\, 4} }{2^{7} K}  \langle 12\rangle^4 [ 34]^4  \int_{\cF} \hspace{0mm}\frac{d^2\tau}{\tau_2^{\, 2}}
\prod_{a=1}^4 \int_\Sigma\hspace{-1.8mm}  \frac{d^2z_a}{\tau_2}   \tau_2 \delta^\ord{2}(z_4) e^{-\frac{\alpha^\prime}{2} \! \sum_{a>b}\! \mathcal{G}(z_a-z_b) k_a \cdot  k_b}  \\
\times\Biggl( 4\pi^2 \Gamma_{\sLambda_{6,6}} +\hspace{-3mm} \sum_{\substack{r , s\, {\rm mod} K\\ (r,s)\ne (0,0)} }\hspace{-3mm}  (2\sin \! \tfrac\pi K)^2  \Gamma_{\sLambda_{2,2}\oplus \mathds{H}(K)}[^s_r]  \Bigl(  \wp(\tfrac{r + s \tau}{ K}) + \tfrac18 \bigl( \mathcal{Y}_{12} + \mathcal{Y}_{34}- \mathcal{Y}_{13}- \mathcal{Y}_{14} - \mathcal{Y}_{23}- \mathcal{Y}_{24}\bigr)  \Bigr) \Biggr)   \end{multline}
where $\Gamma_{L}[^s_r] = \tau_2^3 \Lambda_L[^s_r]$ is the Narain partition function associated to the lattice $L$, with $\Lambda_L[^s_r]$ defined in \eqref{UntwistNarain} and \eqref{TwistNarain}. The normalisation has been fixed such that it reproduces the supergravity one-loop amplitude \cite{Bern:2011rj} in the tropical limit, see Appendix \ref{Sugra1loop}. Using the property that $\mathcal{Y}(z)$ and $\mathcal{G}(z)$ are modular invariant and applying the same reasoning as in Section \ref{CharZK}, one can write the amplitude as  an integral over the $\Gamma_0(K)$ fundamental domain 
\begin{multline} \label{M4grav1-loopFK}
 \mathcal{M}^{\scalebox{0.6}{1-loop}}_{\scalebox{0.6}{type II}}(1^-,2^-,3^+,4^+) =  i \frac{\alpha^{\prime\, 4}g_{\scalebox{0.5}{4}}^{\, 4} (2\sin \! \tfrac\pi K)^2 }{2^{7}}  \langle 12\rangle^4 [ 34]^4  \\
 \int_{\cF_K} \hspace{-2mm}\frac{d^2\tau}{\tau_2^{\, 2}}  \,\Gamma_{\sLambda_{1,1}\oplus \sLambda_{1,1}[K]\oplus \mathds{H}(K)} \prod_{a=1}^4 \int_\Sigma\hspace{-1.8mm}  \frac{d^2z_a}{\tau_2}   \tau_2 \delta^\ord{2}(z_4) e^{-\frac{\alpha^\prime}{2} \! \sum_{a>b}\! \mathcal{G}(z_a-z_b) k_a \cdot  k_b}  \\
\hspace{-2mm}\times \Bigl(  \wp(\tfrac1 K) + \tfrac18 \bigl( \mathcal{Y}_{12} + \mathcal{Y}_{34}- \mathcal{Y}_{13}- \mathcal{Y}_{14} - \mathcal{Y}_{23}- \mathcal{Y}_{24}\bigr)  \Bigr) \; .  \end{multline}
Albeit the explicit dependence in the dilaton field through $g_{\scalebox{0.5}{4}}^{\, 4}$, this amplitude only depends on the Narain moduli through the Narain partition function of the lattice \eqref{LatticeUntwisted}
\be \Gamma_{\sLambda_{1,1}\oplus \sLambda_{1,1}[K]\oplus \mathds{H}(K)}(t,\bar t) = \Gamma_{\mathds{F}_2(K)}(t,\bar t)  =  \tau_2^3  \sum_{{\mathcal{Q}} \in \mathds{F}_2(K) } e^{\pi i \tau p_L({\mathcal{Q}})^2 - \pi i \bar \tau p_R({\mathcal{Q}})^2 } \; . \ee
This expression is manifestly invariant under the automorphism group of the perturbative lattice \eqref{PertF2}. This is the starting point for the low-energy expansion and extraction of the lowest higher-derivative terms a.k.a. threshold corrections.

%%%%%%%%%%%%%%%%%%%%%%%%
\subsection{${\cal R}^4$ and $D^2{\cal R}^4$ couplings}
\label{R41loop}
In this section we discuss the ${\cal R}^4$ and $D^2{\cal R}^4$ couplings, by which we mean their respective completions to supersymmetry invariants discussed in Sections \ref{22Harmonic} and \ref{20Harmonic}.  In ${\cal N} = 6$ supergravity, obtained as asymmetric
orbifolds of  tori, tree-level scattering amplitudes of untwisted
states such as gravitons or graviphotons are identical to the corresponding
amplitudes in the parent ${\cal N} = 8$ theory. This implies that the tree-level effective action for the massless field is the truncation of the maximally supersymmetric tree-level effective action to the fields of ${\cal N} = 6$ supergravity. 

This implies that the ${\cal R}^4$ term receives a tree-level contribution with a `famous' $2\zeta(3)$ coefficient, while no such a term appears for $D^2{\cal R}^4$ (since $s+t+u=0$) nor for ${\cal F}^2 {\cal R}^2$ (since this violates the continuous $SO(2,6)$ symmetry at tree-level). 

The situation changes at one-loop where even `projections' of the untwisted sector ($s=0$, $r\neq 0$) contribute to the threshold functions for other higher-derivative couplings. In particular the one-loop threshold corrections to ${\cal R}^4$ and $D^2{\cal R}^4$ couplings can be extracted from the 4-graviton amplitude \eqref{M4grav1-loopFK} computed above. The expansion of \eqref{M4grav1-loopFK} gives 
\bea  \label{M4grav1-loopWilson}
&&  \mathcal{M}^{\scalebox{0.6}{1-loop W}}_{\scalebox{0.6}{type II}, L}(1^-,2^-,3^+,4^+) \\
&=&  i \frac{\alpha^{\prime\, 4}g_{\scalebox{0.5}{4}}^{\, 4} (2\sin \! \tfrac\pi K)^2 }{2^{7}}  \langle 12\rangle^4 [ 34]^4  
 \int_{\cF_K^L} \hspace{-2mm}\frac{d^2\tau}{\tau_2^{\, 2}}  \,\Gamma_{\sLambda_{1,1}\oplus \sLambda_{1,1}[K]\oplus \mathds{H}(K)} \prod_{a=1}^4 \int_\Sigma\hspace{-1.8mm}  \frac{d^2z_a}{\tau_2}   \tau_2 \delta^\ord{2}(z_4)  \CR
&& \qquad \times \Biggl( \sum_{n=0}^\infty  \frac{ ( \frac{\alpha^\prime}{4} )^n}{n!} \Bigl( ( \mathcal{G}_{12} + \mathcal{G}_{34})  s +( \mathcal{G}_{14} + \mathcal{G}_{23} ) t +  ( \mathcal{G}_{13} + \mathcal{G}_{24} ) u  \Bigr)^n  \Biggr) \CR
&& \qquad \qquad \times  \Bigl(  \wp(\tfrac1 K) + \tfrac18 \bigl( \mathcal{Y}_{12} + \mathcal{Y}_{34}- \mathcal{Y}_{13}- \mathcal{Y}_{14} - \mathcal{Y}_{23}- \mathcal{Y}_{24}\bigr)  \Bigr) \CR
& =&   i \frac{\alpha^{\prime\, 4}g_{\scalebox{0.5}{4}}^{\, 4} (2\sin \! \tfrac\pi K)^2 }{2^{7}}  \langle 12\rangle^4 [ 34]^4  
 \int_{\cF_K^L} \hspace{-2mm}\frac{d^2\tau}{\tau_2^{\, 2}}  \,\Gamma_{\sLambda_{1,1}\oplus \sLambda_{1,1}[K]\oplus \mathds{H}(K)} \Bigl( \wp(\tfrac1 K) + \frac{\alpha^\prime s}{8} \int_\Sigma\frac{d^2z}{\tau_2} \mathcal{G}(z) \mathcal{Y}(z) + \mathcal{O}(\alpha^{\prime 2}) \Bigr)\nonumber   \eea
where one peruses $s+t+u=0$, \eqref{Gzero} and \eqref{Yzero} to simplify the elliptic integrals. Here $L>\hspace{-1.8mm} > 1$ is a cutoff of the fundamental domain $\cF_K^L =\cup_{\gamma \in P\hspace{-0.2mm} S\hspace{-0.2mm}L(2,\mathds{Z})/ \Gamma_0(K)} \cF^L\bigr|_\gamma $ with  $\cF^L$ the component of the standard fundamental domain with $\tau_2< L$. The coupling is then defined by the `renormalised' coupling in which one subtracts the divergence in $L$. When the divergence is power-low this renormalisation is unambiguous, and we shall simply drop the dependence in $L$. In our convention this gives the $\mathcal{R}^4$ perturbative coupling 
\be \gFour^4 \mathcal{E}_\grad{0}{0}{0}^{\scalebox{0.6}{pert}} = \frac{2\zeta(3)}{\gFour^2} + \frac{(2\sin \frac{\pi}{K})^2}{2\pi}  \int_{\cF_K} \hspace{-1.2mm}\frac{d^2\tau}{\tau_2^{\, 2}}  \wp(\tfrac1 K)  \Gamma_{\sLambda_{1,1}\oplus \sLambda_{1,1}[K]\oplus \mathds{H}(K)}  \; , \ee
where we define 
\bea&&\frac{(2\sin \frac{\pi}{K})^2}{2\pi}  \int_{\cF_K} \hspace{-1.2mm}\frac{d^2\tau}{\tau_2^{\, 2}}  \wp(\tfrac1 K)\Gamma_{\sLambda_{1,1}\oplus \sLambda_{1,1}[K]\oplus \mathds{H}(K)}   \CR
&=&   \lim_{L\rightarrow \infty} \Biggl( \frac{(2\sin \frac{\pi}{K})^2}{2\pi}  \int_{\cF_K^L} \hspace{-1.2mm}\frac{d^2\tau}{\tau_2^{\, 2}}  \wp(\tfrac1 K)\Gamma_{\sLambda_{1,1}\oplus \sLambda_{1,1}[K]\oplus \mathds{H}(K)}   - \frac{5\pi}{6} L^2   \Biggr)  \; .  \eea
Note that the divergent term in $L$ does not depend on $K$, which follows from the property that it matches by construction the corresponding divergence of the supergravity amplitude in Schwinger parameter space regularisation. In practice one can simply forget about this divergence because supersymmetry Ward identities \eqref{WardoneloopR4} do not allow for the freedom of adding a constant.

For $K=2$ we have 
\be \frac{(2\sin \frac{\pi}{2})^2}{2\pi}  \wp(\tfrac12) = \frac{4\pi}{3} \Lambda_{D_4} = \frac{4\pi}{3} \bigl( 2 E_2(2\tau) - E_2(\tau)\bigr)\; , \label{P2E2}  \ee
and for $K=3$
\be \frac{(2\sin \frac{\pi}{3})^2}{2\pi}  \wp(\tfrac13) = \frac{3\pi}{2} \Lambda_{A_2\oplus A_2} = \frac{3\pi}{4} \bigl( 3 E_2(3\tau) - E_2(\tau)\bigr)\; .  \label{P3E2}  \ee
As required by supersymmetry \eqref{R4twotwo}, we verify that $\wp(\tfrac1 K)$ is a holomorphic modular form. Moreover $\wp(\tfrac1 K)$ is precisely the holomorphic modular form that appears in the helicity supertrace $B^{\mathds{Z}_K}_8$ in \eqref{HelicityBn}, consistently with the property that only BPS states preserving at least one-third of the supersymmetries contribute to the $\mathcal{R}^4$ coupling.

The $D^2 \mathcal{R}^4$ coupling involves the integral 
\be \frac{1}{4\pi}   \int_\Sigma\frac{d^2z}{\tau_2} \mathcal{G}(z) \mathcal{Y}(z)  = - \frac{\pi^2}{45} E_{1,2}(\tau,\bar\tau) \; , \label{GYintegral} \ee
that we compute in Appendix \ref{calI}, with the definition \eqref{Eisensteinsw} for the weight 2 non-holomorphic Eisenstein series $E_{1,2}$. The corresponding coupling at one-loop is therefore 
\bea \mathcal{E}_\grad{1}{0}{0}^{\scalebox{0.6}{1-loop}} &=& -(2\sin \tfrac{\pi}{K})^2  \frac{\pi^2}{45} \int_{\cF_K} \hspace{-1.2mm}\frac{d^2\tau}{\tau_2^{\, 2}}    E_{1,2}(\tau,\bar\tau) \Gamma_{\sLambda_{1,1}\oplus \sLambda_{1,1}[K]\oplus \mathds{H}(K)} \label{ED2R4Regul}  \\
&=&  \lim_{L\rightarrow \infty} \Biggl( -(2\sin \tfrac{\pi}{K})^2  \frac{\pi^2}{45} \int_{\cF_K^L} \hspace{-1.2mm}\frac{d^2\tau}{\tau_2^{\, 2}}    E_{1,2}(\tau,\bar\tau) \Gamma_{\sLambda_{1,1}\oplus \sLambda_{1,1}[K]\oplus \mathds{H}(K)}   + \frac{2\pi^2}{135} L^2 - \frac{\zeta(3)}{\pi} \log L \Bigr) \; . \nonumber \eea 
The power low divergence depends on the choice of regulator and is not physically relevant, but the logarithmic divergence corresponds to the supergravity form factor logarithmic divergence 
\be \mathcal{M}_4\Bigl[ t_8 t_8 \mathcal{R}^4(k=0) \Bigr]^{\scalebox{0.6}{1-loop} } \sim \frac{\kappa^2}{8\pi^2} \log L\,  \mathcal{M}_4\Bigl[ t_8 (st^s_8 + t t_8^t + u t_8^u) \mathcal{R}^4 (k=0) \Bigr]^{\scalebox{0.6}{tree} } \; , \ee
that affects the Callan-Symanzik equation independently of the renormalisation scheme. We confirm in particular the logarithmic divergence anticipated in \eqref{LogDivD2R4}, and the non-holomorphic Eisenstein series $E_{1,2}$ indeed verifies the Laplace equation \eqref{E12LaplaceEquation}.

\subsection{$\bar{\cal F}^2 {\cal R}^2$ at one-loop}

In the following, we will focus on the two-graviphoton two-graviton amplitude corresponding to the 1/2 BPS supersymmetry invariant described in Section \ref{secF2R2}. This amplitude vanishes in $\mathcal{N}=8$ and therefore does not receive contributions from the trivial orbifold sector $(r,s)= (0,0)$.

The graviphoton vertex in the $(0,0)$ superghost picture reads
\begin{equation}
V^\gra{0}{0}_a = a_{\mu} (\partial{X}^{5+i6} + i \tfrac{k_\rho}{2}\psi^\rho
\psi^{5+i6}) (\bar\partial\tilde{X}^\mu + i \tfrac{k_\sigma}{2}\tilde\psi^\sigma \tilde\psi^\mu )
e^{i\tfrac{k}{2} (X+\tilde{X})}\; ,
\end{equation}
with the convention that $X^{5 + i 6} = \frac{ X^5 + i X^6}{\sqrt{2}}$. 
For a negative polarisation, this is the graviphoton $\bar{F}_{\dot{\alpha}\dot{\beta}}$ of $U(1)$ weight $w=1$, while for a positive polarisation this is the component $F_{\alpha\beta\hspace{0.2mm}  56}$ of the 15 graviphotons $F_{\alpha\beta ij}$ of $U(1)$ weight $w=- \frac13$ in \eqref{LinearisedFields}.\footnote{Note that $i=5,6$ correspond to the indices of  the R-symmetry $SU(6) \supset S(U(4)\times U(2))$, while $I = 5,6$ in $X^I$ and $\psi^I$ correspond to the indices of $SO(6) \supset SO(4)\times SO(2)$ where $SO(4)$ of $T^4$ is  broken by the orbifold.}
The `charge' conservation associated to the rotation in the plan $56$ implies that there is no contribution at tree-level. Also at one-loop terms with one or two insertions of $\psi^{5+i6}$ vanish  
 \be \langle{{\partial}}X^{5+i6} k_2\psi
\psi^{5+i6} k_3\psi \psi^{\mu_3} k_4\psi \psi^{\mu_4} \rangle = 0 \quad , \quad 
\langle k_1\psi
\psi^{5+i6} k_2\psi
\psi^{5+i6} k_3\psi \psi^{\mu_3} k_4\psi \psi^{\mu_4} \rangle = 0
\ee since there is no way to contract $\psi^{5+i6}$. More generally there is no contraction between bilinear in fermions $k \psi \psi^{5 + i 6}$, and two fermion bilinears only produce the term
\cite{Bianchi:2006nf} \be
\big\langle{{\partial}}X^{5+i6} {{\partial}}X^{5+i6} k_{3 \rho} \psi^\rho  \psi^{\mu_3} k_{4 \sigma}\psi^\sigma 
\psi^{\mu_4} \big\rangle = p_L^{5+i6} p_L^{5+i6} \bigl( k_3 \cdot k_4 \eta^{\mu_3\mu_4} -
k_3^{\mu_4} k_4^{\mu_3} \bigr)   S_\alpha(z_3-z_4)^2 \ee 
for even spin structures, where $p_L^{5+i6}$ denotes the zero-mode momentum component
\be p_L^{5 + i 6} = \frac{p_{L 5}({\mathcal{Q}}) + i p_{L 6}({\mathcal{Q}})}{\sqrt{2}} \; , \ee
for the charge ${\mathcal{Q}}$ in the perturbative lattice. The Lorentz structure eventually boils down to $(f_3^L f_4^L)$. 

For the odd spin structure we need the $(-1,0)$ superghost picture vertex operator 
\begin{equation}
V^\gra{-1}{0}_a = a_{\mu} \psi^{5+i6} e^{- \varphi}  (\bar\partial\tilde{X}^\mu + i \tfrac{k_\sigma}{2}\tilde\psi^\sigma \tilde\psi^\mu )
e^{i\tfrac{k}{2} (X+\tilde{X})}\; , 
\end{equation}
and the picture changing operator 
\be P_L^\ord{1} = G_{MN}  \partial X^M \psi^N e^\varphi + \ldots \; . \ee
This gives the non-zero contribution  
\be  \bigl\langle   \partial X^M(z_0) G_{MN} \psi^N(z_0)   \psi^{5+i 6}(z_1)  \partial X^{5 + i 6}(z_2)  f_{3} \psi \psi(z_3) f_{4} \psi \psi(z_4) \bigr\rangle = i   (p_L^{5 + i 6})^2 \varepsilon^{\mu\nu\sigma\rho} f_{3\mu\nu} f_{4\sigma\rho} \ee
where the six fermions give the six-dimensional Levi-Civita tensor and the bosons the zero mode. Here we have not been careful with the normalisation, but one knows from supersymmetry that the correct combination is 
\be ( p_L^{5 + i 6})^2 ( f_3^L f_4^L  + \tfrac i 2 \varepsilon f_3^L f_4^L   )  \ee
such that only the positive helicity amplitude is non zero.\footnote{The corresponding open-string amplitude is a correction to the $\mathcal{N}=2$ prepotential, and the holomorphic part only couples to the positive helicity in the effective Lagrangian Re$[ \cF^{\prime\prime}(\phi) ( F^2 + \tfrac i 2 \varepsilon F^2)]$.}

The complete amplitude eventually combines the right and left sectors to give the kinematic factor for the $[--++]$ helicities 
\be ( f_3^L f_4^L  + \tfrac i 2 \varepsilon f_3^L f_4^L   ) \times t_8 f_1^Rf_2^Rf_3^Rf_4^R = -8  \langle 12 \rangle^2 [34]^4 \ee
corresponding to $ \bar{\cal F}^2 {\cal R}^2$, while it vanishes for the helicities $[++--]$, as required by supersymmetry. Factorising out this kinematic structure, one obtain the scalar amplitude
\begin{multline}  \mathcal{A}^{\scalebox{0.6}{1-loop}}_{\scalebox{0.6}{type II}} = - i  \pi   \alpha^{\prime 3}  \int_{\cF_K} \hspace{0mm}\frac{d^2\tau}{\tau_2^{\, 2}}
\prod_{a=1}^4 \int_\Sigma\hspace{-1.8mm}  \frac{d^2z_a}{\tau_2}   \tau_2 \delta^\ord{2}(z_4) e^{-\frac{\alpha^\prime}{2} \! \sum_{a>b}\! \mathcal{G}(z_a-z_b) k_a \cdot  k_b}  \\
\times\Biggl( 8 \pi  (2\sin \! \tfrac\pi K)^2 \frac{1}{K} \sum_{\substack{r , s\, {\rm mod} K\\ (r,s)\ne (0,0)} }\hspace{-3mm}   \Gamma_{\sLambda_{2,2}\oplus \mathds{H}(K)}[^s_r][ (p_L^{5 + i 6})^2 ] \Biggr) \; . \end{multline}
Once again one can simplify this amplitude as an integral over the $\Gamma_0(K)$ fundamental domain 
\be \mathcal{A}^{\scalebox{0.6}{1-loop}}_{\scalebox{0.6}{type II}} = - i 8 \pi^2  (2\sin \! \tfrac\pi K)^2    \alpha^{\prime 3}  \int_{\cF_K} \hspace{0mm}\frac{d^2\tau}{\tau_2^{\, 2}}
\prod_{a=1}^3 \int_\Sigma\hspace{-1.8mm}  \frac{d^2z_a}{\tau_2}   e^{-\frac{\alpha^\prime}{2} \! \sum_{a>b}\! \mathcal{G}(z_a-z_b) k_a \cdot  k_b}   \Gamma_{\mathds{F}_2(K)}[ (p_L^{5 + i 6})^2 ]  \; , \ee
where 
\be \Gamma_{\mathds{F}_2(K)}[ (p_L^{5 + i 6})^2 ] = \tau_2^3 \sum_{{{\mathcal{Q}}}\in \mathds{F}_2(K)} \frac{( p_{L5}({{\mathcal{Q}}}) + i p_{L6}({{\mathcal{Q}}}))^2}{2}  e^{ i\pi \tau p_{L}({{\mathcal{Q}}})^2 - i \pi \bar \tau p_R({{\mathcal{Q}}})^2 }\; . \ee
From this expression one deduces directly the 1/2 BPS coupling 
\be \cF^{\scalebox{0.6}{1-loop}}_\gra{0}{0} = 8 \pi  (2\sin \! \tfrac\pi K)^2  \int_{\cF_K} \hspace{0mm}\frac{d^2\tau}{\tau_2^{\, 2}} \Gamma_{\mathds{F}_2(K)}[ (p_L^{5 + i 6})^2 ] \; . \ee
Note that there is no need to regularise the integral over $\cF_K$ in this case since the corresponding supergravity amplitude vanishes and the integral converges. At the next order in $\alpha^{\prime 2}$ one gets the integral
\be   \int_{\Sigma}\frac{d^2 z}{\tau_2} \mathcal{G}(z)^2  = \frac{\pi^2}{45} E_{2,0}(\tau,\bar \tau) \; ,  \ee
where $E_{2,0}$ is the real analytic Eisenstein series \eqref{Eisensteinsw}. The corresponding coupling is
 \be \cF^{\scalebox{0.6}{1-loop}}_\gra{1}{0} = \frac{4 \pi^3}{45}  (2\sin \! \tfrac\pi K)^2  \int_{\cF_K} \hspace{0mm}\frac{d^2\tau}{\tau_2^{\, 2}} E_{2,0}(\tau,\bar \tau)\,  \Gamma_{\mathds{F}_2(K)}[ (p_L^{5 + i 6})^2 ] \; . \ee
This coupling is related to $\cE_\grad{1}{0}{0}$ by supersymmetry as explained in Section \eqref{20Harmonic}. Applying the $SO(2,6)$ invariant operator $\det \cD$ of $U(1)$ weight two to $\cE_\grad{1}{0}{0}$ one obtains the coupling $ \cF^{\scalebox{0.6}{1-loop}}_\gra{1}{0}$ as follows 
\bea &&\varepsilon^{abcd} \cD_{ab} \cD_{cd}\cE^{\scalebox{0.6}{1-loop}}_\grad{1}{0}{0}\CR
&=& -  \frac{ \pi^2(2\sin \! \frac\pi K)^2}{45}    \int_{\cF_K} \hspace{-2mm}\frac{d^2\tau}{\tau_2^{\, 2}} E_{1,2}(\tau,\bar \tau)\,  \tau_2^3  \hspace{-3mm} \sum_{{\mathcal{Q}} \in \mathds{F}_2(K) }  \varepsilon^{abcd} \cD_{ab} \cD_{cd} e^{\pi i \tau p_L({\mathcal{Q}})^2 - \pi i \bar \tau p_R({\mathcal{Q}})^2 }  \CR
&=& -  \frac{4 \pi^3(2\sin \! \frac\pi K)^2}{45}  \int_{\cF_K} \hspace{-2mm}\frac{d^2\tau}{\tau_2^{\, 2}} E_{1,2}(\tau,\bar \tau)\,  \tau_2^4  \hspace{-3mm} \sum_{{\mathcal{Q}} \in \mathds{F}_2(K) }   (p_L^{5+i6})^2 \bigl( 2\pi \tau_2 p_R^2 - 3 \bigr)  e^{\pi i \tau p_L({\mathcal{Q}})^2 - \pi i \bar \tau p_R({\mathcal{Q}})^2 } 
\CR
&=& -  \frac{4 \pi^3(2\sin \! \frac\pi K)^2}{45}  \int_{\cF_K} \hspace{-2mm} d^2\tau E_{1,2}(\tau,\bar \tau) \,2 i \bar \partial_\tau  \Biggl( \tau_2^3 \hspace{-3mm}  \sum_{{\mathcal{Q}} \in \mathds{F}_2(K) }   (p_L^{5+i6})^2  e^{\pi i \tau p_L({\mathcal{Q}})^2 - \pi i \bar \tau p_R({\mathcal{Q}})^2 } \Biggr)\CR
&=& -  \frac{4 \pi^3(2\sin \! \frac\pi K)^2}{45}  \int_{\cF_K} \hspace{-2mm}\frac{d^2\tau}{\tau_2^{\, 2}} E_{2,0}(\tau,\bar \tau)\,  \tau_2^3  \hspace{-3mm} \sum_{{\mathcal{Q}} \in \mathds{F}_2(K) }  e^{\pi i \tau p_L({\mathcal{Q}})^2 - \pi i \bar \tau p_R({\mathcal{Q}})^2 }\CR
&=& -  \cF^{\scalebox{0.6}{1-loop}}_\gra{1}{0}\; .
\eea
Note that we did not take care of the regularisation of $\cE^{\scalebox{0.6}{1-loop}}_\grad{1}{0}{0}$ in this computation because all divergent terms cancel thanks to the insertion of the left momentum $p_L$.

%%%%%%%%%%%%%%%%%%%%%%%%
\subsection{Minimal theta series}
\label{MinTheta}
In this section we prove that the $\bar \cF^2 \cR^2$ coupling $\cF_\gra{0}{0}^{\scalebox{0.6}{1-loop}}$ can be written as a rank one theta series for $SO(2,6)$ defined as the group of four by four symplectic matrices over the quaternions \cite{Krieg}. For $K=2$ and $3$  one finds  
{\allowdisplaybreaks \bea &&  \cF_\gra{0}{0}^{\scalebox{0.6}{1-loop}} \CR
&=& \frac{4\pi K}{K-1} \int_{\cF_K} \hspace{-1mm} \frac{ d^2 \tau}{\tau_2^{\; 2}} \tau_2^{\; 3} \hspace{-2mm}  \sum_{{{\mathcal{Q}}}\in \mathds{F}_2(K)} \frac{( p_{L5}({{\mathcal{Q}}}) + i p_{L6}({{\mathcal{Q}}}))^2}{2}  e^{ i\pi \tau p_{L}({{\mathcal{Q}}})^2 - i \pi \bar \tau p_R({{\mathcal{Q}}})^2 }\CR
&=& \frac{2\pi K \RS}{K-1} \int_{\cF_K}  \hspace{-1mm}\frac{ d^2 \tau}{\tau_2^{\; 2}} \tau_2^{\; \frac52}\hspace{-2mm} \sum_{(m,n)\in \mathds{Z}^2} e^{-\frac{\pi}{\tau_2} \Rs^2 | m + n \tau |^2 } \CR
&& \hspace{-4mm}\times\hspace{-4mm}  \sum_{Q\in \mathds{M}_2(K)}\hspace{-4mm} e^{2\pi i m( Q + \frac12 na ) \cdot a} \Bigl({\scriptstyle p_{L5}(Q+na)^2 - \frac{1}{4\pi \tau_2} + \frac{ \sqrt{2} \Rs (m + \bar \tau n)}{\tau_2} p_{L5}(Q+na)+ \frac{ \Rs^2 (m+ \bar \tau n)^2}{2\tau_2^{\; 2} }}\Bigr) e^{ i\pi \tau p_{L}(Q+na)^2 - i \bar \tau p_R(Q+na)^2 }\nonumber \\ 
&=&  \frac{2\pi K \RS}{K-1} \int_{\cF_K} \frac{ d^2 \tau}{\tau_2^{\; 2}} \tau_2^{\; \frac52}   \sum_{Q\in \mathds{M}_2(K)} \Bigl(  p_{L5}(Q)^2 - \frac{1}{4\pi \tau_2} \Bigr)   e^{ i\pi \tau p_{L}(Q)^2 - i \bar \tau p_R(Q)^2 } \CR
&&+ \frac{4\pi K \Rs}{K-1} \int_0^\infty \hspace{-2mm} d\tau_2  \tau_2^{\frac12}  \int_{-\frac12}^{\frac12} d\tau_1 \sum_{m\ge 1}   \sum_{Q\in \mathds{M}_2(K)}\Bigl({\scriptstyle  p_{L5}(Q)^2 - \frac{1}{4\pi \tau_2} + \frac{ \sqrt{2} \Rs m }{\tau_2} p_{L5}(Q)+\frac{ (\Rs m)^2}{2\tau_2^{\; 2} }  }\Bigr) \CR
&& \hspace{80mm} \times e^{-\frac{\pi}{\tau_2} (\RS m)^2  +i\pi \tau p_{L}(Q)^2 - i \bar \tau p_R(Q)^2 +2\pi i m  Q \cdot a}\CR
&&- \frac{4\pi  \Rs}{K(K-1)} \int_0^\infty \hspace{-2mm}d\tau_2  \tau_2^{\frac12} \int_{-K}^{K} \hspace{-2mm}d\tau_1 \sum_{m\ge 1}   \sum_{Q\in \mathds{M}_2(K)^*}\Bigl({\scriptstyle  p_{L5}(Q)^2 - \frac{1}{4\pi \tau_2} + \frac{ \sqrt{2} \Rs m }{\tau_2} p_{L5}(Q)+\frac{ (\Rs m)^2}{2\tau_2^{\; 2} } }\Bigr) \CR
&& \hspace{80mm} \times e^{-\frac{\pi}{\tau_2} (\Rs m)^2+  i\pi \tau p_{L}(Q)^2 - i \bar \tau p_R(Q)^2 +2\pi i m  Q \cdot a}\nonumber \\
&=&  - \frac{\pi \RS^2}{3}+\frac{4\pi K \RS^2}{K-1} \sum_{\substack{Q\in \mathds{M}_2(K) \\ Q^2=0}}^\prime \biggl( \sum_{ \substack{ d\ge 1  \\ \frac{Q}{d} \in \mathds{M}_2(K)}} d \biggr)  \Bigl( 1+ \frac{ p_{L5}(Q)}{|p_{L5}(Q)|} \Bigr) e^{- 2\pi \Rs \sqrt{ 2 p_{L5}(Q)^2}  + 2 \pi i Q \cdot a} \CR
&& - \frac{4\pi  \RS^2}{K-1} \sum_{\substack{Q\in \mathds{M}_2(K)^{*} \\ Q^2=0}}^\prime \biggl( \sum_{ \substack{ d\ge 1  \\ \frac{Q}{d} \in \mathds{M}_2(K)^{*}}} d \biggr)  \Bigl( 1+ \frac{ p_{L5}(Q)}{|p_{L5}(Q)|} \Bigr) e^{- 2\pi \Rs \sqrt{ 2 p_{L5}(Q)^2}  + 2 \pi i Q \cdot a} \CR
&=& -\frac{\pi \RS^2}{3} \Biggl( 1-\frac{24}{K-1}  \sum_{\substack{Q\in \mathds{M}_2(K)^{*} \\ Q^2=0\\ p_{L5}(Q)>0}} \biggl( K\hspace{-2mm}\sum_{ \substack{ d\ge 1  \\ \frac{Q}{d} \in   \mathds{M}_2(K) }} \hspace{-2mm} d- \hspace{-2mm} \sum_{ \substack{ d\ge 1  \\ \frac{Q}{d} \in  \mathds{M}_2(K)^{*} }} \hspace{-2mm} d \biggr) e^{- 2\pi \Rs \sqrt{ 2} p_{L5}(Q) + 2 \pi i Q \cdot a} \Biggr) \; . 
\eea}
 In the first step we used Poisson summation formula. In the second we used the orbit method to unfold the domain of integration,  with the two orbits of $\Gamma_0(K)$ of doublet $(m,n)\in \mathds{Z}^2$ depending of the congruence of  $n / {\rm gcd}(m,n)$ modulo $K$.  In the third step  we used integration by part to show that the first term vanishes and to simplify the two others using that the insertion of $p_{L5}(Q)^2 - \frac{1}{4\pi \tau_2}$ can be written as the derivative $\frac{1}{i\pi} (  \partial_\tau +\frac{i}{4\tau_2}) $. In the last step we absorbed $m$ in $Q$ and carried out the integrals. 
 
 The condition $p_{L5}(Q)>0$  is moduli independent because $p_{L5}(Q)\ne 0$ for all Narain moduli. For $Q$ defined as a two by two Hermitian matrix $Q = \big(\colvec[0.8]{ m  & q \\ q^*& n }\big) $ over the quaternions $\mathds{H}(K)$ as in \eqref{pLpRFive}, $Q^2=2 \det Q= 0$ implies that $mn = |q|^2 \ge 0$ and $p_{L5}(Q)> 0$  reduces to the condition that ${\rm tr} \, Q = m+n> 0$. Equivalently one can write that $Q\ge 0$ as a matrix. One can therefore conclude that only 1/2 BPS states of a given orientation, {\it i.e.} both $m$ and $n$ positive,  contribute to this coupling, and the complex conjugate coupling will get contributions from the 1/2 BPS states with the opposite orientation, with $m$ and $n$ both negative. 

Using the parametrisation of $SO^*(8)/U(4)= SO(2,6)/SO(2) \times SO(6)$ in terms of a complex Hermitian matrix over the quaternions $t = a + i  \RS v^\dagger v$, one finds  
 \be - 2\pi \RS \sqrt{2} p_{L5}(Q)  + 2 \pi i Q \cdot a = 2\pi i {\rm tr}  Q  ( a + i \RS v^\dagger v ) = 2\pi i \tr\, Q t\; ,  \ee
and the perturbative threshold function can be written as 
 \bea \cF_\gra{0}{0}^{\scalebox{0.6}{1-loop}} &=& -\frac{\pi \det[t_2]}{3}\biggl( 1+\frac{24}{K-1}  \sum_{\substack{Q\in \mathds{M}_2(K)^{*} \\ Q>0 ,\,  Q\times Q=0}} \biggl( \hspace{-1mm} \sum_{ \substack{ d\ge 1  \\ \frac{Q}{d} \in  \mathds{M}_2(K)^{*}  }} \hspace{-1mm} d-\hspace{-2mm}\sum_{ \substack{ d\ge 1  \\ \frac{Q}{Kd} \in \mathds{M}_2(K)^{*}  }} \hspace{-2mm}K d \biggr)  e^{2\pi i {\rm tr}\,  t \, Q}\biggr) \ , \CR
 &=&  -\frac{\pi \det[t_2]}{3}\biggl( 1+\frac{24}{K-1}  \sum_{\substack{Q\in \mathds{M}_2(K)^{*} \\ Q>0 ,\,  Q\times Q=0}} \biggl( \hspace{-1mm} \sum_{ \substack{ d\ge 1   \\ d\ne 0 {\rm \, mod\, }K \\ \frac{Q}{d} \in  \mathds{M}_2(K)^{*}  }} \hspace{-1mm} d \biggr)  e^{2\pi i {\rm tr}\,  t \, Q}\biggr) \eea
 where we used that a matrix $Q\in \mathds{M}_2(K)$ satisfying $\det Q=0$ must be in $K\mathds{M}_2^{*}  $ because $\det Q = mn -|q|^2=0$ implies $q \in \alpha \mathds{H}(K)$ for $m=0$ mod $K$. One can always write a degenerate Hermitian matrix as
 \be  Q = \left(\begin{array}{cc} m \; & q \\ q^* \; & n\end{array}\right)  =\left(\begin{array}{c}  q_1  \\  q_2\end{array}\right)  (q_1^*,q_2^*)    \ , \ee
 with $q_1 \in \mathds{H}(K)$ and $q_2\in \frac{1}{\alpha} \mathds{H}(K)$. 
The greatest common divisor of $Q$ in $\mathds{M}_2(K)^{*}$ is the norm square $|{\rm gcrd}(q_1,\alpha q_2)|^2$  of the greatest common right divisor ${\rm gcrd}(q_1, \alpha q_2)$ in $\mathds{H}(K)$. The measure factor simply counts the number of   quaternions in $\mathds{H}(K)$ with norm square $p=|{\rm gcrd}(q_1, \alpha q_2)|^2$ such that 
 \be \cF_\gra{0}{0}^{\scalebox{0.6}{1-loop}}   = -\frac{\pi \det[t_2]}{3} \sum_{{\bf q} \in   \mathds{H}(K) \oplus \frac{1}{\alpha} \mathds{H}(K)}  e^{2\pi i    {\bf q}^\dagger t {\bf q}}  \ . \ee
This rank one theta series is in the minimal automorphic representation of $SO(2,6)$ \cite{Krieg} and is manifestly invariant under the T-duality group $\Cong{D}{4}_{0*}(\alpha)$.  
 
By supersymmetry, the non-perturbative coupling must be a linear combination of rank one theta series  of $SO^*(12)$ defined as the group of six by six symplectic matrices over the quaternions. We will argue that the non-perturbative coupling is the theta series    
 \be  \Theta_{2} = -\frac{\pi \det[T_2]}{3} \sum_{{\bf q} \in  \mathds{H}(K)^2 \oplus\frac{1}{\alpha} \mathds{H}(K) }  e^{2\pi i{\bf    q}^\dagger T {\bf q}}  \ ,\ee
with
\be T = T_1+ i R V^{-1} V^{-1 \dagger} = \left( \begin{array}{cc} b  + \tilde{c}^\dagger c  + c \tilde{c}^\dagger + \frac{i}{\gFour^{\; 2}} + c^\dagger t  c  \;\;   & \tilde{c}^\dagger + c^\dagger t   \\ \tilde{c} +  t c   & t \end{array}\right) \ ,  \ee 
in terms of the NS-NS moduli $t$, the axio-dilaton $b + \frac{i}{\gFour^2}$ and the Ramond-Ramond moduli $(c,\tilde{c})$. The perturbative expansion of $ \Theta_{2}$ rescaled to the string frame gives 
\begin{multline}\gFour^{\, 2}   \Theta_{2} = - \frac{\det[t_2]}{24} \Biggl( \sum_{{\bf q} \in  \mathds{H}(K) \oplus \frac{1}{\alpha}  \mathds{H}(K)}  e^{2\pi i   {\bf q}^\dagger t {\bf q}} \\ +\sum^\prime_{p\in   \mathds{H}(K)} \sum_{{\bf q} \in  \mathds{H}(K) \oplus \frac{1}{\alpha}  \mathds{H}(K)}   e^{-2 \pi \frac{|p|^2}{\gfour^{\; 2}} +2\pi  i |p|^2 ( b+\tilde{c}^\dagger c  + c \tilde{c}^\dagger + c^\dagger t c)  +2\pi  i  \bigl(  {\bf q}^\dagger (\tilde{c}+t c)  p + p^* ( \tilde{c}^\dagger + c^\dagger t) {\bf q} \bigr)  +2\pi  i    {\bf  q}^\dagger t {\bf q} } \Biggr)   \ , \label{ThetaExpand}  \end{multline} 
which is by construction consistent with the perturbative computation. The corrections for ${\bf q}=0$ and $p\ne 0$ can be interpreted as Euclidean NS5 brane instantons with the NS5 charge $k = |p|^2$ and the action 
\be S_{\rm NS5} = \frac{2\pi k}{\gFour^2}- 2\pi i k b  = \frac{2\pi}{g_{\rm s}^2} \frac{\sqrt{G} \tilde{\RS} k}{K}- 2\pi i \frac{k}{K} B_{5678910} \; , \ee
where $G$ is the metric on $T^4$ at the symmetric point, i.e. one-half the Cartan matrix of $D_4$ or $A_2\oplus A_2$, and $\tilde{\RS}$ is the radius of the twisted circle $S^1_{\rm t}$. The Euclidean NS5 brane can therefore be interpreted as wrapping $k$ times the six-torus $T^6 / \mathds{Z}_K$ and $k$ must be an integer. As one may have expected, the instanton measure depends explicitly on the divisibility of $k$ by $K$, since the number of $p\in \mathds{H}(K)$ with $|p|^2=k$ gives the instanton measure
\be \mu(k) = \frac{24}{K-1} \sum_{\substack{d|k\\ d\ne 0 {\rm \, mod\, } K}} d \; . \ee
The perturbative coupling allows in principle the possibility that there be an additional contribution for $p\in \alpha \mathds{H}(K)$. However, the same coupling in the S-dual frame would then have non-perturbative corrections with $p\in \frac{1}{K} \mathds{H}(K)$ and therefore an NS5 charge $k \in \mathds{Z}/K^2$, whereas we expect the coupling $\cF_\gra{0}{0}$ in the S-dual frame to have Euclidean NS5 branes with charge $k \in \mathds{Z}/K$. This is the situation one observes for  NS5 instanton corrections in CHL orbifolds of the heterotic string \cite{Harvey:1996ir,Gregori:1997hi,David:2006ud,Bachas:2008jv} and the dual orbifold of the type II string on $T^2\times K3$ \cite{Bossard:2018rlt}.\footnote{The NS5 instanton measure of the $\mathcal{R}^2$ coupling in heterotic CHL orbifolds $\mu(k) =  \frac{24}{K+1} \bigl( \sum_{d|k} d+\sum_{Kd|k} Kd \bigr)$ is  indeed similar to the one we find here.}

Note that this coupling does not receive corrections associated to Euclidean D-brane with vanishing NS5 charge. This is consistent with the property that there is no 1/2 BPS Euclidean Brane in the asymmetric orbifold theory and the only 1/2 BPS instantons necessarily carry a non-zero NS5 charge \cite{Bianchi:2010aw}.

Using the Poisson summation formula one obtains that $ \Theta_{2} $ is invariant under the theta congruent subgroup $\Cong{D}{6}_{0}(\alpha) \subset Sp(6,\mathds{H}(K))$ of symplectic matrices with a lower-triangular matrix $C= 0$ mod $\alpha$ after conjugation by $\sigma_{36}$, the symplectic matrix that exchange the third and the sixth component, i.e. $g \in \Cong{D}{6}_{0}(\alpha)$ is of the form
\be g = \left(\begin{array}{cccccc} 1\, &\, 0\, &\, 0\, &\, 0\, &\, 0\, &\, 0\\ 0\, &\,  1\, &\, 0\, &\, 0\, &\, 0\, &\, 0\\0& 0&0&0&0&1\\0& 0&0&1&0&0\\0& 0&0&0&1&0\\0\, &\,  0\, &\! -1\, &\, 0\, &\, 0\, &\, 0 \end{array}\right) \left( \begin{array}{cc} A& B \\ C & D\end{array}\right)  \left(\begin{array}{cccccc} 1\, &\, 0\, &\, 0\, &\, 0\, &\, 0\, &\, 0\\ 0& 1&0&0&0&0\\0\, &\,  0\, &\, 0\, &\, 0\, &\, 0\, &\! -1\\0& 0&0&1&0&0\\0& 0&0&0&1&0\\0& 0&1&0&0&0 \end{array}\right)\; , \quad \left( \begin{array}{cc}\; D^\dagger \; & -B^\dagger \; \\  -C^\dagger \; & \; A^\dagger \; \end{array}\right) \left( \begin{array}{cc}\; A \; & \; B\; \\ \; C\; & \; D\; \end{array}\right)  = \left( \begin{array}{cc}\; \mathds{1}  \; & \; 0\; \\ \; 0\; & \; \mathds{1} \; \end{array}\right) \ee
with $A,B,D$ three by three matrices over $\mathds{H}(K)$ and $C$ over  $\alpha \mathds{H}(K)$. 
It is moreover invariant under the Fricke  transformation $\gamma_F$ \eqref{Fricke}
\bea t&\rightarrow&  -\biggl( \begin{array}{cc} \frac{1}{\alpha^{*}}  & 0 \\ 0 & \alpha \end{array}\biggr) t^{-1}   \biggl( \begin{array}{cc}  \frac{1}{\alpha}  & 0 \\ 0 & \alpha^* \end{array}\biggr) \; , \qquad \tilde{c}\rightarrow  \biggl( \begin{array}{cc} \frac{1}{\alpha^{*}}  & 0 \\ 0 & \alpha \end{array}\biggr) c\; , \qquad {c}\rightarrow -  \biggl( \begin{array}{cc} \alpha & 0 \\ 0 & \frac{1}{\alpha^{*}}   \end{array}\biggr) \tilde{c}\; , \CR
b&\rightarrow& b + c^\dagger \tilde{c} + \tilde{c}^\dagger c \; , \qquad \gFour\rightarrow \gFour\; , \eea
using Poisson summation over $\big(\colvec[0.8]{ \alpha^{-1}   & 0 \\ 0& \alpha^* }\big) {\bf q} \in \frac{1}{\alpha} \mathds{H}(K) \oplus  \mathds{H}(K)$ with 
\be  \Theta_{2} \rightarrow \frac{\det t}{\det \bar t}  \,  \Theta_2 \; ,\ee
with 
\be \det t = \det \left[\left( \begin{array}{ccc} 0\; &\; 0& 0 \\0&  \alpha & 0 \\ 0& 0 & \frac{1}{\alpha^{*}} \end{array}\right) T + \left( \begin{array}{ccc} 1\; &\; 0\; &\;  0 \\0& 0 & 0 \\ 0& 0 & 0 \end{array}\right)  \right]  \; . \ee
We  define the Fricke theta group $\Cong{D}{6}_{0 *}(\alpha)$ as the group generated by $\Cong{D}{6}_{0}(\alpha) $ and the Fricke transformation $\gamma_{\rm F}$. $\Theta_2$ is the unique rank one theta series consistent with the congruent subgroup symmetry $\Cong{D}{6}_{0*}(\alpha) $ that we conjecture to be the U-duality group of the theory. We shall find further evidence for this proposal by looking at the non-perturbative coupling $\cE_\grad{0}{0}{0}$ and $\cE_\grad{1}{0}{0}$  in \cite{GBinprep}. 
%%%%%%%%%%%%%%%%%%%%%%%%
\subsection{Large radius expansions of  ${\cal R}^4$ and $D^2{\cal R}^4$ couplings}
In order to write the one-loop couplings more explicitly one can perform an expansion at large radius $\RS$, with $\RS$ the radius of the circle untouched by the orbifold action that goes to infinity. From \eqref{E00Z2R} one finds
{\allowdisplaybreaks \bea
\mathcal{E}^{\scalebox{0.6}{1-loop}}_\gra{0}{0} &=& \frac{\pi^3}{27} \RS^4+  \RS \frac{(2\sin \frac{\pi}{K})^2}{2\pi}  \int_{\cF_K} \hspace{-1.2mm}\frac{d^2\tau}{\tau_2^{\, 2}}  \wp(\tfrac1 K)\Gamma_{\mathds{M}_2(K) }   \CR
&&\hspace{-8mm}  +\; \frac{\RS}{3(K-1)}   \sum^\prime_{\substack{Q \in\mathds{M}^*_2(K) \\ \det Q=0}}  \Biggl( \sum_{Q/d \in \mathds{M}^*_2(K) } \hspace{-5mm} d^3 + K^2 \hspace{-3mm}\sum_{Q/d \in \mathds{M}_2(K)   }  \hspace{-5mm} d^3\, \Biggr)  \frac{1 + 2\pi  \RS \mathcal{M}(Q) }{ \mathcal{M}(Q)^3} e^{- 2\pi \Rs \mathcal{M}(Q)\mathcal{M}(Q)+  2\pi i (Q,B)} \nonumber \\
&&  +\frac{8\RS }{(K-1)^2}   \sum_{\substack{Q \in  \mathds{M}^*_2(K)  \\ \det Q < 0}}  \Biggl( \sum_{Q/d \in  \mathds{M}^*_2(K) } \hspace{-5mm} d^3  \hspace{-0.8mm} \sum_{\substack{\ell | \frac{ K \det Q}{d^2}\\ \ell \ne 0 {\, \rm mod\, } K   }} \hspace{-1mm} \ell+ K^2 \hspace{-3mm}\sum_{Q/d \in  \mathds{M}_2(K)  }  \hspace{-3mm} d^3  \hspace{-2.2mm} \sum_{\substack{\ell | \frac{\det Q}{ d^2}\\ \ell \ne 0 {\, \rm mod\, } K }} \hspace{-1mm} \ell\, \Biggr) \CR
&& \hspace{60mm} \times  \frac{1 + 2\pi  \RS \mathcal{M}(Q)  }{ \mathcal{M}(Q)^3} e^{- 2\pi \Rs \mathcal{M}(Q) +  2\pi i (Q,B)}\; ,  \eea}
where $\mathcal{M}(Q) = \sqrt{2 p_R(Q)^2}$ and $\mathds{M}_2(K)$ is the lattice $\sLambda_{1,1}[2]\oplus \mathds{H}(K)$. The factor in $\frac{\pi^3}{27} \RS^4$ comes from the Kaluza--Klein tower of massive spin two supermultiplets and therefore does not depend on $K$. The holomorphicity  of $\wp(\frac1K)$ ensures that there is no contributions from 1/6 BPS world-sheet instantons with $\det Q > 0$. The measure factor for 1/3 BPS charges can be written in terms of the degeneracy $d_j^\ord{\frac13}(Q)$ of 1/3 BPS states of charge $Q$ and spin $j +2$ according to  \eqref{OneThird2u}, \eqref{OneThird2t}, \eqref{OneThird3u} and \eqref{OneThird3t}
\be \frac{4 }{(K-1)^2}  \Biggl( \sum_{Q/d \in  \mathds{M}^*_2(K) } \hspace{-5mm} d^3  \hspace{-0.8mm} \sum_{\substack{\ell | \frac{ K \det Q}{d^2}\\ \ell \ne 0 {\, \rm mod\, } K   }} \hspace{-1mm} \ell+ K^2 \hspace{-3mm}\sum_{Q/d \in  \mathds{M}_2(K)  }  \hspace{-3mm} d^3  \hspace{-2.2mm} \sum_{\substack{\ell | \frac{\det Q}{ d^2}\\ \ell \ne 0 {\, \rm mod\, } K }} \hspace{-1mm} \ell\, \Biggr) = \hspace{-4mm}\sum_{\substack{ d\in \mathds{N} \\ Q/d \in  \mathds{M}^*_2(K)}}  \hspace{-4mm} d^3 \sum_{j \in \frac{\mathds{N}}{2}} (-1)^{2j}(2j+1) d_j^\ord{\frac13}(Q/d)\ee
which allows for the standard interpretation of instantons as solitons on the Euclidean time circle \cite{Bossard:2016hgy}. 

Equivalently one can take the expansion at large twisted radius $\tilde{\RS}$, 
\bea
\mathcal{E}^{\scalebox{0.6}{1-loop}}_\gra{0}{0} &=& \frac{\pi^3}{27 K^2} \tilde{\RS}^4+ 2\pi  \tilde{\RS}   \int_{\cF} \hspace{-1.2mm}\frac{d^2\tau}{\tau_2^{\, 2}}  \Gamma_{\sLambda_{6,6}}  \CR
&&\hspace{-15mm}  +\; \frac{\tilde{\RS}}{3K^2(K-1)}   \sum^\prime_{\substack{Q \in \frac12 \mathds{M}^0_2(K) \\ \det Q=0}}  \Biggl( \sum_{Q/d \in\frac12 \mathds{M}^0_2(K) } \hspace{-5mm} d^3 + K^2 \hspace{-5mm}\sum_{Q/d \in \mathds{M}^{0*}_2(K)   }  \hspace{-5mm} d^3\, \Biggr)  \frac{1 + 2\pi  \tilde{\RS} \mathcal{M}(Q) }{ \mathcal{M}(Q)^3} e^{- 2\pi \tilde{\Rs} \mathcal{M}(Q)+  2\pi i (Q,\tilde{B})} \CR
&&  +\frac{8\tilde{\RS}
 }{K^2(K-1)^2}   \sum_{\substack{Q \in   \frac12 \mathds{M}^0_2(K) \\ \det Q < 0}}  \Biggl( \sum_{Q/d \in  \frac12 \mathds{M}^0_2(K)  } \hspace{-5mm} d^3  \hspace{3mm} \sum_{\substack{\ell | \frac{ K^2 \det Q}{d^2}\\ \ell \ne 0 {\, \rm mod\, } K   }} \hspace{0mm} \ell+ K^2 \hspace{-3mm}\sum_{Q/d \in  \mathds{M}^{0*}_2(K)  }  \hspace{-5mm} d^3  \hspace{3mm} \sum_{\substack{\ell | \frac{K\det Q}{ d^2}\\ \ell \ne 0 {\, \rm mod\, } K }} \hspace{-0mm} \ell\, \Biggr) \CR
&& \hspace{50mm} \times  \frac{1 + 2\pi  \tilde{\RS} \mathcal{M}(Q) }{ \mathcal{M}(Q)^3} e^{- 2\pi \Rs \mathcal{M}(Q) +  2\pi i (Q,\tilde{B})}\; ,  \eea
where $\mathds{M}^0_2(K) = \sLambda_{1,1}\oplus \mathds{H}(K)$. The term linear in $\tilde{\RS}$ is the maximally supersymmetric $\mathcal{R}^4$ coupling restricted to the $\mathcal{N}=6$ Narain moduli.  Quite reassuringly for consistency, one gets back the maximally supersymmetric theory at $\tilde{\RS} \rightarrow \infty$. The factor in $\frac{\pi^3}{27 K^2} \tilde{\RS}^4$ comes from the Kaluza--Klein tower of massive spin two and spin three-half supermultiplets, which mass spectrum depends on the orbifold order $K$. The world-sheet instantons are again at most 1/3 BPS, but they cannot be interpreted as  string solitons in this case. 

The $D^2 \mathcal{R}^4$ coupling expands similarly at large $\RS$ using \eqref{D2R4Appendix}
 \bea \cE_\grad{1}{0}{0}^{\scalebox{0.6}{1-loop}}  &=& -  \frac{8 \pi^5 \RS^6}{42525} - \frac{2\zeta(3)}{\pi} \log \RS- \RS \frac{\pi^2}{45}  (2\sin \tfrac \pi K)^2  \int_{\cF_K} \frac{d^2 \tau}{\tau_2^{\, 2}} E_{1,2}(\tau) \Gamma_{\mathds{M}_2(K)}  \CR
 &&  -\frac{16 \pi^2}{45(K-1)} \RS^{\frac{7}{2}}\hspace{-1.8mm} \sum_{\substack{ Q \in \mathds{M}_2^*(K)\\ Q^2 = 0 }} \hspace{-1mm} \Biggl(K \hspace{-3mm}\sum_{Q/d \in \mathds{M}_2(K)   }  \hspace{-5mm} d^5- \sum_{Q/d \in \mathds{M}^*_2(K) } \hspace{-5mm} d^5 \, \Biggr)\frac{K_{\frac52}(2\pi \RS \mathcal{M}(Q) )}{\mathcal{M}(Q)^{ \frac52}} e^{2\pi i (Q,B)} \CR
  &&  \hspace{-2mm}+ \frac{4\zeta(3)}{\pi (K-1) }    \sum_{\substack{ Q \in \mathds{M}_2^*(K)\\ Q^2 = 0 }} \hspace{-1mm}  \Biggl(K \hspace{-3mm}\sum_{Q/d \in \mathds{M}_2(K)   }  \hspace{-5mm} d^{-1}- \sum_{Q/d \in \mathds{M}^*_2(K) } \hspace{-5mm} d^{-1} \, \Biggr) e^{-2\pi \Rs \mathcal{M}(Q)+2\pi i (Q,B)} \CR
  %%%%%%%%%%%%%%%%%%%%%%%%%%%%%%%%%%%%%%%%%
  &&  +  \frac{2}{\pi(K-1)} \hspace{-2mm} \sum_{\substack{ Q \in \mathds{M}_2^*(K)\\ \det \! Q >0\\ \det\! Q = 0 {\, \rm mod\, } K}} \hspace{-1mm} \Biggl(K \hspace{-3mm}\sum_{Q/d \in \mathds{M}_2(K)   }  \hspace{-5mm} d^5\sigma_{3}(  \tfrac{\det Q}{d^2}) - \hspace{-4mm} \sum_{Q/d \in \mathds{M}^*_2(K) } \hspace{-5mm} d^5\sigma_{3}(  \tfrac{\det Q}{d^2})  \, \Biggr)   \frac{e^{-2\pi \Rs \mathcal{M}(Q) +2\pi i (Q,B)} }{\det Q^3} \CR
    %%%%%%%%%%%%%%%%%%%%%%%%%%%%%%%%%%%%%%%%%
  && + \frac{1}{2\pi(K-1)} \hspace{-2mm} \sum_{\substack{ Q \in \mathds{M}_2^*(K)\\ \det \! Q <0\\ \det\! Q = 0 {\, \rm mod\, } K}} \hspace{-1mm} \Biggl(K \hspace{-3mm}\sum_{Q/d \in \mathds{M}_2(K)   }  \hspace{-5mm} d^5\sigma_{3}(  \tfrac{\det Q}{-d^2}) - \hspace{-4mm} \sum_{Q/d \in \mathds{M}^*_2(K) } \hspace{-5mm} d^5\sigma_{3}(  \tfrac{\det Q}{-d^2})  \, \Biggr)   \frac{e^{-2\pi \Rs \mathcal{M}(Q) +2\pi i (Q,B)} }{-\det Q^3} \CR
  && \hspace{30mm} \times  \biggl( 1  - \frac{4\pi \RS \det Q}{\mathcal{M}(Q)} \Bigl( 1 - \frac{ \det Q}{\mathcal{M}(Q)} \Bigr) + \frac{2\pi^2 \RS^2 \det Q^2}{ \mathcal{M}(Q)^2}\biggr) \; . 
 \eea
The first two terms come from the  Kaluza--Klein tower of massive spin two supermultiplets and do not depend on $K$. The logarithmic term is of course related to the divergence in \eqref{ED2R4Regul}. The third term is the coupling in five dimensions. As expected we have contributions from both 1/3 BPS world-sheet instantons  with $\det Q<0$ and 1/6 BPS world-sheet instantons  with $\det Q>0$. The 1/6 BPS world-sheet instantons measure is related to the degeneracy $d_j^\ord{\frac16}(Q)$ of 1/6 BPS states of charge $Q$ and spin $j +\frac52$ according to  \eqref{OneSixth2u}, \eqref{OneSixth2t}, \eqref{OneSixth3u} and \eqref{OneSixth3t}
\be \frac{-4 }{K-1
} \Biggl(K \hspace{-3mm}\sum_{Q/d \in \mathds{M}_2(K)   }  \hspace{-5mm} d^5\sigma_{3}(  \tfrac{\det Q}{d^2}) - \hspace{-4mm} \sum_{Q/d \in \mathds{M}^*_2(K) } \hspace{-5mm} d^5\sigma_{3}(  \tfrac{\det Q}{d^2})  \, \Biggr)  = \hspace{-4mm}\sum_{\substack{ d\in \mathds{N} \\ Q/d \in  \mathds{M}^*_2(K)}}  \hspace{-4mm} d^5 \sum_{j \in \frac{\mathds{N}}{2}} (-1)^{2j}(2j+1) d_j^\ord{\frac16}(Q/d)\ee
and the 1/3 BPS world-sheet instantons measure is related to the degeneracy $d_j^\ord{\frac13}(Q)$ of 1/3 BPS states of charge $Q$ and spin $j +2$ according to  \eqref{OneThird2uB10}, \eqref{OneThird2tB10}, \eqref{OneThird3uB10} and \eqref{OneThird3tB10}
\be \frac{6}{K-1
} \Biggl(K \hspace{-3mm}\sum_{Q/d \in \mathds{M}_2(K)   }  \hspace{-5mm} d^5\sigma_{3}(  \tfrac{\det Q}{d^2}) - \hspace{-4mm} \sum_{Q/d \in \mathds{M}^*_2(K) } \hspace{-5mm} d^5\sigma_{3}(  \tfrac{\det Q}{d^2})  \, \Biggr)  = \hspace{-4mm}\sum_{\substack{ d\in \mathds{N} \\ Q/d \in  \mathds{M}^*_2(K)}}  \hspace{-4mm} d^5 \sum_{j \in \frac{\mathds{N}}{2}} (-1)^{2j}(2j+1) j(j+1)d_j^\ord{\frac16}(Q/d)\ee
that appears in the helicity supertrace $B_{10}^{\mathds{Z}_K}$ in \eqref{BSMult}. This can be understood from the relations  
\be E_4 = \frac{2i}{3}\bigl( \partial_\tau - \tfrac{i}{\tau_2}\bigr) E_{1,2}\; , \qquad \overline{E_4} =- \frac{1}{12}\bigl( \bar\partial_\tau + \tfrac{i}{\tau_2}\bigr) \bar \partial_\tau \bigl( \tau_2^2 \bar \partial_\tau E_{1,2} \bigr) \ee
and so the perturbative coupling \eqref{ED2R4Regul} is directly related to the 1/6 BPS helicity supertrace $B^{\mathds{Z}_K}_{10}$ in \eqref{HelicityBn}, consistently with the property that only BPS states preserving at least one-sixth  of the supersymmetries contribute to the $D^2 \mathcal{R}^4$ coupling. 

%%%%%%%%%%%%%%%%%%%%%%%%

%
%%%%%%%%%%%%%%%%%%%%%%%%%
%%%%%%%%%%%%%%%%%%%%%%%%
%%%%%%%%%%%%%%%%%%%%%%%%
%%%%%%%%%%%%%%%%%%%%%%%%

%%%%%%%%%%%%%%%%%%%%%%%%
%%%%%%%%%%%%%%%%%%%%%%%%
%%%%%%%%%%%%%%%%%%%%%%%%
\section{BPS Branes in the asymmetric orbifold}
\label{NonPerturbativeBranes} 
%%%%%%%%%%%%%%%%%%%%%%%%

In standard toroidal compactifications `physical' D-brane charges (R-R charges) transform in spinorial representations of the T-duality group. For instance on $\mathds{R}^{1,3}\times T^6$  they transform in the ${\bf 32}$ Majorana--Weyl spinor representation of $Spin(6,6)$ that combines with the NS-NS charges in the $({\bf 2},{\bf 12})$ of $SL(2)\times SO(6,6)$ to produce the ${\bf 56}$ of $E_{7(7)}$. The first $SO(6,6)$ vector of charges corresponds to the momentum and winding of the perturbative string along $T^6$, while the second corresponds to the NS5 and KK(6,1) charges. Similarly on $\mathds{R}^{1,4}\times  T^5$, after dualizing all antisymmetric tensors to vectors, they transform in the ${\bf 16}$ Majorana--Weyl spinor representation  of $Spin(5,5)$ that combines with the NS-NS charges for the perturbative string in the ${\bf 10}$ and for the NS5 charge in the singlet to produce the ${\bf 27}$ of $E_{6(6)}$. Finally on $\mathds{R}^{1,5}\times T^5$ they transform in the ${\bf 8}_S$ Majorana--Weyl spinor representation of $Spin(4,4)$ that combines with the NS-NS charges in the ${\bf 8}_V$  to produce the ${\bf 16}$ of $Spin(5,5)$. 

Euclidean D-branes transform similarly in the spinor representation of opposite chirality since they can be thought of as the `physical' D-branes of the formal T-dual along Euclidean time \cite{Bianchi:2012ud, Bianchi:2012kt}.

In the asymmetric orbifold case the continuous `T-duality' group  $SO(6,6)$ is broken to  $SO(2,6)\sim SO^*(8)$ in $D=4$ dimensions and  from $SO(5,5)$ to $SO(1,5)$ in $D=5$. Decomposing $T^6 = T^2_{\rm t}\times T^4_{\rm a}$ yields  $SO(6,6)\supset SO(2,6)\times SU(2)_1\times_{\mathds{Z}_2} SU(2)_2$ with
${\bf 32}\rightarrow ({\bf 8}_S, {\bf 2}_2) + ({\bf 8}_C, {\bf 2}_1)$ in $D=4$.  Decomposing $T^5 = S^1_{\rm t}\times T^4_{\rm a} $ yields  $SO(5,5)\supset SO(1,5)\times SU(2)_1\times_{\mathds{Z}_2} SU(2)_2$ with
${\bf 16}\rightarrow ({\bf 4}, {\bf 2}_2) + (\bar{\bf 4}, {\bf 2}_1)$ in $D=5$.

For Type IIB on $S^1 \times S^1_{\rm t} \times T^4_{\rm a}$, the `physical' D-branes, that couple to the 16 `gravi-photons' in the R-R sector (8 `electric' and 8 `magnetic' dual), are given by bound-states of D1-, D3- and D5-branes. They only preserve 8 out of the 32 original supercharges and are thus 1/3 BPS in ${\cal N}=6$ supergravity \cite{Bianchi:2010aw}.   

When the asymmetric $\mathds{Z}_2$ orbifold acts as a reflection  on the left-movers on $T^4_{\rm a}$ and a shift of  $S^1_{\rm t}$ one finds: two bound-states  of D1 on $S^1\subset T^2_{\rm t}$ with D5 on $S^1\times T^4_{\rm a}$, four bound-states of D1 on $S^1\subset T^4_{\rm a}$ with D3 on dual $T^3_\perp\subset T^4_{\rm a}$, four bound-states of  D3 on $T^2 \times S^1\subset T^2_{\rm t}\times T^4_{\rm a}$ with D5 on $T^2\times T^3_\perp\subset T^2_{\rm t}\times T^4_{\rm a}$, three bound-states of D3 on $S^1\times T^2 \subset T^2_{\rm t}\times T^4_{\rm a}$ with D3 on $S^1\times T^2_\perp\subset T^2_{\rm t}\times T^4_{\rm a}$. Altogether they transform in the $({\bf 8}_S, {\bf 2}_2)$ of  $Spin(2,6)\times SU(2)_1\times SU(2)_2$, while $({\bf 8}_C, {\bf 2}_1)$ is projected out by the orbifold, as we will discuss more in detail below.

The low-energy excitations can be described in terms of open strings. For a bound-state of $N$ D-branes and as many ($N'=N$) `images' under the asymmetric orbifold action, open strings starting and ending on the D-branes of the same kind give rise to  
${\cal N}=4$ vector multiplets in the adjoint of $U(N)$ while open string starting on the D-brane of one kind and ending on their images give rise to ${\cal N}=2$ hypermultiplets in the bi-fundamental. Due to the orbifold identification $N=N'$, the latter end up in the adjoint of the diagonal group and the two ${\cal N}=4$ vector multiplets are to be identified. In the end one has an ${\cal N}=4$ vector multiplet, equivalent to an ${\cal N}=2$  vector multiplet and a hypermultiplet in the adjoint,  coupled to an additional adjoint hypermultiplet for `regular' branes of this kind. Moreover one expects to have `fractional' branes wrapping only half of the circle acted on by the shift. The main new feature with respect to the `regular' branes is that they also couple to twisted RR fields. Since in the asymmetric orbifolds we consider twisted RR fields are all massive due to the shift, `fractional' branes do not seem to require a different low-energy description. The only macroscopic difference with respect to regular branes is their tension that affects the value of the gauge coupling and the strength of the self-interactions among open strings.

A similar analysis applies to the ${\mathds{Z}}_3$ case. Each brane has two images with $N'=N''=N$. In addition the three ${\cal N}=4$ vector multiplets that are identified with one another, there are three hyper-multiplets in the bi-fundamentals that are also identified with one another. One gets in the end an ${\cal N}=2$  vector multiplet and two adjoint hypermultiplets for `regular' branes of this kind, as for $\mathds{Z}_2$. 

The Euclidean D-branes (instantons) in the Type IIB frame are in one-to-one correspondence with the `physical' D-branes in the  Type IIA description. 

According to \cite{Bianchi:2010aw} one has schematically the following bound states of Euclidean branes in five dimensions that are invariant under the $\mathds{Z}_2$ orbifold action:  \bea [1]\: {\rm D({-}1)} + {\rm D3}_{{T}_{\rm a}^4} \; ,  \quad [4]\:{\rm D1}_{S^1_{\rm t} \times S^1_{\rm a}} + {\rm D3}_{S^1_{\rm t}\times T^3_{{\rm a}\bot}}  \, , \quad [3]\:{\rm D1}_{T^2_{\rm a}} + {\rm D1}_{T^2_{{\rm a} \bot}}\; .  \label{SchematicDp} \eea
We shall explain in the following section that these eight fundamental bound states generate a spinor representation of $Spin(1,5) \sim SL(2,\mathds{H}(\mathds{R}))$ with the structure of $\frac{1}{\alpha} \mathds{H} \oplus \mathds{H}$. In four dimensions the Euclidean D-brane can moreover wrap the additional circle and we have similarly
\bea [1]\: {\rm D1}_{T^2_{\rm t}} + {\rm D5}_{T^2_{\rm t}\times {T}_{\rm a}^4} \; ,  \quad [4]\:{\rm D1}_{S^1 \times S^1_{\rm a}} + {\rm D3}_{S^1 \times T^3_{{\rm a}\bot}}  \, , \quad [3]\: {\rm D3}_{T^2_{\rm t} \times T^2_{\rm a}} + {\rm D3}_{T^2_{\rm t} \times  T^2_{{\rm a} \bot}} \; . \eea

\subsection{Euclidean branes at the $D_4$ symmetric point}
Le us discuss the asymmetric orbifold projection of the $T^4$ Narain lattice in some more details. For the string zero modes of momentum $\vec{m}$ and winding number $\vec{n}$ along $T^4$, we have the left and right projections 
\bea p_L(\vec{m},\vec{n})^2 &=& \frac{1}{2} G^{-1} ( \vec{m} + (G + B) \vec{n} , \vec{m} + (G + B) \vec{n}) \; , \CR
 p_R(\vec{m},\vec{n})^2 &=& \frac{1}{2} G^{-1} ( \vec{m} + (-G + B) \vec{n} , \vec{m} + (-G + B) \vec{n}) \; .\eea
A $D_4$ symmetric point is obtained at \footnote{Recall that we define the left and right mementa without dimensions, such that they appear as $e^{\pi i \tau p_L^2 - \pi i \bar \tau p_R^2}$ in the Narain partition function.}
\be G = \frac{1}{2} A = \left( \begin{array}{cccc} 1 \; & - \frac12 \; &\; 0\; &\; 0\\ - \frac12 \; & \; 1\; &-\frac12\; & - \frac12 \\
0\; &- \frac12 & \; 1\; &\; 0\\ 0\; &- \frac12 & \; 0 \; &\; 1\end{array}\right)\; , \qquad B = \left( \begin{array}{cccc} 0\; & \;  \frac12 \; &\; 0\; &\; 0\\ - \frac12 \; & \; 0\; &\; \frac12\; & \;  \frac12 \\
0\; &- \frac12 & \; 0\; &\; 0\\ 0\; &- \frac12 & \; 0 \; &\; 0\end{array}\right)  \; , \label{GBZ2}\ee
where $A$ is the $D_4$ Cartan matrix and $G + B \in SL(4,\mathds{Z})$. The condition $p_L(\vec{m},\vec{n})=0$ can be solved for integral coefficients $\vec{m} = - (G + B) \vec{n}$, and 
\be p_R( - (G + B) \vec{n},\vec{n})^2 =  A(\vec{n},\vec{n})\; ,  \ee
The projected zero modes are therefore in the $D_4$ lattice.

The asymmetric orbifold projection, i.e.  the $O(4,4)$ reflection $p_L\rightarrow -p_L$, can be lifted to a $Spin(4,4)$ reflection by choosing $-\mathds{1}\in SU(2) \subset Spin(4,4)$. We choose $-\mathds{1}\in SU(2)_1$, with the  convention for the three trial fundamental representations 
\bea p_L &\in& ({\bf 2}_1,{\bf 2}_2) \; , \qquad p_R \in ({\bf 2}_3,{\bf 2}_4) \;, \CR
p^S_L &\in& ({\bf 2}_1,{\bf 2}_3) \; , \qquad p^S_R \in ({\bf 2}_2,{\bf 2}_4)\;  , \CR
p^C_L &\in& ({\bf 2}_1,{\bf 2}_4) \; , \qquad p^C_R \in ({\bf 2}_2,{\bf 2}_3)\; , 
\eea
under $SU(2)_1\times SU(2)_2\times SU(2)_3 \times SU(2)_4  \subset  Spin(4,4)$, so that the orbifold reflection for the spinor representations is $p^S_L \rightarrow - p^S_L$ and $p^C_L\rightarrow -p^C_L$ for the positive and negative chirality Majorana--Weyl spinors. Note that the `left projections' $p^S_L$ and $p^C_L$ are not associated in any way to the left mover strings. We use the notation left and right simply because these left projections do transform under $-\mathds{1}\in SU(2)_1$ of the $\mathds{Z}_2$ orbifold action while the right projections are invariant.

The Euclidean D-brane charges along $T^4$ are the D(-1) charge $q\in \mathds{Z}$, the D1 charge  $Q^{ab} \in \wedge^2 \mathds{Z}^4$, and the D3 brane charge $p\in \mathds{Z}$, with the `left' and `right' projections 
\bea  p_L^S(q,Q,p)^2 &=& \frac{1}{2 |G|^\frac12} \bigl( q- \tfrac12 \tr[ B Q]  - \tfrac14 \tr[ B^\star B] p  - |G|^\frac12 p\bigr)^2\CR
&& \qquad - \frac14 \tr\bigl[ ( Q + B^\star p) \cdot \bigl(  |G|^{-\frac12} G (Q+B^\star p ) G +( Q+B^\star p)^\star \bigr)\bigr] \CR
p_R^S(q,Q,p)^2 &=& \frac{1}{2 |G|^\frac12} \bigl( q- \tfrac12 \tr[ B Q]  - \tfrac14 \tr[ B^\star B] p  + |G|^\frac12 p\bigr)^2\CR
&& \qquad - \frac14 \tr\bigl[ ( Q + B^\star p) \cdot \bigl(  |G|^{-\frac12} G (Q+B^\star p ) G -( Q+B^\star p)^\star \bigr)\bigr] \; , \label{PSLR} \eea 
where $|G| = \det G$ and $B^\star_{ab} = \frac{1}2 \varepsilon_{abcd} B^{cd}$. The $\mathds{Z}_2$ orbifold condition $p_L^S(q,Q,p)=0$ can be solved for integral coefficients at the $D_4$ symmetric point and for the appropriate parametrisation 
\be q = q_1 + q_2 + q_4\, , \quad Q =  \left( \begin{array}{cccc} 0\; &\;   - q_2 -\sum_{i=1}^4 q_i\; &\; -q_1-q_2\; &\; -\sum_{i=1}^4 q_i \\  q_2 +\sum_{i=1}^4 q_i \; & \; 0\; &\; q_3\; & -q_1 \\
q_1+q_2 \; &- q_3 & \; 0\; &\; q_2\\ \sum_{i=1}^4 q_i \; &q_1 & -q_2   \; &\; 0\end{array}\right)  \; ,  \quad p = q_4\; , \label{DbraneChargesT4} \ee
one obtains
\be p_R^S(q)^2 = 2 A^{-1}(\vec{q},\vec{q})\; . \ee
This shows that the branes localised in the twisted circle, combining a brane and its image under the $\mathds{Z}_2$ action, have R-R charges in $ \mathds{H}$.  Note that the image of a single D(-1) brane under $\mathds{Z}_2$ is not exactly a single D3 wrapping $T^4$ as suggested in \eqref{SchematicDp}, because the metric $G$ is not Euclidean ($G_{ij} \neq \delta_{ij}$)  and the $B$-field is not zero. 

\vskip 4mm

The Euclidean D-brane charges for brane wrapping both the twisted circle $S^1$ and cycles in  $T^4$ are the D1 charges $\vec{q}\in \mathds{Z}^4$ and the D3 charges  $\vec{p} \in \mathds{Z}^4$, with the left and right projections
\bea p^C_L(\vec{q},\vec{p})^2 &=& \frac{1}{2|G|^\frac12} G ( \vec{q} - (|G|^\frac12 G^{-1} + B^\star) \vec{p} , \vec{q} - (|G|^\frac12 G^{-1} + B^\star)  \vec{p}) \; , \CR
p^C_R(\vec{q},\vec{p})^2 &=& \frac{1}{2|G|^\frac12} G ( \vec{q} + (|G|^\frac12 G^{-1} -B^\star) \vec{p} , \vec{q} + (|G|^\frac12 G^{-1} - B^\star)  \vec{p}) \; . \label{PCLR} \eea
The solution to $p^C_L(\vec{q},\vec{p})= 0$ is integral at the $D_4$ symmetric point and for the appropriate parametrisation 
\bea q_1 &=& -\sum_{i=1}^4 p^\prime_i\, , \quad q_2 = -p_2^\prime\; , \quad q_3 = p_1^\prime + p_2^\prime, \quad q_4 = p_2^\prime + p_4^\prime \; , \CR
p_1 &=&-\sum_{i=1}^3 p^\prime_i\, , \quad  p_2 = -p_2^\prime -p_3^\prime\, , \quad  p_3 =-p_3^\prime\, , \quad  p_4 =p_4^\prime\; , \label{pprimepara} \eea
one has 
\be p_R^C(p^\prime)^2 = 2 A^{-1}(\vec{p}{}^{\, \prime},\vec{p}{}^{\, \prime})\; . \ee
The lattice of {\it invariant} R-R charges can therefore be defined as integral quaternions in $ \mathds{H}$. We shall now argue that there are also `fractional branes' with R-R charges in  $ \frac{1}{\alpha} \mathds{H}$.

The quaternion ring structure is preserved by the $\gamma$-matrix product 
\be \left( \begin{array}{c} \vec{q} \\ \vec{p} \end{array}\right) =  \left( \begin{array}{cc} Q & q\mathds{1} \\-p\mathds{1}  & \; Q^\star \end{array}\right)
\left( \begin{array}{c} \vec{m} \\ \vec{n} \end{array}\right) \ee
whose norm is one-half the product of the even norms. We found that the `geometric' Euclidean D-brane charges introduced in  \cite{Bianchi:2008cj} can be parametrised in terms of  Hurwitz quaternions $\tilde{p}$, $p$ in $\mathds{H}$ such that $p_R^C(\tilde{p})^2= 2|\tilde{p}|^2 $ and $p_R^S(p)^2= 2|p|^2 $. With the parametrisation of a $D_4$ lattice vector with a Hurwitz quaternions $q \in  \mathds{H}$, with norm $p_R(q)^2 = 2 |q|^2$, one finds that the $\gamma$-matrix product reduces to the quaternion product 
\be {p} \rightarrow  q \tilde{p} \; . \ee
For a twisted state the string charge $q\in  \frac{1}{\alpha} \mathds{H}$, and this product suggests that $p$ must generally be valued in $\frac{1}{\alpha} \mathds{H}$. This is indeed required by the T-duality $\Cong{A}{3}_{0*}(\alpha)$ transformation 
\be (m,q,n)\rightarrow ( m , q + b m,n + b^* q + q^* b + |b|^2 m)\; , \qquad (\tilde{p},p)\rightarrow (\tilde{p},p + b \tilde{p})\; , \ee
for $b \in \frac{1}{\alpha} \mathds{H}$. We find evidence that $\Cong{A}{3}_{0*}(\alpha)$ is a symmetry of the non-perturbative couplings, and therefore that 1/3 BPS `fractional branes'  with $p\in \frac{1}{\alpha} \mathds{H}$  exist.

\vskip 4mm

The  dimensionless action of the Euclidean D-brane instanton in five dimensions reads \footnote{We use $\frac{1}{\gfive^2} = \frac{|G|^{\frac{1}{2}} \tilde{\RS}}{2  g_{\rm s}^2} = \frac{\tilde{\RS}}{4  g_{\rm s}^2} $ at the $D_4$ symmetric point. }
\be S_{\scalebox{0.6}{D}}= \frac{4\pi}{\gFive} \sqrt{ \tilde{\RS} |p|^2 + \tilde{\RS}^{-1} |\tilde{p} - a^* p|^2} = \frac{2\pi}{ g_{\rm s}} \sqrt{  \tilde{\RS}^2 |p|^2 +  |\tilde{p} - a^* p|^2} \; , \ee
where $\tilde{\RS}$ is the radius modulus in string length, such that the twisted circle radius is $R(S^1_{\rm t}) = \tilde{\RS} \sqrt{\alpha^\prime}$. At vanishing Wilson line $a=0$, we find therefore that the smallest tension for brane wrapped over the twisted circle with $\tilde{p}=0$ and $p=\frac{1}{\alpha}$ is $ \frac{ \pi \sqrt{2} \tilde{\RS}}{g_{\rm s}} $, whereas one has $ \frac{2\pi}{g_{\rm s}}  $ for the branes localised on two image points on the twisted torus with $\tilde{p}=1$ and $p=0$. 

In an asymmetric orbifold such that $G, B \sim 1$ as in \eqref{GBZ2} (i.e. a `string-size' torus) one also expects to have `non-geometric' (fractional) branes  \cite{Gaberdiel:2002jr, Kawai:2007qd} that admit a consistent CFT description in terms of boundary states and string amplitudes on the disk with mixed boundary conditions  \cite{Bianchi:2016bgx, Bianchi:2017sds} but lack a simple `large volume' (supergravity) limit. The non-perturbative corrections \eqref{ThetaExpand} indeed also include Euclidean D-branes with $p\in \frac{1}{\alpha} \mathds{H}$, for which the minimal tension is $\frac{\pi \sqrt{2} \tilde{\RS}}{g_{\rm s}}$. Note that this is not the result one would obtain from the naive expectation that `fractional' branes in the asymmetric orbifold correspond to branes wrapping only half of the twisted circle, because their charges would then be half so as to give $p\in \frac{1}{2} \mathds{H}$ and a minimal tension of $ \frac{\pi \tilde{\RS}}{g_{\rm s}} $. This factor of $\frac{1}{\sqrt{2}}$ (instead of $\frac12$) between the minimal tensions of non-geometrical fractional branes and geometrical ones is reminiscent of the analysis in \cite{Gaberdiel:2002jr} for a similar asymmetric orbifold of the bosonic string. The lack of a clear geometric interpretation of these fractional branes was stressed in \cite{Gaberdiel:2002jr, Kawai:2007qd} in connection with the peculiar value of the mass-shift in the low-lying open string spectrum.\footnote{The latter suggests an interpretation in terms of a `small' angle as in \cite{Anastasopoulos:2011hj}.}
One may be able nonetheless to interpret them as branes wrapping only half of the twisted circle with the additional requirement that $|p|^2 \in \mathds{N}$, which should follow from consistency of the boundary states in string theory. This issue deserves further study.

\vskip 4mm

The same reasoning applies to the four-dimensional theory. The Euclidean D-brane wrapping both $T^2_{\rm t}$ and cycles in  $T^4_{\rm a}$ are the D1 charge $q\in \mathds{Z}$, the D3 charge  $Q \in \wedge^2 \mathds{Z}^4$, and the D5 brane charge $p\in \mathds{Z}$, with the same left and right projections as in \eqref{PSLR}. One concludes that the corresponding Euclidean brane charges $p$ are valued in $ \frac{1}{\alpha} \mathds{H}$. The Euclidean D-branes wrapping the additional (untwisted) circle $S^1$ and cycles in the asymmetric orbifolded $T^4$ are the D1 charges $\vec{q}\in \mathds{Z}^4$ and the D3 charges  $\vec{p} \in \mathds{Z}^4$, with the same left and right projections as in \eqref{PCLR} and the corresponding R-R charges are valued in $\mathds{H}$. 

\vskip 4mm

This concludes the determination of the Euclidean D-brane charge lattice $(\frac{1}{\alpha} \mathds{H})^2 \oplus \mathds{H}^2$. We have done the same analysis for the $\mathds{Z}_3$ orbifold at the $A_2\oplus A_2$ symmetric point 
\be G = \frac{1}{2} A = \left( \begin{array}{cccc} 1 \; & - \frac12 \; &\; 0\; &\; 0\\ - \frac12 \; & \; 1\; &\; 0\; &\; 0  \\
0\; &\; 0  & \; 1\; &- \frac12 \\ 0\; &\; 0  & - \frac12  \; &\; 1\end{array}\right)\; , \qquad B = \left( \begin{array}{cccc} 0\; & \;  \frac12 \; &\; 0\; &\; 0\\ - \frac12 \; & \; 0\; &\; 0\; & \;  0 \\
0\; &\; 0  & \; 0\; &\;  \frac12 \\ 0\; &\; 0  & - \frac12 \; &\; 0\end{array}\right)  \; , \ee
and one concludes in the same way that the  D-brane charges are valued in $(\frac{1}{\alpha} \mathds{H}(3))^2 \oplus \mathds{H}(3)^2$.

The expansion of the non-perturbative $\bar{\mathcal{F}}^2 \mathcal{R}^2$ coupling in \eqref{ThetaExpand} is consistent with this definition, since the R-R charges appearing in the Fourier expansion are indeed valued in $(\frac{1}{\alpha} \mathds{H}(K))^2 \oplus \mathds{H}(K)^2$. In \cite{GBinprep} we will determine the non-perturbative $\mathcal{R}^4$ coupling assuming the existence of a non-trivial U-duality group including the `small' T-duality group $\Cong{D}{4}_{0}(\alpha)$ and a minimal congruent subgroup $\Cong{D}{6}_1(\alpha)\subset Sp(6,\mathds{H}(K))$.\footnote{$\Cong{D}{6}_1(\alpha)$ is the subgroup of elements $g = \mathds{1}$ mod $\alpha$ in $Sp(6,\mathds{H}(K))$.} This automorphic function indeed turns out to be invariant under the full T-duality group $\Cong{D}{4}_{0*}(\alpha)$ and admits D-brane instanton corrections with R-R charges in $(\frac{1}{\alpha} \mathds{H}(K))^2 \oplus \mathds{H}(K)^2$.

The construction of the low energy effective theory for such Euclidean D-brane instantons is rather subtle in general. Consider for simplicity a `bound state' of Euclidean D-branes wrapping cycles in $T^4$, which is invariant under the orbifold $\mathds{Z}_2$ action and preserves eight supersymmeries.  Its R-R charge satisfies then  \eqref{DbraneChargesT4}. Fixing moreover $q_1=q_2=0$ without loss of generality, one gets the configuration\footnote{The `extra' Euclidean D-strings are due to the off-diagonal components of the metric and B-field.} 
\be N \bigl(  {\rm D(-1)} +  {\rm D3} \bigr) + N^\prime {\rm D1}_{23} + (N + N^\prime) \bigl( {\rm D1}_{41} +{\rm D1}_{21}  \bigr) \label{Dexample}\ee
with the number of intersections
\be N^2 + N^\prime (N + N^\prime)\; . \ee
Before the orbifold projection, this configuration is T-dual to \footnote{Any vector $Q$ in $\sLambda_{4,4}$ is in the $O(4,4,\mathds{Z})$ orbit of a canonical vector with the same gcd($Q)$ and the same invariant norm $(Q,Q)$ equal twice the number of intersections.}
\be N_1  {\rm D1}_{12} + N_2 {\rm D1}_{34} \ee
with 
\be {\rm gcd}(N,N^\prime) = {\rm gcd}(N_1,N_2)\; , \qquad N^2 + N^\prime (N + N^\prime) = N_1 N_2 \; . \ee
Starting from the original brane configuration \eqref{Dexample},  tachyon condensation should lead to a brane recombination  \cite{Gava:1997jt, Angelantonj:2011hs} such that the field content of the corresponding low energy effective theory consists in a $U(N_1)\times U(N_2)$ $\mathcal{N}=4$ vector multiplet and a bi-fundamental hyper-multiplet. For $N^\prime=0$ one has $N_1=N_2=N$ and the $\mathds{Z}_2$ orbifold action should act on the gauge theory by exchange of the two types of Chan-Paton labels such that the orbifolded theory is a $U(N)$ gauge theory including one 
$\mathcal{N}=2$ vector multiplet and two adjoint hypermultiplets. The same is true for $N^\prime = - N$, that corresponds to taking Euclidean D-strings with opposite orientation, but generically $N_1\ne N_2$ and one must a priori analyse  tachyon condensation in the orbifolded theory to derive the corresponding low energy effective gauge theory.  Even for $N^\prime =0$ or $- N$, it would be important to derive the low energy effective theory from first principle as in \cite{Gava:1997jt, Angelantonj:2011hs}, but we expect  the argument above to apply, at least for $N$ odd (or $N$ not divisible by $K$ in general).

\subsection{NS5 and KK(6,1)}
The solitonic branes on $\mathds{R}^{1,3}\times T^6$ involve NS5 branes wrapping a five-cycle in $T^6$ and KK(6,1) branes that wrap $T^6$ with a Taub-NUT fibration along one circle in $T^6$. For the $\mathds{Z}_K$ orbifold, one may distinguish two cases. If the NS5 five-cycle only wraps a three-cycle in $T^4_{\rm a}$, then its image under the T-duality reflection on $T^4_{\rm a}$ is a KK(6,1) brane with a fibration along the dual circle in $T^4_{\rm a}$, so we have schematically 
\be [4]\: {\rm NS5}_{T^2_{\rm t} \times T^3_{\rm a}} + {\rm KK(6,1)}_{T^6 , S^1_{{\rm a}\bot}} \; . \ee
On the contrary if the NS5 brane wraps entirely $T^4_{\rm a}$, it is mapped to itself under T-duality, and so is a KK(6,1) brane with a fibration along $T^2_{\rm t}$, so we have schematically 
\be [2]\: {\rm NS5}_{S^1_{\rm t} \times T^4_{\rm a}} \; , \qquad [2]\:  {\rm KK(6,1)}_{T^6 , S^1_{{\rm t}}} \; . \ee
Such solitonic branes invariant under the orbifold action must wrap $K$ times $T^4_{\rm a}$ and have therefore charges divisible by $K$. Only the KK(6,1) brane with a fibration along the twisted circle can have integer charge because the Taub-NUT charge can be fractional since the periodicity of the twisted circle is effectively divided by $K$. This is consistent with the property that solitonic branes invariant under the orbifold $\mathds{Z}_K$ action generally have charges in 
\be K \mathds{F}_2(K)^* = K\sLambda_{1,1} \oplus \sLambda_{1,1}[K] \oplus \alpha \mathds{H}(K) \; . \ee

With the same argument, NS5 Euclidean branes must wrap a multiple of $K$ times $T^5$ to be invariant under the $\mathds{Z}_K$ orbifold action. One expects to have fractional NS5 Euclidean branes wrapping only $T^5 / \mathds{Z}_K$. Writing the tension for an NS5 brane wrapping $k$ times $T^5 / \mathds{Z}_K$  (i.e. $\frac{k}{K}$ times $T^5$) one gets  
\be \frac{2\pi k}{\gFour^2} = \frac{2\pi}{g_{\rm s}^2} \frac{\sqrt{G} \tilde{\RS} k}{K} \; . \ee
This justifies that $k$ must be integer for the non-perturbative coupling $\cF_\gra{0}{0}$. 

%%%%%%%%%%%%%%%%%%%%%%%%

%%%%%%%%%%%%%%%%%%%%%%%%

%%%%%%%%%%%%%%%%%%%%%%%%
%%%%%%%%%%%%%%%%%%%%%%%%
\section{Summary and conclusions}

Let us try and summarise our results, draw some conlcusions and indicate some lines for future investigation. 

We have shown how to embed ${\cal N}=6$ supergravity in string theory using asymmetric orbifolds, thus proving that only $K=2,3$ produce consistent models in $D=4,5$ while $K=4,6$ don't, contrary to naive expectation.\footnote{At the classical level ${\cal N}=6$ supergravity can be defined also in $D=6$ (which enjoys ${\cal N} =(2,1)$ local supersymmetry) but it is inconsistent at the quantum level due to chiral anomalies. Consistent asymmetric orbifolds in string theory include massless gravitino multiplets in the twisted sectors that enhance supersymmetry to ${\cal N}=8$ (i.e. ${\cal N} =(2,2)$ non-chiral supersymmetry in $D=6$).} We have then discussed how supersymmetry constrains the lowest higher-derivative corrections to the low-energy effective action. Relying on chiral factorisation of the integrands and on the results of \cite{Bianchi:2015vsa} for 4-point amplitudes of open-string vector bosons at one-loop in (unoriented) D-branes configurations with ${\cal N}\le 4$ supersymmetry in $D=4$ we have computed the one-loop threshold corrections to $\mathcal{R}^4, D^2\mathcal{R}^4$ and $\bar{\mathcal{F}}^2\mathcal{R}^2$ couplings. We have determine the T-duality group $\Cong{D}{4}_{0*}(\alpha)$ of symmetries of the one-loop perturbative coupling and of the spectrum of the theory.

Assuming that there is a non-trivial U-duality group and that the symmetry of the perturbative coupling is a symmetry of the non-perturbative theory we have been able to determine a unique non-pertubative $\bar{\mathcal{F}}^2\mathcal{R}^2$ coupling consistent with supersymmetry and the expected D-brane and NS5-brane instanton corrections. This conjecture will be strengthen in \cite{GBinprep}, in which the non-perturbative $\mathcal{R}^4$ coupling will be determined upon similar assumptions. 

These non-perturbative couplings provide a precise prediction for the spectrum of 1/3 BPS D-branes and their instanton analogues, as well as the corresponding integration measures (fluctuation determinants). But this approach is largely based on symmetries and the very rigid structure of Hurwitz quaternions. It would be interesting to explicitly construct the set of 1/3 BPS D-branes as boundary states of the conformal field theory, derive precisely the corresponding low-energy effective theories and compute their partition functions. A first analysis suggests that the effective theory for $N$ such D-branes is $U(N)$ $\mathcal{N}=2$  superYang--Mills coupled to two adjoint hypermultiplets when the R-R charge ${\bf q}$ is such that  $|{\rm gcrd}({\bf q})|^2= N^2$ for an integer $N$ not divisible by the orbifold order $K$.

 Another point that might be worth confirming is the uniqueness of string models with ${\cal N}= 6$ supersymmetry. We have been able to exclude asymmetric orbifolds with $K=4,6$ and in fact higher $K$ as well as to prove that two known consistent constructions with $K=3$ are indeed equivalent in that the full perturbative (massless and massive) spectra exactly coincide. The extension of these results to ${\cal N}= 5$ supersymmetry, for which there is no massless matter supermultiplet (except for gravitini that would enhance supersymmetry) may also be worth pursuing.   

%%%%%%%%%%%%%%%%%%%%%%%%
\subsection*{Acknowledgements}

We would like to thank Costas Bachas, Emilian Dudas, Axel Kleinschmidt, Francisco Morales, Boris Pioline, Gianfranco Pradisi and Oliver Schlotterer for useful discussions. GB gratefully acknowledges the warm hospitality of the University of Roma ``Tor Vergata'' where this work was initiated.

%%%%%%%%%%%%%%%%%%%%%%%%
\appendix 

\section{Hurwitz Quaternions and generalisations}

One can identify the root lattice $D_4$ with the set of Hurwitz quaternions that we write $\mathds{H}(2) = \mathds{H}$. To distinguish the quaternions imaginary units from the complex imaginary $i$, we write them $e_i$, with the quaternion algebra  
\be e_i e_j = -\delta_{ij} + \varepsilon_{ij}{}^k e_k \ . \ee
The Hurwitz quaternions is the ring over $\mathds{Z}$ generated by the unit norm quaternions  
\be 1\ ,  \quad u_1 =  e_1 \ , \quad u_2 = e_2 \; ,  \quad  u_3 =- \frac{1+e_1+e_2+e_3}{2} \ . \label{D4basis}  \ee
One verifies easily that the bilinear form 
\be (q,p) = q^* p + p^* q \; , \ee
with $1^* = 1$, $e_i^* = - e_i$ defines the $D_4$ Cartan matrix $\eta$ for the simple roots \eqref{D4basis}, so that $\mathds{H}$ defined as the set of vectors $q = n^0 + n^i u_i$ with $n^\mu \in \mathds{Z}$ can be identified with the $D_4$ lattice with the even norm
\be \eta_{\mu\nu} = u_\mu^* u_\nu + u_\nu^* u_\mu = \left(\begin{array}{cccc} 2 & 0& 0&-1\\ 0 & 2 & 0 & -1\\ 0 & 0& 2 & -1\\ -1&-1&-1& 2\end{array}\right) \; . \ee
 It follows that 
\be \sum_{q \in \mathds{H}} e^{2\pi i \tau |q|^2} = 2 E_2(2\tau) - E_2(\tau) = 1 + 24 \sum_{n=1}^\infty \sum_{\substack{d|n\\ d \ne 0 \, {\rm mod}\,  2}} \hspace{-4mm} d\hspace{2mm}  e^{2\pi i n \tau}= 
\eta^4 O_8 = \frac{2 \pi^2}{3} \wp(\tfrac{1}{2})
 \; .    \ee
can be identified with the $D_4$ Siegel--Narain theta series. 

The ring $\mathds{H}$ defines a maximal order as a ring of integers in  $\mathds{H}(\mathds{Q}) = \mathds{Q}[1,e_1,e_2,e_3]$. Importantly for us, there is a notion of prime decomposition in $\mathds{H}$ \cite{Krieg}. The 24 unit norm Hurwitz quaternions $u$ satisfying $|u|^2 =1$ are invertible of inverse $u^*$. So all Hurwitz quaternions are left and right divisible by any unit norm Hurwitz quaternion. They are $\pm 1$, $\pm e_i$ and $\frac{\pm1 \pm e_1\pm e_2\pm e_3}{2}$ and they form the order 24 finite group $SL(1,\mathds{H}) =\mathds{H}^\times $, which is also called the binary tetrahedral subgroup of  $SU(2)$, the double cover of the alternative group on four elements Alt$_4\in SO(3)$. 

There are also 24 Hurwitz quaternions of norm $2$: $\pm 1 \pm e_i$ and $\pm e_i \pm e_j$ for $i\ne j$.  They are all related by $SL(1,\mathds{H})$ and we shall write $\alpha = 1 + e_1$ for the representative of the unique element of norm square two in $\mathds{H} / \mathds{H}^\times$.

One can define the notion of left and right divisibility by a Hurwitz quaternion and we write respectively gcrd$({\bf q})$ and gcld$({\bf q})$ the greatest common right and left divisor of some vector of quaternions ${\bf q}$. It is only fixed up to right and left multiplication by a unit norm Hurwitz quaternion. One says that a Hurwitz quaternion is prime if it is only divisible by unit norm Hurwitz quaternions or by itself multiplied by a unit norm Hurwitz quaternion. Any Hurwitz quaternion not divisible by an integer $N\ne 1$ admits a unique ordered factorisation in primes in $\mathds{H} / \mathds{H}^\times$. Each real prime factor of $N\ne 1$ admits however an ambiguous prime factorisation in all quaternions $p$ of norm square $N$. The prime factorisation is induced by norm factorisation, i.e. for any element $q\in \mathds{H}$ such that $|q|^2$ is divisible by $n$ in $\mathds{N}^+$,  there exists a left (respectively right) divisor $p$ of $q$ in $\mathds{H}$ of norm square $|p|^2 = n$. 

One defines the Hurwitz zeta function 
\be  \zeta_\mathds{H}(s)  = \sum_{q\in \mathds{H}}^\prime |q|^{-2s} =  24 \sum_{n=1}^\infty \sum_{\substack{d|n\\ d \ne 0 \, {\rm mod}\,  2}} \hspace{-4mm} d\hspace{2mm}  n^{-s} =24 (1-2^{1-s}) \zeta(s) \zeta(s-1) \ . \label{ZetaH} \ee

One will also be interested in divisor sums over the quaternions
\be  \sum_{p|q \in \mathds{H}}  |p|^{2s}=\sum_{\substack{p \in \mathds{H}\\ p^{-1} q \in \mathds{H}}} |p|^{2s} =\sum_{\substack{p \in \mathds{H}\\  q p^{-1} \in \mathds{H}}} |p|^{2s}  \; . \ee
This sum can be simplified by decomposing $q$ over the real primes $\mathpzc{p}$ dividing $|q|^2$
\be |q|^2 = \prod_{\mathpzc{p}} \mathpzc{p}^{n_{\mathpzc{p}}} \; , \qquad   q = \prod_{\mathpzc{p}} \varpi_{ \mathpzc{p}^{n_{\mathpzc{p}}},a_\mathpzc{p} } \;  ,  \ee
by realising that the divisor sum is Eulerian, i.e. is the product over real primes $\mathpzc{p}$ dividing $|q|^2$ of the formula obtained for $\varpi_{ \mathpzc{p}^{n_{\mathpzc{p}}},a_\mathpzc{p} }$, up to the overall factor of $24$ counting the unite norm quaternions in $\mathds{H}^\times$. The formula is trivial for the powers of $2$ because $\alpha^n$ is the unique element in $\mathds{H} / \mathds{H}^{\times}$ of norm square $2^n$ and
\be  \sum_{p|\alpha^n \in \mathds{H}}  |p|^{2s} = 24 \sum_{k=0}^n |\alpha^k|^{2s} = 24    \sum_{d|2^n } d^s\; . \ee
For a prime $\mathpzc{p} \ne 2$ one writes a quaternion $ \mathpzc{p}^n  \varpi_{ \mathpzc{p}^{m},a }$ of norm square $\mathpzc{p}^{2n+m}$ with $  \varpi_{ \mathpzc{p}^{m},a }$ not divisible by ${\mathpzc{p}}$. By construction $\varpi_{ \mathpzc{p}^{m},a }$ admits a unique decomposition in prime quaternions up to left multiplication by $\mathds{H}^\times$, and so its divisor sum simplifies to 
\be  \sum_{p|\varpi_{ \mathpzc{p}^{m},a } \in \mathds{H}}  |p|^{2s} = 24 \sum_{k=0}^m \mathpzc{p}^{s k} = 24    \sum_{d|\mathpzc{p}^{m} } d^s\; . \ee
A generic left divisor of $ \mathpzc{p}^n  \varpi_{ \mathpzc{p}^{m},a }$ can be written as $\varpi_{\mathpzc{p}^k,b} \mathpzc{p}^{l} \varpi_{ \mathpzc{p}^{r},a }$ where $k + l \le n$, $r\le m$ and $\varpi_{ \mathpzc{p}^{r},a }$ divides $ \varpi_{ \mathpzc{p}^{m},a }$, and the index $b$ runs over $\sigma_1( \mathpzc{p}^k)$ (the sigma divisor sum) independent values for the independent quaternions of norm square $\mathpzc{p}^k$ up to left multiplication by $\mathds{H}^{\times}$. This expression is however redundant because $\varpi_{\mathpzc{p}^k,b}$ can include powers of $\mathpzc{p}$ and possibly $\varpi_{ \mathpzc{p}^{r},a }$. To enumerate them one considers the increasing norm square
\begin{align}&\varpi_{\mathpzc{p}^k,b} \; , \quad    0\le k\le n \; , \quad  &1\le |\varpi_{\mathpzc{p}^k,b} |^2 \le \mathpzc{p}^{n} \;  , \CR
& \varpi_{\mathpzc{p}^n,b}\varpi_{\mathpzc{p}^k,a}\; , \quad  1\le k\le m \; , \quad  &\mathpzc{p}^{n+1}\le |\varpi_{\mathpzc{p}^n,b}\varpi_{\mathpzc{p}^k,a}|^2 \le \mathpzc{p}^{n+m} \; , \CR
& \varpi_{\mathpzc{p}^k,b} \mathpzc{p}^{n-k} \varpi_{\mathpzc{p}^m,a} \; , \quad    0\le k\le n-1 \; , \quad  &\mathpzc{p}^{n+m+1}\le |\varpi_{\mathpzc{p}^k,b} \mathpzc{p}^{n-k} \varpi_{\mathpzc{p}^m,a}  |^2 \le \mathpzc{p}^{2n+m} \; . \end{align}
This decomposition is unique because the factor of $\varpi_{\mathpzc{p}^k,a}$ can always be absorbed in $\varpi_{\mathpzc{p}^k,b}$ for a norm square less than $ \mathpzc{p}^{n} $, and the factor of $ \mathpzc{p}^{k}$ can always be absorbed in $\varpi_{\mathpzc{p}^n,b}\varpi_{\mathpzc{p}^k,a}$ for a norm square less than $\mathpzc{p}^{n+m}$. As a result one obtains 
\bea   \sum_{p|\mathpzc{p}^n  \varpi_{ \mathpzc{p}^{m},a }  \in \mathds{H}}  |p|^{2s}  &=&24\left(   \sum_{k=0}^n  \frac{1-\mathpzc{p}^{k+1}}{1-\mathpzc{p}} \mathpzc{p}^{sk}+\frac{1-\mathpzc{p}^{n+1}}{1-\mathpzc{p}}\sum_{k=1}^m \mathpzc{p}^{s(n+k)}  + \sum_{k=0}^{n-1}  \frac{1-\mathpzc{p}^{k+1}}{1-\mathpzc{p}} \mathpzc{p}^{s(2n-k+m)} \right) \CR
&=& 24\sum_{k=0}^n \hspace{-0.5mm}\sum_{l=0}^{2n+m-2k}  \hspace{-2mm} \mathpzc{p}^{k(s+1)+l s }  =24 \sum_{d| \mathpzc{p}^n} d^{s+1} \sigma_s( \tfrac{\mathpzc{p}^{2n+m}}{d^2}) \; .\eea
In this way we obtain that for a general quaternion $q\in \mathds{H}$ 
\be  \sum_{p|q  \in \mathds{H}}  |p|^{2s} =  24   \hspace{-3mm} \sum_{\substack{d|q\\ d \ne 0\, {\rm mod}\, 2}}   \hspace{-2mm} d^{s+1} \sigma_s( \tfrac{|q|^2}{d^2}) \; .  \label{SigmaH2} \ee

Any Hurwitz quaternion of even norm square is both left and right divisible by $\alpha$, and we can therefore define the principal two-sided ideal  $\alpha \mathds{H} =  \mathds{H} \alpha$ of Hurwitz quaternions divisible by $\alpha$. Left and right divisibility are not equivalent for other non-real prime Hurwitz quaternions, and the only two-sided ideals in $\mathds{H}$ are  
\be \alpha N \mathds{H}  \; , \qquad   N \mathds{H} \; ,\label{RingProjection} \ee
for any $N \in \mathds{N}^+$. 

One can define the general linear group $GL(n,\mathds{H})$ over the Hurwitz quaternions as the set of invertible $n$ by $n$ matrices and the special linear group $SL(n,\mathds{H})$ as the set of  $n$ by $n$ matrices with unit Dieudonn\'e determinant. The Dieudonn\'e determinant is the homomorphism $GL(n,\mathds{H})\rightarrow \mathds{R}^+$ that is defined as the norm of the product of the diagonal components for an upper triangular (respectively lower triangular) $n$ by $n$ matrix over the quaternions. For example for $n=2$
\be \gamma = \left( \begin{array}{cc} a \; & \; b\;  \\ c & d \end{array}\right) \ ,\ee
the real determinant $\det \gamma=1$ is 
\be  \det \gamma =  \bigl| a  d -  a c  a^{-1} b  \bigr| \label{det} \ , \ee
if $a\ne 0$, and $\det \gamma  = | c \, b |$ otherwise. Extended over the real quaternions $x = x^0  + x^i e_i $ with $x^\mu \in \mathds{R}$, that we write $\mathds{H}(\mathds{R})$, one has the isomorphism $SL(n,\mathds{H}(\mathds{R})) \cong SU^*(2n)$. For $n=2$ $SU^*(4) = Spin(1,5)$. Because $\alpha \mathds{H} $ is a two-sided ideal one can define the congruent subgroup $\Cong{A}{3}_0(\alpha)\subset SL(2,\mathds{H})  $ of matrices $\gamma$ with $c\in \alpha \mathds{H} $,  which will be identified as (part of) the T-duality group in five dimensions.

One can also define the symplectic group  $Sp(2n,\mathds{H})$ as the set of $2n$ by $2n$ matrices $\gamma$ over the Hurwitz quaternions with 
\be \gamma = \left( \begin{array}{cc}\; A \; & \; B\; \\ \; C\; & \; D\; \end{array}\right) \ , \qquad \left( \begin{array}{cc}\; D^\dagger \; & -B^\dagger \; \\  -C^\dagger \; & \; A^\dagger \; \end{array}\right) \left( \begin{array}{cc}\; A \; & \; B\; \\ \; C\; & \; D\; \end{array}\right)  = \left( \begin{array}{cc}\; \mathds{1}  \; & \; 0\; \\ \; 0\; & \; \mathds{1} \; \end{array}\right)\; .  \ee
Extended over the reals one has the isomorphism $Sp(2n,\mathds{H}(\mathds{R})) \cong SO^*(4n)$, with a symplectic vector in $\mathds{H}^{2n}$ corresponding to a pseudo-real vector of $SO^*(4n)$. For $n=2$ its double cover is $Spin^*(8) = Spin(2,6)$. Because $\alpha \mathds{H} $ is a two-sided ideal one can define the congruent subgroup $\Cong{D}{4}_0(\alpha)\subset Sp(4,\mathds{H})  $ of matrices $\gamma$ with $C$ a two by two matrix in $\alpha \mathds{H} $, that will be identified as (part of) the T-duality group in four dimensions. 

\subsection*{Quaternions orders for $\mathds{Z}_K$ orbifolds with $K=2$, $3$, $4$, $6$}

These properties generalise for other rings of quaternions over $\mathds{Z}$. The ones relevant for the $\mathds{Z}_K$ orbifold theories with $K=2$, $3$, $4$ and $6$ are
{\allowdisplaybreaks
\bea  \mathds{H}=\mathds{H}(2) &=& \mathds{Z}\Bigl[ 1,e_1 , e_2 , \frac{1+e_1+e_2+e_3}{2}\Bigr] \; ,  \nonumber\\
 \mathds{H}(3) &=& \mathds{Z}\Bigl[ 1,e_1 , \frac{e_1+\sqrt{3} \, e_2}{2} , \frac{1 + \sqrt{3} \, e_3}{2}\Bigr] \; , \nonumber\\
  \mathds{H}(4) &=& \mathds{Z}\Bigl[ 1,e_1 , e_2 , e_3 \Bigr] \; ,  \nonumber\\
  \mathds{H}(6) &=&  \mathds{Z}\Bigl[ 1,e_1-e_2 , e_2-e_3 , \frac{1+e_1+e_2+e_3}{2} \Bigr] \; .\label{HKquaternions} \eea}
One straightforwardly checks that they define rings over $\mathds{Z}$ by computing the multiplication table of the generators. The ring $\mathds{H}(K) = \mathds{Z}[ 1,u_1,u_2,u_3]$ defines an even lattice with bilinear form $\eta_{\mu\nu}$ defined  as
\be q^* p + p^* q = \eta_{\mu\nu} n^\mu m^\nu \;  , \ee
for 
\be q = n^0  + n^i u_i\, , \quad p = m^0  + m^i u_i\, , \ee
with coefficients $n^\mu$ and $m^\mu$ in $\mathds{Z}$. One identifies $\mathds{H}(K)$ with the lattices 
\be \mathds{H}(2) = D_4\;, \quad \mathds{H}(3) = A_2\oplus A_2\, , \quad  \mathds{H}(4) = A_1\oplus A_1\oplus A_1\oplus A_1\, , \quad \mathds{H}(6) = A_2\oplus A_2[2]\, , \ee
where $A_2[2]$ is the lattice with bilinear form twice the Cartan matrix of $A_2$. One  defines the reduced discriminant of $\mathds{H}(K)$ as $\sqrt{\det \eta} = K \in \mathds{N}^+$. Both $\mathds{H}(4)$ and $\mathds{H}(6)$ are non-maximal orders inside the Hurwitz quaternions, while $\mathds{H}(3)$ is the unique maximal order of reduced discriminant $\sqrt{\det \eta} = 3$  \cite{Vigneras}. 

These four rings have the property that factorisation in $\mathds{H}(K)$ is induced by norm factorisation  \cite{FNF}, i.e. the property for any element $q\in \mathds{H}(K)$ such that $|q|^2$ is divisible by $n$ in $\mathds{N}^+$,  there exists a left (respectively right) divisor $p$ of $q$ in $\mathds{H}(K)$ of norm square $|p|^2 = n$. This property implies that the discrete groups $SL(n,\mathds{H}(K))$ and $Sp(2n,\mathds{H}(K))$ satisfy arithmetic properties that are essential for the definition of modular forms \cite{Krieg}. According  to \cite{FNF} there are only twelve rings of quaternions over $\mathds{Z}$ that define Euclidean lattice with this property of factorised induced by norm factorisation. Among them, only the four above are relevant to orbifold theories. 

The Siegel--Narain partition function 
\be \Lambda_{\mathds{H}(K)}(\tau) =\sum_{q\in \mathds{H}(K)} e^{2\pi i \tau |q|^2} \ee
can be written in terms of the weight two Eisenstein series $E_2(\tau)$ as
\bea  \Lambda_{\mathds{H}(2)}(\tau) &=& 2E_2(2\tau) - E_2(\tau)  = 1  + 24 \sum_{n=1}^\infty \Bigl( \sum_{\substack{ d | n\\ 2  / \hspace{-1.2mm} | \hspace{0.6mm}d}} d\Bigr) e^{2\pi i n \tau} = \eta^4 O_8 \; , \label{QuaternionNpartition} \CR
 \Lambda_{\mathds{H}(3)}(\tau) &=& \frac{1}{2} \bigl( 3 E_2(3\tau) - E_2(\tau)\bigr)  = 1  + 12 \sum_{n=1}^\infty \Bigl( \sum_{\substack{ d | n\\ 3  / \hspace{-1.2mm} | \hspace{0.6mm}d}} d\Bigr) e^{2\pi i n \tau}= \eta^4 \chi_0^{SU(3)} \chi_0^{SU(3)} \CR
 \Lambda_{\mathds{H}(4)}(\tau) &=& \frac{1}{3} \bigl( 4 E_2(4\tau) - E_2(\tau)\bigr)  = 1  + 8 \sum_{n=1}^\infty \Bigl( \sum_{\substack{ d | n\\ 4  / \hspace{-1.2mm} | \hspace{0.6mm}d}} d\Bigr) e^{2\pi i n \tau}\CR
 \Lambda_{\mathds{H}(6)}(\tau) &=& \frac{1}{4} \bigl( 6 E_2(6\tau) -3 E_2(3\tau) + 2 E_2(2\tau) - E_2(\tau) \bigr) \;  ,  \eea
which shows explicitly that they are weight 2 modular forms with respect to the congruent subgroup $\Gamma_0(K)\subset SL(2,\mathds{Z})$. One can extract the number of quaternions of norm square $|q|^2=n$ in $\mathds{H}(K)$ from these formula.

By construction, all the elements of norm square one are invertible so one defines the group $\mathds{H}(K)^\times= SL(1,\mathds{H}(K))$ of unit quaternions in $\mathds{H}(K)$. They are given by \be \mathds{H}(2)^\times  = \mathds{Z}_2 \ltimes {\rm Alt}_4 \supset \mathds{Z}_2 \, , \quad \mathds{H}(3)^\times  ={\rm Alt}_4   \supset \mathds{Z}_3  \, , \quad \mathds{H}(4)^\times = Q_8 \supset \mathds{Z}_4    \, ,\,  \quad \mathds{H}(6)^\times  =  \mathds{Z}_6 \; , \ee
where we write Alt$_n$ for the alternating group to avoid confusion with the Cartan notation $A_n$, and $Q_8$ is the order eight quaternion group. The fact that $\mathds{Z}_K\subset \mathds{H}(K)^\times$ ensures that the cyclic group of order $K$ acts through the left multiplication in $ \mathds{H}(K)$ on the lattice $ \mathds{H}(K)^\times$. This distinguishes the four rings of quaternions listed here in the twelve rings with factorisation induced by norm factorisation.\footnote{The eight others also define $K$-modular lattices for $K=5$, $6$, $7$, $10$, $12$, $13$, $22$, but do not admit a $\mathds{Z}_K$  automorphism defined by the left multiplication by unites.} This action allows to define the  asymmetric orbifold action $\mathds{Z}_K \subset SU(2) \subset SO(4)$ on $T^4$.

For each of these quaternion rings one can define principal two-sided ideals for non-real quaternions of norm square dividing $K$. These two-sided ideal determine the possible congruent subgroups of $SL(n,\mathds{H}(K))$ and $Sp(2n,\mathds{H}(K))$.

In $\mathds{H}(3)$ the only element of norm square $3$ in $\mathds{H}(3) / \mathds{H}(3)^\times$ is $\alpha = \sqrt{3}\, e_1$. One has the principal two-sided ideal $\alpha \mathds{H}(3)$ and the dual lattice can be identified with the elements in $\frac{1}{\alpha} \mathds{H}(3)$. The zeta function can be defined as 
\be  \zeta_{\mathds{H}(3)}(s)  = \sum_{q\in \mathds{H}(3)}^\prime |q|^{-2s} =  12 (1-3^{1-s}) \zeta(s) \zeta(s-1) \ , \label{ZetaH3} \ee
and the divisor sum simplifies to 
\be  \sum_{p|q  \in \mathds{H}(3)}  |p|^{2s} =  12   \hspace{-3mm}\sum_{\substack{d|q\\ d \ne 0\, {\rm mod}\, 3}}   \hspace{-2mm} d^{s+1} \sigma_s( \tfrac{|q|^2}{d^2}) \; .  \label{Sigma3} \ee

In $\mathds{H}(4)$ there are three elements of norm square $2$ in $\mathds{H}(4) / \mathds{H}(4)^\times$, that one chooses as $1+e_i$, and three elements of norm square $4$ in $\mathds{H}(4) / \mathds{H}(4)^\times$, that one represents as $2$,  $1+e_1+e_2+e_3$ and its conjugate.  One has the three principal two-sided ideals $(1+e_i)\mathds{H}(4)$ that all include the principal ideal $2\mathds{H}(4)$. The dual lattice can be identified with  $\frac12 \mathds{H}(4)$.

In $\mathds{H}(6)$ there is only one  element of norm square $2$ in $\mathds{H}(6) / \mathds{H}(6)^\times$, that one chooses as $e_2-e_3$, and seven elements of norm square $3$ in $\mathds{H}(6) / \mathds{H}(6)^\times$. Among the seven, only $e_1+e_2+e_3$ defines a two-sided ideal. There are seven elements of norm square $6$ in $\mathds{H}(6) / \mathds{H}(6)^\times$, but only $\alpha  =  2e_1-e_2-e_3$ defines a two sided ideal, which is included in the two ideal defined above. The  two-sided ideals are are therefore 
\be (e_2-e_3)  N \mathds{H}(6)\; , \quad (e_1+e_2+e_3) N  \mathds{H}(6)\; ,  \quad  \alpha N \mathds{H}(6) =   (2e_1-e_2-e_3)  N \mathds{H}(6)\; . \ee
The dual lattice can be identified with $\frac{1}{\alpha}   \mathds{H}(6)$.

\section{Explicit helicity supertraces and BPS states indices}
\label{HelicitySupertraces23}
In this Appendix we give the helicity supertraces for the $\mathds{Z}_2$ and $\mathds{Z}_3$ orbifold theories and derive from it the supersymmetry protected BPS indices. 
\subsection{ $\mathds{Z}_2$ case}
The lattice of  zero modes is $\mathds{F}_2(2)^*$ with
\be \mathds{F}_2(2) = \sLambda_{1,1}\oplus \sLambda_{1,1}[2]\oplus D_4\; , \ee
and
\bea B^{\mathds{Z}_2}_6 &=& \frac{45}{4} \Bigl( 4 \int_{-\frac12}^{\frac12}\hspace{-2mm} d\tau_1\;  \Lambda_{\sLambda_{1,1}\oplus \sLambda_{1,1}[2]\oplus D_4 }- \int_{-1}^{1}\hspace{-3mm}  d\tau_1 \; \Lambda_{\sLambda_{1,1}\oplus \sLambda_{1,1}[\frac12]\oplus D_4^*} \Bigr)\; ,  \CR 
B^{\mathds{Z}_2}_8 &=& \frac{315}{8} \Bigl( \frac{8}{3}  \int_{-\frac12}^{\frac12}\hspace{-2mm}  d\tau_1\, \Lambda_{\sLambda_{1,1}\oplus \sLambda_{1,1}[2]\oplus D_4} \bigl(2+2E_2(2\tau) - E_2(\tau)\bigr)\CR
&& \hspace{20mm}-\frac{1}{3} \int_{-1}^{1}\hspace{-3mm}  d\tau_1 \, \Lambda_{\sLambda_{1,1}\oplus  \sLambda_{1,1}[\frac12]\oplus D_4^*} \bigl(4-2E_2(\tau)+ E_2(\tfrac \tau 2)\bigr)\Bigr) \; ,   \eea
which is consistent  with the property that for each untwisted charge ${\cal Q}\in  \mathds{F}_2(2)$ with ${\cal Q}^2=0$ there is a single 1/2 BPS spin two supermultiplet, and two  1/2 BPS spin three half supermultiplets for each twisted charge   ${\cal Q}\in \mathds{F}_2(2)^*\smallsetminus  \mathds{F}_2(2)$ with ${\cal Q}^2=0$. 

Let us define $d_j^\ord{\frac{k}{6}}({\cal Q})$ as the number of supermultiplets $\psi \in \mathcal{E}_j^{k/6}$ with zero mode ${\cal Q} \in \mathds{F}_2(2)^*$. The comparison of $B^{\mathds{Z}_2}_8$ with \eqref{BSMult} gives for the untwisted charge ${\cal Q}\in  \mathds{F}_2(2)$ with ${\cal Q}^2<0$
\be \sum_{j\in \mathds{N}/2} (-1)^{2j} (2j+1) d_j^\ord{\frac13}({\cal Q}) = \; 20 \hspace{-3mm}  \sum_{\substack{d | \frac{{\cal Q}^2 }{2} \\ d = 1 \, {\rm mod} \, 2}} \hspace{-3mm} d \; , \label{OneThird2u} \ee
and for the twisted states ${\cal Q}\in  \mathds{F}_2(2)^* \smallsetminus  \mathds{F}_2(2) $ with ${\cal Q}^2<0$
\be \sum_{j\in \mathds{N}/2} (-1)^{2j} (2j+1) d_j^\ord{\frac13}({\cal Q}) = \; 4 \hspace{-3mm}  \sum_{\substack{d | {\cal Q}^2 \\ d = 1 \, {\rm mod} \, 2}} \hspace{-3mm} d \; .  \label{OneThird2t} \ee
For the 1/3 BPS states $B^{\mathds{Z}_2}_{10}$ gives for the untwisted charge ${Q}\in  \mathds{F}_2(2)$ with ${\cal Q}^2<0$
\be \sum_{j\in \mathds{N}/2} (-1)^{2j} (2j+1) j (j+1) d_j^\ord{\frac13}({\cal Q}) = \; 6\sum_{2d | {\cal Q}^2 }  d^3  \; ,  \label{OneThird2uB10}\ee
and for the twisted states ${\cal Q}\in\mathds{F}_2(2)^*\smallsetminus \mathds{F}_2(2)$ with ${\cal Q}^2<0$ 
\be \sum_{j\in \mathds{N}/2} (-1)^{2j} (2j+1) j (j+1)  d_j^\ord{\frac13}({\cal Q}) = -6 \sum_{2d | {\cal Q}^2  }  d^3  \; , \label{OneThird2tB10}\ee
and therefore vanishes if ${\cal Q}^2$ is odd.

The states with charge ${\cal Q}\in\mathds{F}_2(2)^*$ with ${\cal Q}^2>0$ are 1/6 BPS, and one obtains from the term in $E_4(-\bar \tau)$ in $B^{\mathds{Z}_2}_{10}$
\be \sum_{j\in \mathds{N}/2} (-1)^{2j} (2j+1) d_j^\ord{\frac16}({\cal Q}) = - 2 \sum_{2d | {\cal Q}^2}  d^3 \; ,  \label{OneSixth2u} \ee
for untwisted charges and 
\be \sum_{j\in \mathds{N}/2} (-1)^{2j} (2j+1) d_j^\ord{\frac16}({\cal Q}) =  2 \sum_{2d | {\cal Q}^2 }  d^3 \; , \label{OneSixth2t}\ee
for twisted charges, and therefore vanishes if ${Q}^2$ is odd.

\subsection{ $\mathds{Z}_3$ case}
In this case 
\be \mathds{F}_2(3) = \sLambda_{1,1}\oplus \sLambda_{1,1}[3]\oplus A_2\oplus A_2\; , \ee
and there is only one massless graviton multiplet at ${\cal Q}=0$, with $d_\frac12^\ord{\frac12}=1$,  one spin 2 multiplet for each untwisted charge and one spin $3/2$ multiplet for each twisted charge. This is indeed consistent with $B_6$ and $B_8$
\bea B_6 &=& \frac{45}{4} \Bigl( 3 \int_{-\frac12}^{\frac12}\hspace{-2mm} d\tau_1\;  \Lambda_{\sLambda_{1,1}\oplus \sLambda_{1,1}[3]\oplus \mathds{H}(3)} -\frac13 \int_{-\frac32}^{\frac32}\hspace{-2mm}  d\tau_1 \; \Lambda_{\sLambda_{1,1}\oplus \sLambda_{1,1}[\frac13]\oplus \frac 1 \alpha \mathds{H}(3)}\Bigr)\; ,  \CR
B_8 &=& \frac{315}{8} \Bigl(   \int_{-\frac12}^{\frac12}\hspace{-2mm}  d\tau_1\, \Lambda_{\sLambda_{1,1}\oplus \sLambda_{1,1}[3]\oplus \mathds{H}(3)} \Bigl( 4 + 3 \frac{3E_2(3\tau)- E_2(\tau)}{2} \Bigr) \CR 
&& \hspace{20mm} -\frac19 \int_{-\frac32}^{\frac32}\hspace{-2mm}  d\tau_1 \, \Lambda_{\sLambda_{1,1}\oplus  \sLambda_{1,1}[\frac13]\oplus  \frac 1 \alpha  \mathds{H}(3)}  \Bigl( 4 - \frac{3E_2(\tau)- E_2(\frac{\tau}{3})}{2} \Bigr) \; .   \eea

For the 1/3 BPS states, the expansion of $B_8$ gives for the untwisted charge ${\cal Q}\in \mathds{F}_2(3) $ with ${\cal Q}^2<0$
\be \sum_{j\in \mathds{N}/2} (-1)^{2j} (2j+1) d_j^\ord{\frac13}({\cal Q}) =  10  \sum_{\substack{d | \frac{{\cal Q}^2}{2} \\  3  / \hspace{-1.2mm} | \hspace{0.6mm}d  }}  d \; ,\label{OneThird3u} \ee
and for the twisted states ${\cal Q}\in\mathds{F}_2(3)^* \smallsetminus \mathds{F}_2(3) $ with ${\cal Q}^2<0$
\be \sum_{j\in \mathds{N}/2} (-1)^{2j} (2j+1) d_j^\ord{\frac13}({\cal Q}) =     \sum_{\substack{d |\frac{3 {\cal Q}^2}{2} \\ 3  / \hspace{-1.2mm} | \hspace{0.6mm}d  }}  d \; . \label{OneThird3t} \ee
For the same states $B_{10}$ gives for the untwisted charge ${\cal Q}\in \mathds{F}_2(3) $ with ${\cal Q}^2<0$
\be \sum_{j\in \mathds{N}/2} (-1)^{2j} (2j+1) j (j+1) d_j^\ord{\frac13}({\cal Q}) = \; 6\sum_{d | \frac{{\cal Q}^2 }{2} }  d^3  \; , \label{OneThird3uB10}\ee
and for the twisted states ${\cal Q}\in \mathds{F}_2(3)^* \smallsetminus  \mathds{F}_2(3) $ with ${\cal Q}^2<0$ and ${\cal Q}^2/2 \in \mathds{N}$
\be \sum_{j\in \mathds{N}/2} (-1)^{2j} (2j+1) j (j+1)  d_j^\ord{\frac13}({\cal Q}) = -3 \sum_{d | \frac{{\cal Q}^2 }{2} }  d^3  \; , \label{OneThird3tB10}\ee
and zero if ${\cal Q}^2/2$ is fractional, {\it i.e.} if $3{\cal Q}^2/2 = \pm 1 $ mod $3$.

The states with charge ${\cal Q}\in \mathds{F}_2(3)^*  $ with ${\cal Q}^2>0$ are 1/6 BPS according to supergravity black hole solutions, and one has then from $B_{10}$ for ${\cal Q}\in \mathds{F}_2(3) $ with ${\cal Q}^2>0$
\be \sum_{j\in \mathds{N}/2} (-1)^{2j} (2j+1) d_j^\ord{\frac16}({\cal Q}) =  -2\sum_{d | \frac{{\cal Q}^2 }{2} }  d^3 \; , \label{OneSixth3u}\ee
and for ${\cal Q}\in  \mathds{F}_2(3)^*  \smallsetminus  \mathds{F}_2(3)  $ with ${\cal Q}^2>0$ and ${\cal Q}^2/2\in \mathds{N}$
\be \sum_{j\in \mathds{N}/2} (-1)^{2j} (2j+1) d_j^\ord{\frac16}({\cal Q}) =   \sum_{d | \frac{{\cal Q}^2 }{2} }  d^3 \; , \label{OneSixth3t} \ee
and zero if ${\cal Q}^2/2$ is fractional.

\section{Elliptic functions and other useful formulae} 

In this Appendix we recall the definitions of some elliptic functions and their properties that are useful in the calculation of helicity supertraces in Section \ref{CharZK} and the worldsheet integrals in Section \ref{String1-loop}.

We start from the Jacobi elliptic $\vartheta(z|\t)$ functions. They are solutions of the heat equation on the torus $(\partial_z^2- 4 \p i \partial_\t)\vartheta(z|\t)=0$
\begin{equation}
\vartheta [\substack{\alpha \\ \beta}] (z|\tau)=\sum_{k\in \mathbb{Z}} q^{(k-\alpha)^2/2} e^{2\pi i(z-\beta)(k-\alpha)}\; . 
\end{equation}
We will denote $\partial_z f$ by $f^\prime$. For our purposes, we are interested in four particular $\vartheta$ functions
\begin{align}
\vartheta [\substack{1/2 \\ 1/2}] (z|\tau)&=\vartheta_1 (z|\t)= 2q^{1/8} \sin (\pi z) \prod_{n=1}^\infty (1-q^n) (1-e^{2\pi i z} q^n) (1-e^{-2\pi i z} q^n) \; , \\
\vartheta [\substack{1/2 \\ 0}] (z|\tau)&=\vartheta_2 (z|\t)= 2q^{1/8} \cos (\pi z) \prod_{n=1}^\infty (1-q^n) (1+e^{2\pi i z} q^n) (1+e^{-2\pi i z} q^n) \; , \\
\vartheta [\substack{0 \\ 0}] (z|\tau)&=\vartheta_3 (z|\t)= \prod_{n=1}^\infty (1-q^n) (1+e^{2\pi i z} q^{n-1/2}) (1+e^{-2\pi i z} q^{n-1/2})\; ,  \\
\vartheta [\substack{0 \\ 1/2}] (z|\tau)&=\vartheta_4 (z|\t)= \prod_{n=1}^\infty (1-q^n) (1-e^{2\pi i z} q^{n-1/2}) (1-e^{-2\pi i z} q^{n-1/2})\; . 
\end{align}
It is useful to notice that $\vartheta_1'(0|\t)=2 \pi \eta(\tau)^3$, where $\eta(\tau)$ is the Dedekind function
\begin{equation}
\eta(\tau) = q^{ 1 \over 24} \prod_{k=1}^\infty (1-q^k)\; . 
\end{equation}
Another relevant function is the Weierstrass function
\begin{equation}
\begin{aligned}
\wp (z|\tau)=& \,4 \pi i \partial_\tau \log \eta(\tau)-\partial_z^2 \log \vartheta_1 (z|\tau)=
-\frac{\vartheta_1''(z)}{\vartheta_1(z)} + \frac{[\vartheta_1'(z)]^2}{ \vartheta_1(z)^2} +\frac{1}{3}\frac{\vartheta_1'''(0)}{\vartheta_1'(0)} \\
=& \, \frac{1}{z^2}+ \left. \sum_{m,n}\right.^\prime \left[\frac{1}{(z+n + m \t)^2}-\frac{1}{(n + m \t)^2} \right]
\end{aligned}
\label{wp_def}
\end{equation}
where the primed sum means that the term $(m,n)=(0,0)$ is excluded.

We also consider general Eisenstein series
\begin{equation} \label{Eisensteinsw}
E_{s,w}(\tau) = \frac{1}{2\zeta(2s+w)} \left. \sum_{m,n}\right.^{\prime} \frac{ \tau_2^{\; s}}{| m \tau+n|^{2s} (m\tau+n)^{w}}
\end{equation}
where $E_{0,w}(\tau)= E_{w}(\tau)$ are the holomorphic Eisenstein series for $w\ge 4$ integer and $E_{s,0}(\tau)$ the real analytic Eisenstein series. The series $E_{s,2}(\tau)$ satisfy the identity
\begin{equation}
E_{s,2}(\tau)  = \frac{2i}{s+1} \partial_{\tau} E_{s+1,0}(\tau)
\label{Es2_Es0}
\end{equation}
We also use the following relations\footnote{With $A_{\rm \scriptscriptstyle GK}$ the Glaisher-Kinkelin constant with $\log A_{\rm \scriptscriptstyle GK}  = \frac{1}{12} - \zeta^\prime(-1)$.}
\begin{equation}
{E}_{1+\epsilon,0}(\tau) =  \frac{3}{\pi \epsilon} +  \frac{72 \log (A_{\rm \scriptscriptstyle GK})-6 \log(4\pi) }{\pi} + \hat{E}_{1,0}(\tau) + \mathcal{O}(\epsilon)\ , \qquad 
\hat{E}_{0,2}(\tau) = 2i \partial_\tau  \hat{E}_{1,0}(\tau) 
\label{E10_relations}
\end{equation}
%\begin{equation}
%\zeta(2\epsilon) E_{\epsilon,0}(\tau) =  \frac{ \Gamma(1-\epsilon)}{\pi^{1-2\epsilon} \Gamma(\epsilon)} \zeta(2-2\epsilon) E_{1-\epsilon,0}(\tau)
%\end{equation}
where $\hat{E}_{0,2}$ is the standard quasi-holomorphic Eisenstein series $\hat{E}_{2}$. More explicitly we have 
\begin{equation}
\hat{E}_{1,0}(\tau) = - \frac{3}{\pi} \log( \tau_2 | \eta(\tau)|^4)
\quad, \quad
\hat{E}_{0,2}(\tau) = \hat{E}_2 (\tau) =
- \frac{6 i}{\pi} \, \partial_\tau \log [\tau_2 \eta^2(\tau) ] \; . 
\end{equation}
The $q$ expansion of $\hat{E}_{0,2}$ and $E_{1,2}$ can be recasted in the form
\begin{align}
\hat{E}_{0,2} & = 1 - 24 \sum_{n=1}^\infty \frac{q^n}{1-q^n} - \frac{3}{\pi \tau_2} \ ,\\
E_{1,2} & = \tau_2 - \frac{45}{2\pi^3 \tau_2^{\; 2}} \sum_{n=1}^\infty \frac{1}{n^3} \frac{1-q^n \bar q^n}{(1-q^n)(1-\bar q^n)} - \frac{180}{\pi} \sum_{n=1}^\infty \frac{1}{n} \frac{q^n(1+q^n) }{(1-q^n)^3}- \frac{90}{\pi^2 \tau_2}  \sum_{n=1}^\infty \frac{1}{n^2} \frac{q^n}{(1-q^n)^2} \nn
\end{align}

The Weierstrass function satisfies the identity
\begin{equation}
(\wp ')^2 = 4 \wp^3-  \frac{4}{3} \pi^4 E_4(\tau)   \wp-  \frac{8}{27}\pi^6 E_6(\tau)=4(\wp -e_1)(\wp -e_2)(\wp -e_3)\; , 
\label{wp_poly}
\end{equation}
where  $-e_i$ are the values of the Weierstrass function at the half-periods 
\begin{equation}
e_1 = -\wp(\tfrac{1}{2}) \quad, \quad
e_2 = -\wp(\tfrac{1+\tau}{2})\quad, \quad
e_2 = -\wp(\tfrac{\tau}{2})\; . 
\end{equation}
The functions $e_i$ are related to the Jacobi functions by the identity
\begin{equation} \label{eiJacobi} 
e_i(\tau) = 4 \pi i \partial_\tau \log \frac{\vartheta_{i+1}(0|\tau)}{\eta(\tau)}\; ,
\end{equation}
and satisfy 
\begin{equation}
e_i^2 = -i \pi \partial_\tau e_i + 4 \pi i e_i \partial_\tau \log \eta +  \frac{2\pi^4}{9}  E_4(\tau)\; . 
\end{equation}
Deriving \eqref{wp_poly} with respect to $z$ we can write the second derivative of $\wp$ in terms of $\wp$ itself
\begin{equation}
\wp''(z)= 6\wp^2(z)- \frac{2\pi^4}{3}  E_4(\tau)\; . 
\end{equation}
We will also use the following relation between Weierstrass and Jacobi functions
\cite{Weiertheta} 
\be  \label{WeierDiffTheta} \wp(z_1)-\wp(z_2)= 4\pi^2 \eta(\tau)^6\frac{  \vartheta_1(z_1+z_2)\vartheta_1(z_2-z_1)}{  \vartheta_1(z_1)^2  \vartheta_1(z_2)^2} \; . \ee

For the purpose of computing the helicity supertraces, it is also useful to express derivatives of $\vartheta_1$ in terms of the first derivative of $\vartheta_1$ and the Weierstrass function using \eqref{wp_def}
\begin{align}
\vartheta_1''(z)&=
\frac{\vartheta_1'(z)^2}{\vartheta_1(z)}+
\frac{\vartheta_1^{(3)}(0)}{3 \vartheta_1'(0)}\vartheta_1(z)
-\vartheta_1(z) {\wp(z)} \\
\vartheta_1^{(3)}(z)&=
\frac{\vartheta_1'(z)^3}{\vartheta_1(z)^2}+
\frac{\vartheta_1^{(3)}(0)}{\vartheta_1'(0)}\vartheta_1'(z)
-\vartheta_1(z)  \wp'(z)-3 {\wp(z)} \vartheta_1'(z)
\\
\vartheta_1^{(4)}(z)&=
\frac{\vartheta_1'(z)^4}{\vartheta_1(z)^3}+
\frac{2\vartheta_1^{(3)}(0)}{\vartheta_1'(0)}\frac{\vartheta_1'(z)^2}{\vartheta_1(z)}
+\frac{\vartheta_1^{(3)}(0)^2 }{3 \vartheta_1'(0)^2}\vartheta_1(z) -\vartheta_1(z) {\wp}''(z) -4 \vartheta_1'(z) \wp'(z) \nonumber \\
&~~-\frac{6 {\wp(z)} \vartheta_1'(z)^2}{\vartheta_1(z)}
-\frac{2 \vartheta_1^{(3)}(0) }{\vartheta_1'(0)} \vartheta_1(z) {\wp(z)}+3 \vartheta_1(z) {\wp(z)}^2
\end{align}
and also
\begin{equation}
\vartheta_1'(0) =  2 \pi \eta^3 \quad, \quad
\vartheta_1^{(3)}(0) =  - \pi ^3 E_2 \eta^3 \quad, \quad
\vartheta_1^{(5)}(0) =  \frac{10}{3} \pi^5 E_2^2 \eta ^3-\frac{4}{3} \pi ^5 E_4 \eta ^3
\end{equation}
In Section \ref{StringOrbifolds} we define
\begin{equation}
\xi(v) = \frac{\sin\pi v }{ \pi}  \frac{\vartheta_1'(0) }{\vartheta_1(v)} 
\end{equation}
which is an even function satisfying $\xi(0) =1$. One shows that its first two non-vanishing derivatives give
\begin{align}
\xi''(0) = & \, - \frac{1}{3} \left[ \frac{\vartheta_1'''(0)}{\vartheta_1'(0)} + \pi^2\right] = \frac{\pi^2}{3} (E_2-1) \; , \\
\xi^{(4)}(0) & \,  = \frac{\pi^4}{5} + \frac{2\pi^2}{3} \frac{\vartheta_1'''(0)}{ \vartheta_1'(0)} 
{+}\frac{2}{3}\Bigg[ \frac{\vartheta_1'''(0)}{\vartheta_1'(0)} \Bigg]^2 {-} \frac{1}{5} \frac{\vartheta_1^{(5)}(0)}{ \vartheta_1'(0)}
= {\pi^4\over 15} (3-10 E_2+2E_4 +5 E_2^2)\; . 
\end{align}

Let us now give some formulae about the torus propagator, defined as 
\begin{equation}
{{G}}(z_1,z_2|\tau)= {{G}}(z_{12}|\tau) = -\log \left|\frac{\vartheta_1(z_{12}|\tau)}{\vartheta_1^\prime(0|\tau)} \right|^2 +2 \pi \frac{{\rm Im}^2 z_{12}}{{\rm Im} \tau}\; . 
\end{equation}
It will be convenient to consider the Green function without zero mode as in \cite{DHoker:2015gmr}
\be \mathcal{G}(z|\tau) = G(z|\tau)-2\log 2\pi |\eta(\tau)|^2 = -\log \left|\frac{\vartheta_1(z|\tau)}{\eta(\tau)} \right|^2 +2 \pi \frac{{\rm Im}^2 z}{{\rm Im} \tau}\; , \ee
which admits the regularised Fourier expansion 
\be
{{\calG}}_\epsilon(z|\tau) = \frac{1}{\pi} \left. \sum_{m,n}\right.^\prime  \frac{\tau_2^{1+\epsilon}}{|m \tau-n|^{2+2\epsilon}} e^{2 \pi i (m x+n y)}\; ,
\label{propagator_fourier}
\ee
such that ${{\calG}}(z|\tau) = \lim_{\epsilon\rightarrow 0} {{\calG}}_\epsilon(z|\tau) $. 
We will often refer to the torus propagator as the one without the zero mode, since the zero mode drops out from the Koba--Nielsen factor  by momentum conservation. 

In Section \ref{String1-loop} we also mention the annulus propagator $G_A$, which is defined from the torus propagator as
\begin{equation}
{{G}}_A(z_1,z_2|\tau_A) = \frac{1}{2} [
{G}(z_1,z_2|\tau_A){+}
{G}(z_1,1{-}\bar{z}_2|\tau_A){+}
{G}(1{-}\bar{z}_1,z_2|\tau_A){+}
{G}(1{-}\bar{z}_1,1{-}\bar{z}_2|\tau_A)
]
\end{equation}
where $\tau_A = i T/2 $ and it reduces to 
\begin{equation}
{{G}}_A(z_1,z_2|\tau_A) = -4 \log \frac{\vartheta_1(z_{12}|\tau)}{\vartheta_1^\prime(0|\tau)}  +4 \pi \frac{{\rm Im}^2 z_{12}}{{\rm Im} \tau}
\label{annulus_prop}
\end{equation}
where we used that the annulus can be obtained from the torus from the involution $\tilde{z}=1{-}z$.

In the one-loop amplitude we meet the function $\calY(z|\tau)$ defined as
\begin{align}
\calY(z|\tau) &=- 2 [\wp(z|\tau)-(\partial_z {{\calG}})^2 (z|\tau)] \label{def_calY_1} \\
& = 2 (\partial_z {{\calG}})^2 (z|\tau)- 2 \partial_z^2 {{\calG}} (z|\tau)-4 \pi i \partial_\tau \log [\tau_2\,\eta^2 (\tau) ]\nonumber 
\end{align}
where in the latter equation we used the identity
\begin{equation}
\wp(z|\tau) - \partial_z^2 {{\calG}} (z|\tau) = 
4 \pi i \partial_\tau \log [\eta(\tau) \sqrt{\tau_2}]\; . 
\label{wp_prop_identity}
\end{equation}
Using the definition of the torus propagator and the heat equation for $\vartheta_1$, the function $\calY$ can be written in a form in which the absence of double pole is manifest 
\begin{equation}
\calY(z|\tau) = 8 \pi i \left[\partial_\tau \log \frac{\vartheta_1(z| \tau)}{\eta(\tau)} +\frac{{\rm Im} z}{\tau_2}\partial_z  \log \vartheta_1(z| \tau)  + i \pi \frac{{\rm Im}^2 z}{\tau_2^2}\right]
\label{calY_def2}
\end{equation}
since the terms $[\vartheta_1'(z)]^2$, which should yield the double pole, cancel.

One  defines the Szeg\"{o} kernel as
\begin{gather}
S_\alpha(z_1,z_2;\tau)=S_\alpha(z_{12})=
\begin{cases}
S(z_{12})= -\partial_{z_1} \mathcal{G}(z_{12}) & \text{if $\alpha=1$,} \\
\dfrac{\vartheta_\alpha(z_{12}) \vartheta_1'(0)}{\vartheta_1(z_{12}) \vartheta_\alpha (0)}   & \text{if $\alpha > 1$.}
\end{cases}
\end{gather}
When $\alpha>1$ the Szeg\"{o} kernel satisfies the identities
\begin{align}
S_{i+1}^2(z) & = \wp (z) + e_i \\
S_{i+1}(z_{12}) S_{i+1}(z_{23}) & =  S_{i+1}(z_{13}) \,\omega_{123} - \partial_{z_1} S_{i+1} (z_{13})
\end{align}
where
\begin{equation}
\omega_{123} = S(z_{12})+S(z_{23})+S(z_{31})=
\partial_{z_1} \log \vartheta_1(z_{12}){+}\partial_{z_2} \log \vartheta_1(z_{23}){+}\partial_{z_3} \log \vartheta_1(z_{31})\; . 
\end{equation}

A variant of Fay's trisecant identity gives 
\begin{equation}
\Omega_{1 2 3} \equiv S(z_{12}) S(z_{23})+S(z_{23}) S(z_{31})+S(z_{31}) S(z_{12}) = - \frac{1}{4} \left( \mathcal{Y}(z_{12})+ \mathcal{Y}(z_{23})+ \mathcal{Y}(z_{31}) \right)\; . 
\label{fay}
\end{equation}

\section{Details on the 1-loop open string amplitude}
\label{open_details}
In this appendix with simplify the one-loop correlators  \eqref{G4conn},\eqref{G4disc},\eqref{G2} for a choice of polarisations $[--++]$ such that 
\be a^-_1\cdot a^-_2 = a^+_3\cdot  a^+_4 = k_2 \cdot a_3^+ = k_2 \cdot a_4^+ = k_3 \cdot a_1^-= k_3 \cdot a_2^- = 0 \; . \ee 
One can then simplify $\mathbb{G}^L_\textup{2-bil}[^r_s]$ in \eqref{G2} as
\begin{equation}
a_1 {\cdot} {\cal P}_1\, a_2 {\cdot} {\cal P}_2 = \frac{1}{4} (f_1 f_2) (2 S_{12}^2 +\Omega_{123}+\Omega_{412})\; . 
\end{equation}
We then compute the field strength traces using spinor-helicity formalism, obtaining that the only non vanishing traces of $f$'s are
\begin{equation}
(f_1 f_2 f_3 f_4) = (f_1 f_3 f_2 f_4) = (f_1 f_3 f_4 f_2) = \frac{1}{4}(f_1 f_2) (f_3 f_4)= \frac{1}{4} \langle 12 \rangle^2 [34]^2\; . 
\end{equation}
$\mathbb{G}^L_\textup{4-bil,conn}[^r_s]$ in \eqref{G4conn} can be simplified using the Fay trisecant identity \eqref{fay} and
\be
\omega_{123}\omega_{341}+\omega_{124}\omega_{431}+
\omega_{132}\omega_{241}+\omega_{234}\omega_{412}+
\omega_{243}\omega_{312}+ \omega_{324}\omega_{413}=- \sum_{i<j} S_{ij}^2 - \sum_i \Omega_i \; , \ee
where we introduced for short  $\Omega_i = \Omega_{i+1,i+2,i+3}$. Finally we can write almost all the contributions in terms of $\mathcal{Y}$ functions
\begin{align}
\mathbb{G}^L_\textup{4-bil,conn}[^r_s]&=
-3 \langle 12 \rangle^2 [34]^2 \mathcal{F}[^r_s]
\\
\mathbb{G}^L_\textup{4-bil,disc}[^r_s]&=
\langle 12 \rangle^2 [34]^2
\left[\mathcal{F}[^r_s]-\frac{1}{2}\mathcal{E}[^r_s](\wp_{12}+\wp_{34})\right]
\\
\mathbb{G}^L_\textup{2-bil}[^r_s]&=
\langle 12 \rangle^2 [34]^2 \mathcal{E}[^r_s] \left(\frac{1}{2}S_{12}^2+\frac{1}{2} S_{34}^2-\frac{1}{8} \sum_{i<j} \mathcal{Y}_{ij} \right)
\end{align}
Combining the three contributions the double poles cancel and using that $\mathcal{F}[^r_s]= - \tfrac{1}{2} \wp (u_{r,s}) \mathcal{E}[^r_s]$, we finally get \eqref{G4L}.

\section{Elliptic integrals}
\label{integrals_1loop}

In this section we compute the relevant integrals that appear in Section \ref{String1-loop}.

\subsection{Few simple integrals}

We start with vanishing integrals. Using the Fourier expansion of the propagator \eqref{propagator_fourier} it is straightforward to prove that the integral of ${{\calG}}$  and its derivatives vanish
\begin{equation} \label{Gzero}
\int \frac{d^2 z}{\tau_2} \, {{\calG}}(z| \tau) = 0
\quad , \quad
\int \frac{d^2 z}{\tau_2} \, \partial_z^k {{\calG}}(z| \tau) = 0\; .  
\end{equation}
One can also prove that the integral of $\calY$ vanishes using \eqref{def_calY_1}. The integral of $\wp$ can be computed using the identity \eqref{wp_prop_identity}
\begin{equation}
\int \frac{d^2 z}{\tau_2} \, \wp(z| \tau) =
4 \pi i \, \partial_\tau \log [\eta(\tau) \sqrt{\tau_2}] =
-\frac{\pi^2}{6} \hat{E}_2 (\tau)\; . 
\end{equation}
The integral of $(\partial_z {{\calG}})^2$ can be performed using the Fourier expansion \eqref{propagator_fourier}
\bea
\int \frac{d^2 z}{\tau_2} \, (\partial_z {{\calG}}_\epsilon)^2 & = &
\left. \sum_{m,n}\right.^\prime  \left. \sum_{m',n'}\right.^\prime 
\int_0^1 dx \, \int_0^1 dy \,
\frac{e^{2 \pi i [(m+m') x+(n+n') y]}\, \tau_2^{2\epsilon}}{(m \tau{-}n) (m '\tau{-}n')|m \tau{-}n|^{2\epsilon} |m '\tau{-}n'|^{2\epsilon}} \CR
& = &
-\tau_2^{2\epsilon}
\left. \sum_{m,n}\right.^\prime 
\frac{1}{(m \tau{-}n)^2 |m \tau{-}n|^{4\epsilon} } =
-2\zeta(2+4 \epsilon)\,
E_{2 \epsilon,2}(\tau) \CR
& \underset{\epsilon\rightarrow 0}{=}&  -\frac{\pi^2}{3}\, \hat{E}_{2}(\tau)\; . 
\eea
Combining these results with \eqref{def_calY_1} we get
\begin{equation}
\int \frac{d^2 z}{\tau_2} \, \calY(z|\tau)  = 0\; .  \label{Yzero}
\end{equation}

\subsection{The integral of ${{\calG}}(z) \calY(z) $}
\label{calI}
In this section we compute the integral \eqref{GYintegral}
\begin{equation}
\calI(\tau) \equiv \int \frac{d^2 z}{\tau_2} \, {{\calG}}(z) \calY(z) \; . 
\end{equation}
Using \eqref{calY_def2} one finds that $\calI(\tau)$ converges, and it will be convenient to use
\begin{align}
\calI (\tau) &= 
%\int \frac{d^2 z}{\tau_2} \,
%{{\calG}}\, [\,\partial_z^2 {{\calG}}-(\partial_z {{\calG}})^2 -4\p i \partial_\t \log(\h \sqrt{\t_2})\,] \\
%&= 
\int \frac{d^2 z}{\tau_2} \,
{{\calG}}(z)\, \left[\,
-8 \pi^2 \frac{{\rm Im}^2 z}{\t_2^2}+
8 \pi i \partial_\t \log \frac{\vartheta_1(z)}{\h} +
8 \pi i \frac{{\rm Im} z}{\t_2}\partial_z  \log \vartheta_1(z)
\, \right]\; ,
\label{calI_no_poles}
\end{align}
which is manifestly free of double pole divergence. 
We then write $\partial_z \log \vartheta_1$ and $\partial_\tau \log \vartheta_1/\eta$ as $q$-expansions 
\begin{align}
\partial_z \log \vartheta_1(z|\tau) & =
%-i \pi \frac{1+e^{2 \pi i z}}{1-e^{2 \pi i z}} - \sum_{k=1}^\infty \frac{2\pi i}{1-e^{2 \pi i z} q^k}+\sum_{k=1}^\infty \frac{2\pi i}{1-e^{-2 \pi i z} q^k}\\
%&=-i \pi \frac{1+e^{2 \pi i z}}{1-e^{2 \pi i z}} + 2\pi i \sum_{\ell=1}^\infty 
%\left[- \sum_{k=1}^\infty  e^{2 \pi i \ell z} q^{\ell k} +\sum_{k=1}^\infty e^{-2 \pi i \ell z} q^{\ell k} 
%\right]\\
%&=
-i \pi-\frac{2 \pi i e^{2 \pi i z}}{1-e^{2 \pi i z}}- 2\pi i \sum_{\ell=1}^\infty 
\frac{q^\ell}{1-q^\ell}
\left(e^{2 \pi i \ell z}- e^{-2 \pi i \ell z}\right)
\label{derz_theta1_expansion}
\\
\partial_\t \log \frac{\vartheta_1(z| \t)}{\h(\t)} & =
% 2 \pi i q \partial_q \log \frac{\vartheta_1(z| \t)}{\h(\t)} =
%\frac{\pi i}{6}
%-\sum_{k=1}^\infty \frac{k e^{2 \pi i z} q^k}{1-e^{2 \pi i z} q^k}-\sum_{k=1}^\infty \frac{k e^{-2 \pi i z} q^k}{1-e^{-2 \pi i z} q^k}\\
%&= \frac{\pi i}{6}
%- 2 \pi i \sum_{\ell=1}^\infty
%\sum_{k=1}^\infty k q^{k \ell} ( e^{2 \pi i z \ell} + e^{-2 \pi i z \ell} ) \\
%&= 
\frac{\pi i}{6}
- 2 \pi i \sum_{\ell=1}^\infty \frac{q^\ell}{(1-q^\ell)^2} ( e^{2 \pi i z \ell} + e^{-2 \pi i z \ell} ) 
\label{dertau_theta1_expansion}
\end{align}
Considering that the integral of ${{\calG}}$ vanishes and also that
\begin{equation}
\int \frac{d^2 z}{\tau_2} {{\calG}}(z) {\rm Im} (z) =
\int \frac{d^2 z}{\tau_2} {{\calG}}(1+\tau-z) {\rm Im} (1+\tau-z) =
-\int \frac{d^2 z}{\tau_2} {{\calG}}(z) {\rm Im} (z) = 0
\end{equation}
when we insert the expansions \eqref{derz_theta1_expansion} and \eqref{dertau_theta1_expansion} in \eqref{calI_no_poles} we get
\begin{equation}
\begin{aligned}
\calI (\tau) &= 
16 \pi^2\!\! \int \! dx\,dy\, {{\calG}}(x,y)
\Bigg\{
\sum_{\ell=1}^\infty \frac{q^\ell}{(1{-}q^\ell)^2}
(e^{2 \pi i z \ell} {+} e^{-2 \pi i z \ell})
+
\sum_{\ell=1}^\infty \frac{y q^\ell}{1{-}q^\ell} (e^{2 \pi i \ell z}{-} e^{-2 \pi i \ell z})+\\
&
~~ +\frac{y \, e^{2 \pi i z}}{1{-}e^{2 \pi i z}}
-\frac{y^2}{2}
\Bigg\} = 16 \pi^2 [\calI^{(1)} (\tau)+\calI^{(2)} (\tau)+\calI^{(3)} (\tau)+\calI^{(4)} (\tau)]\; . 
\end{aligned}
\end{equation}
We compute the contributions $\calI^{(i)}$ separately using the Fourier expansion of the propagator in \eqref{propagator_fourier}, where we set $\epsilon=0$ directly because all the integrals converge. We first compute the following integrals
\begin{equation}
\begin{aligned}
\int \frac{d^2 z}{\tau_2} {{\calG}}(z) e^{2 \pi i z \ell} & =
%\left. \sum_{m,n}\right.^\prime \frac{1}{|m \tau+n|^2}
%\int_0^1 dx\, e^{2 \pi i (\ell-m) x}
%\int_0^1 dy\, e^{2 \pi i (n+\ell\tau) y} = \\
%&=
\frac{\tau_2}{2 \pi^2 i}
\sum_{n \in \mathbb{Z}}\frac{q^\ell-1}{|n+\ell \tau|^{2}(n+\ell \tau)} \\
\int \frac{d^2 z}{\tau_2} {{\calG}}(z) \, y\,e^{2 \pi i z \ell} & =
%\left. \sum_{m,n}\right.^\prime \frac{1}{|m \tau+n|^{2}}
%\int_0^1 dx\, e^{2 \pi i (\ell-m) x}
%\int_0^1 dy\, y\,e^{2 \pi i (n+\ell\tau) y} = \\
%&=
\frac{\tau_2}{2 \pi^2 i}
\sum_{n\in \mathbb{Z}} 
\left[
\frac{q^\ell}{|n+\ell \tau|^{2}(n+\ell \tau)}
+
\frac{1-q^\ell}{2 \pi i|n+\ell \tau|^{2}(n+\ell \tau)^2}
\right]
\label{int_G_y}
\end{aligned}
\end{equation}
Using the above formulae we can easily compute $\calI^{(1)} (\tau)$ and $\calI^{(2)} (\tau)$ obtaining
\begin{align}
\calI^{(1)} (\tau) +\calI^{(2)} (\tau) =
\frac{\tau_2}{2 \pi^2 i}
\sum_{\ell=1}^\infty\sum_{n\in \mathbb{Z}}
\left[
- \frac{q^\ell}{|n+\ell \tau|^{2}(n+\ell \tau)}
+
\frac{1+q^\ell}{2 \pi i|n+\ell \tau|^{2}(n+\ell \tau)^2}
\right]\; . 
\end{align}
The third term can be computed similarly to \eqref{int_G_y} expanding the denominator $1-e^{2 \pi i z}$
\begin{equation}
\begin{aligned}
\calI^{(3)} (\tau) & =
\frac{\tau_2}{2 \pi^2 i}
\sum_{\ell=1}^\infty
\sum_{n\in \mathbb{Z}} 
\left[
\frac{q^\ell}{|n+\ell \tau|^{2}(n+\ell \tau)}
+
\frac{1-q^\ell}{2 \pi i|n+\ell \tau|^{2}(n+\ell \tau)^2}
\right]\; . 
\end{aligned}
\end{equation}
The last integral reads
\begin{equation}
\begin{aligned}
\calI^{(4)} (\tau) &= 
-\frac{1}{2} \int dx\,dy\, {{\calG}}(x,y)\, y^2 =
\frac{\tau_2}{\pi (2 \pi i)^2} \sum_{n \neq 0} \frac{1}{ |n+0 \tau|^{2}(n+0 \tau)^2}\; . 
\end{aligned}
\end{equation}
Combining all the $\calI^{(i)}$ we get an Eisenstein series
\begin{equation}
\sum_{i=1}^4 \calI^{(i)} (\tau) =
-\frac{1}{4 \pi^3}
\left. \sum_{\ell,n} \right.^\prime
\frac{\tau_2}{|n+\ell \tau|^{2}(n+\ell \tau)^2}=
-\frac{\pi}{180 } E_{1,2}(\tau)
\end{equation}
and finally the result
\begin{equation}
\calI
 (\tau)=- \frac{4 \pi^3}{45} \, E_{1,2}(\tau)\; . 
\end{equation}

\section{Lattices and Eisenstein series} 

In this appendix we shall derive some properties of the $SL(2,\mathds{H})$ and $Sp(4,\mathds{H})$ real analytic Eisenstein series that appear as one-loop couplings defined as Siegel--Narain theta lifts. The discrete groups and their congruent subgroups are identified as automorphism groups of even lattices. For simplicity we consider only the Hurwitz quaternions, but the formulae extend straightforwardly to the Eisenstein quaternions $\mathds{H}(3)$. All computations and proofs are indeed valid for $K=3$. In this section we write $R$ for the radius $\RS$ or $\tilde{\RS}$ in string units for short. 

\subsection{\texorpdfstring{$ SL(2,\mathds{H})$ as $O(\sLambda_{1,1}\oplus D_4)$}{Gamma(alpha) in SL(2,H) as O(II_1,1[2] + D4} }
A pseudo-real Weyl spinor of $Spin(1,5)$, or equivalently a real $SU(2)$ doublet of Weyl spinor of $Spin(1,5)$, can be realised as a doublet of quaternions through the homomorphism $Spin(1,5)=SU^*(4)=SL(2,\mathds{H}(\mathds{R}))$. A vector of $SO(1,5)$ is then a Hermitian two by two matrix over the quaternions. 

The even lattice $\sLambda_{1,1}\oplus D_4$ can be realised in this way as the set the Hermitian two by two matrix over the Hurwitz quaternions
\be Q = \left( \begin{array}{cc} m\,  &\,  q\\ q^* & n \end{array}\right)   \in \mathds{M}_2^0 \qquad \mbox{with}\;  n\in \mathds{Z}\, ,\;  m\in \mathds{Z}\, , \;  q\in \mathds{H}\; .  \ee
The even bilinear form is $(Q,Q)=2\det Q= 2 mn - 2 |q|^2$. The automorphism group of $\mathds{M}_2^0 =\sLambda_{1,1}\oplus D_4$ is $SL(2,\mathds{H})/\mathds{Z}_2$, with the transformation 
\be Q\rightarrow \gamma^\dagger Q \gamma \; . \ee
Any vector $Q$ of vanishing norm factorises in two spinors as $Q={\bf q} {\bf q}^\dagger$ for a two-vector ${\bf q}$ of Hurwitz quaternions. One can always find $\gamma \in SL(2,\mathds{H})$ for $\det Q=0$ such that 
\be \gamma Q =  \left( \begin{array}{cc} {\rm gcd}(Q)\,  &\,  0 \\ 0  & 0 \end{array}\right) \ee
and $\gamma{\bf q}=({\rm gcrd}({\bf q}),0) $ with $|{\rm gcrd}({\bf q})|^2= {\rm gcd}(Q)$. The property of factorisation induced by norm factorisation summarised in \eqref{QuaternionNpartition} implies that for a given ${\rm gcd}(Q)\in \mathds{N}$ there are 
\be  24 \sum_{\substack{ d | {\rm gcd}(Q) \\ 2  / \hspace{-1.2mm} | \hspace{0.6mm}d}} d  \ee 
distinct ${\rm gcrd}({\bf q}) \in \mathds{H}$ such that $|{\rm gcrd}({\bf q})|^2= {\rm gcd}(Q)$. For a homogeneous function of degree $-s$, $f(\lambda Q)= \lambda^{-s} f( Q)$, one has therefore the factorisation 
\be  \sum^\prime_{{\bf q} \in \mathds{H}^2 } f({\bf q} {\bf  q}^\dagger)= \sum^\prime_{\substack{Q\in\mathds{M}_2^0 \\ \det Q= 0 }}   \Biggl(24 \sum_{\substack{ d |{\rm gcd}(Q)\\ 2  / \hspace{-1.2mm} | \hspace{0.6mm}d}} d\Biggr)   f(Q) =\frac{\zeta_\mathds{H}(s)}{2\zeta(s)} \sum^\prime_{\substack{Q\in\mathds{M}_2^0 \\ \det Q= 0 }}  f(Q)\label{factorSum}  \ee
where $\zeta_\mathds{H}(s)$ is the zeta function over the Hurwitz quaternions \eqref{ZetaH}. It is convenient to introduce complete zeta functions satisfying functional relations. For the Riemann zeta function one one the standard 
\be \xi(s) = \pi^{-\frac{s}{2}} \Gamma(\tfrac{s}{2}) \zeta(s)\; , \ee 
and we introduce the completed zeta function over the Hurwitz quaternions as
\be \xi_\mathds{H}(s) = ( \sqrt{2} \pi)^{-s} \Gamma(s) \zeta_\mathds{H}(s) \ . \ee
It satisfies the functional identity $ \xi_\mathds{H}(2-s)  =    \xi_\mathds{H}(s)$ and is a meromorphic function of $s$ with simple poles at $s=0,2$ of residues $-1$ and $1$. Its zero's are located on the two lines Re$[s]=\frac12$ and Re$[s]=\frac32$ according to Riemann hypothesis.\footnote{There is the generalisation $ \xi_{\mathds{H}(K)}(s) = \bigl( \tfrac{K}{4}\bigr)^{\frac{s}{2}} \pi^{-s} \Gamma(s) \zeta_{\mathds{H}(K)}(s)$.}

The perturbative lattice is $ \sLambda_{1,1}[2]\oplus D_4$, it is defined by the set of Hermitian two by two matrices over the Hurwitz quaternions with 
\be Q = \left( \begin{array}{cc} m\,  &\,  q\\ q^* & n \end{array}\right) \in \mathds{M}^1_2 \qquad \mbox{with}\;  n\in \mathds{Z}\, ,\;  m\in 2\mathds{Z}\, , \;  q\in \mathds{H}\; , \ee
that we write $\mathds{M}_2$ or $\mathds{M}_2(2)$ in the main text. We also introduce notations for the dual lattices 
\be Q = \left( \begin{array}{cc} m\,  &\,  q\\ q^* & n \end{array}\right)   \in \mathds{M}^2_2 = 2 \mathds{M}^{1*}_2 \qquad \mbox{with}\;  n\in  \mathds{Z}\, ,\;  m\in 2\mathds{Z}\, , \;  q\in \alpha \mathds{H}\; ,  \ee
\be Q = \left( \begin{array}{cc} m\,  &\,  q\\ q^* & n \end{array}\right)   \in \mathds{M}^3_2=2 \mathds{M}^{*}_2 \qquad \mbox{with}\;  n\in2  \mathds{Z}\, ,\;  m\in 2\mathds{Z}\, , \;  q\in \alpha \mathds{H}\; ,  \ee
with $\alpha$ any Hurwitz quaternion of norm square  $2$, e.g. $\alpha = 1+e_1$. 
Checking the representation on $ \mathds{M}^1_2$,  one finds that the congruent subgroup   of  $\Cong{A}{3}_0(\alpha) \subset SL(2,\mathds{H})$ preserving  $ \sLambda_{1,1}[2]\oplus D_4$ is  defined as
\be \gamma = \left( \begin{array}{cc} a\,  &\,  b\\  c\, & \, d\end{array}\right) \in SL(2,\mathds{H}) \; , \qquad c = 0 \;  \mbox{mod} \, \alpha \; . \ee
As explained in Appendix A, this condition is consistent with the group multiplication because $\alpha \mathds{H}$ is a two-sided ideal and both $bc$ and $ca $ are in $\alpha \mathds{H}$. The automorphism group of the lattice $ \sLambda_{1,1}[2]\oplus D_4$  is in fact the conjugate $SL(2,\mathds{H}) $ group 
\be \gamma = \left( \begin{array}{cc} a\,  &\,  b\\  c\, & \, d\end{array}\right) \in  \left( \begin{array}{cc} 1\,  &\,  0\\  0\, & \, \alpha\end{array}\right)  SL(2,\mathds{H}) \left( \begin{array}{cc} 1\,  &\,  0\\  0\, & \, \frac{1}{\alpha} \end{array}\right) \; , \qquad c \in \alpha \mathds{H}\, , \quad b \in \frac1{\alpha} \mathds{H} \; . \ee
We write this group $\Cong{A}{3}_{0*}(\alpha)$, as the group generated by $\Cong{A}{3}_{0}(\alpha)$ and the element $\scalebox{0.6}{$ \left( \begin{array}{cc} 0  & - \frac1{\alpha}\\  \alpha \, & \, 0\end{array}\right)$}$. 

The integral spinors of  $\Cong{A}{3}_0(\alpha) $ are the vectors in $\alpha \mathds{H}\oplus  \mathds{H}$, and \eqref{factorSum} generalises to
\be  \sum^\prime_{{\bf q} \in\alpha \mathds{H}\oplus  \mathds{H} } f({\bf q} {\bf  q}^\dagger)= \frac{\zeta_\mathds{H}(s)}{2\zeta(s)} \sum^\prime_{\substack{Q\in\mathds{M}^1_2 \\ \det Q= 0 }}  f(Q)\label{factorSumC}\; .  \ee

\subsection{$SL(2,\mathds{H})$ Eisenstein series}

One defines the $SL(2,\mathds{H})$ Eisenstein series for ${\rm Re}[s]>4$ as the absolutely convergent sums 
\be E_{ \alpha \mathds{H} \oplus  \mathds{H},s }^{\mathfrak{sl}_2 \mathds{H}}  = \frac{1}{\zeta_\mathds{H}(s)} \sum_{{\bf q} \in   \alpha \mathds{H} \oplus  \mathds{H}}^\prime |v^{-1\dagger}({\bf q})|^{-2s} \; , \qquad E_{ \mathds{H}^2,s}^{\mathfrak{sl}_2 \mathds{H}}  = \frac{1}{\zeta_\mathds{H}(s)} \sum_{{\bf q}\in   \mathds{H} \oplus \mathds{H}}^\prime |v^{-1\dagger}({\bf q})|^{-2s} \ , \ee  
with the zeta function over the  the Hurwitz quaternions \eqref{ZetaH}. By construction $E_{ \mathds{H}^2,s}^{\mathfrak{sl}_2 \mathds{H}} $ is invariant under $SL(2,\mathds{H})$, while $E_{ \alpha \mathds{H} \oplus  \mathds{H},s }^{\mathfrak{sl}_2 \mathds{H}}$ is invariant under the congruent subgroup $\Cong{A}{3}_0(\alpha) $. The Eisenstein series have analytic continuations to meromorphic functions of $s\in \mathds{C}$, with poles at $s\ne 4$ and a countable set of points on the lines ${\rm Re}[s] = \frac12$ and $\frac32$ corresponding to the non-trivial zeros of the Riemann zeta function. They are normalised such that $E_{ \alpha \mathds{H} \oplus  \mathds{H},0 }^{\mathfrak{sl}_2 \mathds{H}} = E_{ \mathds{H}^2,0}^{\mathfrak{sl}_2 \mathds{H}}  =1$.   

Writing 
\be  |v^{-1\dagger}({\bf q})|^2 = R^{-1}|q + a p |^2 + R |p|^2 \; , \ee
for $R \in \mathds{R}^+$ and $a\in \mathds{H}(R)$, one finds the relation \footnote{This is the same analogy as for the Eisenstein series $E_{s,w}(\frac{\tau}{2})$ for the sum over $m\in 2 \mathds{Z}$.}
\be E_{ \alpha \mathds{H} \oplus  \mathds{H},s }^{\mathfrak{sl}_2 \mathds{H}} ( R^2 , a) = 2^{-\frac{s}{2}} E_{\mathds{H}^2,s }^{\mathfrak{sl}_2 \mathds{H}} ( R^2/2 , \alpha^{-1} a)\; . \ee
Using Poisson summation formula one computes that 
\begin{multline}   E_{ \alpha \mathds{H} \oplus  \mathds{H},s }^{\mathfrak{sl}_2 \mathds{H}} =2^{-s} R^s + \frac{\xi_\mathds{H}(s-2)}{ 4\xi_\mathds{H}(s)} R^{4-s} \\ + 2^{-\frac{s}{2}}\frac{24  R^2}{ \xi_\mathds{H}(s)}  \sum_{q\in \mathds{H}}^\prime \sum_{\substack{d|q\\ d= 1\, {\rm mod}\, 2}} \Bigl( \frac{d^{s-1}}{|q|^{s-2}}  \sigma_{s-2}( \tfrac{|q|^2}{d^2}) \Bigr) K_{s-2}( 2\pi R |q|)  e^{ i \pi    ( q^* a + a^* q )}  \end{multline}
where one uses \eqref{SigmaH2}. In this form the analytic continuation is manifest and this asymptotic expansion is absolutely convergent for almost all $s$. 
 It satisfies that $2^{\frac{s}2} \xi_\mathds{H}(s)  E_{ \alpha \mathds{H} \oplus  \mathds{H},s }^{\mathfrak{sl}_2 \mathds{H}} $ is symmetric under $s\rightarrow 4-s$.

 The same function is obtained by the Siegel--Narain theta series 
 \be \int_{\cF_2} \frac{d^2 \tau}{\tau_2^{\, 2}} E_{\frac{s-3}{2},2}(\tau)  \Gamma_{\sLambda_{1,1}[2]\oplus D_4}  =( 2 - 2^s) \xi(s)  E_{ \alpha \mathds{H} \oplus  \mathds{H},s }^{\mathfrak{sl}_2 \mathds{H}}  \; . \ee
 The even quadratic form is then defined as $(Q,Q) = 2 \det Q$ and
\bea p_L(Q) &=& \frac1{\sqrt{2}}\Bigl( \frac{ m + q^* a + a^* q + |a|^2 n}{R} + R n \Bigr) \; , \CR
  p_R(Q)^2 &=& \frac12\Bigl( \frac{ m + q^* a + a^* q + |a|^2 n}{R} - R n \Bigr)^2 +2 |q + a n |^2 \; . \label{pLpRFive} \eea
One computes  the Fourier expansion using the Rankin--Selberg unfolding method, the formula
 \be \int_0^\infty \frac{1}{4\pi \sqrt{\tau_2 x}} e^{- \frac{\pi}{\tau_2} x - \frac{\pi}{2}  \tau_2 x } \Bigl( \pi \tau_2 x K_{\frac{s}{2}}(\tfrac{\pi \tau_2 x}{2}) + \bigl( \pi \tau_2 x - s + 1\bigr) K_{\frac{s-2}{2}} (\tfrac{\pi \tau_2 x}{2}) \Bigr) = K_{s-2}(2\pi x) \; ,\ee
 and the property that the regularised integral of the product of two $SL(2,\mathds{Z})$ Eisenstein series vanishes \cite{MR656029}. Recall that $ \int_{\cF_2} \frac{d^2 \tau}{\tau_2^{\, 2}}$ must be regularised to take into account the split of the one-loop string amplitude in the sum of the Wilsonian effective action and the non-analytic component of the amplitude. 
 
 For ${\rm Re}[s]>4$ one can unfold the domain $\cF_2$ using the Poincar\'e sum expression for $E_{\frac{s-3}{2},2}(\tau)$ 
\bea \int_{\cF_2} \frac{d^2 \tau}{\tau_2^{\, 2}} E_{\frac{s-3}{2},2}(\tau)  \Gamma_{\sLambda_{1,1}[2]\oplus D_4} &=& \pi^{-\frac{s}{2}} \Gamma(\tfrac{s}{2}) \Biggl( \sum^\prime_{\substack{Q\in \sLambda_{1,1}[2] \oplus D_4\\ Q^2= 0 }} \frac{1}{(2 p_R(Q)^2)^{\frac{s}{2}}}- \frac12\hspace{-3mm} \sum^\prime_{\substack{Q\in \sLambda_{1,1}[\frac12] \oplus D^*_4\\ Q^2= 0 }} \frac{1}{(2 p_R(Q)^2)^{\frac{s}{2}}}\Biggr) 
\CR
 &=& \pi^{-\frac{s}{2}} \Gamma(\tfrac{s}{2}) (1-2^{s-1}) \sum^\prime_{\substack{Q \in  \mathds{M}^2_2 \\ \det Q = 0 }} \frac{1}{ \bigl( \tr (v^{-1\dagger} Q v^{-1})^2\bigr)^\frac{s}{2} }  \CR
  &=&(2-2^s) \xi(s) \frac{1}{\zeta_\mathds{H}(s)} \sum^\prime_{{\bf q} \in \alpha \mathds{H}\oplus  \mathds{H}} \frac{1}{|v^{-1\dagger}({\bf q})|^{2s}} \; . \label{UnfoldingSL2}
  \eea
In this last step we use the property that  any $Q\in \mathds{M}_2^1$ with $\det Q=0$ belongs to $\mathds{M}_2^2$, and any such $Q$ factorises into $Q = {\bf q} {\bf q}^\dagger$, with ${\bf q}$ of right greatest common divisor  over $\mathds{H}$  a Hurwitz quaternion gcrd$({\bf q})$ of norm square $|{\rm gcrd}({\bf q})|^2 = {\rm gcd}(Q)$, see equation \eqref{factorSumC}. 
 Applying the same steps one computes that 
 \be  \int_{\cF_2} \frac{d^2 \tau}{\tau_2^{\, 2}} E_{\frac{s-3}{2},2}(2\tau)  \Gamma_{\sLambda_{1,1}[2]\oplus D_4}  =0  \; ,\ee
for ${\rm Re}[s]>4$, and therefore for all $s$ by analytic continuation. This is a consequence of the transformation property of the Siegel--Narain theta series under inversion  $\tau\rightarrow - \frac{1}{\tau}$. 

The lattice $\sLambda_{1,1}\oplus D_4$ can be identified in the same way as $\mathds{M}_2^0$ and one obtains that 
 \be \int_{\cF_2} \frac{d^2 \tau}{\tau_2^{\, 2}} E_{\frac{s-3}{2},2}(2\tau)  \Gamma_{\sLambda_{1,1}\oplus D_4}  =2^{\frac{1- s}2} ( 2^{s-1}-1) \xi(s)  E_{\mathds{H}^2,s }^{\mathfrak{sl}_2 \mathds{H}}\; . \ee
while 
 \be \int_{\cF_2} \frac{d^2 \tau}{\tau_2^{\, 2}} E_{\frac{s-3}{2},2}(\tau)  \Gamma_{\sLambda_{1,1}\oplus D_4}  = 0 \; . \ee
 This equation is true for any $SL(2,\mathds{Z})$ modular form of weight two, and is ensures that the large twisted circle limit of the orbifold theory amplitude gives back the maximally supersymmetric amplitude.

 One can also relate these Eisenstein series to Poincar\'e sums. An element $(q,p)\in \alpha \mathds{H}\oplus \mathds{H}$ divided on the left by its left greatest common divisor  $d\in \mathds{H}$ is  in the orbit of $(0,d)$ if $d^{-1} q \in\alpha  \mathds{H}$, and in the orbit of $(d,0)$ if $d^{-1} q \notin \alpha  \mathds{H}$. Using this one obtains that 
 \bea E_{ \alpha \mathds{H} \oplus  \mathds{H},s }^{\mathfrak{sl}_2 \mathds{H}}  &=&\sum_{\gamma \in \Cong{A}{3}_0(\alpha) / P_1} R^s |_\gamma +2^{-s} \sum_{\gamma \in \Cong{A}{3}_0(\alpha) / \bar P_1} \frac{R^s}{(R^2 + |a|^2)^s} \bigg|_\gamma   \; , \CR
  E_{ \mathds{H}^2,s}^{\mathfrak{sl}_2 \mathds{H}}  &=&  \sum_{\gamma \in \Cong{A}{3}_0(\alpha) / P_1} R^s |_\gamma + \sum_{\gamma \in \Cong{A}{3}_0(\alpha) / \bar P_1} \frac{R^s}{(R^2 + |a|^2)^s} \bigg|_\gamma  \ , \eea  
 where $ R_{\scalebox{0.5}{T}} = \frac{R}{R^2 + |a|^2}$ is the T-dual radius and $P_1$ (respectively $ \bar P_1$)  is the parabolic stabiliser of $(1,0)$ (respectively  $(0,1)$) in $ \Cong{A}{3}_0(\alpha) $.

 \subsection{\texorpdfstring{$\Cong{D}{4}_0(\alpha) \subset Sp(4,\mathds{H})$ as $O(\sLambda_{1,1}\oplus \sLambda_{1,1}[2]\oplus D_4)$}{Gamma in Sp(4,H) as O(II_1,1 + II_1,1[2] + D4}}
 \label{OrbitsSp4}
Before to define Eisenstein series, we will describe the three fundamental representations of $Spin(2,6)$ and the corresponding discrete representations of the double cover of $Sp(4,\mathds{H})$. For short we will use the same notation $Sp(4,\mathds{H})$ for the double cover.

One defines $g \in Sp(4,\mathds{H})$ from the  two by two matrices $A$, $B$, $C$ and $D$ over $\mathds{H}$ such that 
\be g = \left( \begin{array}{cc}\; A \; & \; B\; \\ \; C\; & \; D\; \end{array}\right) \ , \qquad \left( \begin{array}{cc}\; D^\dagger \; & -B^\dagger \; \\  -C^\dagger \; & \; A^\dagger \; \end{array}\right) \left( \begin{array}{cc}\; A \; & \; B\; \\ \; C\; & \; D\; \end{array}\right)  = \left( \begin{array}{cc}\; \mathds{1}  \; & \; 0\; \\ \; 0\; & \; \mathds{1} \; \end{array}\right)\; .  \ee
A symplectic vector ${\bf k}$ of $Sp(4,\mathds{H})$ is a Weyl spinor of $Spin(2,6)$ that can be defined as a real $SU(2)$ doublet of Weyl spinors. We write the vector ${\bf k}\in \mathds{H}^4$ as a doublet  ${\bf k}= ({\bf q},{\bf p})$ of vector ${\bf q}$ and ${\bf p}$ in $ \mathds{H}^2$, consistently with the decomposition of $g$ in two by two matrices introduced above. The Weyl spinor representation admits the pure imaginary invariant bilinear form 
 \be {\bf k}^\dagger \omega {\bf k} = {\bf q}^\dagger {\bf p} - {\bf p}^\dagger {\bf q} \in {\rm Im}[\mathds{H}] \ee
 that corresponds to the $SU(2)$ triplet of $SO^*(8)$ singlets obtained by applying the $SO^*(8)$ bilinear form to the $SU(2)$ doublet of vectors in the real representation $({\bf 2},{\bf 8})$ of $SU(2)\times SO^*(8)$. 
 
  The negative chirality spinor $\widetilde{{\bf k}}$ transforms in a representation of a trial $SO^*(8)\subset Spin(2,6)$. The algebraic relation between the two representations can only be written algebraically for the  block diagonal matrix and the upper and lower triangular matrices separately, as 
 \be  \left( \begin{array}{cc}\; A \; & \; 0\; \\ \; 0\; & \; A^{-1\dagger} \; \end{array}\right) \rightarrow  \left( \begin{array}{cc}\; A^{-1\dagger} \; & \; 0\; \\ \; 0\; & \; A \; \end{array}\right) \; , \quad  \left( \begin{array}{cc}\; \mathds{1} \; & \; B \; \\ \; 0\; & \; \mathds{1} \; \end{array}\right) \rightarrow  \left( \begin{array}{cc}\; \mathds{1} \; & \; \widetilde{B} \; \\ \; 0\; & \; \mathds{1} \; \end{array}\right) \; , \quad  \left( \begin{array}{cc}\; \mathds{1} \; & \; 0 \; \\ \; C\; & \; \mathds{1} \; \end{array}\right) \rightarrow  \left( \begin{array}{cc}\; \mathds{1} \; & \; 0 \; \\ \; \widetilde{C}\; & \; \mathds{1} \; \end{array}\right)   \; ,   \label{HomoWeyl}\ee
with $B$ and $C$ Hermitian and the tilde involution defined for  $Q =  \big(\colvec[0.8]{ \tilde{m} & q \\ q^* & \tilde{n} }\big)$ as
  \be \widetilde Q  \equiv \left( \begin{array}{cc}  \tilde{n} &-q\\ - q^* & \tilde{m} \end{array}\right) \; . \ee
  There is no algebraic relation for the general matrix $ \big(\colvec[0.8]{A& B \\C& D }\big)$ because the two representations  are defined for distinct double covers of the adjoint group $  Sp(4,\mathds{H})/\mathds{Z}_2$. The tilde involution satisfies 
\be \label{Minor}Q \widetilde Q  = \widetilde Q Q = \det Q \, \mathds{1} \ , \ee
and
\be Q\rightarrow A^{-1\dagger} Q A^{-1} \; , \qquad \widetilde{Q} \rightarrow A \widetilde{Q} A^\dagger \; , \ee
for $A\in SL(2,\mathds{H})$, which ensures that \eqref{HomoWeyl} is locally a homomorphism. 

One defines the $SO(2,6)$ vector $\mathcal{Q}$ in 
 \be \mathds{F}^0_2 = \mathds{Z}\oplus \mathds{M}_2^0 \oplus \mathds{Z}  =\sLambda_{2,2}\oplus D_4 \ee
with $Q\in  \mathds{M}_2^0$ transforming as above under $SL(2,\mathds{H})$ and the two integers $m$, $n$, such that the unipotent elements act on $\mathcal{Q} =  (n,Q,m)$ as
\be (n,Q,m)\rightarrow \bigl(n  , Q + \widetilde{B} n , m - \tr B Q - \det B n \bigr)\; , \quad  (n,Q,m)\rightarrow \bigl(n + \tr C \widetilde{Q} - \det C m  , Q -C m  , m \bigr)\; . \ee
This action is consistent with the action of Gamma matrices acting on Dirac spinors $({\bf k},\widetilde{{\bf k}})$ as the matrix
\be \Gamma(\mathcal{Q})=  \left(\begin{array}{cccc} 0\, &\, 0\,  & \, \widetilde{Q}  \, & \, m\\ 0\, &\, 0 \, &\,  n\,  & -Q\\ Q \, & \, m\,  &\,  0\, &\, 0 \\ n &- \widetilde{Q} \, & \, 0\,  & \, 0 \, \end{array}\right)\; , \ee 
which satisfies 
\be \bigl\{ \Gamma(\mathcal{Q}) ,  \Gamma(\mathcal{Q}^\prime) \bigr\} =  (\mathcal{Q},\mathcal{Q}^\prime )  \mathds{1} \; ,\ee
for the bilinear form
\be  (\mathcal{Q},\mathcal{Q} )   = 2(m n + \det Q)\; . \ee

As proved in \cite{Krieg}, a null symplectic vector $
{\bf k} \in \mathds{H}^4$ satisfying ${\bf k}^\dagger \omega {\bf k}=0$ is in the $Sp(4,\mathds{H})$ orbit of a canonical symplectic vector $\gamma {\bf k} =  ({\rm gcrd}({\bf k}) ,0,0,0)$ where ${\rm gcrd}({\bf k})$ is the greatest right divisor of ${\bf k}$. For any null vector $\mathcal{Q}\in \mathds{F}^0_2$, there exists a null symplectic vector ${\bf k}$ such that $\mathcal{Q}\cdot {\bf k}=0$. Using the constraint on $\gamma {\bf k}$ one obtains that 
\be \gamma \mathcal{Q} =  \left( 0,   \left( \begin{array}{ccc}0  \, &  \, 0 \\ 0 \, &  \,n \end{array}\right), m \right)\; , \ee  
and using then the $\gamma {\bf k}$  Levi stabiliser $SL(2,\mathds{Z})\subset Sp(2,\mathds{H}) \ltimes (\mathds{H}^2 \oplus \mathds{Z}) $ 
\be \left( \begin{array}{cccc} 1 \; \, &\, \;  0\; \,  &\, \;  0\;  \, &\; \,   0\\ 0 & a & 0 & b \\ 0 & 0 & 1 & 0 \\ 0 & c & 0 & d \end{array}\right) \left( \begin{array}{cccc} 1 \; \, &\, \;  -q^*\; \,  &\, k +q^* p  \, &\; -p^*\\ 0 & 1 & -p& 0 \\ 0 & 0 & 1 & 0 \\ 0 & 0 & q & 1 \end{array}\right)  \ee
on $\gamma \mathcal{Q}$ one obtains that there exists $\gamma^\prime$ such that 
\be \gamma^\prime \mathcal{Q} =  \left( 0,   \left( \begin{array}{ccc}0  \, &  \, 0 \\ 0 \, &  \,0 \end{array}\right), {\rm gcd}(\mathcal{Q}) \right)\; . \ee  

\medskip

We define the theta congruent subgroup $\Cong{D}{4}_0(\alpha) \subset Sp(4,\mathds{H})$ such that \footnote{Here the $*$ state for any Hurwitz quaternion, just like one  would write $(^a_c{}^b_d) = (^*_0{}^*_*)$ mod $N$ for the standard congruent subgroup $\Gamma_0(N) \subset SL(2,\mathds{Z})$.}
\be g = \left( \begin{array}{cccc} * \; \, &\, \;  *\; \,  &\, \;  *\;  \, &\; \,   *\\ 0 & * & * & 0 \\ 0 & * & * & 0 \\ * & * & * & * \end{array}\right) \; \mbox{mod}\; \alpha \ .\ee 
It is conjugate to the congruent subgroup with  $C=0$ mod $\alpha$, but it will be more convenient to use this basis in which the block diagonal subgroup of parameter $A$ is the five-dimensional (restricted) T-duality group $\Cong{A}{3}_0(\alpha) \subset  SL(2,\mathds{H})$. 

One defines then the sublattices 
  \bea \mathds{F}^1_2 &=& \mathds{F}_2= \mathds{Z}\oplus \mathds{M}^1_2 \oplus \mathds{Z} = \sLambda_{1,1}\oplus \sLambda_{1,1}[2]\oplus D_4\; , \CR
 \mathds{F}^2_2 &=& 2 \mathds{F}_2^* = 2\mathds{Z}\oplus \mathds{M}^{2}_2 \oplus 2\mathds{Z} = 2\sLambda_{1,1}\oplus \sLambda_{1,1}[2]\oplus 2D^*_4\;  \eea
and
  \be \mathds{F}^3_2 = 2 \mathds{F}_2^{0*} = 2\mathds{Z}\oplus \mathds{M}^{3}_2 \oplus 2\mathds{Z} = 2\sLambda_{2,2}\oplus 2D^*_4\; .  \ee
The lattice $\mathds{F}^0_2 $ and its dual are invariant under $Sp(4,\mathds{H})$, while $\mathds{F}^1_2 =\mathds{F}_2$ and its dual 
  are invariant under the congruent subgroup $\Cong{D}{4}_0(\alpha) \subset Sp(4,\mathds{H})$. The $\Cong{D}{4}_0(\alpha)$ symplectic vector can be defined in $\mathds{H}^4$, $\mathds{H}^2 \oplus (\alpha \mathds{H})^2$ or $(\alpha \mathds{H})^4$. 
  
  One can moreover define a Fricke duality of the lattice $\mathds{F}_2$, because it is 2-modular, i.e. that its dual  is $\sqrt{2}$ times $\mathds{F}_2$ up to an irrational rotation in $SO(2,6)$   \cite{Persson:2015jka}. This reflects into the rational symplectic matrix 
  \be \gamma_{\rm F} =  \left(\begin{array}{cccc} 0\, &\, 0\,  & - \frac{1}{\alpha^*}  \, & \, 0 \\ 0\, &\, 0 \, &\, 0 \,  & -\alpha  \\ \alpha  \, & \, 0\,  &\,  0\, &\, 0 \\ 0 \, &\, \frac{1}{\alpha^*} \, & \, 0\,  & \, 0 \, \end{array}\right) \in Sp(4,\mathds{H}(\mathds{Q})) \label{Fricke} \ee
that preserves $\mathds{H}^2 \oplus (\alpha \mathds{H})^2$ and that acts on $\mathcal{Q}$ as
\bea   \left(\begin{array}{cccc} \tilde{m}\, &\, q\,  & m   \, & \, 0 \\ q^*\, &\, \tilde{n} \, &\, 0 \,  & m \\ n  \, & \, 0\,  &- \tilde{n} \, &\, q \\ 0 & n  \, & \, q^* \,  & -\tilde{m} \, \end{array}\right) &\rightarrow&   \left(\begin{array}{cccc} 0\, &\, 0\,  & - \alpha^*  \, & \, 0 \\ 0\, &\, 0 \, &\, 0 \,  & -\frac1{ \alpha } \\ \frac1{\alpha }  \, & \, 0\,  &\,  0\, &\, 0 \\ 0\,  &\, \alpha^* \, & \, 0\,  & \, 0 \, \end{array}\right)   \left(\begin{array}{cccc} \tilde{m}\, &\, q\,  & m   \, & \, 0 \\ q^*\, &\, \tilde{n} \, &\, 0 \,  & m \\ n  \, & \, 0\,  &- \tilde{n} \, &\, q \\ 0 & n  \, & \, q^* \,  & - \tilde{m} \, \end{array}\right)  \left(\begin{array}{cccc} 0\, &\, 0\,  & - \frac{1}{\alpha^*}  \, & \, 0 \\ 0\, &\, 0 \, &\, 0 \,  & -\alpha  \\ \alpha  \, & \, 0\,  &\,  0\, &\, 0 \\ 0 \, &\,\frac{1}{\alpha^*} \, & \, 0\,  & \, 0 \, \end{array}\right)  \CR
&=&  \left(\begin{array}{cccc} 2\tilde{n}\, &-\alpha^{-1} q\alpha \,  & n   \, & \, 0 \\ -(\alpha^{-1} q\alpha)^*\, &\, \frac{\tilde{m}}{2} \, &\, 0 \,  & n \\ m  \, & \, 0\,  &- \frac{\tilde{m}}{2} \, &-\alpha^{-1} q\alpha \\ 0 & m  \, & - (\alpha^{-1} q\alpha)^* \,  & -2\tilde{n} \, \end{array}\right) \; .   \eea
Note that $q\rightarrow p^{-1} q p$ is an automorphism of $\mathds{H}$ for $p = \alpha$, but not for a Hurwitz quaternion of norm  $|p|^2 > 2$ \cite{Krieg}.  This rational symplectic transformation  is an automorphism of $\mathds{F}^1_2$ and $\mathds{F}^2_2$, as well as the symplectic module $\mathds{H}^2 \oplus (\alpha \mathds{H})^2$. Although $\gamma_{\rm F} \notin Sp(4,\mathds{H})$. One checks that $ \gamma_{\rm F} \Cong{D}{4}_0(\alpha) \gamma_{\rm F} = \Cong{D}{4}_0(\alpha) $ and we define the Fricke theta group $\Cong{D}{4}_{0 *}(\alpha)$ as the group generated by  $\Cong{D}{4}_0(\alpha)$ and $\gamma_{\rm F} $.  We find that the T-duality group of the theory is the Fricke theta group   $\Cong{D}{4}_{0 *}(\alpha)$.

Fricke duality acts on the projective coordinates \footnote{  In terms of the torus complex structure $U$ and K\"{a}hler modulus  $T$ one has  \be t =\left(\begin{array}{cc} U  \; &\;b+ U a\\ b^* +U a^*  \; &\; T +U |a|^2 \end{array} \right) \; .  \ee}
\be t  = B+i R v^{-1} v^{-1\dagger} =\left(\begin{array}{cc} b_1+i R \tilde{R}^{-1} \; &\; \beta + i R \tilde{R}^{-1} a\\ \beta^* + i  R \tilde{R}^{-1} a^*  \; &\; b_2+i R (\tilde{R} + \tilde{R}^{-1}   |a|^2 )\end{array} \right)  \ee
as
\be t\rightarrow  -\biggl( \begin{array}{cc} \frac{1}{\alpha^{*}}  & 0 \\ 0 & \alpha \end{array}\biggr) t^{-1}   \biggl( \begin{array}{cc}  \frac{1}{\alpha}  & 0 \\ 0 & \alpha^* \end{array}\biggr) \; .\ee

  There are two orbits for the $\Cong{D}{4}_0(\alpha)$ symplectic vector. Either ${\bf k}\, {\rm gcrd}({\bf k})^{-1}  = ( a_{11} , a_{12} , c_{11} , c_{12})$ with $a_{12}$ and $c_{11}$ in $\alpha \mathds{H}$ and ${\bf k} = \gamma ({\rm gcrd}({\bf k}) ,0,0,0)$, or  ${\bf k}\, {\rm gcrd}({\bf k})^{-1}  = ( a_{21} , a_{22} , c_{21} , c_{22})$ with $a_{22}$ and $c_{21}$ not both in $\alpha \mathds{H}$ and ${\bf k} = \gamma (0 ,{\rm gcrd}({\bf k}),0,0)$.

  Once again, any null vector $ \mathcal{Q} $ in $\mathds{F}_2^1$ will satisfy $\mathcal{Q}\cdot {\bf k} = 0$ for some ${\bf k} $. Using the two different orbits for ${\bf k} $ one finds then the two possibilities for $ \mathcal{Q} $
  \be  \mathcal{Q} = \gamma  \left( 0,   \left( \begin{array}{ccc}0  \, &  \, 0 \\ 0 \, &  \,n \end{array}\right), m \right)\; , \quad  \mathcal{Q} = \gamma  \left( 0,   \left( \begin{array}{ccc}n  \, &  \, 0 \\ 0 \, &  \,0 \end{array}\right), m \right) \; . \ee  
In the first case the Levi stabiliser is $\Gamma_0(2) \subset \Cong{D}{2}_0(\alpha) \ltimes (\mathds{H}^2 \oplus \mathds{Z}) $ and there are two orbits, $m=0$ and $n=0$. In the second the Levi stabiliser is $\Gamma_0(2) \subset \Cong{D}{2}_0(\alpha) \ltimes (\mathds{H} \oplus (\alpha \mathds{H}) \oplus 2 \mathds{Z}) $ \footnote{We use the notation $\Cong{A}{$2n\,$-$1$}_0(\alpha) \subset SL(n,\mathds{H})$ for $n\ge 2$, for $n=1$ $SL(1,\mathds{H})$ is the order 24 finite group of unite Hurwitz quaternions and there is no congruent subgroup. We use $\Cong{D}{$2n$}_0(\alpha) \subset Sp(2n,\mathds{H})$ for $n\ge1$ in the same way and for $n=1$ $Sp(2,\mathds{H}) = SL(1,\mathds{H})\times SL(2,\mathds{Z})$ while $\Cong{D}{2}_0(\alpha)= SL(1,\mathds{H})\times \Gamma_0(2)$, where  $ \Gamma_0(2) \subset SL(2,\mathds{Z})$.}
\be \left( \begin{array}{cccc} a \; \, &\, \;  0\; \,  &\, \;  b\;  \, &\; \,   0\\ 0 & 1 & 0 & 0 \\ c & 0 & d & 0 \\ 0 & 0 & 0 & 1 \end{array}\right) \left( \begin{array}{cccc} 1 \; \, &\, \;  0\; \,  &\, \, 0\, \, &\; p \\ q^* & 1 & p^*& k +q^* p   \\ 0 & 0 & 1 & -q \\ 0 & 0 & 0 & 1 \end{array}\right)  \ee
and there are also two orbits. The two orbits with $n=0$ and $m\ne 0$ are the same $\Cong{D}{4}_0(\alpha)$ orbit, as can be checked using the subgroup $\Gamma_0(2)\times \Gamma_0(2)$ acting on the four real integers in $\mathcal{Q}$. The two orbits with $m=0$ and $n\ne 0$ are related by $\gamma_{\rm F} $. We conclude that there are two $\Cong{D}{4}_{0*}(\alpha)$ orbits for  null vectors in $\mathds{F}_2^1$ and $\mathds{F}_2^2$.

The lattice of perturbative charges is $\mathds{F}^*_2 = \frac{1}{2} \mathds{F}_2^2$ and $\mathds{F}_2 = \mathds{F}_2^1$. The same analysis generalises straightforwardly to $K=3$ and one obtains that there are two $\Cong{D}{4}_{0*}(\alpha)$ orbits of vector of vanishing norm in $\mathds{F}^*_2(K) $, 
  \be  \mathcal{Q} = \gamma  \left( 0,   \left( \begin{array}{ccc}0  \, &  \, 0 \\ 0 \, &  \,0 \end{array}\right), m \right)\; , \quad  \mathcal{Q} = \gamma  \left( 0,   \left( \begin{array}{ccc}m  \, &  \, 0 \\ 0 \, &  \,0 \end{array}\right), 0 \right) \; . \ee  
In the first case $\mathcal{Q}$ must be in  $\mathds{F}_2(K)$, while in the second $\mathcal{Q}\in \mathds{F}_2(K)$ if and only if $m = 0$ mod $K$. 

\subsection{$Sp(4,\mathds{H})$ and $\Cong{D}{4}_0(\alpha)$ Eisenstein series}
One can now describe Eisenstein series for $\mathds{F}_2^k$ with $k=0,1,2,3$. We define the Eisenstein series as the convergent sum for Re$[s]> 3$
  \be E^{\mathfrak{sp}_4 \mathds{H}}_{ \mathds{F}^k_2,s} = \frac{1}{2\zeta(2s)} \sum^\prime_{\substack{\mathcal{Q} \in \mathds{F}^k_2\\ (\mathcal{Q},\mathcal{Q})=0}} \frac{1}{|V(\mathcal{Q})|^{2s}} \; ,   \ee
and as meromorphic functions of $s$ on $\mathds{C}$ by analytic continuations with a pole at the real value $s=3$. They are normalised to $1$ at $s=0$. The definition of $|V(\mathcal{Q})|^2$ and the $\sLambda_{1,1}\oplus \sLambda_{1,1}[2]\oplus D_4$ bilinear form are 
  \bea |V(\mathcal{Q})|^2  &=& R^{-2} ( m - \tr B Q - \det B n )^2 + \tr ( v^{-1\dagger } ( Q + \widetilde{B}) v^{-1})^2 + R^2 n^2\CR
  (\mathcal{Q},\mathcal{Q}) &=& 2 mn + 2\det Q\; ,\eea
  where $R$ is the circle radius, $v$ parametrises the five-dimensional NS moduli (i.e. the twisted circle radius and the Wilson line along $T^4$), while $B$ parametrises  NS axions. We will also use the notations 
  \be (Q,Q) = 2 \det Q \; , \qquad G(Q,Q) =  \tr ( v^{-1\dagger }  Q  v^{-1})^2 \; . \ee
These Eisenstein series are related to the Siegel--Narain theta lifts 
\bea  \int_{\cF_2} \frac{d^2 \tau}{\tau_2^{\, 2}} E_{s-2,2}(2\tau) \Gamma_{ \sLambda_{1,1}\oplus \sLambda_{1,1}[2]\oplus D_4}  &=& 2^{s-1}  \xi(2s)  E^{\mathfrak{sp}_4 \mathds{H}}_{ \mathds{F}^1_2,s}  - 2^{s}  \xi(2s)   E^{\mathfrak{sp}_4 \mathds{H}}_{ \mathds{F}^2_2,s}  \; , \CR
 \int_{\cF_2} \frac{d^2 \tau}{\tau_2^{\, 2}} E_{s-2,2}(\tau) \Gamma_{ \sLambda_{1,1}\oplus \sLambda_{1,1}[2]\oplus D_4}  &=&2 \xi(2s)  E^{\mathfrak{sp}_4 \mathds{H}}_{ \mathds{F}^1_2,s}  -  2^{2s} \xi(2s)  E^{\mathfrak{sp}_4 \mathds{H}}_{ \mathds{F}^2_2,s}\; ,  \CR
 \int_{\cF_2} \frac{d^2 \tau}{\tau_2^{\, 2}} E_{s-2,2}(2\tau) \Gamma_{ \sLambda_{2,2}\oplus D_4}  &=& 2^{s-1}  \xi(2s)  E^{\mathfrak{sp}_4 \mathds{H}}_{ \mathds{F}^0_2 ,s}  - 2^{s+1}  \xi(2s)   E^{\mathfrak{sp}_4 \mathds{H}}_{ \mathds{F}^3_2,s} \; ,  \CR
 \int_{\cF_2} \frac{d^2 \tau}{\tau_2^{\, 2}} E_{s-2,2}(\tau) \Gamma_{ \sLambda_{2,2} \oplus D_4}  &=&2 \xi(2s)  E^{\mathfrak{sp}_4 \mathds{H}}_{ \mathds{F}^0_2 ,s}  -2^{2s+1}  \xi(2s)  E^{\mathfrak{sp}_4 \mathds{H}}_{ \mathds{F}^3_2,s} \; .  \eea

Using the Rankin--Selberg unfolding method one can compute the large radius limit of the  Siegel--Narain theta lifts. One obtains in this way that the last function vanishes identically 
\be  \int_{\cF_2} \frac{d^2 \tau}{\tau_2^{\, 2}} E_{s-2,2}(\tau) \Gamma_{ \sLambda_{2,2} \oplus D_4}  = 0 \; . \ee
This is due to the property that any null vector in $ \mathds{F}^3_2$ must be in $2 \mathds{F}^0_2$, so that 
\be E^{\mathfrak{sp}_4 \mathds{H}}_{ \mathds{F}^3_2,s}  = 2^{-2s} E^{\mathfrak{sp}_4 \mathds{H}}_{ \mathds{F}^0_2,s} \; . \ee
It will nonetheless be useful to define the sublattice $ \mathds{F}^3_2$ for the non-perturbative completion \cite{GBinprep}.

For the lattice $\mathds{F}_2^1$ one computes using Rankin--Selberg unfolding method that 
{\allowdisplaybreaks \bea &&  \int_{\cF_2} \frac{d^2 \tau}{\tau_2^{\, 2}} E_{s,2}(2\tau) \Gamma_{ \sLambda_{1,1}\oplus \sLambda_{1,1}[2]\oplus D_4} \\
 &=& (2^{s+1} - 2^{-s-2})  \xi(2s+4) R^{2s+4} +(2^{s-1} - 2^{-s}) \frac{s}{s+1} \frac{\xi(2s+1)\xi(2s-1)}{\xi(2s+2) } R^{2-2s} \CR
 &&  - \frac{R^{s+\frac{5}{2}} }{2^{2s+1}} \sum_{\substack{ Q \in \sLambda_{1,1}[\frac12] \oplus D_4^*\\ Q^2 = 0 }} e^{2\pi i (Q,B)}  \hspace{-4mm} \sum_{\substack{ d^{-1} Q \in \sLambda_{1,1}[\frac12] \oplus D_4^*\\ d = 1 \, {\rm mod}\, 2}} \hspace{-3mm} d^{2s+3}\hspace{2mm}  \frac{K_{s+\frac32}(2\pi R \sqrt{ 2 p_R(Q)^2})}{\sqrt{2 p_R(Q)^2}^{s + \frac32}} \nonumber \\
  &&  +  \frac{2^{2s+1} s\xi(2s+1)}{(s+1)\xi(2s+2) }  R^{\frac{3}{2}-s}  \hspace{-5mm}\sum_{\substack{ Q \in \sLambda_{1,1}[\frac12] \oplus D_4^*\\ Q^2 = 0 }} e^{2\pi i (Q,B)} \hspace{-4mm} \sum_{\substack{ d^{-1} Q \in \sLambda_{1,1}[\frac12] \oplus D_4^*\\ d = 1 \, {\rm mod}\, 2}} \hspace{-3mm} d^{1-2s} \hspace{2mm}  \frac{K_{s-\frac12}(2\pi R \sqrt{ 2 p_R(Q)^2})}{\sqrt{2 p_R(Q)^2}^{ \frac12-s}}  \nonumber \\
  &&  -  \frac{R}{2\xi(2s+2)}  \sum_{\substack{ Q \in \sLambda_{1,1}[\frac12] \oplus D_4^*\\ Q^2 \ne 0}} \hspace{-2mm} e^{2\pi i (Q,B)} \hspace{-7mm}  \sum_{\substack{ d^{-1} Q \in \sLambda_{1,1}[\frac12] \oplus D_4^*\\ d = 1 \, {\rm mod}\, 2}} \hspace{-6mm} \frac{d^{2s+3}  \sigma_{2s+1}( | \frac{Q^2}{d^2}|)}{|Q^2|^s} \int_0^\infty \hspace{-3mm} d t  \sqrt{t} W_{s,2}(\pi Q^2  t) e^{- \frac{\pi}{t } R^2 - \pi t G(Q,Q)}\nn
 \eea}
 where 
 \be W_{s,2}(2\pi x) = \frac{1}{(s+1) \sqrt{|x|}} \Bigl( (s+1-2\pi x) K_{s+\frac12}(2\pi | x|) - 2\pi |x| K_{s+\frac32}(2\pi |x|)\Bigr) \ee
 is defined such that 
 \be E_{s,2}(\tau) = \tau_2^s - \frac{s}{s+1} \frac{ \xi(2s+1)}{\xi(2s+2)} \tau_2^{-1-s} + \frac{2}{\xi(2s+2)} \sum_{n \in \mathds{Z} }^\prime \frac{\sigma_{2s+1}(|n|)}{|n|^s} W_{s,2}(2\pi  n \tau_2) e^{2\pi i \tau_1} \; . \ee
The same computation for $E_{s,2}(\tau)$ gives 
 \bea &&  \int_{\cF_2} \frac{d^2 \tau}{\tau_2^{\, 2}} E_{s,2}(\tau) \Gamma_{ \sLambda_{1,1}\oplus \sLambda_{1,1}[2]\oplus D_4} \\
 &=&  \xi(2s+4) R^{2s+4} -  \frac{s}{s+1} \frac{\xi(2s+1)\xi(2s-1)}{\xi(2s+2) } R^{2-2s}  + R \int_{\cF_2} \frac{d^2 \tau}{\tau_2^{\, 2}} E_{s,2}(\tau) \Gamma_{\sLambda_{1,1}[2]\oplus D_4}  \CR
 &&  +2^{s+1} R^{s+\frac{5}{2}} \sum_{\substack{ Q \in \sLambda_{1,1}[\frac12] \oplus D_4^*\\ Q^2 = 0 }} e^{2\pi i (Q,B)}  \hspace{-4mm} \sum_{d^{-1} Q \in \sLambda_{1,1}[\frac12] \oplus D_4^*} \hspace{-3mm} (-1)^{\frac{2n}{d}} d^{2s+3}\hspace{2mm}  \frac{K_{s+\frac32}(2\pi R \sqrt{ 2 p_R(Q)^2})}{\sqrt{2 p_R(Q)^2}^{s + \frac32}} \CR
  &&  - \frac{2^{s+1} s\xi(2s+1)}{(s+1)\xi(2s+2) }  R^{\frac{3}{2}-s}  \hspace{-6mm}\sum_{\substack{ Q \in \sLambda_{1,1}[\frac12] \oplus D_4^*\\ Q^2 = 0 }} e^{2\pi i (Q,B)} \hspace{-6mm} \sum_{ d^{-1} Q \in \sLambda_{1,1}[\frac12] \oplus D_4^*} \hspace{-5mm} (-1)^{\frac{2n}{d}}d^{1-2s} \hspace{2mm}  \frac{K_{s-\frac12}(2\pi R \sqrt{ 2 p_R(Q)^2})}{\sqrt{2 p_R(Q)^2}^{ \frac12-s}}  \CR
  &&  \hspace{-4mm} +  \frac{2^{s+1}R}{\xi(2s+2)}  \sum_{\substack{ Q \in \sLambda_{1,1}[\frac12] \oplus D_4^*\\ Q^2 \ne 0}} \hspace{-4mm} e^{2\pi i (Q,B)} \hspace{-9mm} \sum_{\substack{ d^{-1} Q \in \sLambda_{1,1}[\frac12] \oplus D_4^*\\  \frac{Q^2}{d^2} = 0 \, {\rm mod} \, 2}} \hspace{-8mm}\frac{(-1)^{\frac{2n}{d}}  d^{2s+3}  \sigma_{2s+1}( | \frac{Q^2}{2d^2}|)}{|Q^2|^s} \int_0^\infty \hspace{-3mm} d t  \sqrt{t} W_{s,2}(\pi Q^2  t) e^{- \frac{\pi}{t } R^2 - \pi t G(Q,Q)}\nn
 \eea
where $Q = \big(\colvec[0.8]{ m  & q \\ q^*& n }\big) $ with $n \in\frac{\mathds{Z}}{2}$ and $q\in  \mathds{H}$ and $2n = 1$ mod $2$ if $q \notin \alpha \mathds{H}$. 

One can use these formulae in the limit $s\rightarrow0$ to obtain the large radius expansion of the one-loop $\mathcal{R}^4$ threshold function, but in practice it is more convenient to use directly the  Rankin--Selberg unfolding method and the Fourier expansion of the holomorphic modular form $2 {E}_{2}(2\tau)- {E}_{2}(\tau)$. One obtains 
{\allowdisplaybreaks\bea && \frac{4\pi}{3}  \int_{\cF_2} \frac{d^2 \tau}{\tau_2^{\, 2}} \bigl( 2 {E}_{2}(2\tau)- {E}_{2}(\tau)\bigr) \Gamma_{ \sLambda_{1,1}\oplus \sLambda_{1,1}[2]\oplus D_4}\CR
&=& \frac{\pi^3}{27} R^4 - \frac{4\pi}{3} R \int_{\cF_2} \frac{d^2 \tau}{\tau_2^{\, 2}} \hat{E}_{2}(\tau) \Gamma_{\sLambda_{1,1}[2]\oplus D_4}\nonumber\\
&&  +\frac{R}{3} \sum^\prime_{\substack{Q \in \sLambda_{1,1}[\frac12] \oplus D_4^*\\ Q^2 =0}}  \Biggl( \sum_{Q/d \in  \sLambda_{1,1}[\frac12] \oplus D_4^* } \hspace{-5mm} d^3 + 4 \hspace{-3mm}\sum_{Q/d \in  \sLambda_{1,1}[2] \oplus D_4 }  \hspace{-5mm} d^3\, \Biggr)  \frac{1 + 2\pi  R \sqrt{ 2 p_R(Q)^2} }{ \sqrt{2 p_R(Q)^2}^3} e^{- 2\pi R\sqrt{ 2 p_R(Q)^2} +  2\pi i (Q,B)} \nonumber\\
&&  +8 R  \sum_{\substack{Q \in \sLambda_{1,1}[\frac12] \oplus D_4^*\\ Q^2 < 0}}  \Biggl( \sum_{Q/d \in  \sLambda_{1,1}[\frac12] \oplus D_4^* } \hspace{-5mm} d^3 \sum_{\substack{\ell | \frac{Q^2}{d^2}\\\ell = 1 \, {\rm mod} \, 2}} \hspace{-3mm} \ell+ 4 \hspace{-3mm}\sum_{Q/d \in  \sLambda_{1,1}[2] \oplus D_4 }  \hspace{-5mm} d^3 \sum_{\substack{\ell | \frac{Q^2}{2d^2}\\\ell = 1 \, {\rm mod} \, 2}} \hspace{-3mm} \ell\, \Biggr) \CR
&& \hspace{50mm} \times  \frac{1 + 2\pi  R \sqrt{ 2 p_R(Q)^2} }{ \sqrt{2 p_R(Q)^2}^3} e^{- 2\pi R\sqrt{ 2 p_R(Q)^2} +  2\pi i (Q,B)} \label{E00Z2R}\eea}

For the one-loop ${D}^2 \mathcal{R}^4$ threshold function we need the limit  $s\rightarrow 1$. The limit diverges and the regularised limit gives
 \bea \label{D2R4Appendix} && \lim_{s\rightarrow 1} \Bigl(  \int_{\cF_2} \frac{d^2 \tau}{\tau_2^{\, 2}} E_{s,2}(\tau) \Gamma_{ \sLambda_{1,1}\oplus \sLambda_{1,1}[2]\oplus D_4} + \frac{s}{s+1} \frac{\xi(2s+1)\xi(2s-1)}{\xi(2s+2)}  \Bigr) \\
 &=&  \frac{2\pi^3}{945} R^{6} + \frac{45 \zeta(3)}{2\pi^3} \log R  + R \int_{\cF_2} \frac{d^2 \tau}{\tau_2^{\, 2}} E_{1,2}(\tau) \Gamma_{\sLambda_{1,1}[2]\oplus D_4}  \CR
 &&  +4 R^{\frac{7}{2}} \sum_{\substack{ Q \in \sLambda_{1,1}[\frac12] \oplus D_4^*\\ Q^2 = 0 }} e^{2\pi i (Q,B)}  \hspace{-4mm} \sum_{d^{-1} Q \in \sLambda_{1,1}[\frac12] \oplus D_4^*} \hspace{-3mm} (-1)^{\frac{2n}{d}} d^{5}\hspace{2mm}  \frac{K_{\frac52}(2\pi R \sqrt{ 2 p_R(Q)^2})}{\sqrt{2 p_R(Q)^2}^{ \frac52}} \CR
  &&  \hspace{-2mm}- \frac{45\zeta(3)}{\pi^3 }    \hspace{-6mm}\sum_{\substack{ Q \in \sLambda_{1,1}[\frac12] \oplus D_4^*\\ Q^2 = 0 }}  \sum_{ d^{-1} Q \in \sLambda_{1,1}[\frac12] \oplus D_4^*} \hspace{-5mm} (-1)^{\frac{2n}{d}}d^{-1} \hspace{2mm}  e^{-2\pi R \sqrt{ 2 p_R(Q)^2}+2\pi i (Q,B)} \CR
  &&  -  \frac{180}{\pi^3}  \sum_{\substack{ Q \in \sLambda_{1,1}[\frac12] \oplus D_4^*\\ Q^2 > 0}}  \sum_{\substack{ d^{-1} Q \in \sLambda_{1,1}[\frac12] \oplus D_4^*\\  \frac{Q^2}{d^2} = 0 \, {\rm mod} \, 2}} \hspace{-8mm}\frac{(-1)^{\frac{2n}{d}}  d^{5}  \sigma_{3}(  \frac{Q^2}{2d^2})}{Q^6} e^{-2\pi R \sqrt{ 2 p_L(Q)^2}+2\pi i (Q,B)} \CR
  && - \frac{180}{\pi^3}   \sum_{\substack{ Q \in \sLambda_{1,1}[\frac12] \oplus D_4^*\\ Q^2 < 0}} \sum_{\substack{ d^{-1} Q \in \sLambda_{1,1}[\frac12] \oplus D_4^*\\  \frac{Q^2}{d^2} = 0 \, {\rm mod} \, 2}} \hspace{-8mm}\frac{(-1)^{\frac{2n}{d}}  d^{5}  \sigma_{3}( - \frac{Q^2}{2d^2})}{-Q^6} e^{-2\pi R \sqrt{ 2 p_R(Q)^2} + 2\pi i (Q,B) }\CR
  && \hspace{30mm} \times  \biggl( 1  - \frac{2\pi R Q^2}{\sqrt{2 p_R(Q)^2}} \Bigl( 1 - \frac{Q^2}{4p_R(Q)^2} \Bigr) + \frac{\pi^2 R^2 Q^4}{ p_R(Q)^2}\biggr) \nn
 \eea

\medskip

As a final comment we will relate the Eisenstein series defined as constrained lattice sums to Poincar\'e sums associated to the three $\Cong{D}{4}_0(\alpha)$ orbit of  null vectors in $\mathds{F}^0_2$ and $\mathds{F}_2^1$. We denote them $\mathcal{O}_1$ for the orbit of the momentum along the $S^1$,  $\mathcal{O}_2$ for the winding along the twisted $S^1$ and $\mathcal{O}_3$ for the momentum along the twisted $S^1$, and respectively $R$, $\tilde{R}_{\scalebox{0.5}{T}}$ and $\tilde{R}$ for the associated radius. These orbits are discrete cosets of $\Cong{D}{4}_0(\alpha)$ 
 \be \mathcal{O}_1 = \Cong{D}{4}_0(\alpha) / ( \Cong{A}{3}_0(\alpha) \ltimes \mathds{M}_2^1)\,, \;   \mathcal{O}_2 = \Cong{D}{4}_0(\alpha) / ( SL(2,\mathds{H}) \ltimes \mathds{M}_2^0) \, , \;   \mathcal{O}_3 = \Cong{D}{4}_0(\alpha) / ( SL(2,\mathds{H}) \ltimes \mathds{M}^3_2) \; , \ee
by the corresponding stabilisers 
\bea  \Cong{A}{3}_0(\alpha) \ltimes \mathds{M}_2^1 \; : \; g &=& \left( \begin{array}{cccc} a_{11} \; \, &\, \;  a_{12} \; \,  &\, \;  b_{11} \;  \, &\; \,   b_{12} \\ a_{21} &  a_{22} &  b_{21}  & b_{22}  \\ 0 & 0   & d_{11}   & d_{12}  \\ 0  & 0  & d_{21}  & d_{22}  \end{array}\right) \; , \CR
SL(2,\mathds{H}) \ltimes \mathds{M}_2^0 \; : \; g &=& \left( \begin{array}{cccc} a_{11} \; \, &\, \;  b_{12} \; \,  &\, \;  b_{11} \;  \, &\; \,   a_{12} \\  0  &  d_{22} &  d_{21}  & 0 \\ 0 & d_{12}   & d_{11}   & 0   \\ a_{21}  & b_{22}  & b_{21}  & a_{22}  \end{array}\right) \; , \CR
SL(2,\mathds{H}) \ltimes \mathds{M}^3_2 \; : \; g &=& \left( \begin{array}{cccc} d_{11} \; \, &\, \;  0  \; \,  &\, \;  0  \;  \, &\; \,   d_{12} \\  b_{21}  &  a_{22} &  a_{21}  & b_{22} \\ b_{21}  & a_{12}   & a_{11}   & b_{12}    \\ d_{21}  & 0   & 0  & d_{22}  \end{array}\right) \; , \eea 
with $D = A^{-1 \dagger}$ and $A^{-1} B$ Hermitian. 
One has in this way 
\bea E^{\mathfrak{sp}_4 \mathds{H}}_{ \mathds{F}^0_2,s} &=& \sum_{\gamma \in \mathcal{O}_1} R^{2s} +\sum_{\gamma \in \mathcal{O}_2} \tilde{R}_{\scalebox{0.5}{T}}^{2s} +\sum_{\gamma \in \mathcal{O}_3} \tilde{R}^{2s} \CR
E^{\mathfrak{sp}_4 \mathds{H}}_{ \mathds{F}^1_2,s} &=& \sum_{\gamma \in \mathcal{O}_1} R^{2s} +\sum_{\gamma \in \mathcal{O}_2} \tilde{R}_{\scalebox{0.5}{T}}^{2s} +2^{-2s} \sum_{\gamma \in \mathcal{O}_3} \tilde{R}^{2s} \CR
E^{\mathfrak{sp}_4 \mathds{H}}_{ \mathds{F}^2_2,s} &=& 2^{-2s} \sum_{\gamma \in \mathcal{O}_1} R^{2s} +\sum_{\gamma \in \mathcal{O}_2} \tilde{R}_{\scalebox{0.5}{T}}^{2s} +2^{-2s} \sum_{\gamma \in \mathcal{O}_3} \tilde{R}^{2s} \; . 
\eea
The relation can be used to define the Poincar\'e sums in function of the constrained lattice sums for almost all $s$. The limit $s\rightarrow 0$ is singular, and the Poincar\'e sums are not constant at $s=0$. Instead, they depend individually on the regularised Eisenstein series at $s=3$. Nevertheless, the sum of the three  Poincar\'e sums tends to one as $s\rightarrow 0$. More generally, for any three coefficients $c_i(s)$ function of $s$ with  $c_i(s) = 1 + \mathcal{O}(s)$ at $s=0$, one has 
 \be \lim_{s\rightarrow 0} \biggl( c_1(s) \sum_{\gamma \in \mathcal{O}_1} R^{2s}+c_2(s) \sum_{\gamma \in \mathcal{O}_2} \tilde{R}_{\scalebox{0.5}{T}}^{2s}+c_3(s) \sum_{\gamma \in \mathcal{O}_3} \tilde{R}^{2s} \biggr) = 1\; .  \ee

%%%%%%%%%%%%%%%%%%%%%%%%

\section{Supergravity one-loop amplitude}
\label{Sugra1loop}
The tree-level supergravity four-graviton amplitude is \footnote{We use the vielbein convention for $\kappa$ commonly used in string theory amplitudes that is $\kappa =\frac{\kappa_h}{2}$ with the metric  $\kappa_h$ used in field theory amplitudes \cite{Bern:2011rj}. Moreover we use the mostly plus signature for the Minkowski metric. We use otherwise the amplitude normalisation, that differs from the standard string theory normalisation by a factor of $4$.}
\be \mathcal{M}^{\scalebox{0.6}{tree}}_{\scalebox{0.6}{sugra}}(1^-,2^-,3^+,4^+) = -i \kappa^2 \langle 12\rangle^4 [ 34]^4  \frac{1}{s  t u}= -i\frac{ \kappa^2 }{2^{10}} t_8 t_8 \mathcal{R}^4 \frac{64}{s  t u}\; .  \ee
It is the low energy limit of the string theory amplitude 
\be \mathcal{M}^{\scalebox{0.6}{tree}}_{\scalebox{0.6}{type II}}(1^-,2^-,3^+,4^+)  = \frac i {2^{11}}\frac{ (2\pi)^7 \alpha^{\prime 7}  g_{\scalebox{0.5}{D}}^{\, 2}}{(2\pi \sqrt{\alpha^\prime})^{10-D}} t_8 t_8 \mathcal{R}^4 \frac{\Gamma(-\frac{\alpha^\prime s}{4})\Gamma(-\frac{\alpha^\prime t}{4})\Gamma(-\frac{\alpha^\prime u}{4})}{\Gamma(1+\frac{\alpha^\prime s}{4})\Gamma(1+\frac{\alpha^\prime t}{4})\Gamma(1+\frac{\alpha^\prime u}{4})} \ee
with the identification \cite{Green:2005ba}
\be \frac12 \alpha^{\prime \frac{D-2}{2}} (2\pi)^{D-3} g_{\scalebox{0.5}{D}}^{\, 2} = \kappa^2 \; ,\label{kappaalpha} \ee
in $D$ dimensions. 

The one-loop supergravity $\mathcal{N}=6$ amplitude in $D=4-2\epsilon$ dimensions takes the form \cite{Bern:2011rj}
\begin{multline} \mathcal{M}^{\scalebox{0.6}{1-loop}}_{\scalebox{0.6}{sugra}}(1^-,2^-,3^+,4^+) =- i \kappa^4 \langle 12\rangle^4 [ 34]^4   \Biggl( I_4(s,t) + I_4(t,u) + I_4(u,s)\\
 - \frac{1}{s^2}\frac{D-2}{D-3} \Bigl(\frac{t u}{2} I_4(t,u)  +t I_3(t) + u  I_3(u) \Bigr)\\ +  \frac{1}{s u}\frac{D-4}{D-3} \Bigl(\frac{s t }{2} I_4(s,t)  +s I_3(s) + t  I_3(t) \Bigr) +  \frac{1}{s t}\frac{D-4}{D-3} \Bigl(\frac{s u }{2} I_4(s,u)  +s I_3(s) + u  I_3(u) \Bigr) \Biggr) \end{multline}
where 
\bea I_4(s,t) &=&\int \frac{d^D p}{(2\pi)^D}  \frac{1}{p^2 (p-k_1)^2 (p-k_1-k_2)^2 (p+k_4)^2} \CR
 I_3(s) &=&\int \frac{d^D p}{(2\pi)^D}  \frac{1}{p^2 (p-k_1-k_2)^2 (p+k_4)^2} \eea
are the integral over the Euclidean momentum $p$. The first line is the $\mathcal{N}=8$ supergravity amplitude and the two others line remove the contribution from the two additional gravitini multiplets. In Schwinger parameter space one obtains after some integrations by part
\bea && \frac{-1}{u} \frac{1}{D-3}  \Bigl(\frac{s t }{2} I_4(s,t)  +s I_3(s) + t  I_3(t) \Bigr)  \CR
&=&\frac{1}{2^4(2\pi)^{D-4}} \int_0^\infty \hspace{-3mm}d\tau_2 \, \tau_2^{\, 3-\frac{D}{2}}  \int_0^1 \hspace{-2mm} dx_3 \int_0^{x_3} \hspace{-2mm} dx_2\int_0^{x_2} \hspace{-2mm} dx_1  \, e^{\pi \tau_2 [(x_2-x_1)(1-x_3) s + x_1 (x_3-x_2) t]}\CR
&& \hspace{5mm}  \Biggl(  \frac{1}{D-3} \frac{s t}{s+t} \bigl( \tfrac12 - 1+x_3-x_1\bigr) + \frac{2}{D-6} \bigl(  (x_2-x_1)(1-x_3) s + x_1 (x_3-x_2) t\bigr) \Biggr)  \CR
&=&\frac{1}{2^4(2\pi)^{D-4}} \int_0^\infty \hspace{-3mm}d\tau_2 \, \tau_2^{\, 3-\frac{D}{2}}  \frac{1}{\pi \tau_2} \int_0^1 \hspace{-2mm} dx_3 \int_0^{x_3} \hspace{-2mm} dx_2\int_0^{x_2} \hspace{-2mm} dx_1  \, e^{\pi \tau_2 [ (x_2-x_1)(1-x_3) s + x_1 (x_3-x_2) t]}\eea
where one recognises the box integral in $D+2=6-2\epsilon $ dimensions as is used in \cite{Bern:2011rj}. After several integrations by part, one obtains the representation of the amplitude 
\bea && i \mathcal{M}^{\scalebox{0.6}{1-loop}}_{\scalebox{0.6}{sugra}}(1^-,2^-,3^+,4^+) \nonumber\\
&=&  \frac{\kappa^4}{2^4 (2\pi)^{D-4}}   \langle 12\rangle^4 [ 34]^4   \int_0^\infty \hspace{-3mm}d\tau_2 \, \tau_2^{\, 3-\frac{D}{2}}   \int_0^1 \hspace{-2mm} dx_3 \int_0^{x_3} \hspace{-2mm} dx_2\int_0^{x_2} \hspace{-2mm} dx_1  \CR
&&\times \Biggl( \Bigl( 1 - \frac{( 1-x_1 + x_2  - x_3)(x_1 - x_2 + x_3) }2  \Bigr) e^{\pi \tau_2 [ (x_2-x_1)(1-x_3) t + x_1 (x_3-x_2) u]}\CR
&&\hspace{-4mm} +\Bigl( 1 - \frac{( 1-x_1 + x_2  - x_3)(x_1 - x_2 + x_3)}2 - 2( x_2-x_1) (1-x_3)   \Bigr) \Bigl( e^{\pi \tau_2 [ (x_2-x_1)(1-x_3) s + x_1 (x_3-x_2) t]}\CR
&& \hspace{80mm} + e^{\pi \tau_2 [ (x_2-x_1)(1-x_3) s + x_1 (x_3-x_2) u]}  \Bigr)  \Biggr)  \eea
which is more appropriate for comparison with the string theory amplitude. In particular the dependence in the dimension $D$ is only in the power of $(2\pi)^2 \tau_2$, and this formula is also valid in $D=5-2\epsilon$ for four-dimensional polarisations.

We will now find that this matches the one-loop string theory amplitude \eqref{M4grav1-loopFK} at low energy, fixing in this way the normalisation. For this purpose we consider the tropical limit  \cite{Tourkine:2013rda}, such that $z_a \sim \tau y_a$ and $\tau \sim i \tau_2$ with $\tau_2>\hspace{-2mm}>1$ and $0\le y_a\le 1$. Using \eqref{def_calY_1}, one computes the tropical limits \footnote{We have for $z = x + i \tau_2 y$
\bea \mathcal{G}(z) &=& 2\pi \tau_2 ( \tfrac16 + y^2) - \log\bigl( | e^{\pi \tau_2 y- i \pi  x} {-} e^{-\pi \tau_2 y+i \pi  x} |^2\bigr)  + \mathcal{O}(e^{-2\pi \tau_2})  \; , \CR
\mathcal{Y}(z) &=& - 8\pi^2  \Bigl(  \tfrac16  -  y \frac{e^{\pi \tau_2 y- i \pi  x} {+} e^{-\pi \tau_2 y+i \pi  x}}{e^{\pi \tau_2 y- i \pi  x} {-} e^{-\pi \tau_2 y+i \pi  x} } +y^2 \Bigr) + \mathcal{O}(e^{-2\pi \tau_2})\; . \eea} 
\be \mathcal{G}(z_a-z_b) \sim 2\pi \tau_2 \bigl( \tfrac16 - |y_a-y_b| + (y_a-y_b)^2 \bigr) \; , \quad \mathcal{Y}(z_a-z_b) \sim - 8\pi^2  \bigl( \tfrac16 - |y_a-y_b| + (y_a-y_b)^2 \bigr) \; . \ee
The tropical limit is obtained up to exponentially suppressed term in $e^{-\pi \tau_2}$ at $y_a\ne y_b$.
Using then \eqref{P2E2} and \eqref{P3E2} one shows that 
\bea 
&& \sum_{\gamma \in PSL(2,\mathds{Z})/ \Gamma_0(K)} (2\sin \! \tfrac\pi K)^2   \wp(\tfrac1 K)  \Gamma_{\sLambda_{1,1}\oplus \sLambda_{1,1}[K]\oplus \mathds{H}(K)}  \label{BPSorbitsSugraP} \\
&\underset{K=2}{\sim}& \frac{2\pi^2}{3} \tau_2^3 \Biggl( 5 \sum_{\substack{ \mathcal{Q} \in \sLambda_{1,1}\oplus \sLambda_{1,1}[2]\oplus \mathds{H}\\ \mathcal{Q}^2=0}}+\sum_{\substack{ \mathcal{Q} \in \sLambda_{1,1}\oplus \sLambda_{1,1}[\frac12]\oplus \frac1\alpha \mathds{H}\smallsetminus\sLambda_{1,1}\oplus \sLambda_{1,1}[2]\oplus \mathds{H} \\ \mathcal{Q}^2=0}} \Biggr) e^{-4\pi \tau_2  p_R(\mathcal{Q})^2} \CR
&\underset{K=3}{\sim}& \frac{2\pi^2}{3} \tau_2^3 \Biggl( 5 \sum_{\substack{ \mathcal{Q} \in \sLambda_{1,1}\oplus \sLambda_{1,1}[3]\oplus \mathds{H}(3)\\ \mathcal{Q}^2=0}}+ \hspace{8mm} \tfrac12\hspace{-10mm} \sum_{\substack{ \mathcal{Q} \in \sLambda_{1,1}\oplus \sLambda_{1,1}[\frac13]\oplus \frac1\alpha \mathds{H}(3)\smallsetminus\sLambda_{1,1}\oplus \sLambda_{1,1}[3]\oplus \mathds{H}(3) \\ \mathcal{Q}^2=0}} \Biggr) e^{-4\pi \tau_2  p_R(\mathcal{Q})^2} \nonumber
\eea
and 
\bea 
&&\frac12 \sum_{\gamma \in PSL(2,\mathds{Z})/ \Gamma_0(K)} (2\sin \! \tfrac\pi K)^2   \Gamma_{\sLambda_{1,1}\oplus \sLambda_{1,1}[K]\oplus \mathds{H}(K)} \CR
&\underset{K=2}{\sim}& \tau_2^3 \Biggl( \sum_{\substack{ \mathcal{Q} \in \sLambda_{1,1}\oplus \sLambda_{1,1}[2]\oplus \mathds{H}\\ \mathcal{Q}^2=0}}-\sum_{\substack{ \mathcal{Q} \in \sLambda_{1,1}\oplus \sLambda_{1,1}[\frac12]\oplus \frac1\alpha \mathds{H}\smallsetminus\sLambda_{1,1}\oplus \sLambda_{1,1}[2]\oplus \mathds{H} \\ \mathcal{Q}^2=0}} \Biggr) e^{-4\pi \tau_2  p_R(\mathcal{Q})^2} \CR
&\underset{K=3}{\sim}&  \tau_2^3 \Biggl(  \sum_{\substack{ \mathcal{Q} \in \sLambda_{1,1}\oplus \sLambda_{1,1}[3]\oplus \mathds{H}(3)\\ \mathcal{Q}^2=0}}- \hspace{8mm} \tfrac12\hspace{-10mm} \sum_{\substack{ \mathcal{Q} \in \sLambda_{1,1}\oplus \sLambda_{1,1}[\frac13]\oplus \frac1\alpha \mathds{H}(3)\smallsetminus\sLambda_{1,1}\oplus \sLambda_{1,1}[3]\oplus \mathds{H}(3) \\ \mathcal{Q}^2=0}} \Biggr) e^{-4\pi \tau_2  p_R(\mathcal{Q})^2} \; , \label{BPSorbitsSugra} 
\eea
which gives the supergravity limit, including the Kaluza--Klein tower of spin two and spin three-half supermultiplets. Keeping only the massless graviton multiplet with $\mathcal{Q}=0$, one gets 
\bea && i  \mathcal{M}^{\scalebox{0.6}{1-loop}}_{\scalebox{0.6}{type II}}(1^-,2^-,3^+,4^+) \\
&\sim&  \frac{\alpha^{\prime\, 2 + \frac{D}{2}}g_{\scalebox{0.5}{D}}^{\, 4}  (2\pi)^{D-4} }{2^{8}}  \langle 12\rangle^4 [ 34]^4    \int_{0}^\infty \hspace{-2mm} d\tau_2 \tau_2^{3-\frac{D}{2}}  \prod_{a=1}^4 \int_0^1  \hspace{-2mm}  dy_a \delta(y_4) \; e^{-\frac{\alpha^\prime}{2}  \sum_{a>b} \mathcal{G}(z_a-z_b) k_a \cdot  k_b}  \CR
&&\hspace{-2mm}\times \Bigl(  \frac{20\pi^2}{3} +  \tfrac18 \bigl( \mathcal{Y}(z_{12}) + \mathcal{Y}(z_{34})- \mathcal{Y}(z_{13})- \mathcal{Y}(z_{14}) - \mathcal{Y}(z_{23})- \mathcal{Y}(z_{24}) \bigr)\Bigr) \CR
&=&  \frac{\alpha^{\prime\, 2+\frac{D}{2}} g_{\scalebox{0.5}{D}}^{\, 4}  (2\pi)^{D-2} }{2^{6}} \langle 12\rangle^4 [ 34]^4 \int_0^\infty \hspace{-3mm}d\tau_2 \, \tau_2^{\, 3-\frac{D}{2}}    \int_0^1 \hspace{-2mm} dx_3 \int_0^{x_3} \hspace{-2mm} dx_2\int_0^{x_2} \hspace{-2mm} dx_1  \CR
&&\times \Biggl( \Bigl( 1 - \frac{( 1-x_1 + x_2  - x_3)(x_1 - x_2 + x_3) }2  \Bigr) e^{\pi \tau_2  \alpha^\prime [ (x_2-x_1)(1-x_3) t + x_1 (x_3-x_2) u]}\CR
&& \hspace{-6mm} +\Bigl( 1 - \frac{( 1-x_1 + x_2  - x_3)(x_1 - x_2 + x_3)}2 - 2( x_2-x_1) (1-x_3)   \Bigr) \Bigl( e^{\pi \tau_2   \alpha^\prime[ (x_2-x_1)(1-x_3) s + x_1 (x_3-x_2) t]}\CR
&& \hspace{78mm} + e^{\pi \tau_2   \alpha^\prime[ (x_2-x_1)(1-x_3) s + x_1 (x_3-x_2) u]}  \Bigr)  \Biggr)  \eea
consistently with the supergravity amplitude using \eqref{kappaalpha}. In the computation we have decomposed the integral over the three $y_a$ variables into the 6 ordered integrals, that give two times the three integrals with the orders $0\le y_1\le y_2\le y_3$, $0\le y_2\le y_3\le y_1$ and $0\le y_3\le y_1\le y_2$. This gives in total two times the three corresponding integrals over $0\le x_1\le x_2\le x_3\le 1$ with the three respective changes of variables 
\bea y_1 &=& 1-x_3 \; , \qquad y_2 = 1-x_2 \; , \qquad y_3 = 1-x_1 \; , \\
 y_1 &=&1-x_1\; , \qquad y_2 = 1-x_3 \; , \qquad y_3 = 1-x_2\; ,  \\
  y_1 &=& x_2 \; , \qquad y_2 = x_3 \; , \qquad y_3 =x_1 \; .
\eea
Note as a cross check that the contribution from the one-half BPS states in \eqref{BPSorbitsSugraP} and \eqref{BPSorbitsSugra} gives also the expected contribution from spin two and spin three-half supermultiplets in supergravity. 

\section{The Dabholkar--Harvey $\mathds{Z}_3$ orbifold}
In this appendix we discuss another string theory construction of $\mathcal{N}=6$ supergravity introduced in \cite{Dabholkar:1998kv}, which is defined as the  asymmetric $\mathds{Z}_3$ orbifold acting  on $T^6$ at the $A_2\oplus A_2\oplus A_2$ symmetric point. The untwisted sector breaks supersymmetry down to $\mathcal{N}=5$, but the twisted sector includes one massless $\mathcal{N}=5$ gravitino multiplet. We want to argue that this theory is identical to the $\mathds{Z}_3$ orbifold theory introduced in Section \ref{StringOrbifolds} at the point in Narain moduli space where the left and right projections of the lattice $\sLambda_{1,1}\oplus \sLambda_{1,1}[3]\oplus A_2 \oplus A_2$ factorise into the left $A_2$ lattice and the right $A_2\oplus A_2\oplus A_2$ lattice, i.e. at zero Wilson lines and for the metric $G$ and $B$ field on $T^2$
\be G = \left( \begin{array}{cc} 3 \; & - \frac{3}{2}\\ - \frac{3}{2}\; & \; 1 \end{array}\right) \; , \qquad B  =  \left( \begin{array}{cc} 0 \; & - \frac{3}{2}\\  \frac{3}{2}\; & \; 0 \end{array}\right)\; . \label{A2fourPoint} \ee
We will show that the spectrum of the two theories are identical. Using \cite{Bianchi:2015vsa}, one obtains for the character valued partition function with the insertion of the $\mathds{Z}_3$ generator $g$ 
\be \frac{\vartheta_3(v) \vartheta_3(\frac13)^2 \vartheta_3(-\frac23) - \sum_{\alpha = 1,2,4} \vartheta_\alpha(v) \vartheta_\alpha(\frac13)^2 \vartheta_\alpha(-\frac23)  }{2 \eta(\tau)^3 \vartheta_1(\frac13)^3} = -\frac{\vartheta_1(\frac{v}{2}) \vartheta_1(\frac{v}{2}-\frac13)^3}{\eta(\tau)^3 \vartheta_1(\frac13)^3} \; , \ee
and the same with $\frac13$ replaced by $-\frac13$ for the insertion of $g^{-1}$, such that the character valued partition function is
 \begin{multline}
{\cal Z}^{\mathds{Z}_3}_{D=4}(v,\bar{v})=\frac{\xi(v)}{3} \overline{ \frac{\xi({v})\vartheta_1(\frac{v}2)^4}{\eta(\tau)^{12}} }\Biggl[  \frac{{\vartheta_1({v\over 2})^4}}{   \eta(\tau)^{12} } \Lambda_{\sLambda_{6,6}}+   ( 2 \sin \tfrac{\pi}{3})^3    \frac{\vartheta_1(\frac{v}{2})\bigl(  \vartheta_1(\frac13-\frac{v}{2})^3- \vartheta_1(\frac13+\frac{v}{2})^3\bigr) }{\eta(\tau)^3 \vartheta_1(\frac13)^3}  \overline{\Lambda}_{A_2}^{\; 3}  \\ -   \frac{ i }{\sqrt{27}} ( 2 \sin \tfrac{\pi}{3})^3   \sum_{r=0}^2  \Bigl( \frac{\vartheta_1(\frac{v}{2}) \bigl( \vartheta_1(\frac \tau 3-\frac{v}{2})^3- \vartheta_1(\frac \tau 3+\frac{v}{2})^3\bigr) }{\eta(\tau)^3 \vartheta_1(\frac \tau 3)^3}  \overline{\Lambda}_{A_2^*}^{\; 3} \Bigr)  \Big|_{\tau\rightarrow \tau + r} \Biggr]  \end{multline}
where the last line is the contribution from the twisted sectors that is obtained by modular transformations. As in \ref{CharZK}, it is convenient to decompose the Lorentzian lattice partition function as a sum over orbits
\be \Lambda_{\sLambda_{6,6}} = \sum_{\gamma \in P\hspace{-0.2mm}SL(2,\mathds{Z}) / \Gamma_0(3) } \Bigl( \bigl( \Lambda_{A_2}^{\; 3} - \tfrac18 (\Lambda_{A^*_2}-\Lambda_{A_2})^3 \bigr)  \overline{\Lambda}_{A_2}^3 \Bigr) \Big|_\gamma \label{Lambda66asPprime}  \ee
where the seed can be identified as  
\be \Lambda_{A_2}^{\; 3} - \tfrac18 (\Lambda_{A^*_2}-\Lambda_{A_2})^3  =  ( 2 \sin \tfrac{\pi}{3})^3  \frac{\eta(\tau)^9}{\vartheta_1(\frac13)^3} \; . \label{A2A2dProduct}  \ee
With the help of this formula one can rewrite the character valued partition function as
\begin{multline}
{\cal Z}^{\mathds{Z}_3}_{D=4}(v,\bar{v})=\sqrt{3} \, \overline{ \frac{\xi({v})\vartheta_1(\frac{v}2)^4}{\eta(\tau)^{12}} }\Biggl[   \frac{\xi(v)\vartheta_1(\frac{v}{2}) \bigl( \vartheta_1(\frac v 2)^3+ \vartheta_1(\frac13-\frac{v}{2})^3- \vartheta_1(\frac13+\frac{v}{2})^3\bigr) }{\eta(\tau)^3 \vartheta_1(\frac13)^3}  \overline{\Lambda}_{A_2}^{\; 3}  \\ -   \frac{ i }{\sqrt{27}}   \sum_{r=0}^2  \Bigl( \frac{ \xi(v)\vartheta_1(\frac{v}{2}) \bigl(\vartheta_1(\frac v 2)^3+ \vartheta_1(\frac \tau 3-\frac{v}{2})^3- \vartheta_1(\frac \tau 3+\frac{v}{2})^3\bigr) }{\eta(\tau)^3 \vartheta_1(\frac \tau 3)^3}  \overline{\Lambda}_{A_2^*}^{\; 3} \Bigr)  \Big|_{\tau\rightarrow \tau + r} \Biggr]  \; . \end{multline}
Using then the theta function identity 
\be \vartheta_1^\prime(0) \Bigl( \vartheta_1(z)^3- \vartheta_1(z+\tfrac13)^3- \vartheta_1(z-\tfrac13)^3\Bigr) = 6 \vartheta_1^\prime(\tfrac13) \vartheta_1(z) \vartheta_1(z+ \tfrac13) \vartheta_1(z-\tfrac13) \; , \label{ThetaFourTerms} \ee
and 
\be \frac{\sqrt{3}}{\pi}  \frac{\vartheta_1^\prime(\frac13)}{\vartheta_1(\frac13)}  = \Lambda_{A_2}\; , \label{A2Theta}  \ee
one finds that 
\be \sqrt{3} \,  \frac{\vartheta_1(\frac{v}{2}) \bigl( \vartheta_1(\frac v 2)^3+ \vartheta_1(\frac13-\frac{v}{2})^3- \vartheta_1(\frac13+\frac{v}{2})^3\bigr) }{\eta(\tau)^3 \vartheta_1(\frac13)^3} =3\frac{\vartheta_1(\frac{v}{2})^2 \vartheta_1(\frac v 2+\frac13) \vartheta_1(\frac v 2-\frac13) }{\eta(\tau)^6 \vartheta_1(\frac13)^2}   \Lambda_{A_2}\; .  \ee
Because the twisted sector is obtained by modular transformations, we conclude that 
\bea 
Z^{\mathds{Z}_3}_{D=4}(y) 
&=& 3  \int_{-\frac12}^{\frac12}\hspace{-2.5mm}d\tau_1   \overline{ \frac{\xi({v})\vartheta_1(\frac{v}{2})^4}{\eta(\tau)^{12}} }  \frac{  \xi(v){\vartheta_1(\frac{v}{2})^2 \vartheta_1(\frac{v}{2}{+}\frac{1}{3})\vartheta_1(\frac{v}{2}{-}\frac{1}{3})
}}{  \vartheta_1(\frac{1}{3})^2 \eta(\tau)^6}  \Lambda_{A_2} \overline{\Lambda}_{A_2}^{\; 3}  \\
&& \hspace{20mm}- \frac{1 }{3}  \int_{-\frac 32}^{\frac 32}\hspace{-2mm}d\tau_1    \overline{ \frac{\xi({v})\vartheta_1(\frac{v}{2})^4}{\eta(\tau)^{12}} }  \frac{  \xi(v){\vartheta_1(\frac{v}{2})^2 \vartheta_1(\frac{v}{2}{+}\frac{\tau}{3})\vartheta_1(\frac{v}{2}{-}\frac{\tau}{3})
}}{  \vartheta_1(\frac{\tau}{3})^2 \eta(\tau)^6} \Lambda_{A_2^*} \overline{\Lambda}_{A_2^*}^{\; 3}    \; , \nonumber \eea
as in \eqref{PartitionHelicity} at the  point \eqref{A2fourPoint} in Narain moduli space where 
\be \Lambda_{\sLambda_{1,1}\oplus \sLambda_{1,1}[3]\oplus \mathds{H}(3)} = \Lambda_{A_2} \overline{\Lambda}_{A_2}^{\; 3} \; . \ee
This proves that the two theories have the same perturbative spectrum, which provides strong evidence that the Dabholkar--Harvey $\mathds{Z}_3$ orbifold theory is just a particular example of the $\mathds{Z}_3$ orbifold theory analysed  in this paper.


\begin{thebibliography}{99}
  
  %\cite{Cremmer:1978ds}
\bibitem{Cremmer:1978ds}
E.~Cremmer and B.~Julia,
``The N=8 Supergravity Theory 1. The Lagrangian,''
\doi{Phys. Lett. B \textbf{80} (1978), 48.}{doi:10.1016/0370-2693(78)90303-9}
%717 citations counted in INSPIRE as of 21 Jan 2022

%\cite{Bianchi:2009wj}
\bibitem{Bianchi:2009wj}
M.~Bianchi, S.~Ferrara and R.~Kallosh,
``Perturbative and non-perturbative $\cN=8$ Supergravity,''
\doi{Phys. Lett. B \textbf{690} (2010), 328-331}{doi:10.1016/j.physletb.2010.05.049}
\eprintN{0910.3674}.
%29 citations counted in INSPIRE as of 10 Mar 2022

%\cite{Bianchi:2009mj}
\bibitem{Bianchi:2009mj}
M.~Bianchi, S.~Ferrara and R.~Kallosh,
``Observations on Arithmetic Invariants and U-Duality Orbits in $\cN =8$ Supergravity,''
\doi{JHEP \textbf{03} (2010), 081}{doi:10.1007/JHEP03(2010)081}
\eprintN{0912.0057}.
%23 citations counted in INSPIRE as of 10 Mar 2022


\bibitem{Hull:1994ys}
C.~M. Hull and P.~K. Townsend, ``Unity of superstring dualities,'' {\em Nucl.
  Phys.} {\bf B438} (1995)  109--137,
\eprint{hep-th/9410167}.
%%CITATION = HEP-TH 9410167;%%.

\bibitem{Witten:1995ex}
E.~Witten, ``{String theory dynamics in various dimensions},''
  \href{http://dx.doi.org/10.1016/0550-3213(95)00158-O}{{\em Nucl.Phys.} {\bf
  B443} (1995)  85--126},
\eprint{hep-th/9503124}.
%%CITATION = HEP-TH/9503124;%%.

%%%%%%%%%%%%%%%%%%%%%%%%%%%%%%%%%%%%%%%%%%%%%%%%%%%%%%%%%%%%%%%%

\bibitem{Green:1981yb}
M.~B. Green and J.~H. Schwarz, ``{Supersymmetrical String Theories},''
\href{http://dx.doi.org/10.1016/0370-2693(82)91110-8}{{\em Phys. Lett.} {\bf
  B109} (1982)  444--448}.
%%CITATION = PHLTA,B109,444;%%.


\bibitem{Green:1997tv}
M.~B. Green and M.~Gutperle, 
``{Effects of D-instantons},''
\doi{Nucl. Phys. B {\bf 498} (1997)  195--227}{doi:10.1016/S0550-3213(97)00269-1}
\eprint{hep-th/9701093}.
%%CITATION = HEP-TH/9701093;%%.

\bibitem{Green:1997di}
M.~B. Green and P.~Vanhove, ``{D-instantons, strings and M-theory},''
  \href{http://dx.doi.org/10.1016/S0370-2693(97)00785-5}{{\em Phys. Lett.} {\bf
  B408} (1997)  122--134},
\eprint{hep-th/9704145}.
%%CITATION = HEP-TH/9704145;%%.


%\cite{Berkovits:1997pj}
\bibitem{Berkovits:1997pj}
N.~Berkovits,
``Construction of $R^4$ terms in $N=2$  $D = 8$ superspace,''
\doi{Nucl. Phys. B \textbf{514} (1998), 191-203}{doi:10.1016/S0550-3213(97)00817-1}
\eprint{hep-th/9709116}.
%46 citations counted in INSPIRE as of 24 Mar 2022

%\cite{Pioline:1998mn}
\bibitem{Pioline:1998mn}
B.~Pioline,
``A note on nonperturbative $R^4$ couplings,''
\doi{Phys. Lett. B \textbf{431} (1998), 73-76}{doi:10.1016/S0370-2693(98)00554-1}
\eprint{hep-th/9804023}.
%76 citations counted in INSPIRE as of 14 Apr 2021


%\cite{Green:1998by}
\bibitem{Green:1998by}
M.~B.~Green and S.~Sethi,
``Supersymmetry constraints on type IIB supergravity,''
\doi{Phys. Rev. D \textbf{59} (1999), 046006}{doi:10.1103/PhysRevD.59.046006}
\eprintN{hep-th/9808061}.
%232 citations counted in INSPIRE as of 14 Apr 2021



%\cite{Obers:1999um}
\bibitem{Obers:1999um}
N.~A.~Obers and B.~Pioline,
``Eisenstein series and string thresholds,''
\doi{Commun. Math. Phys. \textbf{209} (2000), 275-324}{doi:10.1007/s002200050022}
\eprint{hep-th/9903113}.
%105 citations counted in INSPIRE as of 21 Jan 2022

\bibitem{Green:1999pv}
M.~B. Green and P.~Vanhove, ``{The low energy expansion of the one-loop type II
  superstring amplitude},''
  \href{http://dx.doi.org/10.1103/PhysRevD.61.104011}{{\em Phys. Rev.} {\bf
  D61} (2000)  104011},
\eprint{hep-th/9910056}.
%%CITATION = HEP-TH/9910056;%%.





%\cite{Obers:2001sw}
\bibitem{Obers:2001sw}
N.~A.~Obers and B.~Pioline,
``Exact thresholds and instanton effects in string theory,''
Fortsch. Phys. \textbf{49} (2001), 359-375 
\eprint{hep-th/0101122}.
%5 citations counted in INSPIRE as of 01 Apr 2021


%\cite{Kazhdan:2001nx}
\bibitem{Kazhdan:2001nx}
D.~Kazhdan, B.~Pioline and A.~Waldron,
``Minimal representations, spherical vectors, and exceptional theta series,''
\doi{Commun. Math. Phys. \textbf{226} (2002), 1-40}{doi:10.1007/s002200200601}
\eprintN{hep-th/0107222}.
%46 citations counted in INSPIRE as of 14 Apr 2021

\bibitem{Basu:2008cf}
A.~Basu and S.~Sethi, ``{Recursion relations from space-time supersymmetry},''
  \href{http://dx.doi.org/10.1088/1126-6708/2008/09/081}{{\em JHEP} {\bf 09}
  (2008)  081},
\eprintN{0808.1250}.
%%CITATION = ARXIV:0808.1250;%%.



  %\cite{Green:2005ba}
\bibitem{Green:2005ba}
M.~B.~Green and P.~Vanhove,
``Duality and higher derivative terms in M theory,''
\doi{JHEP \textbf{01} (2006), 093}{doi:10.1088/1126-6708/2006/01/093}
\eprintN{hep-th/0510027}.
%149 citations counted in INSPIRE as of 15 Apr 2021





\bibitem{Pioline:2010kb}
B.~Pioline, ``{$R^4$ couplings and automorphic unipotent representations},''
  \href{http://dx.doi.org/10.1007/JHEP03(2010)116}{{\em JHEP} {\bf 03} (2010)
  116},
\eprintN{1001.3647}.
%%CITATION = 1001.3647;%%.


%\cite{Green:2011vz}
\bibitem{Green:2011vz}
M.~B.~Green, S.~D.~Miller and P.~Vanhove,
``Small representations, string instantons, and Fourier modes of Eisenstein series,''
\doi{J. Number Theor. \textbf{146} (2015), 187-309}{doi:10.1016/j.jnt.2013.05.018}
\eprintN{1111.2983}.
%42 citations counted in INSPIRE as of 14 Apr 2021


%\cite{Bossard:2014lra}
\bibitem{Bossard:2014lra}
G.~Bossard and V.~Verschinin,
``Minimal unitary representations from supersymmetry,''
\doi{JHEP \textbf{10} (2014), 008}{doi:10.1007/JHEP10(2014)008}
\eprintN{1406.5527}.
%25 citations counted in INSPIRE as of 14 Apr 2021

%\cite{Bossard:2014aea}
\bibitem{Bossard:2014aea}
G.~Bossard and V.~Verschinin,
``$\mathcal{E} {D}^4 \mathcal{R}^4$ type invariants and their gradient expansion,''
\doi{JHEP \textbf{03} (2015), 089}{doi:10.1007/JHEP03(2015)089}
\eprintN{1411.3373}.
%26 citations counted in INSPIRE as of 14 Apr 2021


%\cite{Gustafsson:2014iva}
\bibitem{Gustafsson:2014iva}
H.~P.~A.~Gustafsson, A.~Kleinschmidt and D.~Persson,
``Small automorphic representations and degenerate Whittaker vectors,''
\doi{J. Number Theor. \textbf{166} (2016), 344-399}{doi:10.1016/j.jnt.2016.02.002}
\eprintNT{1412.5625}.
%9 citations counted in INSPIRE as of 24 Mar 2022


\bibitem{Wang:2015jna}
Y.~Wang and X.~Yin, ``{Constraining higher derivative supergravity with
  scattering amplitudes},''
  \href{http://dx.doi.org/10.1103/PhysRevD.92.041701}{{\em Phys. Rev.} {\bf
  D92} (2015) no.~4, 041701},
\eprintN{1502.03810}.
%%CITATION = ARXIV:1502.03810;%%.

%\cite{Bossard:2015uga}
\bibitem{Bossard:2015uga}
G.~Bossard and V.~Verschinin,
``The two $\nabla^{6} R^{4}$ type invariants and their higher order generalisation,''
\doi{JHEP \textbf{07} (2015), 154}{doi:10.1007/JHEP07(2015)154}
\eprintN{1503.04230}.
%22 citations counted in INSPIRE as of 14 Apr 2021

%\cite{Gourevitch:2019knu}
\bibitem{Gourevitch:2019knu}
D.~Gourevitch, H.~P.~A.~Gustafsson, A.~Kleinschmidt, D.~Persson and S.~Sahi,
``Fourier coefficients of minimal and next-to-minimal automorphic representations of simply-laced groups,''
\doi{Can. J. Math. \textbf{74} (2022) no.1, 122-169}{doi:10.4153/S0008414X20000711}
\eprintNT{1908.08296}.
%3 citations counted in INSPIRE as of 24 Mar 2022



%%%%%%%%%%%%%%%%%%%%%%%%%%%%%%%%%%%%%%%%%%%%%%

 


%\cite{Green:1997as}
\bibitem{Green:1997as}
M.~B.~Green, M.~Gutperle and P.~Vanhove,
``One loop in eleven-dimensions,''
\doi{Phys. Lett. B \textbf{409} (1997), 177-184}{doi:10.1016/S0370-2693(97)00931-3}
\eprint{hep-th/9706175}.
%306 citations counted in INSPIRE as of 21 Jan 2022



\bibitem{Kiritsis:1997em}
E.~Kiritsis and B.~Pioline, ``{On $R^4$ threshold corrections in type IIB
  string theory and (p,q) string instantons},''
  \href{http://dx.doi.org/10.1016/S0550-3213(97)00645-7}{{\em Nucl. Phys.} {\bf
  B508} (1997)  509--534},
\eprint{hep-th/9707018}.
%%CITATION = HEP-TH/9707018;%%.



\bibitem{Pioline:1997pu}
B.~Pioline and E.~Kiritsis, ``{U-duality and D-brane combinatorics},''
  \href{http://dx.doi.org/10.1016/S0370-2693(97)01398-1}{{\em Phys. Lett.} {\bf
  B418} (1998)  61--69},
\eprint{hep-th/9710078}.
%%CITATION = HEP-TH/9710078;%%.


%\cite{Green:1999pu}
\bibitem{Green:1999pu}
M.~B.~Green, H.~h.~Kwon and P.~Vanhove,
``Two loops in eleven-dimensions,''
\doi{Phys. Rev. D \textbf{61} (2000), 104010}{doi:10.1103/PhysRevD.61.104010}
\eprint{hep-th/9910055}.
%168 citations counted in INSPIRE as of 21 Jan 2022

\bibitem{Basu:2007ru}
A.~Basu, ``{The $D^4 R^4$ term in type IIB string theory on $T^2$ and U-duality},'' \href{http://dx.doi.org/10.1103/PhysRevD.77.106003}{{\em Phys.
  Rev.} {\bf D77} (2008)  106003},
\eprintN{0708.2950}.
%%CITATION = 0708.2950;%%.

%\cite{Green:2008uj}
\bibitem{Green:2008uj}
  M.~B.~Green, J.~G.~Russo and P.~Vanhove,
  ``Low energy expansion of the four-particle genus-one amplitude in type II superstring theory,''
 \doi{JHEP {\bf 0802} (2008) 020}{doi:10.1088/1126-6708/2008/02/020}
  \eprintN{0801.0322}.
  %%CITATION = doi:10.1088/1126-6708/2008/02/020;%%
  %65 citations counted in INSPIRE as of 11 Jun 2018


%\cite{Green:2010wi}
\bibitem{Green:2010wi}
M.~B.~Green, J.~G.~Russo and P.~Vanhove,
``Automorphic properties of low energy string amplitudes in various dimensions,''
\doi{Phys. Rev. D \textbf{81} (2010), 086008}{doi:10.1103/PhysRevD.81.086008}
\eprintN{1001.2535}.
%85 citations counted in INSPIRE as of 19 Jan 2022

\bibitem{Green:2010sp}
M.~B. Green, J.~G. Russo, and P.~Vanhove, ``{String theory dualities and
  supergravity divergences},''
  \href{http://dx.doi.org/10.1007/JHEP06(2010)075}{{\em JHEP} {\bf 1006} (2010)
   075},
\eprintN{1002.3805}.
%%CITATION = ARXIV:1002.3805;%%.

%\cite{Green:2010kv}
\bibitem{Green:2010kv}
M.~B.~Green, S.~D.~Miller, J.~G.~Russo and P.~Vanhove,
``Eisenstein series for higher-rank groups and string theory amplitudes,''
\doi{Commun. Num. Theor. Phys. \textbf{4} (2010), 551-596}{doi:10.4310/CNTP.2010.v4.n3.a2}
\eprintN{1004.0163}.
%77 citations counted in INSPIRE as of 21 Jan 2022

%\cite{Green:2014yxa}
\bibitem{Green:2014yxa}
M.~B.~Green, S.~D.~Miller and P.~Vanhove,
``$SL(2, \mathbb{Z})$-invariance and D-instanton contributions to the $D^6 R^4$ interaction,''
\doi{Commun. Num. Theor. Phys. \textbf{09} (2015), 307-344}{doi:10.4310/CNTP.2015.v9.n2.a3}
\eprintN{1404.2192}.
%45 citations counted in INSPIRE as of 21 Jan 2022



%\cite{Bossard:2020xod}
\bibitem{Bossard:2020xod}
G.~Bossard, A.~Kleinschmidt and B.~Pioline,
``1/8-BPS Couplings and Exceptional Automorphic Functions,''
\doi{SciPost Phys. \textbf{8} (2020) no.4, 054}{doi:10.21468/SciPostPhys.8.4.054}
\eprintN{2001.05562}.
%5 citations counted in INSPIRE as of 21 Jan 2022


  %\cite{Harvey:1996ir}
\bibitem{Harvey:1996ir}
J.~A.~Harvey and G.~W.~Moore,
``Five-brane instantons and $R^2$ couplings in $\mathcal{N}=4$ string theory,''
\doi{Phys. Rev. D \textbf{57} (1998), 2323-2328}{doi:10.1103/PhysRevD.57.2323}
\eprint{hep-th/9610237}.
%108 citations counted in INSPIRE as of 05 Mar 2022



%\cite{Bachas:1997mc}
\bibitem{Bachas:1997mc}
C.~Bachas, C.~Fabre, E.~Kiritsis, N.~A.~Obers and P.~Vanhove,
``Heterotic / type I duality and D-brane instantons,''
\doi{Nucl. Phys. B \textbf{509} (1998), 33-52}{doi:10.1016/S0550-3213(97)00639-1}
\eprint{hep-th/9707126}.
%134 citations counted in INSPIRE as of 01 Apr 2021


%\cite{Gregori:1997hi}
\bibitem{Gregori:1997hi}
A.~Gregori, E.~Kiritsis, C.~Kounnas, N.~A.~Obers, P.~M.~Petropoulos and B.~Pioline,
``$R^2$ corrections and nonperturbative dualities of N=4 string ground states,''
\doi{Nucl. Phys. B \textbf{510} (1998), 423-476}{doi:10.1016/S0550-3213(97)00635-4}
\eprint{hep-th/9708062}.
%138 citations counted in INSPIRE as of 05 Mar 2022

%\cite{Kiritsis:1997hf}
\bibitem{Kiritsis:1997hf}
E.~Kiritsis and N.~A.~Obers,
``Heterotic type I duality in $D<10$-dimensions, threshold corrections and D instantons,''
\doi{JHEP \textbf{10} (1997), 004}{doi:10.1088/1126-6708/1997/10/004}
\eprint{hep-th/9709058}.
%93 citations counted in INSPIRE as of 01 Apr 2021


%\cite{Bianchi:1998vq}
\bibitem{Bianchi:1998vq}
M.~Bianchi, E.~Gava, J.~F.~Morales and K.~S.~Narain,
``D strings in unconventional type I vacuum configurations,''
\doi{Nucl. Phys. B \textbf{547} (1999), 96-126}{doi:10.1016/S0550-3213(99)00004-8}
\eprint{hep-th/9811013}.
%33 citations counted in INSPIRE as of 01 Apr 2021

%
%%\cite{Bianchi:2007rb}
%\bibitem{Bianchi:2007rb}
%M.~Bianchi and J.~F.~Morales,
%``Unoriented D-brane Instantons vs Heterotic worldsheet Instantons,''
%\doi{JHEP \textbf{02} (2008), 073}{doi:10.1088/1126-6708/2008/02/073}
%\eprintN{0712.1895}.
%%33 citations counted in INSPIRE as of 01 Apr 2021


  %\cite{Bossard:2018rlt}
\bibitem{Bossard:2018rlt}
G.~Bossard, C.~Cosnier-Horeau and B.~Pioline,
``Exact effective interactions and 1/4-BPS dyons in heterotic CHL orbifolds,''
\doi{SciPost Phys. \textbf{7} (2019) no.3, 028}{doi:10.21468/SciPostPhys.7.3.028}
\eprintN{1806.03330}.
%7 citations counted in INSPIRE as of 19 Apr 2021

\bibitem{Hull:2004in}
C.~M.~Hull,
``A Geometry for non-geometric string backgrounds,''
\doi{JHEP \textbf{10} (2005), 065}{doi:10.1088/1126-6708/2005/10/065}
\eprint{hep-th/0406102}.
%524 citations counted in INSPIRE as of 24 Mar 2022

%%%%%%%%%%%%%%%%%%%%%%%%%%%%%%%%%%%%%%%%%%%%%%%%%%%%%%%%%%%%
%%%%%%%%%%%%%%%%%%%%%%%%%%%%%%%%%%%%%%%%%%%%%%%%%%%%%%%%%%%%

  %\cite{Ferrara:1989nm}
\bibitem{Ferrara:1989nm}
S.~Ferrara and C.~Kounnas,
``Extended supersymmetry in four-dimensional Type {II} Strings,''
\doi{Nucl. Phys. B \textbf{328} (1989), 406-438.}{doi:10.1016/0550-3213(89)90335-0}
%67 citations counted in INSPIRE as of 17 Jan 2022

%\cite{Bianchi:2008cj}
\bibitem{Bianchi:2008cj}
M.~Bianchi,
``Bound-states of D-branes in L-R asymmetric superstring vacua,''
\doi{Nucl. Phys. B \textbf{805} (2008), 168-181}{doi:10.1016/j.nuclphysb.2008.07.008}
\eprintN{0805.3276}.
%11 citations counted in INSPIRE as of 17 Jan 2022
  
  %\cite{Bianchi:2010aw}
\bibitem{Bianchi:2010aw}
M.~Bianchi,
``On $\mathcal{R}^{4}$ terms and MHV amplitudes in $\mathcal{N}$ = 5,6 supergravity vacua of Type II superstrings,''
\doi{Adv. High Energy Phys. \textbf{2011} (2011), 479038}{doi:10.1155/2011/479038}
\eprintN{1010.4736}.
%3 citations counted in INSPIRE as of 22 Mar 2021

%\cite{Narain:1986am}
\bibitem{Narain:1986am}
K.~S.~Narain, M.~H.~Sarmadi and E.~Witten,
``A note on toroidal compactification of Heterotic String Theory,''
\doi{Nucl. Phys. B \textbf{279} (1987), 369-379.}{doi:10.1016/0550-3213(87)90001-0}
%790 citations counted in INSPIRE as of 24 Mar 2022






%\cite{Dabholkar:1998kv}
\bibitem{Dabholkar:1998kv}
A.~Dabholkar and J.~A.~Harvey,
``String islands,''
\doi{JHEP \textbf{02} (1999), 006}{doi:10.1088/1126-6708/1999/02/006}
\eprint{hep-th/9809122}.
%61 citations counted in INSPIRE as of 13 Jan 2022

%\cite{Gunaydin:1983bi}
\bibitem{Gunaydin:1983bi}
M.~G\"{u}naydin, G.~Sierra and P.~K.~Townsend,
``The Geometry of N=2 Maxwell--Einstein Supergravity and Jordan Algebras,''
\doi{Nucl. Phys. B \textbf{242} (1984), 244-268.}{doi:10.1016/0550-3213(84)90142-1}
%581 citations counted in INSPIRE as of 01 Apr 2021

%\cite{Bianchi:2007va}
\bibitem{Bianchi:2007va}
M.~Bianchi and S.~Ferrara,
``Enriques and Octonionic Magic Supergravity Models,''
\doi{JHEP \textbf{02} (2008), 054}{doi:10.1088/1126-6708/2008/02/054}
\eprintN{0712.2976}.
%22 citations counted in INSPIRE as of 10 Mar 2022

  
\bibitem{Krieg}
A.~Krieg, ``Modular forms on half-spaces of quaternions'', Lecture Notes  in Mathematics {\bf 1143} (1985).
 

\bibitem{Vigneras}
Marie-France Vign\'eras , “Arithm\'etique des alg\`ebres de quaternions'' Lecture Notes in Mathematics  (1980) 800.



\bibitem{GreenSchwWitt}
M.B. Green, J.H. Schwarz and E. Witten, ``Superstring theory,'' Cambridge University Press.


%\cite{Bossard:2010bd}
\bibitem{Bossard:2010bd}
G.~Bossard, P.~S.~Howe and K.~S.~Stelle,
``On duality symmetries of supergravity invariants,''
\doi{JHEP \textbf{01} (2011), 020}{doi:10.1007/JHEP01(2011)020}
\eprintN{1009.0743}.
%86 citations counted in INSPIRE as of 21 Jan 2022

%\cite{Bern:2013uka}
\bibitem{Bern:2013uka}
Z.~Bern, S.~Davies, T.~Dennen, A.~V.~Smirnov and V.~A.~Smirnov,
``Ultraviolet properties of $\mathcal{N}=4$ Supergravity at four loops,''
\doi{Phys. Rev. Lett. \textbf{111} (2013) no.23, 231302}{doi:10.1103/PhysRevLett.111.231302}
\eprintN{1309.2498}.
%142 citations counted in INSPIRE as of 21 Jan 2022


%\cite{Bern:2014sna}
\bibitem{Bern:2014sna}
Z.~Bern, S.~Davies and T.~Dennen,
``Enhanced ultraviolet cancellations in $\mathcal N=5$ supergravity at four loops,''
\doi{Phys. Rev. D \textbf{90} (2014) no.10, 105011}{doi:10.1103/PhysRevD.90.105011}
\eprintN{1409.3089}.
%144 citations counted in INSPIRE as of 21 Jan 2022



%\cite{Bossard:2011tq}
\bibitem{Bossard:2011tq}
G.~Bossard, P.~S.~Howe, K.~S.~Stelle and P.~Vanhove,
``The vanishing volume of D=4 superspace,''
\doi{Class. Quant. Grav. \textbf{28} (2011), 215005}{doi:10.1088/0264-9381/28/21/215005}
\eprintN{1105.6087}.
%82 citations counted in INSPIRE as of 21 Jan 2022



%\cite{Bianchi:2015vsa}
\bibitem{Bianchi:2015vsa}
M.~Bianchi and D.~Consoli,
``Simplifying one-loop amplitudes in superstring theory,''
\doi{JHEP \textbf{01} (2016), 043}{doi:10.1007/JHEP01(2016)043}
\eprintN{1508.00421}.
%17 citations counted in INSPIRE as of 13 Jan 2022


%\cite{Bianchi:2010mg}
\bibitem{Bianchi:2010mg}
M.~Bianchi, R.~Poghossian and M.~Samsonyan,
``Precision Spectroscopy and Higher Spin symmetry in the ABJM model,''
\doi{JHEP \textbf{10} (2010), 021}{doi:10.1007/JHEP10(2010)021}
\eprintN{1005.5307}.
%14 citations counted in INSPIRE as of 10 Mar 2022



%\cite{Kiritsis:1997hj}
\bibitem{Kiritsis:1997hj}
E.~Kiritsis,
``Introduction to superstring theory,''
\eprint{hep-th/9709062}.
%223 citations counted in INSPIRE as of 21 Jan 2022




%\cite{Bergshoeff:1997gy}
\bibitem{Bergshoeff:1997gy}
E.~Bergshoeff, B.~Janssen and T.~Ort\'{\i}n,
``Kaluza-Klein monopoles and gauged sigma models,''
\doi{Phys. Lett. B \textbf{410} (1997), 131-141}{doi:10.1016/S0370-2693(97)00946-5}
\eprint{hep-th/9706117}.
%95 citations counted in INSPIRE as of 24 Feb 2022



%\cite{Alvarez-Gaume:1983ihn}
\bibitem{Alvarez-Gaume:1983ihn}
L.~\'{A}lvarez-Gaum\'{e} and E.~Witten,
``Gravitational Anomalies,''
\doi{Nucl. Phys. B \textbf{234} (1984), 269.}{doi:10.1016/0550-3213(84)90066-X}
%1712 citations counted in INSPIRE as of 15 Mar 2022




%\cite{Andrianopoli:2001zh}
\bibitem{Andrianopoli:2001zh}
L.~Andrianopoli, R.~D'Auria and S.~Ferrara,
``Supersymmetry reduction of N extended supergravities in four-dimensions,''
\doi{JHEP \textbf{03} (2002), 025}{doi:10.1088/1126-6708/2002/03/025}
\eprint{hep-th/0110277}.
%83 citations counted in INSPIRE as of 24 Mar 2022

%\cite{Narain:1986qm}
\bibitem{Narain:1986qm}
K.~S.~Narain, M.~H.~Sarmadi and C.~Vafa,
``Asymmetric Orbifolds,''
\doi{Nucl. Phys. B \textbf{288} (1987), 551.}{doi:10.1016/0550-3213(87)90228-8}
%489 citations counted in INSPIRE as of 24 Mar 2022


%\cite{Narain:1990mw}
\bibitem{Narain:1990mw}
K.~S.~Narain, M.~H.~Sarmadi and C.~Vafa,
``Asymmetric orbifolds: Path integral and operator formulations,''
\doi{Nucl. Phys. B \textbf{356} (1991), 163-207.}{doi:10.1016/0550-3213(91)90145-N}
%88 citations counted in INSPIRE as of 24 Mar 2022



%\cite{Anastasopoulos:2009kj}
\bibitem{Anastasopoulos:2009kj}
P.~Anastasopoulos, M.~Bianchi, J.~F.~Morales and G.~Pradisi,
``(Unoriented) T-folds with few T's,''
\doi{JHEP \textbf{06} (2009), 032}{doi:10.1088/1126-6708/2009/06/032}
\eprintN{0901.0113}.
%10 citations counted in INSPIRE as of 26 Mar 2022

%\cite{Bianchi:2012xz}
\bibitem{Bianchi:2012xz}
M.~Bianchi, G.~Pradisi, C.~Timirgaziu and L.~Tripodi,
``Heterotic T-folds with a small number of neutral moduli,''
\doi{JHEP \textbf{10} (2012), 089}{doi:10.1007/JHEP10(2012)089}
\eprintN{1207.2665}.
%3 citations counted in INSPIRE as of 26 Mar 2022

%\cite{Bianchi:1990yu}
\bibitem{Bianchi:1990yu}
M.~Bianchi and A.~Sagnotti,
``On the systematics of open string theories,''
\doi{Phys. Lett. B \textbf{247} (1990), 517-524.}{doi:10.1016/0370-2693(90)91894-H}
%551 citations counted in INSPIRE as of 26 Mar 2022

%\cite{Bianchi:1990tb}
\bibitem{Bianchi:1990tb}
M.~Bianchi and A.~Sagnotti,
``Twist symmetry and open string Wilson lines,''
\doi{Nucl. Phys. B \textbf{361} (1991), 519-538.}{doi:10.1016/0550-3213(91)90271-X}
%430 citations counted in INSPIRE as of 26 Mar 2022




%\cite{Andrianopoli:2006ub}
\bibitem{Andrianopoli:2006ub}
L.~Andrianopoli, R.~D'Auria, S.~Ferrara and M.~Trigiante,
``Extremal black holes in supergravity,''
Lect. Notes Phys. \textbf{737} (2008), 661-727, 
\eprint{hep-th/0611345}.
%131 citations counted in INSPIRE as of 24 Mar 2022





%\cite{Persson:2015jka}
\bibitem{Persson:2015jka}
D.~Persson and R.~Volpato,
``Fricke S-duality in CHL models,''
\doi{JHEP \textbf{12} (2015), 156}{doi:10.1007/JHEP12(2015)156}
\eprintN{1504.07260}.
%28 citations counted in INSPIRE as of 09 Mar 2022


\bibitem{GBinprep}
G.~Bossard, in preparation.



%\cite{Cremmer:1979uq}
\bibitem{Cremmer:1979uq}
E.~Cremmer, J.~Scherk and J.~H.~Schwarz,
``Spontaneously broken $N=8$ supergravity,''
\doi{Phys. Lett. B \textbf{84} (1979), 83-86.}{doi:10.1016/0370-2693(79)90654-3}
%271 citations counted in INSPIRE as of 09 Mar 2022




 %\cite{Drummond:2003ex}
\bibitem{Drummond:2003ex}
J.~M.~Drummond, P.~J.~Heslop, P.~S.~Howe and S.~F.~Kerstan,
``Integral invariants in N=4 SYM and the effective action for coincident D-branes,''
\doi{JHEP \textbf{08} (2003), 016}{doi:10.1088/1126-6708/2003/08/016}
\eprint{hep-th/0305202}.
%76 citations counted in INSPIRE as of 14 Apr 2021



%\cite{Drummond:2010fp}
\bibitem{Drummond:2010fp}
J.~M.~Drummond, P.~J.~Heslop and P.~S.~Howe,
``A Note on N=8 counterterms,''
\eprintN{1008.4939}.
%26 citations counted in INSPIRE as of 14 Apr 2021
  
  
  
%\cite{Howe:1981gz}
\bibitem{Howe:1981gz}
P.~S.~Howe,
``Supergravity in Superspace,''
\doi{Nucl. Phys. B \textbf{199} (1982), 309-364.}{doi:10.1016/0550-3213(82)90349-2}
%174 citations counted in INSPIRE as of 24 Mar 2022


%\cite{Dobrev:1985qv}
\bibitem{Dobrev:1985qv}
V.~K.~Dobrev and V.~B.~Petkova,
``All positive energy unitary irreducible representations of extended conformal supersymmetry,''
\doi{Phys. Lett. B \textbf{162} (1985), 127-132.}{doi:10.1016/0370-2693(85)91073-1}
%311 citations counted in INSPIRE as of 15 Apr 2021

  
  
%\cite{Ferrara:1999zg}
\bibitem{Ferrara:1999zg}
S.~Ferrara and E.~Sokatchev,
``Short representations of $SU(2,2|N)$ and harmonic superspace analyticity,''
\doi{Lett. Math. Phys. \textbf{52} (2000), 247-262}{doi:10.1023/A:1007641619266}
\eprint{hep-th/9912168}.
%47 citations counted in INSPIRE as of 14 Apr 2021


  \bibitem{CollingwoodMcGovern}
D.~H. Collingwood and W.~M. McGovern, {\em Nilpotent orbits in semisimple {L}ie
  algebras}.
\newblock Van Nostrand Reinhold Mathematics Series. Van Nostrand Reinhold Co.,
  New York, 1993.







  %\cite{DHoker:2014oxd}
\bibitem{DHoker:2014oxd}
E.~D'Hoker, M.~B.~Green, B.~Pioline and R.~Russo,
``Matching the $D^{6}R^{4}$ interaction at two-loops,''
\doi{JHEP \textbf{01} (2015), 031}{doi:10.1007/JHEP01(2015)031}
\eprintN{1405.6226}.
%64 citations counted in INSPIRE as of 07 Oct 2021



%\cite{Berg:2016wux}
\bibitem{Berg:2016wux}
M.~Berg, I.~Buchberger and O.~Schlotterer,
``From maximal to minimal supersymmetry in string loop amplitudes,''
\doi{JHEP \textbf{04} (2017), 163}{doi:10.1007/JHEP04(2017)163}
\eprintN{1603.05262}.
%25 citations counted in INSPIRE as of 24 Feb 2022



%\cite{Bianchi:2012ud}
\bibitem{Bianchi:2012ud}
M.~Bianchi and G.~Inverso,
``Unoriented D-brane instantons,''
\doi{Fortsch. Phys. \textbf{60} (2012), 822-834}{doi:10.1002/prop.201200047}
\eprintN{1202.6508}.
%15 citations counted in INSPIRE as of 24 Mar 2022

%\cite{Bianchi:2012kt}
\bibitem{Bianchi:2012kt}
M.~Bianchi, G.~Inverso and L.~Martucci,
``Brane instantons and fluxes in F-theory,''
\doi{JHEP \textbf{07} (2013), 037}{doi:10.1007/JHEP07(2013)037}
\eprintN{1212.0024}.
%26 citations counted in INSPIRE as of 24 Mar 2022



%%%%%%%%%%%%%%%%%%%%%%%%%%%%%%%%%%%%%%%%%%%%%%%%%%%%%%

%\cite{Kawai:1985xq}
\bibitem{Kawai:1985xq}
H.~Kawai, D.~C.~Lewellen and S.~H.~H.~Tye,
``A Relation Between Tree Amplitudes of Closed and Open Strings,''
\doi{Nucl. Phys. B \textbf{269} (1986), 1-23}{doi:10.1016/0550-3213(86)90362-7}
%986 citations counted in INSPIRE as of 24 Mar 2022



%\cite{Bianchi:2006nf}
\bibitem{Bianchi:2006nf}
M.~Bianchi and A.~V.~Santini,
``String predictions for near future colliders from one-loop scattering amplitudes around D-brane worlds,''
\doi{JHEP \textbf{12} (2006), 010}{doi:10.1088/1126-6708/2006/12/010}
\eprint{hep-th/0607224}.
%40 citations counted in INSPIRE as of 31 Mar 2021

%\cite{Bern:2011rj}
\bibitem{Bern:2011rj}
Z.~Bern, C.~Boucher-Veronneau and H.~Johansson,
``$\mathcal{N} \ge  4$ supergravity amplitudes from gauge theory at one loop,''
\doi{Phys. Rev. D \textbf{84} (2011), 105035}{doi:10.1103/PhysRevD.84.105035}
\eprintN{1107.1935}.
%113 citations counted in INSPIRE as of 27 Sep 2021


%\cite{David:2006ud}
\bibitem{David:2006ud}
J.~R.~David, D.~P.~Jatkar and A.~Sen,
``Dyon spectrum in generic $\mathcal{N}=4$ supersymmetric $\mathds{Z}_N$ orbifolds,''
\doi{JHEP \textbf{01} (2007), 016}{doi:10.1088/1126-6708/2007/01/016}
\eprint{hep-th/0609109}.
%84 citations counted in INSPIRE as of 05 Mar 2022

%\cite{Bachas:2008jv}
\bibitem{Bachas:2008jv}
C.~Bachas, M.~Bianchi, R.~Blumenhagen, D.~L\"ust and T.~Weigand,
``Comments on Orientifolds without Vector Structure,''
\doi{JHEP \textbf{08} (2008), 016}{doi:10.1088/1126-6708/2008/08/016}
\eprintN{0805.3696}.
%25 citations counted in INSPIRE as of 10 Mar 2022


%\cite{Bossard:2016hgy}
\bibitem{Bossard:2016hgy}
G.~Bossard and B.~Pioline,
``Exact $\nabla^4 R^4$ couplings and helicity supertraces,''
\doi{JHEP \textbf{01} (2017), 050}{doi:10.1007/JHEP01(2017)050}
\eprintN{1610.06693}.
%10 citations counted in INSPIRE as of 09 Apr 2021


   %\cite{Gaberdiel:2002jr}
\bibitem{Gaberdiel:2002jr}
M.~R.~Gaberdiel and S.~Schafer-Nameki,
``D-branes in an asymmetric orbifold,''
\doi{Nucl. Phys. B \textbf{654} (2003), 177-196}{doi:10.1016/S0550-3213(03)00062-2}
\eprint{hep-th/0210137}.
%24 citations counted in INSPIRE as of 07 Apr 2021

  %\cite{Kawai:2007qd}
\bibitem{Kawai:2007qd}
S.~Kawai and Y.~Sugawara,
``D-branes in T-fold conformal field theory,''
\doi{JHEP \textbf{02} (2008), 027}{doi:10.1088/1126-6708/2008/02/027}
\eprintN{0709.0257}.
%21 citations counted in INSPIRE as of 07 Apr 2021

%\cite{Bianchi:2016bgx}
\bibitem{Bianchi:2016bgx}
M.~Bianchi, J.~F.~Morales and L.~Pieri,
``Stringy origin of $4D$ black hole microstates,''
\doi{JHEP \textbf{06} (2016), 003}{doi:10.1007/JHEP06(2016)003}
\eprintN{1603.05169}.
%18 citations counted in INSPIRE as of 11 Mar 2022

%\cite{Bianchi:2017sds}
\bibitem{Bianchi:2017sds}
M.~Bianchi, D.~Consoli and J.~F.~Morales,
``Probing Fuzzballs with Particles, Waves and Strings,''
\doi{JHEP \textbf{06} (2018), 157}{doi:10.1007/JHEP06(2018)157}
\eprintN{1711.10287}.
%21 citations counted in INSPIRE as of 11 Mar 2022


%\cite{Anastasopoulos:2011hj}
\bibitem{Anastasopoulos:2011hj}
P.~Anastasopoulos, M.~Bianchi and R.~Richter,
``Light stringy states,''
\doi{JHEP \textbf{03} (2012), 068}{doi:10.1007/JHEP03(2012)068}
\eprintN{1110.5424}.
%26 citations counted in INSPIRE as of 10 Mar 2022

%\cite{Gava:1997jt}
\bibitem{Gava:1997jt}
E.~Gava, K.~S.~Narain and M.~H.~Sarmadi,
``On the bound states of $p$-branes and $(p+2)$-branes,''
\doi{Nucl. Phys. B \textbf{504} (1997), 214-238}{doi:10.1016/S0550-3213(97)00508-7}
\eprint{hep-th/9704006}.
%123 citations counted in INSPIRE as of 24 Mar 2022


%\cite{Angelantonj:2011hs}
\bibitem{Angelantonj:2011hs}
C.~Angelantonj, C.~Condeescu, E.~Dudas and G.~Pradisi,
``Non-perturbative transitions among intersecting-brane vacua,''
\doi{JHEP \textbf{07} (2011), 123}{doi:10.1007/JHEP07(2011)123}
\eprintN{1105.3465}.
%10 citations counted in INSPIRE as of 20 Mar 2022



\bibitem{FNF}
Dennis R. Estes and Gordon Nipp, “Factorization in quaternion orders'' Journal of Number Theory {\bf 33}, 224 (1989). 

%\cite{DHoker:2015gmr}
\bibitem{DHoker:2015gmr}
E.~D'Hoker, M.~B.~Green and P.~Vanhove,
``On the modular structure of the genus-one Type II superstring low energy expansion,''
\doi{JHEP \textbf{08} (2015), 041}{doi:10.1007/JHEP08(2015)041}
\eprintN{1502.06698}.
%74 citations counted in INSPIRE as of 25 Mar 2022

 \bibitem{Weiertheta}
F. W. J. Olver et al., “NIST Handbook of Mathematical Functions'' Cambridge University
Press (2010); {{\hypersetup{urlcolor=darkred}\href{http://dlmf.nist.gov}{DLMF, Digital Library of Mathematical Functions}\hypersetup{urlcolor=blue}}}.


\bibitem{MR656029}
D.~Zagier, ``The {R}ankin-{S}elberg method for automorphic functions which are
  not of rapid decay,'' {\em J. Fac. Sci. Univ. Tokyo Sect. IA Math.} {\bf 28}
  (1981) no.~3, 415--437 (1982).  

%\cite{Tourkine:2013rda}
\bibitem{Tourkine:2013rda}
P.~Tourkine,
``Tropical Amplitudes,''
\doi{Annales Henri Poincar\'e \textbf{18} (2017) no.6, 2199-2249}{doi:10.1007/s00023-017-0560-7}
\eprintN{1309.3551}.
%36 citations counted in INSPIRE as of 17 Oct 2021

\end{thebibliography}
\end{document}